\documentclass[12pt,a4paper,twoside,british]{book}
\pdfoutput=1
\usepackage[utf8]{inputenc}
\usepackage[british]{babel}
\usepackage{amsmath}
\usepackage{amsfonts}
\usepackage{amssymb}
\usepackage{graphicx}
\usepackage{lmodern}
\usepackage{fancyhdr}
\usepackage{ccaption}
\usepackage{booktabs}

\usepackage{subfigure}
\usepackage{acronym}
\usepackage[nottoc]{tocbibind}
\usepackage{upgreek}
\usepackage{setspace}

\usepackage{hyperref} 
\hypersetup{
    colorlinks,
    citecolor=black,
    filecolor=black,
    linkcolor=black,
    urlcolor=black
}

\usepackage[left=3cm,right=2.5cm,top=2.5cm,bottom=2.5cm]{geometry}
\author{Carlos García Argos}
\title{PhD Thesis\\A Silicon Strip Detector for the Phase II High Luminosity Upgrade of the ATLAS Detector at the Large Hadron Collider}

\captionstyle{\centerlastline}
\captionnamefont{\bfseries}
\captiontitlefont{\bfseries} 

\def\dedx{{\rm d}E/{\rm d}x}
\def\pt{p_{{\rm T}}}

\def\de_dx{\ensuremath{\frac{{\rm d}E}{{\rm d}x}}}

\def\m{~{\rm m}}
\def\cm{~{\rm cm}}
\def\mm{~{\rm mm}}
\def\mum{~{\rm \upmu m}} 
\def\nm{~{\rm nm}}

\def\ns{~{\rm ns}}
\def\ps{~{\rm ps}}

\def\mrad{~{\rm mrad}}
\def\urad{~{\rm \upmu rad}}

\def\MHz{~{\rm MHz}}

\def\MeV{~{\rm MeV}}
\def\GeV{~{\rm GeV}}
\def\TeV{~{\rm TeV}}

\def\ifb{~{\rm fb^{-1}}}
\def\inb{~{\rm nb^{-1}}}

\def\V{\ensuremath{~{\rm V}}}
\def\mV{\ensuremath{~{\rm mV}}}

\def\A{\ensuremath{~{\rm A}}}

\def\uA{\ensuremath{~{\rm \upmu A}}}

\def\nF{\ensuremath{~{\rm nF}}}
\def\pF{\ensuremath{~{\rm pF}}}

\def\fC{\ensuremath{~{\rm fC}}}

\def\degC{\ensuremath{^{\circ}{\rm C}}}

\def\ENC{\ensuremath{~e^{-} {\rm ENC}}}

\renewcommand{\chaptermark}[1]{\markboth{#1}{}}
\renewcommand{\sectionmark}[1]{\markright{\thesection.\ #1}}
\begin{document}
\pagenumbering{arabic}
\frontmatter
\begin{titlepage}
\headsep 2 cm
    \begin{center}
      \begin{spacing}{2}
        {\Large  UNIVERSIDAD DE VALENCIA}

        {\Large  FACULTAD DE FÍSICA}

\vspace{2cm}
\includegraphics[clip,scale=0.75]{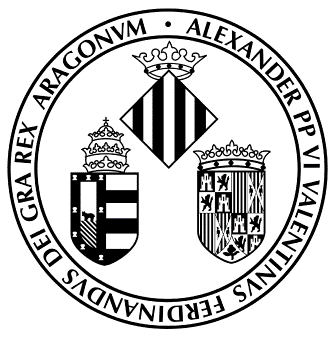}
      \end{spacing}
    \end{center}
\vspace{0.1cm} 
        \begin{center}   
        \begin{spacing}{2} 
                {\Large PhD Thesis}

\vspace{0.75cm}
                {\Large \bf A Silicon Strip Detector for the \\ \vspace{-0.6cm}
                Phase II High Luminosity Upgrade \\ \vspace{-0.6cm}
                of the ATLAS Detector at the \\ \vspace{-0.08cm}
                Large Hadron Collider}

\vspace{1.25cm}
        \end{spacing}
\vspace{.0cm}
\large Author: \\ Carlos García Argos \\
\vspace{0.5cm}
\large Supervisors: \\ Dr. Carlos Lacasta Llácer \\
Dr. Stephen J. McMahon \\
\vspace{0.5cm}
\large Valencia, 2014

        \end{center}
\end{titlepage}      
\thispagestyle{empty} 

\vspace{12cm}

\newpage      

\thispagestyle{empty}
\newpage

\chapter{Abstract}

ATLAS is a particle physics experiment at the Large 
Hadron Collider (LHC) that detects
proton-proton collisions at a centre of mass energy of $14\TeV$. The 
Semiconductor Tracker is part of the Inner Detector, implemented using silicon 
microstrip detectors with binary read-out,
providing momentum measurement of charged particles with excellent 
resolution. The operation of the LHC and the ATLAS experiment started in 2010,
with ten years of operation expected until major upgrades are needed in the
accelerator and the experiments. The ATLAS tracker will need to be completely 
replaced
due to the radiation damage and occupancy of some detector elements and the 
data links at high luminosities. These upgrades 
after the first ten years of operation are named the Phase-II Upgrade and 
involve a re-design of the LHC, resulting in the High Luminosity Large Hadron 
Collider (HL-LHC).

This thesis presents the work carried out in the testing of the ATLAS Phase-II
Upgrade electronic systems in the future strips tracker after 2023, to be
installed for operations in the HL-LHC period. The high
luminosity and number of interactions per crossing that will happen after the
HL-LHC starts require a complete replacement of the ATLAS tracker. The systems
that have been defined for the Phase-II Upgrade will be designed to cope with 
that
increased radiation and have the right granularity to maintain the performance
with higher pile-up.

In this thesis I present 
results on single modules and larger
structures comprising multiple modules. 

The single modules are built using silicon microstrip sensors with four rows of
1280 strips. The read-out of the strips is done using 128 channel chips, glued
and bonded on a hybrid circuit that holds 20 chips. Two hybrids are glued to 
the sensor to read-out all its strips.

In addition to the new sensors and read-out chips, the specifications for the 
ATLAS Phase-II Upgrade programme require a different powering 
scheme in the strips tracker than the current ATLAS Semiconductor Tracker. Two
approaches have been proposed, which are serial powering and Direct Current to
Direct Current (DC-DC) conversion. 
The decision on which will be used is not final yet, pending the results on 
efficiency and performance of the tracker using both of them.

Larger structures are constructed by mounting the single modules on a bus tape
that carries the signals to one end of the structure, which interfaces with the
tracker read-out systems. The bus tape is glued on a structure that
provides mechanical support and cooling. All the modules on a
structure are read-out through the same interface, aggregating multiple signals
in one physical channel. The structures are called staves or stavelets. The 
latter 
typically mount four modules on a side of the structure. Two different 
stavelets have been tested in the context of this thesis, one with serial
powering and one with DC-DC conversion. Both are single-sided objects and 
double-sided objects have been constructed and tested in other institutes.

One full size stave with twelve modules on one side has been constructed. It is 
powered using DC-DC conversion, and tested at the Rutherford Appleton 
Laboratory (RAL) as part of the work for this 
thesis.

In the context of the current ATLAS Semiconductor Tracker studies, I present an
analysis of the data taken by the detector from the beginning of operation
in 2010 until the first Long Shut-down in 2013. The analysis consists of an
energy loss study in the Semiconductor Tracker, a task the detector was not
designed to perform. 

However, the availability of the Time-over-Threshold of the signals
generated by particles traversing the detector elements allows an
estimation of the charge deposited by the particles. This calculation of the
energy loss is typically used to perform particle identification, a feature 
that is usually not required from the tracker. In addition, I present a study
that proposes the use of this energy loss calculation as a means of tracking
radiation damage in the silicon.

\chapter*{Acknowledgements and Remarks}

\thispagestyle{empty}

A lot of other people have contributed in one way or another to this thesis, 
from the various institutes that participate in the ATLAS Collaboration. In 
hope of not forgetting too many of them, I would like to thank, in no 
particular order: Fernando, Ashley, Bruce, Matt, Ingo, Conrad, Tony W., Tony A., 
Phil, Dave, Sergio G., Sergio D., Pippa, Susanne, Ingrid, Uli, 
Miguel, Víctor, Todd, Samer, Bart, Ian, Madalina, Takashi and
Stephen.

Thanks to the colleagues from IFIC in Valencia, for their help and 
collaboration: Carlos, Pepe and Ricardo. Special thanks to Carlos for
agreeing to supervise this thesis, for his thorough reviews and chats about
the various topics in the thesis.

I do not want to miss the opportunity to thank the Science and Technology 
Facilities Council (STFC) in the United Kingdom for their support, both
financial and technical, with the
tests done at Rutherford Appleton Laboratory, in close collaboration with
Peter Phillips. A special thanks to Peter for his help.

A big thanks to Steve for also agreeing to supervise this work and for his 
constant support, his kindness and enthusiasm.

A huge thanks to my family and friends for their support and love in the 
distance.

The author acknowledges the financial support from Spanish grant 
SEII-2010-00101, IFIC-CSIC and CERN.

This PhD Thesis is written following the spelling recommendations from the 
Oxford English Dictionary and Fowler's \textit{A Dictionary of Modern English
Usage}.

The figures, tables, equations and units are written following the ATLAS
Style Guide.

This document was last updated \today.

\newpage 

\thispagestyle{empty}

\mainmatter

\setcounter{page}{5}
\tableofcontents

\listoffigures

\listoftables

\renewcommand{\chaptermark}[1]{\markboth{\chaptername\ \thechapter.\ #1}{}}
\renewcommand{\sectionmark}[1]{\markright{\thesection.\ #1}}
\fancyhf{}
\fancyhead[RO]{\small \rightmark}
\fancyhead[LE]{\small \leftmark}
\fancyfoot[RO,LE]{\thepage}

\chapter{Introduction}

This thesis is focused on silicon strips detectors. In particular, studies
of the ATLAS Phase II Upgrade for the High Luminosity Large Hadron Collider 
(HL-LHC), with special emphasis on the future silicon strips tracker. This 
upgrade of the ATLAS detector will 
take place in the early 2020s and poses several challenges due to the harsh 
conditions that the LHC experiments will have to cope with. This thesis also 
discusses some work done on the performance studies of the current ATLAS 
silicon strips tracker.

This introduction chapter gives a brief overview of the current ATLAS detector.

   \section{The Large Hadron Collider Accelerator and its Experiments at CERN}
   
The Large Hadron Collider is the hadron accelerator 
and collider that was installed in the $26.7$~kilometres long tunnel that 
held the CERN Large Electron-Positron Collider (LEP) and its experiments. 
LEP was decommissioned in 2000 and LHC civil works commenced in 1998, and its 
installation was finished in 2008. 

The layout of the LHC is the same one as that of LEP, with eight straight 
sections. The two high luminosity insertions are located at opposite sides of
the LHC ring, 
Intersection Points~1 and~5 (IP1 and IP5), where the two large general-purpose 
experiments, ATLAS and CMS, are placed. 

The LHC is capable of accelerating and colliding hadrons, 
namely protons, and heavy ions. The injection of protons and heavy ions into 
the LHC is done by the previously existing infrastructure: the Linear 
Accelerators Linac2 (protons) and Linac3 (heavy ions), Booster, the Proton
Synchrotron (PS) and the Super Proton Synchrotron (SPS). They perform the 
pre-acceleration of the particles that are injected into the LHC ring.
   
The main design parameters of the LHC for proton-proton (p-p) operation are the
maximum proton single-beam energy of $7\TeV$, for a $14\TeV$ maximum centre of 
mass energy, and the peak luminosity of 
$10^{34}\cm^{-2} {\rm s}^{-1}$, with 
$1.15\times 10^{11}$ protons per bunch, crossing in the experiments every 
$25\ns$. In 
addition to the p-p collisions, the LHC also collides heavy ions, in particular 
lead nuclei, at a beam energy of $5.5\TeV$ and a design luminosity of 
$10^{27}{\rm cm}^{-2}{\rm s}^{-1}$ \cite{LHC08,ATLAS}.
   
   \begin{figure}[!htb]
    \begin{center}
     \includegraphics[scale=0.38]{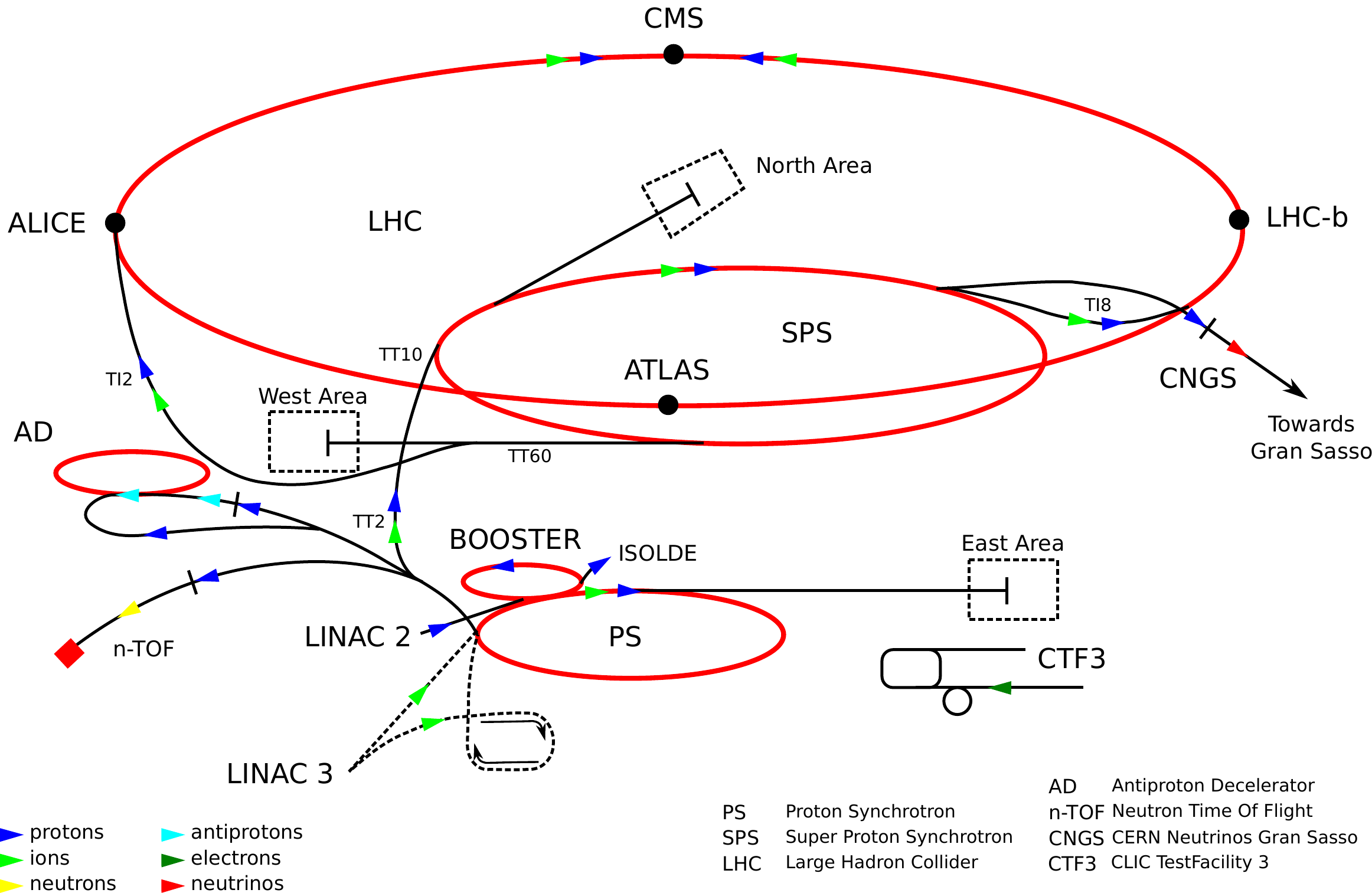}
     \caption{The CERN Accelerator Complex showing all experiments around the 
     LHC.}\label{fig:1-lhc_complex}
    \end{center}
   \end{figure}

The whole CERN accelerator complex, shown in Figure~\ref{fig:1-lhc_complex}, 
comprises 
not only the LHC ring, but also the experiments, injectors and other non-LHC 
experiments. The six LHC experiments, of which four are depicted in the 
figure, are:

\begin{itemize}
  \item \textit{A Large Ion Collider Experiment} (ALICE): it is a specialized 
detector to study heavy ion collisions, which at the LHC are lead-lead (Pb-Pb)
and proton-lead (p-Pb). 
The experiment is located at the intersection point IP2.
  \item \textit{A Toroidal LHC Apparatus} (ATLAS): it is one of the two large 
general-purpose experiments at the LHC, located at intersection point IP1. It 
is used to study a wide range of physics, including the search for 
the Higgs boson, extra dimensions, super-symmetry and particles that could make 
up dark matter.
\item \textit{Compact Muon Solenoid} (CMS): it is the other large 
general-purpose experiment together with ATLAS, located at intersection point 
IP5.
\item \textit{Large Hadron Collider beauty} (LHCb): located at intersection 
point IP8, it is a specialized B-physics experiment, that measures the 
parameters of CP violation in the interactions of b-hadrons (heavy particles 
containing a bottom quark).
\item \textit{Large Hadron Collider forward} (LHCf): it is a special-purpose 
experiment for astroparticle physics, designed to study the particles generated 
in the ``forward'' region of collisions, those almost directly in line with the 
colliding proton beams. It shares the interaction point IP1 with ATLAS.
\item \textit{TOTal Elastic and diffractive cross section Measurement} (TOTEM): 
it shares intersection point IP5 with CMS and has multiple detectors spread 
over $440$ metres. Its purpose is to measure the 
structure and effective size of the proton, as well as precisely measure 
the cross section of proton-proton interactions.
\end{itemize}

This thesis is focused on the ATLAS silicon tracker and the next section
gives a short description of the ATLAS experiment. Further details about the
silicon tracker will be part of chapter 2.

   \section{The ATLAS Experiment at the Large Hadron Collider}\label{sec:atlas}

ATLAS is a multi-purpose detector that was designed to detect charged and most 
neutral particles. The detected particles are the decay products of unstable 
Standard Model 
particles, such as the $Z$ and $W^{\pm}$, the recently observed 
Higgs boson~\cite{Aad20121} and other particles from physics beyond the 
Standard Model. 

\begin{figure}[!htbp]
 \begin{center}
  \includegraphics[scale=0.5]{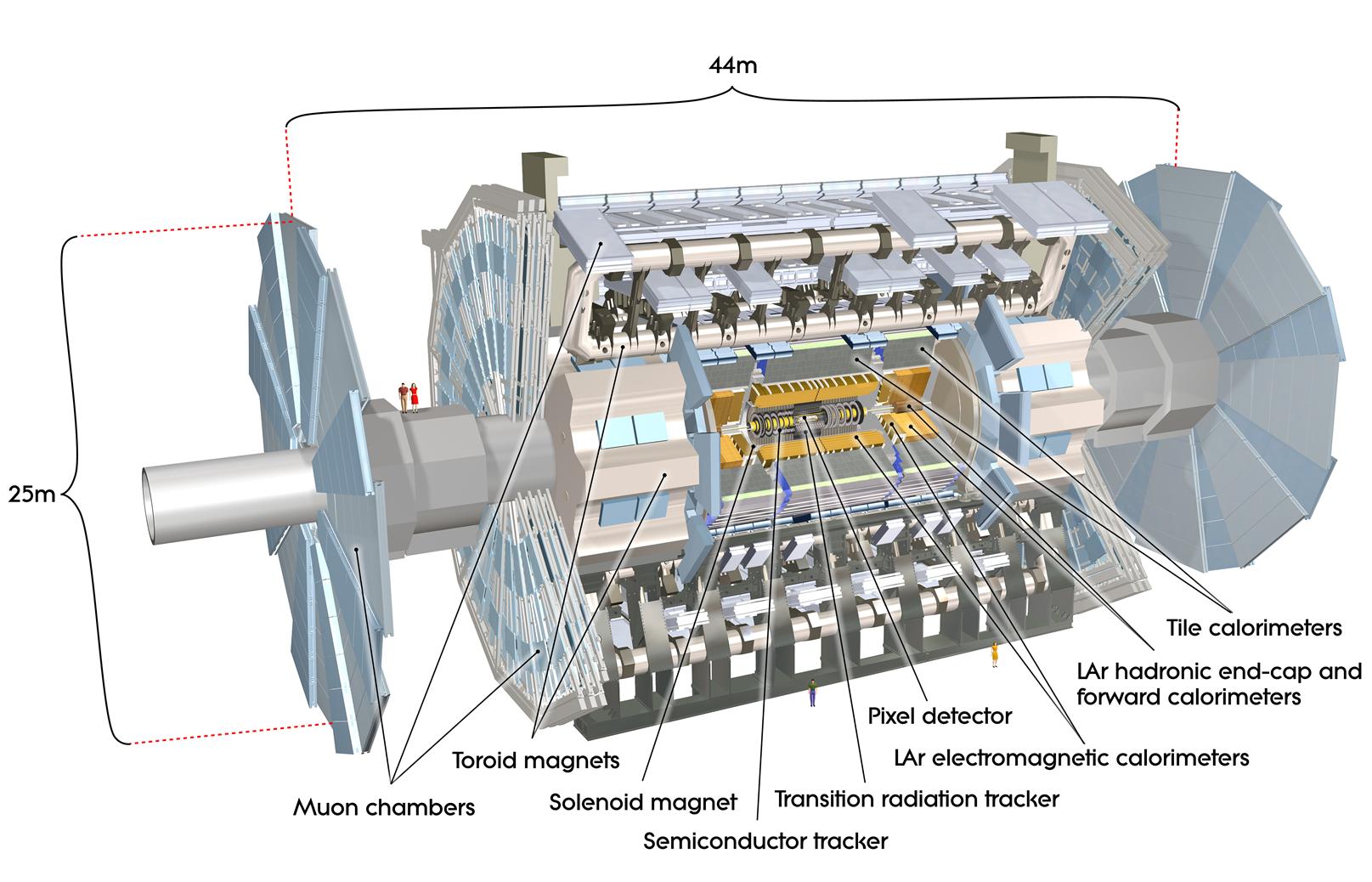}
  \caption{View of the whole ATLAS detector.}\label{fig:1-atlas}
 \end{center}
\end{figure}

A schematic representation of the ATLAS detector is shown in 
Figure~\ref{fig:1-atlas}, where the layout of the sub-detectors can be seen. 
The dimensions are also indicated in the figure. The detector is $44$ metres 
long and 
$25$ metres high, weighting approximately $7000$ tonnes~\cite{ATLAS}. A brief 
description of its sub-detectors follows.

The first element, surrounding the interaction point, is the Inner Detector, 
which performs the tracking role, one of the central functions in 
a particle detector. A discussion on trackers, as well as some additional 
information about the ATLAS Inner Detector, is part of chapter 2.

It 
is designed to do robust pattern recognition, excellent momentum resolution and 
both primary and secondary vertices measurements for charged particles with 
transverse momentum as low as $0.1~{\rm GeV}$. It covers the pseudorapidity 
range $\left| \eta \right| < 2.5$. It is immersed in a $2$ tesla magnetic field 
that bends the trajectory of charged particles, providing the information 
needed to determine the momenta of the particles. 

Three sub-systems form the Inner Detector: a Pixel detector and a 
Semi-Conductor Tracker (SCT) both implemented using silicon sensors, and a 
Transition Radiation Tracker (TRT) that uses ionization caused by charged 
particles in straw tubes filled with gas. The three
are surrounded by a solenoid magnet that generates the $2$ tesla magnetic 
field,
designed to minimize the material thickness before the calorimeters, in order
to have an adequate calorimetry performance.
A more detailed description of the Semi-Conductor Tracker is given in 
chapter~2.

The calorimetry system, surrounding the Inner Detector, is composed of two 
types of calorimeters. The first one, that surrounds
the inner detector, is the electromagnetic calorimeter, dedicated to absorbing
and measuring the energy of electrons and photons that leave the tracker.
The second layer is the hadronic calorimeter, which is responsible for 
performing the equivalent task for hadrons. The calorimeters have a large 
pseudorapidity coverage, $\left| \eta \right| < 4.9$.
The two types are implemented
using different technologies and more detailed information is available 
in~\cite{ATLAS}.

The Muon spectrometer is located at the outer part of the detector and it 
is designed to detect charged particles that are not absorbed in the 
calorimeters and to measure their momenta in the $\left| \eta \right| < 2.7$ 
range. These particles are muons, which are the only detectable particles 
that are not stopped in the calorimeters, both the muons that originated in 
the hard interaction and those that are products of decays inside the detector. 

The muons trajectories are bent by a set of toroid magnets that can reach up 
to $4.7$ tesla, serving the same purpose as the solenoid in the Inner Detector, 
that is, bend the particles trajectories to measure their momenta.

\begin{figure}[!htbp]
 \begin{center}
  \includegraphics[scale=0.27]{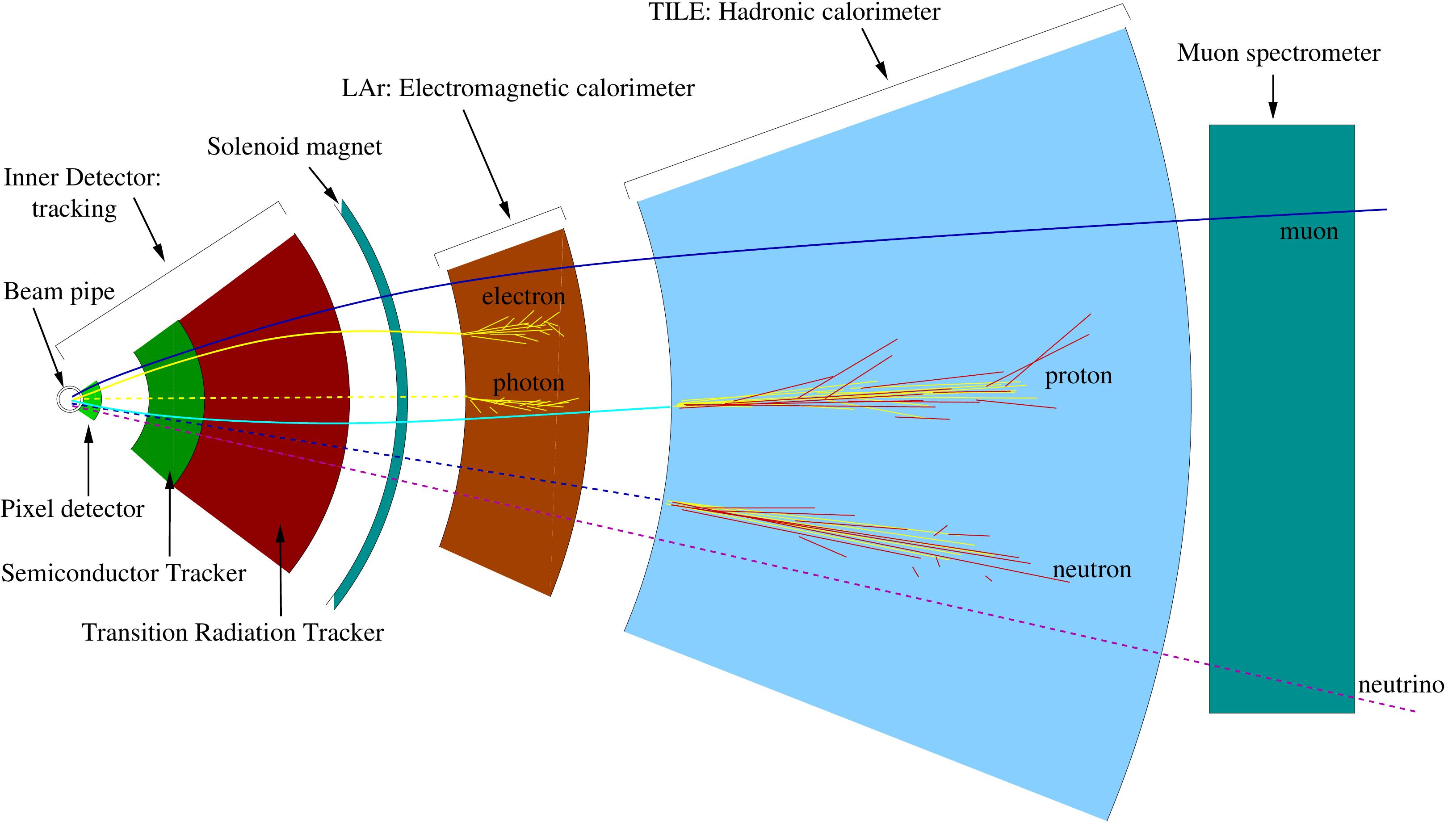}
  \caption[Axial representation of particle detection at the ATLAS 
  sub-detectors.]{Axial representation of particle detection at the ATLAS 
  sub-detectors (to scale): the dashed lines indicate that the particle 
  is not detected in that section of the 
  experiment.}\label{fig:1-atlas_particles}
 \end{center}
\end{figure}

Figure~\ref{fig:1-atlas_particles} shows each of the ATLAS layers with some 
sample particle species and their representative traces in each of the 
sub-detectors. The dashed lines for a given particle type means that the 
particle is not detected in that part of the detector.
The trajectories of charged particles such as protons and electrons are bent 
in the Inner Detector, whereas neutral particles such as neutrons are 
invisible to it. Photons, electrons, protons and neutrons create showers in 
their respective calorimeters, while muons go all the way to the muon chambers, 
and neutrinos cannot be seen by ATLAS because they interact very rarely with 
matter.

\section{Operating Conditions of the ATLAS Detector}

The read-out system of the whole detector has to deal with an enormous amount 
of data. Although the occupancy is never close to $100\%$, the available 
number of channels is still a huge number. The Inner Detector has a total of
more than $86$ million channels, $80$ million in the pixel detector, $6.3$ 
million in the SCT and $351000$ in the TRT. The electromagnetic calorimeter 
has $173304$ channels while the hadronic calorimeter has $19008$ channels. 
Last, the 
muon system has a total of more than $1$ million channels. All together, it is 
more than $87$ million channels.

At the LHC design luminosity, the mean number of collisions is expected to be 
about
$23$ per bunch crossing. This means that an event will consist of $23$ single
minimum bias events superimposed, what is called pile-up. 

The tracker occupancy increases with the number of interactions per bunch
crossing, and the estimated peak values are around $0.1\%$ for the Pixel 
detector, $1\%$ for the SCT and $40\%$ for the TRT, with a maximum of $30$ 
reconstructed vertices in 2012~\cite{ATLAS-CONF-2012-042}. The increase in 
occupancy of the tracker
may eventually cause performance degradation, leading to the impossibility to
resolve close-by tracks.

With respect to the radiation environment, in particular of the tracker, the
radiation levels are very high in this area. The extreme levels of 
radiation lead to damage in both the silicon sensors and the electronics.
The fluences that the tracker sub-detector will accumulate at the LHC design
luminosity are shown in Table~\ref{tab:1-fluences}. These levels of radiation
cause several problems for the Inner Detector: the silicon sensors are 
affected, the electronics performance suffers and the particles contribute to
background hits. All the materials used in the Inner Detector must be qualified
to survive the doses and fluences expected at their 
locations~\cite{ATLASidTDR}.

\begin{table}[!htb]
\begin{center}
\begin{tabular}{c|cc|cc}
\toprule
 & \multicolumn{2}{c}{$1\MeV$ equivalent neutron fluence}  & \multicolumn{2}{c}{Ionizing dose}\\
 & \multicolumn{2}{c}{[$10^{13}\cm^{-2}/{\rm year}$]}  & \multicolumn{2}{c}{[${\rm kGy/year}$]}\\
 \hline
Sub-detector & Typical  & Maximum   & Typical  & Maximum  \\
 Pixel detector & $5$ & $50$ & $30$ & $300$ \\
 SCT & $1.5$ & $2$ & $4$ & $10$ \\
 TRT & $0.7$ & $1$ & $2$ & $6$\\
\bottomrule
\end{tabular}
\caption[Annual fluences and ionizing doses for the tracker at the LHC design 
luminosity.]{Annual fluences and ionizing doses for the tracker at the LHC design 
luminosity~\cite{ATLASidTDR}.}\label{tab:1-fluences}
\end{center}
\end{table}

The effects of radiation damage in silicon detectors will be briefly presented
in chapter 2.

\section{Thesis Outline}

This thesis is structured as follows. Chapter 2 gives an overview of tracking 
in High Energy Physics experiments, focused on modern semiconductor trackers 
and with some insight into the ATLAS Semi-Conductor Tracker. The Phase II 
of the ATLAS Upgrade is presented in chapter 3, with additional specific 
information on the upgrade of the inner tracker in chapter 4. 
The results of the tests performed on the strips structures for the upgrade are
presented in chapter 5. 

Chapter 6 discusses a study of the ATLAS Semiconductor Tracker performance 
over time, using a method to calculate the $\dedx$ in the SCT,
as a means to track the radiation damage in the silicon. 
Finally, chapter 7 is the conclusions chapter that summarizes the 
results of the tracker upgrade and $\dedx$ studies.

A shorter version of the complete thesis has been written in Spanish and 
included as chapter 8.

\chapter{Tracking in High Energy Physics Experiments}

\section{The Role of Trackers}

The main role of a tracker in a High Energy Physics (HEP) experiment, such 
as those at the LHC, is to measure the trajectory of charged particles, by 
recording several points along the track. These points are found by the 
particles ``hitting'' the tracker at known locations, defined by the position
of the tracker components. These hits result in the track
coordinates in radius 
and the longitudinal coordinates.
The hits are 
fitted to a curve and the momentum is measured from the curvature of charged 
particles in a magnetic field. Tracking is central in the experiments event 
reconstruction and analysis~\cite{WellsTracking}.

Trackers also contribute in the process of particle identification (PID). 
It is possible in some cases to measure the rate of energy loss, $\dedx$, 
in the tracker. However, dedicated detectors are normally used for this task 
and the main contribution from the tracker is matching the tracks with:
\begin{itemize}
 \item Showers in the calorimeter to identify electrons from their 
 characteristic shower shape.
\item Muon chamber track segments, improving the muon momentum measurement.
\end{itemize}
 
There are, in general, two types of tracker geometries, depending on the 
physics. 
Classical, central-forward-backward trackers (one cylinder barrel and two 
endcaps) are symmetric about the interaction point and they usually 
are immersed in a magnetic field that is oriented parallel to the beam 
direction. This is the case of ATLAS. There are asymmetric 
trackers like the LHCb tracker, which may not be immersed inside a magnetic 
field. 
This chapter is
focused on the classical trackers, like the ATLAS Inner Detector.

\section{Track Reconstruction}

The process of reconstructing the tracks of an event includes extrapolating 
the tracks from the hit information, also reconstructing the interaction 
point for each track. One event in the ATLAS detector involves, in general, 
one hard interaction, generating multiple vertices, a primary vertex from the
hard interaction and the rest from the secondary interactions. 

As an example, for proton-proton collisions at the LHC instantaneous design 
luminosity of
$10^{34}~{\rm cm}^{-2} {\rm s}^{-1}$ and $E_{cm} = 14~{\rm TeV}$ with $25\ns$
operation, there are approximately 23 hard interactions per crossing on 
average. Therefore, 
about that mean number of primary vertices will be reconstructed for each 
event. 

Apart from the primary interactions in the collision point, there are other 
interactions inside the detector volume. These secondary interactions can be 
identified by reconstructing their vertices, which are called, accordingly, 
secondary vertices. They are reconstructed from tracks left by particles that 
are the decay products of a variety of other particles, such as tau-leptons, 
strange hadrons, photon conversions and nuclear 
interactions. Looking at the charged decay products is one way of detecting 
neutral particles in trackers.

The tracker is immersed in a magnetic field, for which usually strong 
magnitude and 
uniform distribution are desired. For instance, the ATLAS Experiment Tracker 
has a $2~{\rm T}$ magnetic field, while the CMS Experiment Tracker has a 
$4~{\rm T}$ one. A stronger magnetic field usually implies a better 
momentum resolution. However, a strong field requires more material 
for the magnet, which in turn has a negative impact on the global efficiency
of the detector, in particular of the sub-systems that are located outside the
magnet.

Charged particles describe a spiral trajectory when they move inside a magnetic
field, which projects to a circle in the plane transverse to the 
magnetic field. The curvature 
of this circumference is proportional to the transverse momentum:

\begin{equation}
 \pt [{\rm GeV}/c] \approx 0.3 \times B [{\rm T}] \times R [{\rm m}]
\end{equation}

\begin{figure}[!htbp]
 \begin{center}
  \includegraphics{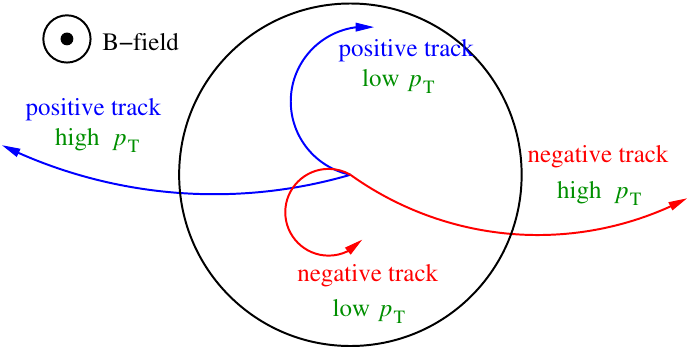}
  \caption{Particle track deflection in the transverse plane due to a perpendicular magnetic field.}\label{fig:2-charge_momentum}
 \end{center}
\end{figure}

Figure~\ref{fig:2-charge_momentum} shows the effect of the magnetic field in 
the deflection of the particles, depending on their charges and the magnitude 
of the transverse momentum. The magnetic field is perpendicular and pointing 
``outside of the paper''. With such orientation of the magnetic field, the 
sense of rotation of the tracks is clockwise for positively charged 
particles and counter-clockwise for negative particles. The radius of the 
curvature is greater for high momentum and vice-versa.

\subsection{Track Parameterization}

\begin{figure}[!htbp]
 \begin{center}
  \includegraphics{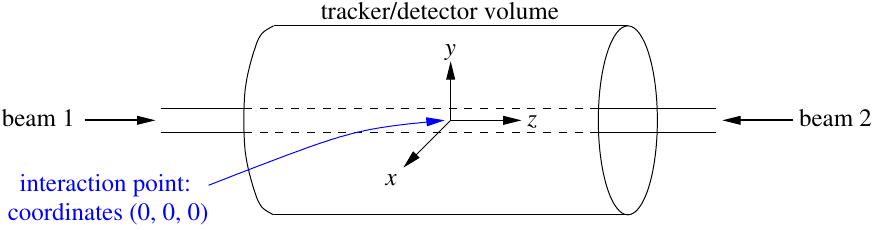}
  \caption{Coordinate system in a tracker.}\label{fig:2-coordinates}
 \end{center}

\end{figure}

The track parameters are described with respect to a given coordinate system. 
Usually the $z$ axis is the longitudinal axis, following the beam line, while 
the $x$ axis points horizontally to the centre of the accelerator ring (in 
circular accelerators, like the LHC) and the $y$ axis points vertically. The 
coordinate system is depicted in Figure~\ref{fig:2-coordinates}.

\begin{figure}[!htbp]
 \begin{center}
  \includegraphics{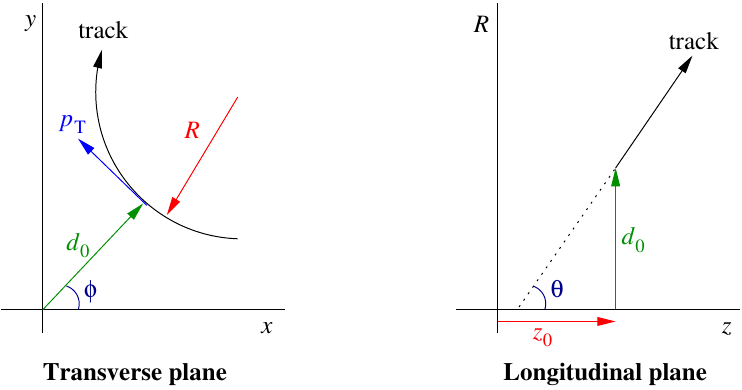}
  \caption{Illustration of the track parameters in the transverse ($xy$) and 
  longitudinal ($Rz$) planes.}\label{fig:2-track_parameters}
 \end{center}

\end{figure}

With the usual approach of a magnetic field orientation parallel to the $z$ 
axis, charged particles describe a circumference in the axial plane. By 
measuring the curvature perpendicular to the magnetic field, the transverse 
momentum of the particle can be measured. The parameters that describe a track 
are also called \textit{perigee parameters} and are depicted in 
Figure~\ref{fig:2-track_parameters}. Their definitions follow:

\begin{itemize}
 \item The transverse impact parameter, $d_0$, is the distance from the origin 
 of the detector to the track's closest approach (point of closest approach, or 
 PCA, to the $z$ axis) in the $xy$ plane. The uncertainty in its resolution 
 depends on the radial position of the active elements, and on the space point 
 precision.

\item The longitudinal impact parameter, $z_0$, is the distance from the origin 
of the detector to the track's closest approach in the $Rz$ plane. Both the
longitudinal and transverse impact 
parameters can be defined with respect to a primary vertex instead of the 
origin. 

\item The azimuthal angle ($xy$ plane) at the point of closest approach, 
$\phi$. It is the angle formed by the $x$ axis and the line between the origin 
and the PCA.

\item The polar angle, $\theta$, is the angle of the track momentum vector with 
the $z$ axis.

\item The pseudorapidity, $\eta = - \ln \tan (\theta/2)$. It is just a 
different way of expressing the polar angle.

\item The transverse momentum, $\pt = p \sin \theta$. It is the projection of 
the momentum vector in the $xy$ plane.

\item The charge, $q$, is defined by the sense of rotation of the helix in the 
$xy$ plane. Usually the charge over the momentum vector magnitude, $q/p$, is 
used.
\end{itemize}

These definitions are used in the ATLAS experiment. Other experiments may use
different notation or definitions.

\subsection{Resolution}

There are a number of uncertainty sources in the aforementioned parameters, 
such as the position measurement precision, the precise knowledge of the 
magnetic field configuration and the contribution that comes from multiple 
scattering in the tracker material. The uncertainties limit the resolution of 
the track 
reconstruction process. The position measurement can be improved through a good 
alignment of the detector, first during the assembly and after by measuring the 
actual position of the detector elements over time. 

In actual trackers, the magnetic field is not homogeneous. Although the 
momentum measurement through curvature is still proportional to the magnetic 
field strength, the field has to be mapped inside the tracker and numerical 
methods are used to interpolate between points in the field and to integrate 
the field 
along the particle trajectory. The parameters are defined locally, depending on 
the coordinate along the trajectory, yet there is still some bias on the 
momentum measurement after these adjustments.

One of the most important sources of uncertainty is multiple scattering in the 
material. When a charged particle traverses a medium, it is deflected by many 
small angle scatters. Most of these deflections are caused by Coulomb 
scattering in the nuclei and, for hadrons, the strong interaction is an 
additional contribution to multiple scattering.

For an incident particle on a layer with thickness $x/X_0$ radiation lengths, 
assuming the projected angular distribution is Gaussian, its width is given by 
the following formula~\cite{Beringer:1481544}.

\begin{equation}
\theta_0 = 13.6~{\rm MeV}\frac{z}{p\beta c} \sqrt{\frac{x}{X_0}}\left( 1 + 0.038\ln\left(\frac{x}{X_0}\right)\right)
\end{equation}

Where $p$ is the momentum, $\beta c$ is the velocity and $z$ is the charge 
number of the incident particle. Scattering is lower with high momentum 
particles.

Minimizing the amount of material is essential in order to reduce multiple 
scattering and other interactions inside the tracker volume. However, since 
there are a number of services inside the tracker, such as supporting 
structures, cooling, electricity and data input/output, the minimization has a 
limit. The amount of material is initially calculated from the engineering 
plans 
and by knowing all the material properties and position of all the pieces. With 
this information, the whole detector can be simulated and, for an accurate 
simulation, the material interactions have to be considered. 

With multiple scattering in mind, there are several parameters resolutions to 
be considered.

The impact parameters resolutions, $\sigma_{d_0}$ and $\sigma_{z_0}$, are 
driven by local misalignments of the tracker, namely the radii and space point 
precision of the layers. For the simplified case of only two measurement 
layers, the resolution on the impact parameter, due to only geometrical 
uncertainty, is~\cite{WellsTracking,Haywood:1994joa}

\begin{equation}
\sigma^2_{d_0} = \frac{r_2^2\sigma_1^2 + r_1^2\sigma_2^2}{(r_2 - r_1)^2}
\end{equation}

Where $r_i$ are the radii of the two layers and $\sigma_i$ are the intrinsic 
measurement errors. In addition, the measurement precision is also degraded by 
multiple scattering:

\begin{equation}
\sigma_{d_0} = \sqrt{\frac{r_2^2\sigma_1^2 + r_1^2\sigma_2^2}{(r_2 - r_1)^2}}
\oplus \frac{r}{p\sin^{3/2}\theta} 13.6\MeV\sqrt{\frac{x}{X_0}}
\end{equation}

At low momentum, the resolution is driven by multiple scattering in the beam 
pipe and the first layers of the tracker. For high momentum 
($\gtrsim 10~{\rm GeV}/c$), there is an asymptotic limit determined by the 
intrinsic tracker resolution and misalignments~\cite{IDCommissioning}.

\begin{figure}[!htbp]
 \begin{center}
  \includegraphics{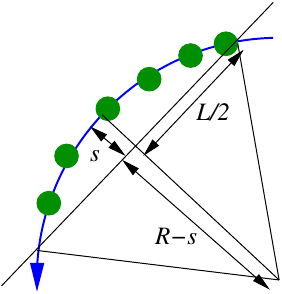}
  \caption{Measurement of the track curvature.}\label{fig:2-measure_momentum}
 \end{center}

\end{figure}

Figure~\ref{fig:2-measure_momentum} depicts a track with several hits and the 
curvature radius is calculated from the sagitta, $s$, and the distance between 
planes or path length, $L$:

\begin{equation}
 R = \frac{L^2}{2s} + \frac{s}{2}
\end{equation}

The relative uncertainty in the transverse momentum is proportional to the 
transverse momentum itself, 
$\pt$, and the sagitta uncertainty, $\sigma_s$. It is also inversely 
proportional to the magnetic field magnitude and the path length:

\begin{equation}
 \frac{\sigma_{\pt}}{\pt} = \frac{8\pt}{0.3BL^2}\sigma_s
\end{equation}

The sagitta uncertainty decreases with increasing number of points, each of 
them with a resolution $\sigma_{r\phi}$:

\begin{equation}
 \sigma_s = \sqrt{\frac{A_N}{N + 4}} \frac{\sigma_{r\phi}}{8}
\end{equation}

Where $A_N = 720$~\cite{Gluckstern1963381} is the statistical factor. The point 
error $\sigma_{r\phi}$ has a constant part that comes from the intrinsic 
resolution and a multiple scattering part. The multiple scattering contribution 
enters the sagitta uncertainty as

\begin{equation}
 \sigma_s \propto \frac{L}{\pt \sin^{1/2} \theta} \sqrt{\frac{L}{X_0}}
\end{equation}

Finally, a precise and reliable reconstruction of the track direction, that is, 
the angular resolution ($\theta$ and $\phi$), is an important contribution to 
the momentum vector determination. Therefore, the angular resolution is vital 
for the search of decay vertices and for matching the tracker signal with 
signals coming from other sub-detectors~\cite{IDCommissioning}. 

\subsection{Track Fitting}

The pattern recognition provides a series of measurements along a particle 
trajectory, together with a first estimation of the track parameters. The track 
fitting is responsible for computing the best possible estimation of those 
track parameters.

There are two widely used track fitting techniques, which are the global least
squares fit and the Kalman filter. They will not be discussed 
in detail here. 

The global least squares fit~\cite{LeastSquaresFitting} 
minimizes the $\chi^2$ function:

\begin{equation}
 \chi^2 = \sum_{i = 1}^{N_{{\rm hits}}} \left(\frac{\Delta_i}{\sigma_i}\right)^2 + 
 \sum_{i = 1}^{N_{{\rm layers}}} \left(\frac{\theta^2_{{\rm scat}}}{\sigma^2_{{\rm scat}}} +
 \frac{\left(\sin\theta_{{\rm local}}\right)^2 \phi^2_{{\rm scat}}}{\sigma^2_{{\rm scat}}}\right)
\end{equation}

The first term of the equation is the sum of all detectors hits of the 
residuals, $\Delta_i$, divided by the measurement error, $\sigma_i$. The second
part of the equation implements the possible changes in the track direction
using the scattering angles, $\phi_{{\rm scat}}$ and $\theta_{{\rm scat}}$. 
$\theta_{\rm local}$ is the angle between the track and the $Z$ axis at the
scattering plane. The factor $\sin\theta_{{\rm local}}$ is needed 
because $\phi_{{\rm scat}}$ is defined in the $x-y$ plane.
The scattering angles enter the algorithm initialized to 
zero and the error, $\sigma_{{\rm scat}}$, is calculated from the traversed 
material thickness~\cite{Limper:1202457,Cornelissen:1005181}.

The other technique is the Kalman filter~\cite{Fruhwirth:1987fm}, which 
actually does least squares fitting, with an iterative process of stepping from 
one measurement to the next, extrapolating the parameters from the previous 
step to the next measurement surface to calculate a prediction of the track 
state at that surface. This method results in a fast algorithm, as it does not
require to invert large matrices. The least squares fitting
involves the inversion of a $N\times M$ matrix with $N$ the number of fit 
parameters and $M$ the number of measurements and the number of scattering 
angles that have been used in the fit~\cite{Limper:1202457}.

There are three steps in the Kalman filter method: prediction using the track 
parameters and their covariance matrix in every step, filtering by updating the 
track parameters with the current hit measurement and smoothing by backwards 
propagation of the previously filtered points~\cite{Main97ranger}.

After the tracks have been fitted, the vertices may be reconstructed, through 
fitting of the tracks to common points of origin. Constrains such as invariant 
mass or momentum vector direction may be applied. The fast vertex fitting 
method, also known as the Billoir algorithm, is described in 
\cite{Billoir1992139}. It uses the fitted tracks parameters and their errors, 
$\Delta q_i = q_i^{{\rm measured}} - F(\mathbf{V}, \vec{p}_i)$, to find 
the vertex position ($\mathbf{V}$) and momentum vectors of the associated 
tracks ($\vec{p}_i$) that minimize

\begin{equation}
\chi^2 = \sum_{i=1}^{N_{{\rm tracks}}} \Delta q_i^T V_i \Delta q_i
\end{equation}

The algorithm operates with small matrices, what leads to a computational 
complexity of order $N_{{\rm tracks}}$, in contrast to the least squares 
method, which requires $N_{{\rm tracks}}^3$ operations.

Additional information on track fitting and vertex fitting techniques is 
available in~\cite{Fruhwirth:429722}.

\subsection{Alignment}\label{sec:2-alignment}

The accuracy with which particle tracks can be reconstructed is highly 
dependant on how well the positions and orientations of the tracker sensors are 
known. This is, in turn, limited by the precision of the assembly process and 
the stability of the structures holding the tracker in place. 

Usually, the intrinsic resolution of the tracker is better than the precision 
of the assembly. Additionally, the position of the detector elements may vary 
over time due to the effect of the magnetic field and other environmental 
effects, such as temperature variations inside the volume~\cite{TrackingLHC}. 
As a consequence, the alignment has to be surveyed periodically, to account for 
possible deformations or movement of the elements in the tracker volume, even 
right after the assembly.

\begin{figure}[!htbp]
 \begin{center}
  \includegraphics{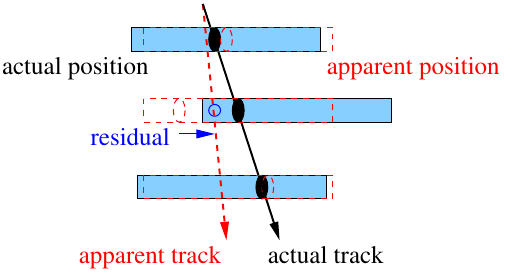}
  \caption{Track hits residuals.}\label{fig:2-residual}
 \end{center}

\end{figure}

The basis of the track based alignment algorithms is the minimization of the 
sum of hit residuals $r$ from high momentum tracks, which have lower multiple 
scattering distortion. A residual is defined as the distance between the 
position of the measurement and the intersection of the fitted track with that 
module, see Figure~\ref{fig:2-residual}. 

The alignment algorithms use six constants for every independent module or 
structure, resulting in six degrees of freedom of a rigid body: three 
translations with respect to the nominal position and three rotations with 
respect to the nominal axis orientations. Module deformations such as twisting 
and bending are ignored. The alignment constants can be determined by 
minimizing the $\chi^2$ function~\cite{IDCommissioning}:

\begin{equation}
 \chi^2 = \sum_{{\rm tracks}} \vec{r}^T V^{-1}\vec{r}
\end{equation}

Where $V$ is the track covariance matrix and $\vec{r}$ is the residuals vector 
for a given track, which is a function of both the track parameters and the 
alignment constants.

The use of several types of tracks may be required, depending on the 
geometrical distortion mode to be constrained. This is because minimization of 
the residuals might not be sufficient to guarantee a correct alignment. Some 
global distortions preserve the helical trajectory of the tracks but bias the 
track parameters, while not affecting the $\chi^2$.

\begin{figure}[!htbp]
 \begin{center}
  \includegraphics{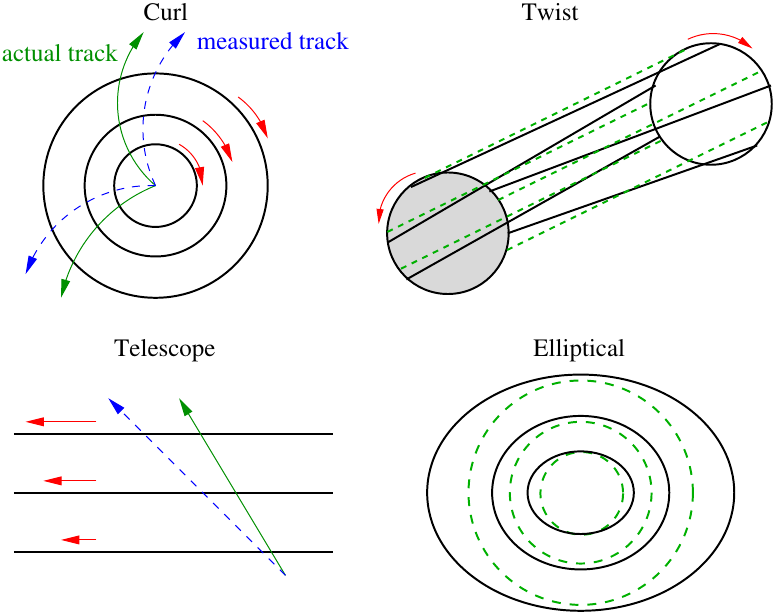}
  \caption[Weak modes of geometrical distortions.]{Weak modes of geometrical distortions~\cite{KollarTracking}.}\label{fig:2-alignment_weak_modes}
 \end{center}
\end{figure}

These are called weak distortion modes, shown in 
Figure~\ref{fig:2-alignment_weak_modes}, and they represent a great danger to 
physics results. For instance, elliptical distortion affects the measurement of 
the vertices masses.

Weak modes are difficult to remove and require different track topologies. Some 
ways to constrain these weak modes are: 

\begin{itemize}
 \item Cosmic ray tracks provide a continuous helical trajectory across the 
 whole tracker. In addition, a large fraction of cosmic rays cross the tracker 
 far from the beam axis, what provides additional constrains not available with 
 the tracks coming from the interaction point.
\item Tracks passing through the overlapping regions of adjacent modules help 
constrain the circumference of cylindrical arrangements, improving the 
determination of the average radial position of the modules.
\item Reconstruction of the invariant mass of track pairs from $Z$ and $J/\Psi$ 
decays provides sensitivity to systematic correlation between separate detector 
elements.
\item Survey measurements provide additional information.
\end{itemize}

See~\cite{ATLAS} and~\cite{KollarTracking} for further details. A brief 
description of the alignment in the ATLAS tracker is given in 
section~\ref{sec:2-ATLASalignment}.

\section{Silicon Trackers}
   
\subsubsection{A Short Introduction to Detectors}   
   
The principle of operation of a detector is the transfer of part or all of the 
radiation energy carried by the particles to the detector mass. This energy 
transfer is a process that converts the radiation energy to some other form of 
energy that is processable by other means, for instance, electric signals. 
Charged particles transfer their energy by means of direct collisions with 
electrons in the atoms of the detector material. The form in which the energy 
appears after its conversion in the detector is dependent of the detector type 
and its design~\cite{Leo1994}.

\begin{figure}[!htbp]
  \begin{center}
 \includegraphics[scale=0.8]{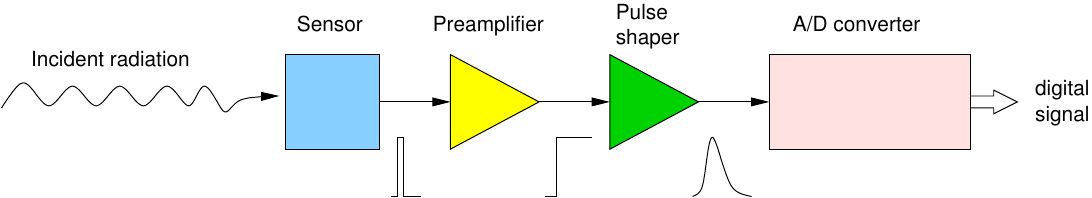}
 
 \caption{Basic detector scheme: from radiation absorption in the sensor to 
 digitization of the event.}\label{fig:2-detector}
 \end{center}
\end{figure}

Semiconductor detectors are based on crystalline semiconductor material and 
generate electric signals that can be processed with electronic technology. The 
most frequent semiconductor materials used for charged particle detection are 
silicon and germanium. Their operation principle is analogous to gas ionization 
detectors: the creation of electron-hole pairs in the material mass during the 
passage of ionizing radiation.

\begin{figure}[!htbp]
  \begin{center}
 \includegraphics{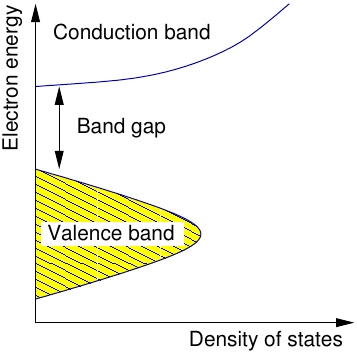}
 \caption{Semiconductor band structure.}\label{fig:2-bandgap}
 \end{center}
\end{figure}

The band gap of any material is the energy difference between the valence and 
the conduction band, as shown in Figure~\ref{fig:2-bandgap}. In conductors, 
such as copper, the separation is very small 
or non existent, while insulators have a very large gap. Semiconductors have a 
smaller band gap, a few electronvolts, which is the 
energy required to create an electron-hole pair. Silicon has a band gap of
$1.12~{\rm eV}$, with an ionization energy of $3.6~{\rm eV}$. In 
comparison, gas detectors have an ionization energy of around 
$30~{\rm eV}$.

\subsubsection{How are semiconductors used in particle detection?} 

\begin{figure}[!htbp]
  \begin{center}
 \includegraphics[scale=0.75]{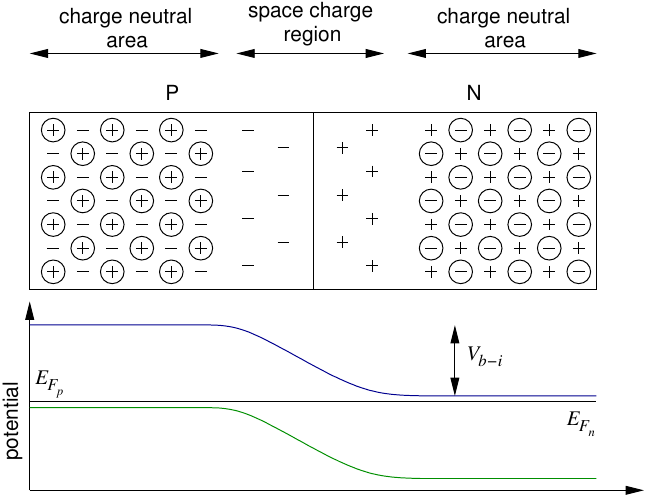}
 \caption{A p-n junction under thermal 
 equilibrium.}\label{fig:2-pn_equilibrium}
 \end{center}
\end{figure}

A semiconductor detector is, essentially, a p-n junction. The p and n regions 
are electrically neutral by themselves, but, when they are part of a p-n 
junction, thermal diffusion drives holes and electrons across the junction. 
This leads to a net positive charge in the n region, which in the end results 
in a built-in potential, $V_{b-i}$, as shown in 
Figure~\ref{fig:2-pn_equilibrium}. In this situation, the Fermi levels of each 
region, $E_{F_n}$ and $E_{F_p}$, determine the built in potential, as $V_{b-i} 
= E_{F_n} - E_{F_p}$.  The diffusion of holes and electrons leads to an area 
free of mobile carriers, named the ``depletion region''. Strictly speaking, the 
depletion area is not completely empty of mobile carriers, as the diffusion 
profile is a gradual transition instead of an abrupt one.

The number of free carriers in the junction when it is under thermal 
equilibrium, that is, unbiased, is several orders of magnitude larger than the 
number of electron-hole pairs that are generated by a Minimum Ionizing Particle 
(MIP) passing through the silicon. For instance, the 
carrier concentration in intrinsic silicon is around 
$1.45\times 10^{10}~{\rm cm}^{-3}$. In a volume 
$1~{\rm cm}\times 1~{\rm cm}\times 300\mum$, there are 
$4.5\times 10^8$ free carriers, while a minimum ionizing particle generates 
$3.2\times 10^4$ electron-hole pairs. The signal to noise ratio (SNR) in this
case is very low.

However, when the junction is reverse biased, the depletion region widens 
across the whole p-n junction. When the bias reaches the ``depletion voltage'', 
the complete volume is devoid of free carriers and all the material is 
sensitive to ionizing particles. In addition, the reverse bias adds an electric 
field 
that helps the collection of the generated electron-hole pairs. A minimum 
ionizing particle traversing the volume will leave some energy behind and 
create a number of electron-hole pairs proportional to that energy. As a 
result, the SNR increases significantly.

\begin{figure}[!htbp]
  \begin{center}
\includegraphics{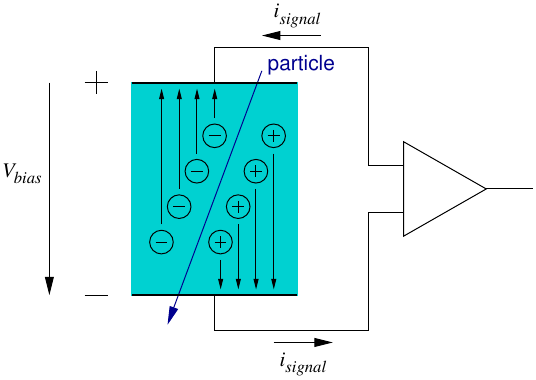}
\caption{Description of the charge collection 
concept.}\label{fig:2-charge_collection}
 \end{center}
\end{figure}

The electron-hole pairs drift due to the electric field present in the 
semiconductor junction, inducing an electrical current. The charge collection 
process is depicted in Figure~\ref{fig:2-charge_collection}~\cite{Spieler2005}.
Because the average energy required to produce an electron-hole pair in 
semiconductors is an order 
of magnitude smaller than gas ionization detectors, semiconductor detectors 
have a greater energy resolution.

The energy may be calculated by integrating the signal current that is induced
in the electrodes.

In order to be able to estimate the energy deposited by the traversing 
particle, some discretization of the collected charge has to be done. Some 
detectors aim at measuring not the energy of the particles, but if a particle 
has gone through, so this information is usually not available as it is not 
necessary. Thus, there are two options when designing the detector:

\begin{itemize}
\item If the goal is to know the path of a particle (tracking) through a 
volume, then the output of the detector would be ``measured some charge above 
threshold''. The threshold is usually set so noise is minimized and efficiency 
is maximized. This is called binary read-out.
\item If the goal is to know the energy deposited by a 
particle traversing the volume, then the amount of charge that is generated 
needs to be measured somehow. The charge generated in the detector volume is 
proportional to the deposited energy: 

\begin{equation}
Q = \frac{E}{E_i}e
\end{equation}

Where $E$ is the absorbed energy and $E_i$ is the ionization energy, which for 
silicon it is $E_i = 3.6~{\rm eV}$. An approximation for the ionization 
energy as a function of the band gap energy is 
\cite{Spieler2005}
$$E_i \approx 2.8 E_g + 0.6~{\rm eV}$$
\end{itemize} 

The absorbed energy is proportional to the amount of material that the particle 
traverses, so the thickness of the semiconductor material is also a parameter 
to be considered for two reasons:

\begin{itemize}
\item Thicker material results in higher absorbed energy, what increases the
energy resolution. Even if energy resolution is not important, charge 
collection efficiency is also increased as more electron-hole pairs are 
generated by minimum ionizing particles for thicker material.
\item But low material is needed in order to reduce multiple scattering in the 
material. Multiple scattering is a major contributor to momentum measurement 
uncertainty for low momentum, up to several tens of ${\rm GeV}/c$ 
transverse momentum~\cite{ValentanScattering09}.
\end{itemize}

Apart from the material, the absorbed energy depends on the type of material, 
which is fixed, and the type of particle: electrons, muons, protons, etc. This 
is related to the $\dedx$ or differential energy loss, described by the 
Bethe-Bloch formula. This topic is discussed in detail in 
chapter~\ref{chap:de_dx}.

\subsubsection{Why Silicon?}

There are several advantages of silicon over other semiconductor materials. To 
begin with, it is very abundant in the Earth's crust ~\cite{Bean1990}, orders 
of magnitude above the other semiconductors' constituents, as can be seen in 
Table~\ref{tab:1-abundance}. Another important advantage of silicon is the fact 
that it is one of the main materials used in everyday electronic technology. 
Therefore, the development of the production techniques is well advanced, 
leading to low production cost, which is always desirable.

\begin{table}[!htbp]
\begin{center}
\begin{tabular}{c c|c c|c c}
\toprule
Element & Fraction & Element & Fraction & Element & Fraction \\ \hline
\textbf{Silicon} & $0.283$ & Zinc & $7\times 10^{-5}$ & Cadmium & $2\times 10^{-7}$ \\
Aluminium & $0.083$ & Gallium & $1.5\times 10^{-5}$ & Indium & $10^{-7}$ \\
Phosphorus & $0.001$ & Germanium & $5\times 10^{-6}$ & Mercury & $8\times 10^{-8}$ \\
Sulfur & $2.6 \times 10^{-4}$ & Arsenic & $1.8 \times 10^{-6}$ & Selenium & $5\times 10^{-8}$ \\
Carbon & $2 \times 10^{-4}$ & Antimony & $2 \times 10^{-7}$ & Tellurium & $1 \times 10^{-9}$ \\
\bottomrule
\end{tabular}
\caption{Abundance of semiconductor constituents in the Earth's 
crust.}\label{tab:1-abundance}
\end{center}
\end{table}

Regarding the physical properties of silicon, they are summarized in the 
following list~\cite{Lutz2007}:

\begin{itemize}
\item It has a small band gap, $E_g = 1.12~{\rm eV}$, with an ionization energy 
of $E_i = 3.6~{\rm eV}$.
\item It has a high specific density, which increases the absorbed energy 
($\dedx$), generating more electron-hole pairs for a minimum ionizing 
particle.
\item Despite that high density, it has a high carrier mobility, which allows 
fast charge collection, in the order of $10$~nanoseconds over hundreds of 
micrometres of fully depleted silicon.
\item It is very pure, with less than $1$ part per million impurities and less 
than $0.1$ parts per billion electrically active impurities.
\item Its rigidity is appropriate for constructing thin, self-supporting 
structures.
\end{itemize}

\begin{table}[!htbp]
\begin{center}
\begin{tabular}{c|c|c|c|c|c}
\toprule
 & Diamond & SiC & GaAs & Ge & \textbf{Si} \\ \hline
 
Z & 6 & 14/6 & 31/33 & 32 & 14 \\
$E_g\ [{\rm eV}]$ & $5.5$ & $3.3$ & $1.42$ & $0.66$ & $1.12$ \\
$E_i\ [{\rm eV}]$ & $13$ & $7.6-8.4$ & $4.3$ & $2.9$ & $3.6$ \\
density $[{\rm g/cm^{3}}]$ & $3.515$ & $3.22$ & $5.32$ & $5.32$ & $2.33$ \\
$\mu_e\ [{\rm cm^2/Vs}]$ & $1800$ & $800$ & $8500$ & $3900$ & $1450$ \\
$\mu_h\ [{\rm cm^2/Vs}]$ & $1200$ & $115$ & $400$ & $1900$ & $450$ \\
\bottomrule
\end{tabular}
\caption{Some physical characteristics of 
semiconductors.}\label{tab:1-semiconductor_characteristics}
\end{center}
\end{table}

Some of the physical properties of silicon are compared to other semiconductor 
materials in 
Table~\ref{tab:1-semiconductor_characteristics}. Although silicon does not have 
the best numbers for all the properties, they are more than adequate for 
detectors. The facts that the fabrication methods production are well known and 
that it is very abundant make it the preferred choice so far. However, 
research on alternatives, such as diamond, silicon carbide (SiC) and other 
compounds as detectors is a very active research field~\cite{Owens:1486869}.

\subsubsection{Characteristics of Silicon Trackers}

    Semiconductor detectors are usually constructed on a substrate which is 
several centimetres long and wide. Depending on how the p-n junctions are 
implanted on the substrate, it is possible to have two types of detectors:
    
    \begin{enumerate}
    \item Pixel detectors are implemented as a matrix of p-n junctions, which 
    are several micrometres long and wide.    
    \item Strip detectors are made by creating a straight pattern of wide p-n 
    junctions separated by several micrometres.
    \end{enumerate}

\begin{figure}[!htbp]
  \begin{center}
\includegraphics[scale=0.75]{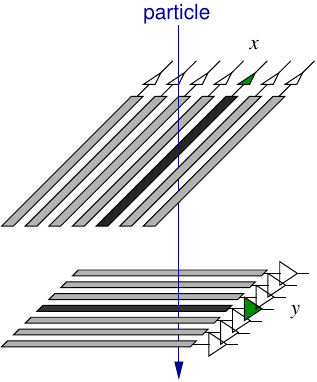} 
\caption{Double sided strip sensor for two dimension 
positioning.}\label{fig:2-double_sided_strips}
 \end{center}
 \end{figure}

For pixel detectors, spatial resolution depends only on the separation between 
the junctions but for strip detectors, a single row of strips will only give 
good resolution in one direction. In order to have spatial resolution in two 
directions, it is necessary to use another set of strips with certain subtended 
angle with respect to the first detector, having a double sided detector as 
shown in Figure~\ref{fig:2-double_sided_strips}.

The configuration shown in the figure has problems at high hit densities, 
because each hit generates an $x$- and a $y$-coordinate. That leads to $n$ 
tracks generating $n$ $x$-coordinates and $n$ $y$-coordinates, simulating $n^2$ 
hits, out of which $n^2 - n$ are fake. Additional information is needed to 
eliminate coordinates not consistent with tracks. The area subtended by two 
sensing strips of length $L$ in a $90^\circ$ angle is $A = L^2$, which provokes 
that a hit in a given strip may form combinations with hits in all the 
transverse strips. The solution to this problem is using ``small-angle 
stereo'', where the subtended angle is (much) lower than $90^\circ$. Thus, if 
the subtended angle is $\alpha$ and the strip pitch is $p$, the capture area is

\begin{equation}
A \approx L^2 \tan \alpha + L p
\end{equation}

The probability of multiple hits within this acceptance area is lower as 
$\alpha$ is made smaller, but resolution in the longitudinal 
coordinate also deteriorates.

Position resolution is the spread of the reconstructed position of the particle
hit with respect to the actual position. There are design parameters that 
influence this resolution. First, the pixel size or strip pitch determines the
area sensitive to one particle hit. Once the size has been fixed, the signal to
noise ratio (SNR) of the detector has an influence on analogue systems. The SNR
is higher if the following conditions are met:

\begin{itemize}
\item Low detector capacitance, achieved with small pixels or short strips. The
noise is directly proportional to the capacitance at the input of the front-end
electronics.
\item Low leakage current, which depends on the quality of the bulk material. 
The noise increases with the square root of 
the leakage current.
\item Large bias resistor, in order to have a good inter-strip isolation and 
increase the integration time.
\item Short and low resistance connection from the detector to the amplifier.
\end{itemize}

When analogue read-out is used, the position resolution is

\begin{equation}
\sigma_x \approx \frac{p}{1.5~{\rm SNR}}
\end{equation}

For binary read-out with threshold counter, the resolution is estimated as

\begin{equation}
\sigma_x \approx \frac{p}{\sqrt{12}}
\end{equation}

Typical SNR values in the electronics are from $15$ to $40$.

\subsubsection{Radiation Damage in Silicon}\label{sec:2-radiationdamage}

Semiconductors in general, and silicon in particular, are relatively sensitive 
to radiation damage. The damage is caused by the incident particles colliding 
with the crystal lattice atoms, causing point defects by ``knocking'' the atoms 
out of their normal positions. 

The defects in the lattice give rise to discrete trapping 
levels in the forbidden band gap, that reduce the number of charge carriers in 
the semiconductor~\cite{Leo1994}. Effects observed from these defects caused by 
radiation include~\cite{Kraner:1984uv}:

\begin{itemize}
 \item Leakage current increase, due to the defects acting as centres that 
 increase the bulk current.
\item Energy resolution decrease, as the defects are discrete trapping centres 
that reduce the number of charge carriers in the semiconductor and, in turn, at 
the electronics input.
\item Reduction of the carrier mobility.
\item Increase in the output pulse rise time or charge collection time, as a 
consequence of the charge trapping and the carrier mobility reduction.
\item Actual material change, what is called ``type inversion'': n type silicon 
turns into p type silicon. This happens because the radiation removes the 
donors and creates acceptors in their place. Type inversion occurs after a
fluence of around $10^{13}~{\rm n_{eq} cm^{-2}}$~\cite{SiTypeInversion}.
\end{itemize}

The leakage current increase can be estimated as~\cite{MollThesis}:

\begin{equation}
 \Delta I = \alpha \Phi V
\end{equation}

Where $V$ is the volume of the silicon, $\Phi$ is the radiation fluence and 
$\alpha$ is the damage constant, which depends on the radiation type and its 
energy. 

Heavy charged particles, protons and neutrons, can cause a large recoil energy 
to the silicon bulk, creating ``clusters'' of damage, apart from some isolated 
single defects. These clusters generate charge traps and 
leakage current increase. Lightly ionizing particles, such as
electrons and minimum ionizing particles, cause primarily single defects.

Refer to~\cite{Leo1994,Kraner:1984uv} for additional information on the damage 
constants.

This model is used to compare with the monitoring of the leakage current
of the silicon detectors. An example of such monitoring during the first three
years of operation is shown in
section~\ref{sec:2-SCTradiationdamage} for the ATLAS Semiconductor Tracker.

In addition to the radiation damage defects described above, heavy charged 
particles and fast neutrons produce vacancy clusters that are not entirely 
stable at room temperature. They ``decay'' or migrate to smaller vacancy 
aggregates, what is called annealing. 

This effect was discovered after irradiating 
detectors, when the observed damage to the detector started decreasing with 
time, depending on what temperature the detector was kept at during the waiting 
period~\cite{Lutz2007,Wunstorf1992149}.

Although this effect might be naively interpreted as a reduction in the 
radiation generated defects in the silicon and true annealing (the crystal 
becoming perfect again) actually exists, in many cases the defects may just be 
transformed into other stable defects~\cite{Lutz2007}.

\section{The ATLAS Semiconductor Tracker}

The Semi-Conductor Tracker (SCT) in the ATLAS Inner Detector covers the radial 
range from $299$ to $514\mm$, and $z$ range up to $\pm 2720.2\mm$, with 
four barrel layers and nine endcap discs to each side of the barrel.

There are $4088$ modules in the SCT, split in $2112$ modules in the barrel 
layers and $988$ on each of the two endcaps. The micro-strip silicon modules 
for the SCT are of six different types, one on
the barrel and five different types on the endcaps. 

The barrel modules are all rectangular, with $768$ read-out strips that have a 
$80\mum$ pitch. The modules are double sided, with the active strips of the two
sides forming a $40\mrad$ stero angle that provides the two-dimensional 
resolution. Each side holds two sensors with $768$ strips, $6\cm$ long, which 
are daisy chained together with wirebonds in order to get $12\cm$ long 
active strips on each module.

The endcap discs layout makes the geometry of the sensors 
more complicated. The modules have a trapezoidal shape, with an interstrip 
angle that varies 
between $161$ and $207\urad$ and a strip pitch that varies from $56.9$ 
to $94.2\mum$. There are five sensor types, 
named W12, W21, W22, W31 and W32. All of them also have $768$ strips, to comply 
with the read-out hybrid assembly that comprises the twelve ABCD3TA 
chips~\cite{Campabadal:994402} (ABC stands for ``ATLAS Binary Chip''). The 
hybrid is folded around the module so each side is read-out by six chips.

All sensors are p-in-n with a thickness of $285\pm 15\mum$, supplied by 
Hamamatsu Photonics and CiS. In principle, all the sensors were meant to be 
fabricated on $\left\langle 111 \right\rangle$ silicon due to availability of 
supply~\cite{SCTsensors07}, but $93$ modules, $2.3\%$ of the total, have 
sensors with $\left\langle 100 \right\rangle$ silicon. The different crystal 
lattice structure has a small but measurable effect on the carrier 
mobility~\cite{Jacoboni197777}. This difference also has an effect on the 
noise performance of the modules~\cite{Silicon111and100}.

The sensors are glued back-to-back
and the 
electronics mounted on a hybrid are glued and wirebonded to build the modules. 
The assembly includes the mounting frame, cooling, powering and data 
input/output.

\begin{figure}[!hbt]
 \begin{center}
  \includegraphics[scale=0.8]{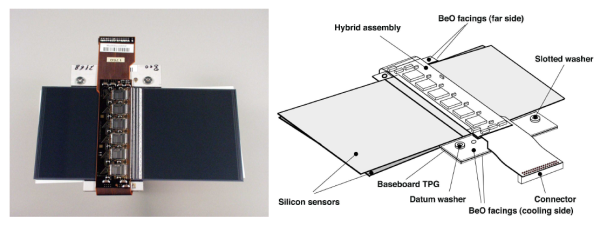}
  \caption{SCT barrel module photograph and illustration, showing the sensors 
  with the built-in stereo angle and the 
  electronics.}\label{fig:2-sct_barrel_module}
 \end{center}
\end{figure}

A barrel module is shown in Figure~\ref{fig:2-sct_barrel_module}, both a photo 
of a fully assembled module and a drawing showing its components: the two 
silicon sensors, the hybrid, facings for the assembly and the connector for the 
electronics.

\begin{figure}[!htb]
 \begin{center}
  \includegraphics[scale=0.8]{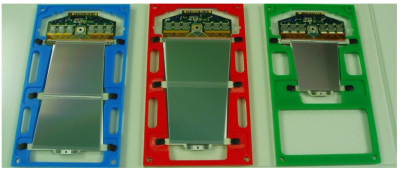}
  \caption{SCT endcap module types (outer, middle and 
  inner).}\label{fig:2-sct_endcap_modules}
 \end{center}
\end{figure}

Three of the endcap module types are shown in 
Figure~\ref{fig:2-sct_endcap_modules}. The outer module is composed of W31 
(top) and W32 (bottom) sensors, the middle module is composed of W22 (top) and 
W21 (bottom) sensors and the inner module has only W12 sensors. A fourth type 
that is not shown is the short-middle, which only has W22 
sensors~\cite{SCTendcap}.

The strips read-out is made through the twelve chips mounted on each hybrid. 
This chip is the ABCD3TA~\cite{Campabadal:994402}, with $128$ channels, 
performing the following functions:

\begin{itemize}
 \item Charge integration.
\item Pulse shaping.
\item Discrimination.
\item Compression and transmission of the data.
\item Configuration of the thresholds and read-out modes.
\end{itemize}

\begin{figure}[!htbp]
 \begin{center}
  \includegraphics[scale=0.8]{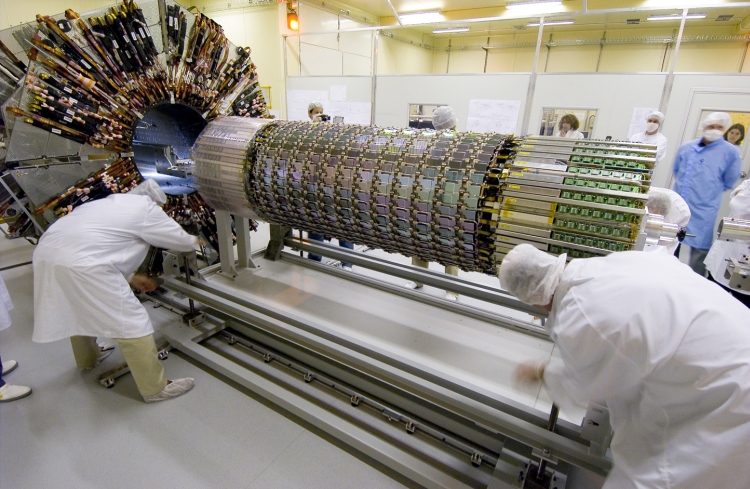}
  \caption{One of the SCT barrels during insertion into the 
  Inner Detector.}\label{fig:2-sct_barrel_assembly}
 \end{center}
\end{figure}

The four SCT barrel layers are numbered from $3$ to $6$, continuing the 
Inner Detector layer numbering, from the three Pixel layers (numbered from
0 to 2). The two endcaps have nine discs, numbered from 
$1$ to $9$, and the two endcap sides are named ``Endcap A'' and ``Endcap C''.

Figure~\ref{fig:2-sct_barrel_assembly} shows one of the SCT 
barrels during the insertion in the Inner Detector.

\begin{figure}[!htbp]
 \begin{center}
  \includegraphics[scale=0.8]{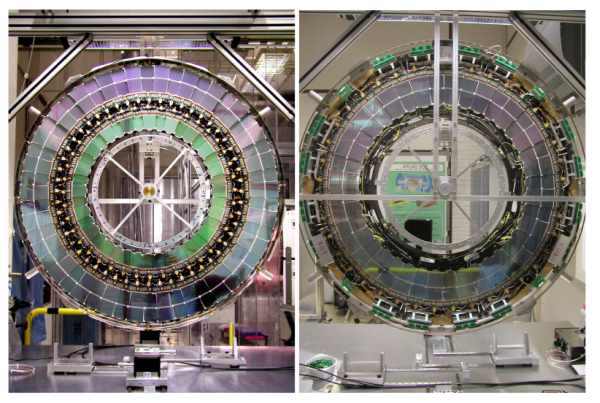}
  \caption{SCT endcap modules mounted on endcap discs: left has outer and inner 
  modules, right shows middle modules.}\label{fig:2-sct_endcap_assembly}
 \end{center}
\end{figure}

Figure~\ref{fig:2-sct_endcap_assembly} shows two sides of an endcap disc, one 
of the sides is populated by outer and inner modules and the other side by 
middle modules.

The SCT sensors are required to operate stably up to $500~{\rm V}$ bias and to 
have less than one percent bad read-out strips, 
before and after a neutron equivalent fluence of 
$F_{{\rm neq}} = 2\times 10^{14}~{\rm cm}^{-2}$.

\subsection{Tracking of the Radiation Damage in the 
SCT}\label{sec:2-SCTradiationdamage}

The model for the leakage current increase with the radiation damage that was 
described before is compared against the actual monitored leakage current of 
the SCT modules. 

\begin{figure}[!htbp]
 \begin{center}
  \includegraphics[scale=0.7]{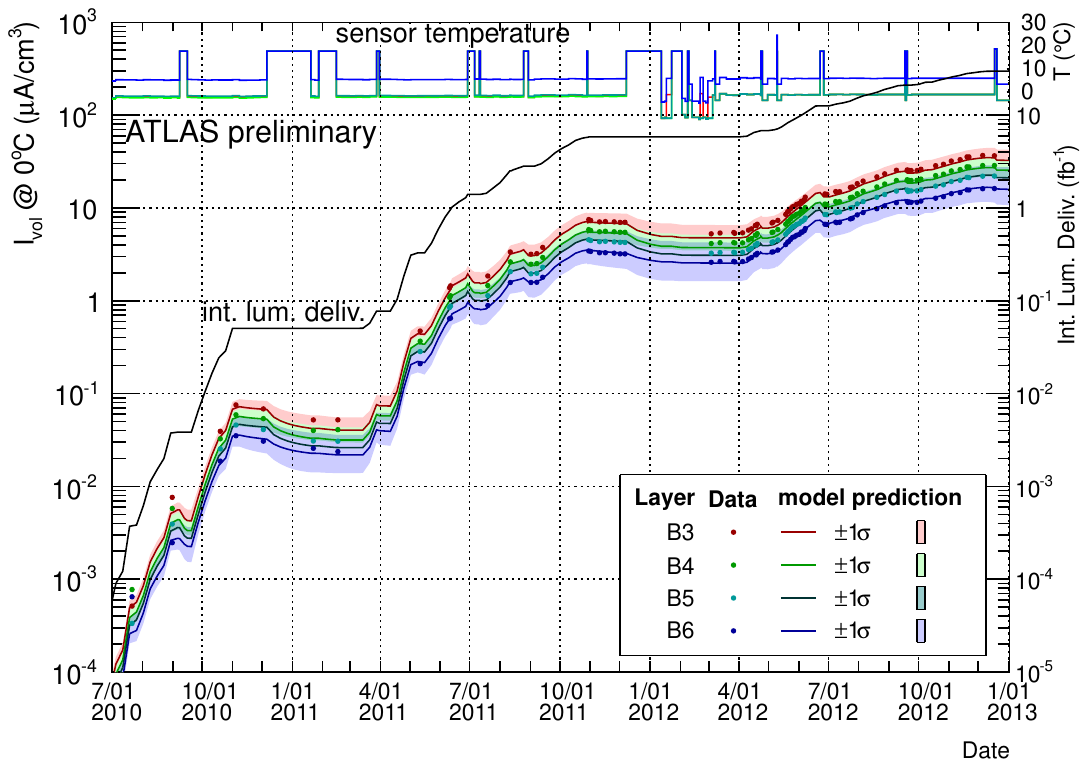}
  \caption[Leakage current evolution in the SCT over 
  time.]{Leakage current evolution in the SCT over 
  time~\cite{Aad:2014mta}.}\label{fig:2-leakcurrent_2010_2012}
 \end{center}
\end{figure}

Figure~\ref{fig:2-leakcurrent_2010_2012}
shows the evolution of the leakage current for 
the four barrel layers, both the measurements and the prediction from the 
model~\cite{Aad:2014mta}. The temperature profile of the sensors is also
plotted, as well as the integrated luminosity. The data ranges from mid-2010 
up to the beginning of 2013.

The plot shows the leakage current density in$\uA/{\rm cm}^3$, normalized to 
$0\degC$.

The temperature dependence of the leakage current follows the following scaling 
formula~\cite{Chilingarov:1511889}:

\begin{equation}
 \frac{I_{{\rm ref}}}{I_T} = \left(\frac{T_{{\rm ref}}}{T}\right)^2 \exp 
 \left( - \frac{E_{\rm gen}}{2k_B} \left(\frac{1}{T_{{\rm ref}}} - \frac{1}{T}\right)\right)
\end{equation}

Where $T_{{\rm ref}}$ is a reference temperature at which the leakage 
current is known, $E_{\rm gen} = 1.21~{\rm eV}$ is the effective generation energy 
in silicon 
and $k_B$ is the Boltzman constant.

\subsection{Noise Performance}

The performance of the silicon modules in terms of noise is usually measured 
through two main parameters, which are not independent: Equivalent Noise Charge 
(ENC) and Noise Occupancy 
(NO). This section only refers to the SCT sub-detector, however, the same 
performance measurements are applicable to the Pixel 
detector~\cite{IDCommissioning}.

The Equivalent Noise Charge is defined as the number of electrons that have to 
be collected from the sensor in order to create a signal equivalent to the 
noise of the sensor. $N$-point gain scans are used to measure this noise, being 
usually $N=3$. In this kind of scan a known, calibrated charge is injected for 
each point to the chip front-end and the discriminator threshold is varied. 
The occupancy of the strips decreases as the threshold is raised.

The voltage at the input of the discriminator is measured over a certain number
of triggers. The noise is assumed to be Gaussian, therefore the voltage at the
discriminator is fitted to a Gaussian distribution, with the mean and sigma
values corresponding to the $V_{t50}$ (named the $50\%$ threshold) and the 
output noise ($\sigma$).

This $50\%$ threshold is plotted against the injected charge, leading to a 
response curve, of which the slope is called the gain ($G$), usually measured 
in ${\rm mV/fC}$. The calibration circuit of the ABCD3TA chip of the SCT 
modules 
is capable of injecting charges between $0.5~{\rm fC}$ and $16~{\rm fC}$.

The aforementioned sigma of the distribution is the output noise, typically 
measured in$\mV$. The input noise is the result of dividing this 
output noise by the gain and, since $1~{\rm fC} \approx 6242~e^{-}$,

\begin{equation}
 \ENC = 6242 [{\rm fC}/e^{-}]\frac{\sigma[{\rm mV}]}{G[{\rm mV/fC}]}
\end{equation}

\begin{figure}[!htbp]
 \begin{center}
 \subfigure[S-curves from a threshold scan for one chip]{
  \includegraphics[scale=0.65]{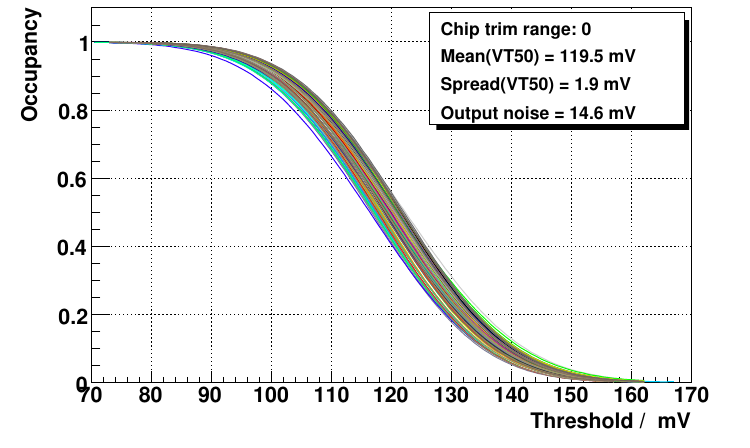}
  \label{fig:2-scurve}
 }
 \subfigure[Response curves]{
   \includegraphics[scale=0.95]{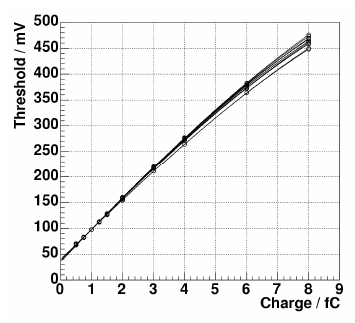}
 }
 \label{fig:2-responsecurve}
  \caption{Sample performance measurements on the SCT modules.}\label{fig:2-scurves_and_responsecurve}
 \end{center}
\end{figure}

When injecting a known charge, plots like those shown in 
Figure~\ref{fig:2-scurve} are generated. This picture shows the s-curves 
from the threshold scan done on 
one chip. The vertical axis is the occupancy and the horizontal axis is the 
threshold setting. The output noise is the standard deviation of the threshold 
from the $50\%$ occupancy and the $V_{t50}$ is the threshold at which the 
occupancy is $50\%$. With $N$ (at least $3$) of these s-curves, the response 
curve can be plotted and the gain of the chip is calculated, leading to the
input noise calculation 
as shown above. The plot in the figure was taken from~\cite{Demirkoz:2007zz}.

In the SCT chip, the ABCD3TA, the gain has an average value of~
$55~{\rm mV/fC}$. The Noise Occupancy in the SCT is defined as the number of 
hits per event for an equivalent threshold setting of $1~{\rm fC}$. 
The ENC requirement of the SCT before irradiation is 
$1500$ electrons r.m.s. and $1800$ electrons r.m.s. for an irradiated module.

With the goal of keeping the noise in the SCT very low, the threshold is set
at a level that suppresses the noise. However, a too high threshold will have
a negative impact on the hit efficiency. The requirement for the SCT system is
comes from a compromise between noise occupancy and efficiency, and it was
established that the noise occupancy has to be below 
$5\times 10^{-4}$ hits per event, while maintaining $99\%$ 
efficiency~\cite{IDCommissioning}.

The noise occupancy is measured without an injected charge, as opposed to
the equivalent noise charge.

As was mentioned before, by fitting the $N$ points to a response curve, such as 
that shown in Figure~\ref{fig:2-responsecurve}, which is a $10$ point gain 
scan. The slope is the gain of the chip and the intersection with the vertical 
axis (zero charge) is called the ``offset''. The plot corresponds to the 12 
chips on one whole SCT module~\cite{SCTDAQ}. When $N=3$, specially in the low 
injected charge region, between $0.5$ and $2.5~{\rm fC}$, the curve is well 
fitted to a straight line.

The noise is dependent on a number of parameters from both the sensor and the 
chip, such as:

\begin{itemize}
\item The amplifier input capacitance.
\item The chip temperature.
\item The sensor temperature.
\item The sensor leakage current.
\end{itemize}

A complete noise model was developed to study the noise of the SCT modules, 
considering these contributions and how they enter the 
equation~\cite{Burdin:1237817}.

\begin{figure}[!htbp]
 \begin{center}
  \includegraphics[scale=1.0]{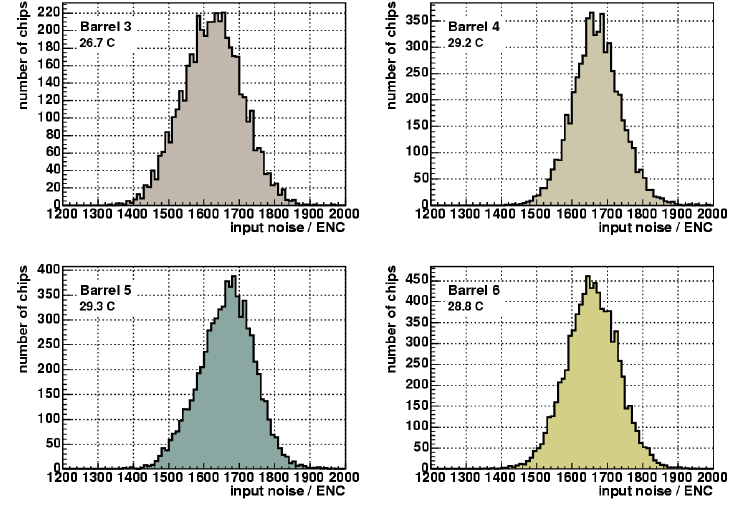}
  \caption{Average input noise per chip for the four SCT 
  barrels.}\label{fig:2-noise_barrel}
 \end{center}
\end{figure}

Before installation, the SCT barrels were tested and characterized in terms of 
noise, both ENC and NO. Figure~\ref{fig:2-noise_barrel} shows the noise 
distributions for each of the four barrels, tested ``warm'', contrary to the 
usual operation temperature of $-7^\circ{\rm C}$. 

The values shown in the plot are conforming to the requirements, since the 
noise at the operation temperature will be about $150\ENC$ lower, accounting 
for a noise variation of
${\rm dENC/d}T \approx 5~e^{-}{\rm ENC/K}$~\cite{SCTDAQ}.

\begin{figure}[!htbp]
 \begin{center}
  \includegraphics{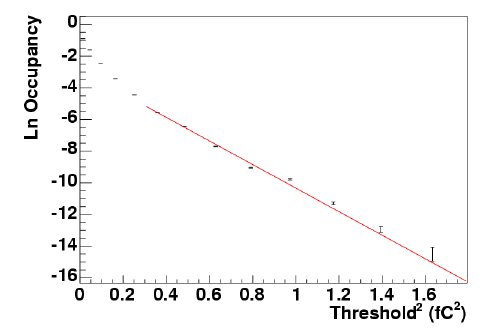}
  \caption{Noise occupancy of one front-end chip as a function of the 
  discriminator threshold.}\label{fig:2-no_sct}
 \end{center}
\end{figure}

Figure~\ref{fig:2-no_sct} represents the natural logarithm of the average noise 
occupancy in one chip, as a function of the threshold squared. This test is 
done by means of threshold scans without charge injection and the linear fit of 
the shown plot allows an estimation of the input noise of the modules.

The full description of the functional testing of the SCT barrels is available 
in~\cite{Phillips:2007db}. The three-point gain test was performed with 
injected charges $1.5$, $2.0$ and $2.5~{\rm fC}$.

The concept of ENC and noise occupancy will be discussed again in chapters 4 
and 7, in the context of the tracker upgrade modules performance.

\subsection{Alignment and Resolution of the Tracker}\label{sec:2-ATLASalignment}

The alignment of the Inner Detector takes place at four levels, using the 
widely extended residual minimization technique mentioned in 
section~\ref{sec:2-alignment}:

\begin{itemize}
 \item Level 0 is the global alignment, in which large structures are aligned 
 with respect to the entire pixel detector.
\item Level 1 is part of the local alignment of the sub-detectors. The pixel 
barrels were split into upper and lower halves plus the two endcaps, while the 
SCT was split into the four barrel layers plus the two endcaps.
\item Level 2 is the next step, in which the pixel and the SCT are aligned 
stave-by-stave. The SCT ``staves'' are not actual staves, but groups of 12 
modules in a row, which are considered rigid bodies.
\item Last, level 3 is the alignment at the module level. Although the pixel 
modules are assembled in staves and these were surveyed after the assembly took 
place, the deformation of the staves is expected to be larger than the survey 
errors. 
\end{itemize}

Each level introduces additional degrees of freedom, but not all alignment 
parameters are used at every step. Also, the expected sizes of the corrections 
at each stage are smaller~\cite{IDCommissioning}.

The goal is that the pixel modules must be aligned with a precision of 
$7\mum$ and the SCT modules with a precision of $12\mum$ in the 
$R\phi$ direction. Both pixel and SCT barrel modules require an alignment 
precision of several tens of micrometres in the $z$ direction, while the endcap 
modules require this precision in the $R$ direction.

\begin{table}[!htb]
 \begin{center}
  \begin{tabular}{l c}
  \toprule
   Parameter & Asymptotic resolution from cosmic ray data 2008 \\ \hline\hline
   $d_0$ [$\upmu {\rm m}$] & $22.1\pm 0.9$ \\
   $z_0$ [$\upmu {\rm m}$] & $112\pm 4$ \\
   $\phi_0$ [mrad] & $0.147\pm 0.006$ \\
   $\theta$ [mrad] & $0.88\pm 0.03$ \\
   $q/p$ [$({\rm GeV}/c)^{-1}$] & $\left(4.83\pm 0.16\right) \times 10^{-4}$\\
   \bottomrule
  \end{tabular}
  \caption{Track parameter resolution for tracks with $\pt > 30~{\rm GeV}/c$ from 
  cosmic ray data.}\label{tab:2-resolution}
 \end{center}
\end{table}

In addition to the minimization of residuals using tracks in the detector, a 
Frequency Scanning Interferometry system is installed in the SCT to monitor 
variations in the shape of the detector. Refer to~\cite{FSIsystem} for further 
details on the FSI system.

The resolution of the track parameters in the Inner Detector were predicted 
using cosmic ray data. Since particles originating from cosmic ray showers 
mostly traverse the detector from top to bottom, it is only possible to assess 
the resolution for the barrel layers. The resolution of the parameters 
excluding multiple scattering can be measured by taking tracks with 
$\pt > 30~{\rm GeV}/c$. These are shown in Table~\ref{tab:2-resolution}, taken 
from~\cite{IDCommissioning}.

The momentum resolution is key in physic analyses. For the Inner Detector, the 
$q/p$ resolution is flat in the pseudorapidity range $|\eta| < 1$, although it 
degrades at high transverse momentum, $\pt \gtrsim 
10~{\rm GeV}/c$~\cite{IDCommissioning}.

\chapter{Upgrades at the LHC}

\section{Motivation}

The Large Hadron Collider operations began in 2010,
with an outstanding performance until early 2013. First, during
2010 and 2011 at a centre of mass energy of $\sqrt{s} = 7\TeV$ and 
after, in 2012, at $\sqrt{s} = 8\TeV$. During this period, it has
delivered a total integrated
luminosity close to $26~{\rm fb}^{-1}$, allowing the two general
purpose experiments, ATLAS and CMS, to record an integrated luminosity
of around $25~{\rm fb}^{-1}$~\cite{Hubaut:1546370}.

Two planned shut-downs will be used to increase the performance of the machine
and to reach its design luminosity and collision energy of $\sqrt{s} = 14\TeV$.
In addition, the four experiments will perform consolidation work to enhance 
their performance and to fix possible problems appeared during operation.
These are called the first long shut-down, LS1,
taking place during 2013 and 2014, and the second long shut-down, LS2, which 
will take place during 2018. The two are part of the Phase-0 Upgrade and 
Phase-I Upgrade, respectively.  

After these two shut-downs, the next step involves a full replacement of the 
accelerator and major updates to the experiments. 

This chapter briefly describes LS1 and LS2, and gives a longer overview of
the major upgrade of the ATLAS experiment at the LHC.

\subsection{Long Shut-down 1}

After the first long shut-down, LS1, from 2013 to 2014, the centre of mass
energy will be increased to $\sqrt{s} = 13$-$14\TeV$, the number of bunches per
beam will increase from 1380 
to 2808 and the bunch spacing will be reduced from $50$ to
$25~{\rm ns}$. With these modifications, the instantaneous luminosity will 
reach $10^{34}~{\rm cm^{-2}s^{-1}}$ from 2015. The expected integrated 
luminosity delivered until the next shut-down will be around 
$100\ifb$~\cite{Zhu:1500464}.

In order to achieve this
goal, consolidation works are taking place during LS1 on the splices at the 
magnets
interconnections~\cite{Rossi:1471000}.

Some experiments will also perform upgrade works of some of their sub-systems 
during LS1.

In the case of ATLAS~\cite{Zhu:1500464}, a new pixel layer, called 
Insertable B-Layer (IBL), is 
going to be installed in the pixel detector. It will be placed between the
existing first pixel layer and the beam pipe, at a radius of $33\mm$. As a 
consequence, the beam pipe is also replaced with a smaller radius one: from
$39\mm$ down to $23.5\mm$~\cite{Flick:1552408}. New front-end electronics, 
optical transceivers and read-out systems have also been developed for the IBL.

In addition to the new layer, the pixel detector will deploy new service 
quarter panels. The initial motivation for it was to move the optical 
transmitters away from the inaccessible parts of the pixel detector and it
finally received some extra improvements for the whole 
sub-system~\cite{Welch:1490708}. These new service panels have been designed
to take services out of the pixel volume, thus reducing material.

The SCT will see a rearrangement of the read-out drivers with an increase
from $90$ to $128$ RODs to cope with the pile-up increase after the shut-down.
The expansion could allow a $\langle\mu\rangle$ up to $80$ at
a level 1 trigger rate of $100~{\rm kHz}$, compared to the SCT design target
of $\langle\mu\rangle = 23$.
In addition, all the off-detector optical transmitters will be replaced with a 
new solution~\cite{LightABLEreview}. These transmitters are used to send the 
Timing, Trigger and 
Control (TTC) signals to the SCT modules and the original devices have 
experienced low reliability~\cite{WeidbergReliability}.

The cooling plant for the pixel and the SCT will be modified to an evaporative
cooling system and the IBL will use CO$_2$ based cooling.

On the calorimeters, all low voltage power supplies will be changed.

\subsection{Long Shut-down 2}

 A second long shut-down will
take place in 2018 to do upgrades to the collimators and injectors, preparing
some systems for the high luminosity LHC~\cite{ICHEPlhc}. After these 
upgrades, the LHC
will pursue the goal of attaining twice its design instantaneous luminosity, 
reaching $2.2\times 10^{34}~{\rm cm^{-2} s^{-1}}$. This will allow the LHC to
deliver between $300$ and $400\ifb$ in its whole operations period,
ending in the early 2020s~\cite{Zhu:1500464}.

To achieve this goal, there will be an injector upgrade, integrating the new 
Linear accelerator, Linac4~\cite{Bartosik:1492996}, into the injector chain, 
also increasing the 
energy of the PS Booster and upgrading the collider 
collimators~\cite{CERN-LHCC-2011-012}.

On the experiments side, the upgrades planned so far for LS2 are the following:

\begin{itemize}
\item ATLAS will upgrade its muon system, with the introduction of a New Small
Wheel (NSW). The Level 1 trigger will also be improved, both in the muon 
spectrometer and the electromagnetic and forward 
calorimeters~\cite{CERN-LHCC-2011-012}. One of the requisites for this upgrade
phase is that they have to be fully compatible with the physics programme of 
the Phase-II upgrades after the third long shut-down.
\item CMS will completely replace its pixel system~\cite{Dominguez:1481838}
 and upgrade the Level 1 
trigger system~\cite{1748-0221-7-01-C01046,Collaboration:1355706}.
\item LHCb will replace the Vertex Locator (VELO) and the silicon strips 
tracker, and upgrade the
read-out system to support a flexible Level-0 trigger. These upgrades will 
cover the operations beyond the third long shut-down~\cite{LHCbUpgrade}.
\item ALICE will install a new silicon tracker around a very small beampipe.
Also, the Time Projection Chamber (TPC) will be upgraded as well as the 
electronics for the other sub-detectors, and a minimum bias trigger will be 
implemented~\cite{ALICEls1ls2}.
\end{itemize}

Depending upon the performance and issues that could arise during operation in 
the 2015-2018 period, other updates may be needed.

\subsection{LHC End of Life}

The expected integrated luminosity delivered by the current Large Hadron 
Collider is around $300$ to $400~{\rm fb}^{-1}$ through the ten years of operation 
that will end in 2022.  By that time, the LHC will be facing the degradation of 
its components due to the accumulated radiation damage~\cite{Rossi:1471000} and 
also parts of the experiments will need replacing.

Figure~\ref{fig:3-lhclumi} shows the expected integrated luminosity of the
LHC until the third long shut-down (LS3), when the LHC replacement will be
deployed. This replacement of the accelerator is described in the following
section.

\begin{figure}[!htbp]
\begin{center}
\includegraphics[scale=0.75]{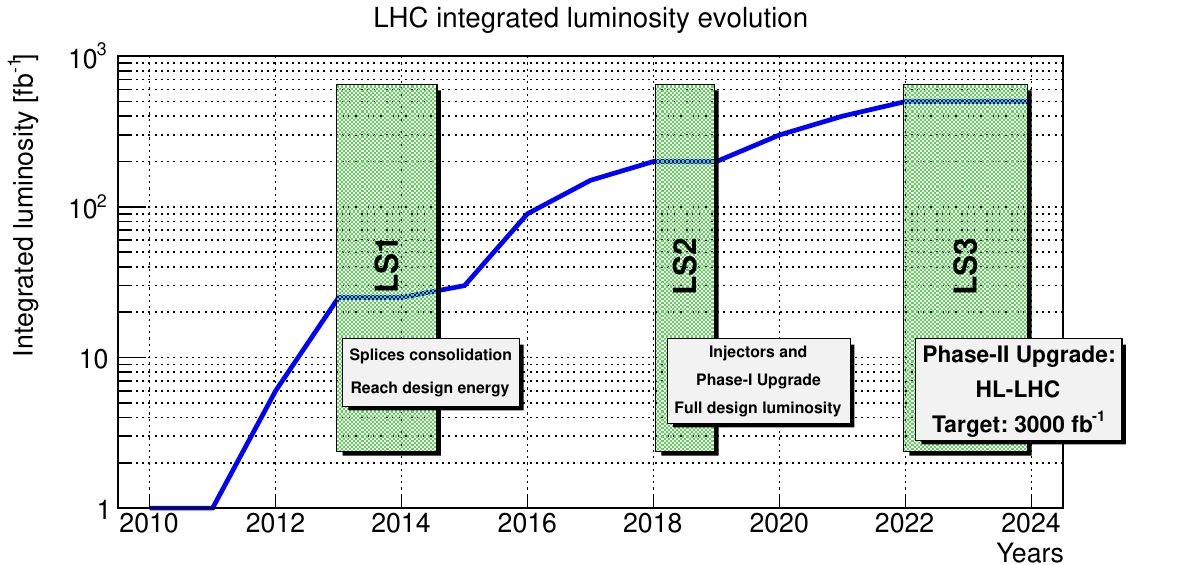}
\caption[Possible integrated luminosity evolution for the LHC with the 
foreseen three long shut-downs 
superimposed.]{Possible integrated luminosity evolution for the LHC with the 
foreseen three long shut-downs 
superimposed~\cite{Rossi:1471000}.}\label{fig:3-lhclumi}
\end{center}
\end{figure}

\section{High Luminosity Large Hadron Collider}

The LHC evolution beyond LS3 is called High Luminosity Large Hadron Collider 
(HL-LHC). The design goal for integrated luminosity over its lifetime is over
$2500\ifb$, up to $3000\ifb$,
with peak luminosity increasing five times with respect
to the LHC design, while maintaining
the $14\TeV$ centre of mass energy. With this goal, the whole LHC+HL-LHC 
programme is expected to deliver over $3000\ifb$.

If the current LHC is not upgraded to a HL-LHC after LS3, maintaining 
the original design
luminosity, the total integrated luminosity that can be expected after 
another ten years of operation would be around 
$1000\ifb$~\cite{PhysicsBriefing}. In any case, the current LHC 
inner triplets have to be replaced after $400\ifb$ integrated
luminosity.

The work for the HL-LHC and the various experiments upgrades has already
begun. Specification and design of the diverse components is well advanced
for some of the systems. Another topic that is being treated are the 
simulations to define the physics cases.

After the discovery of the Standard Model Higgs boson, made by ATLAS and 
CMS in the LHC~\cite{Aad20121}, 
the case study for a High Luminosity Large Hadron Collider 
(HL-LHC) is easier to define and quantify. The Standard Model 
predicts, for a Higgs boson mass around $125\GeV/c^2$, that all the decay 
final states are accessible for exploration with the whole LHC programme,
given a high enough 
integrated luminosity. 

For instance, the detection and measurement of the decays 
$H\rightarrow \mu^+\mu^-$ and $H\rightarrow Z\gamma$ require an integrated 
luminosity of the order of $3000~{\rm fb}^{-1}$, which justifies the need for 
a major upgrade of the accelerator. 

The extra amount of data will also allow better 
precision measurement of the Higgs couplings, with precisions between $5\%$ 
and $30\%$. For instance, the full luminosity should allow studying Higgs 
self-coupling in channels $HH\rightarrow \tau\tau bb$ and 
$HH\rightarrow \gamma \gamma bb$~\cite{ATLAS:1502664}. This factor of ten 
increase in the luminosity 
is beyond the design specifications of the LHC and its experiments.

In addition to the 
characterization of the Standard Model 
Higgs boson decay modes, there are other scenarios of physics Beyond the 
Standard Model (BSM), such as supersymmetric partners of quarks and gluons 
with mass greater than $1.5\TeV/c^2$. 

Another example is the new $Z'$ gauge bosons that would 
imply the existence of new weak interactions, which requires at least
$300~{\rm fb}^{-1}$ but needs the full HL-LHC luminosity if its mass is above 
$2.5\TeV/c^2$. 

Table~\ref{tab:cap2-designgoals} shows a brief comparison between the design 
specifications of the HL-LHC and those of the LHC~\cite{LHC08,Rossi:1471000}.

\begin{table}[!htbp]
\begin{center}
\begin{tabular}{p{8cm}|c|c}
\toprule
Parameter & LHC & HL-LHC \\ \hline
Peak luminosity [${\rm cm}^{-2}{\rm s}^{-1}$] & $10^{34}$ & $5\times 10^{34}$ \\
Maximum integrated yearly luminosity [${\rm fb}^{-1}$] & $40$ & $>250$\\
Beam energy [${\rm TeV}$] & $7$ & $7$ \\
Dipole magnetic field [${\rm T}$] & $8.3$ & $8.3$ \\
Bunches per beam & $2808$ & $2808$ \\
Bunch spacing [${\rm ns}$] & $25$ & $25$ \\
Protons per bunch & $1.15\times 10^{11}$ & $2\times 10^{11}$ \\
Average pile-up & $27$ & $140$ \\
$\beta^{*}$ at the IP [${\rm cm}$] & $55$ & $15$ \\
\bottomrule
\end{tabular}
\caption{Design specifications of the LHC and the HL-LHC.}\label{tab:cap2-designgoals}
\end{center}
\end{table}

In this context of increasing luminosity and pile-up, the upgrade of the 
experiments is mandatory. The current detectors 
elements will have accumulated an amount of radiation that will render them 
unsuitable for further use. 

In particular, the inner tracking systems of
the two general purpose experiments, ATLAS and CMS, will have reached their
end of lifetime. Therefore, they need to be replaced for the future 
HL-LHC and they will also need to meet new requirements:

\begin{itemize}
\item Improved radiation hardness, due to the much higher luminosity and  
fluence.
\item Higher granularity (especially in the tracker) due to the increased 
pile-up and track density.
\item Reduction of material for better $\pt$ resolution, and adequate pattern 
recognition with the higher occupancy.
\item Additional trigger level for a higher rate of at least $500$~kHz.
\end{itemize}

These requirements call for a new design of the trackers in the experiments 
and will also need an upgrade of the electronics for the data acquisition
of the other sub-detectors.

\subsection{Experiments Upgrades in the HL-LHC}

The four experiments in the LHC, ALICE, ATLAS, CMS and LHCb, will need
some upgrades after the LHC end of life. For the first three experiments, these
upgrades are needed in order to cope with the much higher luminosity and 
pile-up that the HL-LHC will deliver. The upgrades in LHCb are aimed at 
triggering at $40~{\rm MHz}$. Also, some of the components, specifically the 
trackers, will have to be replaced due to the radiation damage incurred in 
the first ten years of operation of the LHC.

The two general purpose experiments, ATLAS and CMS, will need similar
upgrades during LS3. These are summarized in the list 
below~\cite{Collaboration:1355706,ATLAS:1502664}:

\begin{itemize}
\item Full replacement of the trackers. The pixel detectors will have 
accumulated their maximum design fluence and the whole tracker will not be 
able to withstand both the fluence in the HL-LHC and the occupancy. In the case
of ATLAS, the SCT and TRT parts will be replaced by an all-silicon tracker.
\item A completely new trigger system is required for the new conditions in
the HL-LHC, with higher particle density in the detectors. An additional
trigger level, called Level-0, is expected to work at a rate of at least
$500~{\rm kHz}$. This is at least a fivefold increase from the current Level-1
trigger rate of $100~{\rm kHz}$.
\item The Trigger and Data Acquisition (TDAQ) systems will need an upgrade for
the new rates in the HL-LHC.
\item Electronics replacements will be needed in various parts of the 
calorimeters and muon spectrometers.
\end{itemize}

The greatest challenge for both experiments is the tracker upgrade. The two 
trackers will be silicon based and have to be designed to cope with a very
high radiation environment and very high pile-up. There will be layout changes 
and a big increase in the number of channels.

The ALICE requirements for the HL-LHC are summarized in the 
following list~\cite{ExpChallengesAllport}:

\begin{itemize}
\item A new inner tracker system to improve the impact parameter resolution,
getting closer to the interaction point, reducing the amount of material and 
the
pixels size. Fast removal and insertion will be needed for modules replacement
during yearly shut-downs.
\item Upgrade of the read-out systems of the TPC, Transition Radiation Detector 
(TRD), Time of Flight detector (TOF), calorimeters and 
muon detectors, to cope with the higher trigger rates.
\end{itemize}

For LHCb, the systems that need an upgrade for the $40~{\rm MHz}$ rate 
are~\cite{LHCbUpgrade}:

\begin{itemize}
\item The VELO and the silicon strips tracker will need a complete replacement.
\item The read-out systems for the tracker straw tubes and the Hybrid Photon
Detectors (HPD) will be
replaced.
\item Also, the present read-out system at a maximum data rate of $1~{\rm MHz}$
will be upgraded to support a flexible Level-0 trigger, reading out all 
detectors at a $40~{\rm MHz}$ rate.
\item In the case of LHCb, the upgrade deployment will take place during LS2 instead 
of LS3.
\end{itemize}

The next section and chapters focus solely on the ATLAS Tracker Upgrade. At
first, I will describe part of the Phase-II Upgrade of the ATLAS Pixel 
Detector, but the main topic of this thesis is the Strips Tracker Upgrade.

   \section{Phase-II Upgrade of the ATLAS Inner Tracker}\label{sec:upgrade_tracker}

\subsection{Tracker Upgrade Overview}

The current ATLAS Inner Detector consists of three layers with three detector 
types: A pixel detector in the innermost layers, a semiconductor tracker (SCT) 
and a transition radiation tracker (TRT) at larger radii. The silicon pixel 
layers provide high-resolution tracking, as does the semiconductor tracker 
built 
with silicon microstrip sensors. After the first long shut-down, a fourth pixel
layer will be installed closer to the interaction point.

The pixel detector was designed to cope with a $1\MeV$ neutron equivalent 
fluence of $10^{15}\cm^{-2}$, which corresponds to an integrated luminosity of
$400\ifb$. The SCT can operate up to a $1\MeV$ neutron equivalent fluence of 
$2\times 10^{14}\cm^{-2}$.

After the ten years of expected lifetime, the 
pixel sub-detector will have accumulated a $1\MeV$ neutron equivalent
fluence of 
around $8\times 10^{14}\cm^{-2}$, while the SCT is expected to be well below 
its design requirement. However, its maximum attainable fluence is much lower 
than the
requirement for the luminosity beyond 2022~\cite{ATLAS,ATLAS:1502664}.

In addition to the radiation damage, the sub-systems were designed to 
accommodate up to $23$ pile-up events. In the first place, the front-end
electronics of both the pixel detector and the SCT cannot cope with the HL-LHC
pile-up without a degradation in their efficiency. In addition, the track 
density 
with that pile-up will also be so high that the TRT would be close to $100\%$ 
occupancy and both the pixel and the SCT would be unable to resolve close-by 
particles.

The bandwidth of the optical links that read-out the data from the tracker 
would also reach saturation in such high pile-up scenario.

As a consequence, after ten years of operation, it will be necessary to 
replace the 
inner tracker with a new one. The innermost layers will be implemented with 
pixel sensors and the outer layers will be all silicon strips. 
All the details can be found
in the Letter of Intent (LoI) for the Phase-II Upgrade of the ATLAS 
Experiment~\cite{ATLAS:1502664}.

The current ATLAS pixel detector has $80$~million channels spread over three 
barrel layers and three discs on each endcap side. The extra layer that was
inserted during LS1 in the pixel detector, called Insertable 
B-Layer (IBL), contributes with an extra $12$~million 
channels~\cite{Flick:1552408}.
The pixel detector after the Phase II upgrade 
will have $638$~million channels, with four 
barrel layers and six discs for each endcap. The sensors for the pixel upgrade 
will be smaller than the sensors in the current detector.

Also, the current ATLAS SCT has $6.3$~million channels with four barrel layers
and nine discs on each endcap. The TRT has $351000$~channels. The new ATLAS 
Strips Tracker will be all-silicon, with $74$~million channels spread over
five barrel layers plus a stub layer and seven endcap discs on each side.

The barrel strips will have different lengths depending on the occupancy of
the region in which they are. The shorter strips will be used in areas with
higher track density.

\begin{figure}[!ht]
\begin{center}
\includegraphics[scale=0.7]{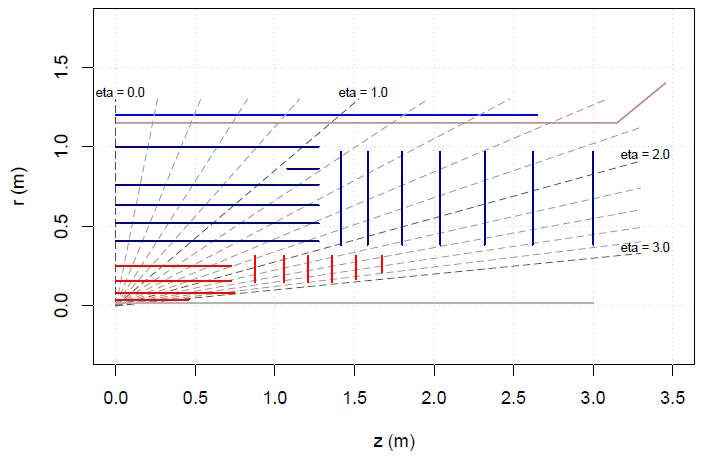}
\caption[Baseline layout of the Inner Tracker
for the ATLAS Phase-II Upgrade.]{Baseline layout of the Inner Tracker
for the ATLAS Phase-II Upgrade~\cite{ATLAS:1502664}.}\label{fig:3-ITk-layout}
\end{center}
\end{figure}

In addition to increasing the capacity of the system, there are other 
requirements to be met. For instance, modularity allows easy replacement of
damaged layers. As an example, it would permit the
removal of the first two pixel layers without removing the beam pipe and also
the whole pixel detector without disturbing the strips tracker. 

The removability 
requirement effectively separates the pixel and strip detectors envelopes,
which needs a Pixel Support Tube (PST) and routing all pixel services within
its own envelope.

The new layout of the Inner Tracker has been defined. There is a baseline
layout, shown in Figure~\ref{fig:3-ITk-layout}, that has four pixel barrel 
layers with six endcap layers on each side. The strip detector has five 
barrel layers and an additional ``stub'' that maintains hermeticity at the
barrel-forward boundary. Finally, seven strips discs compose the endcaps
on each side. 

The beam pipe is assumed to be
$33\mm$ radius and the innermost layer of the pixel detector will be at
a radius of $39\mm$~\cite{ATLAS:1502664}.

This layout was conceived following the requirement of having at least
$14$ hits in the tracker, with a minimum of four hits in the pixel
detector, up to $|\eta| < 2.5$~\cite{Clark:1477118}.

The total coverage of the pixel
detector is close to $|\eta| < 3.0$ and the whole inner tracker covers up to
$|\eta| < 2.5$. The last pixel barrel layer will be placed at a radius of 
$250\mm$ and the maximum length in $z$ of each barrel is $747\mm$. The pixel
barrel will be composed of staves, holding between $22$ and $36$ modules.

The pixel endcaps will cover radial distances between $150.1\mm$ and $315\mm$
with $z$ positions between $877\mm$ and $1675\mm$.

The strips barrel will be surrounding the pixel barrel, with a maximum radius
of around one metre and length of the active area up to $1.25\m$. The endcap
regions start around $z = 1.4\m$ and end at around $3\m$, covering
radii ranges between $405\mm$ and $1000\mm$. Further details on the strip 
detector can be checked in the following section and in the next chapter.

Other alternative layouts have been defined. They are briefly described below,
refer to~\cite{ATLAS:1502664} for more detailed information.

\begin{itemize}
\item  The conical layout is based on a bent stave in the pixel layers, 
reducing the material in the forward region.
\item Five pixel barrel layers, what allows a more robust pattern
recognition and better two-particle separation in high $\pt$ jets.
\item In the alpine layout, pixel modules at high $\eta$ are placed with 
increasing inclination angle. This layout removes the need for endcap discs 
and reduces the total silicon area in the pixel from $8.2\m^2$ to $4.6\m^2$.
\end{itemize}

Although these are the proposals included in the Letter of Intent, the layout
is not complete yet. For instance, the ``stub'' in the strips detector lacks of
modularity.

The next section is focused on the strip detector and further details about
the pixel detector and the other sub-systems upgrades can be found in the 
Phase-II Letter of Intent~\cite{ATLAS:1502664}.

\subsection{The Strip Detector for the Upgrade}\label{sec:3-stripsupgrade}

The strip detector for the ATLAS Phase-II Upgrade will have one barrel region 
covering a radius of $1\m$ and a length of $1.3\m$ to each side of the 
interaction point. The endcaps cover the length up to $3\m$. 
The pseudorapidity range coverage of the strip detector is $|\eta| < 2.5$.

For the barrel, a layout with six layers that include 
five full length cylinders surrounding the pixel
detector along the beam line is conceived. The fifth barrel layer will be a 
short ``stub'' that covers the 
acceptance loss between the endcap and the barrel on each side, as shown in
Figure~\ref{fig:3-ITk-layout}.

\begin{figure}[!htbp]
\begin{center}
\includegraphics[scale=0.19]{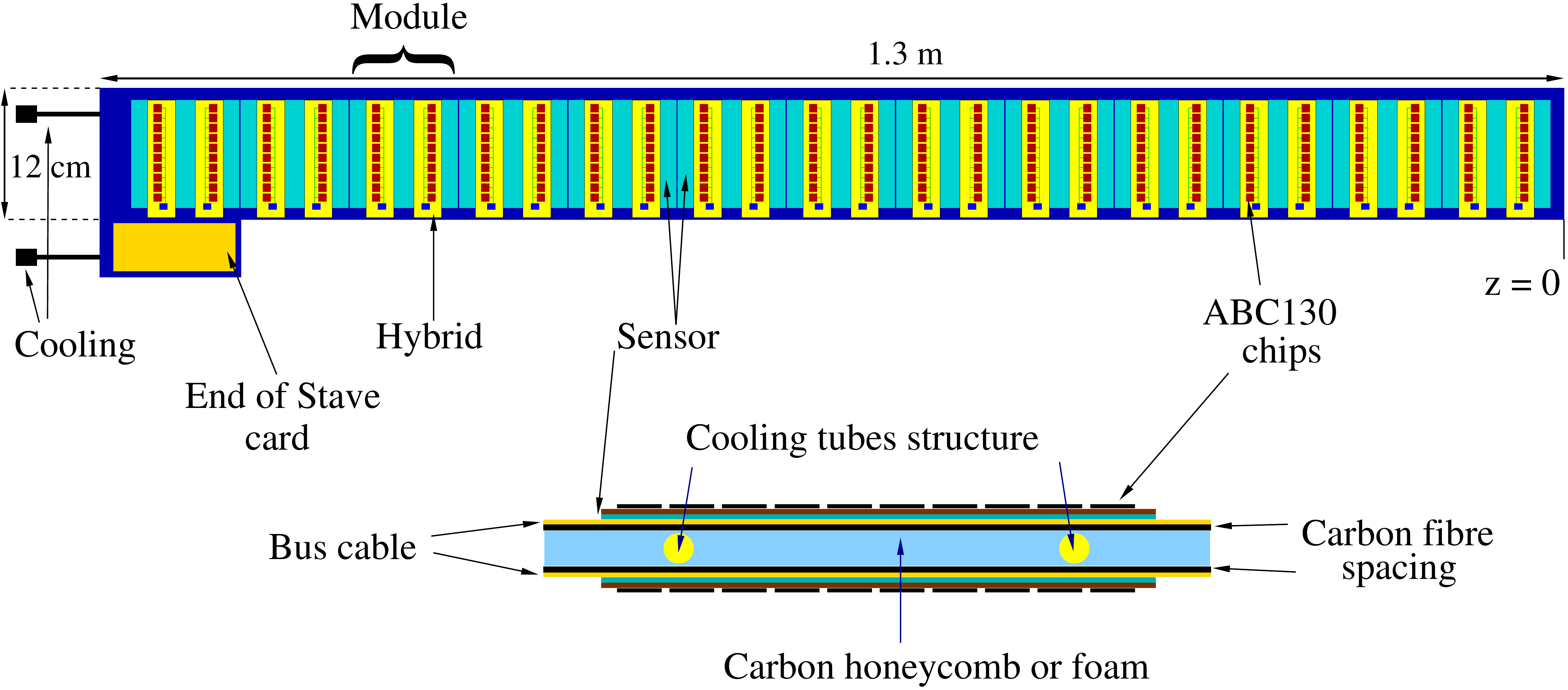}
\caption[Drawing of a barrel stave showing the 13 modules and the End of Stave 
card.]{Drawing of a barrel stave showing the 13 modules and the End of Stave 
card. The drawing does not reflect the actual device 
arrangement.}\label{fig:3-Stave}
\end{center}
\end{figure}

The ``stave'' is the basic mechanical element of the barrel. A drawing is shown 
in Figure~\ref{fig:3-Stave}, it consists of a 
stave core that provides support for the detectors,
powering, read-out electronics and cooling services. 

The detectors are modules made from a silicon microstrip sensor, with two
hybrids holding the read-out chips glued on top of the sensor, and wire-bonded
to the strips. The sensors and the electronics will be described in the next 
chapter.

Each stave consists of two sides for double
dimension position resolution and each side holds 13 modules, for a total of
26 modules in a double sided stave. The End of Stave (EoS) card holds the 
interface to the stave modules, including the read-out electronics and the 
optical transceivers. The total length of a stave including the End of Stave 
card is around $1.3\m$ and its width is around $12\cm$.

The short stub in the fifth layer is expected to have two modules on each 
side of the core, for a total of four modules per stub.

The barrel will be composed of $472$ full length staves, with $236$ 
staves to each side of $z = 0$. The fifth layer will hold $64$ stubs on each
side of $z = 0$.
The first three layers have modules with $23.82\mm$
long strips (``short strips'') and the three outer layers (the two outer 
cylinders and the stub layer) have modules with
$47.755\mm$ long strips (``long strips''). 

\begin{figure}[!ht]
\begin{center}
\includegraphics[scale=0.6]{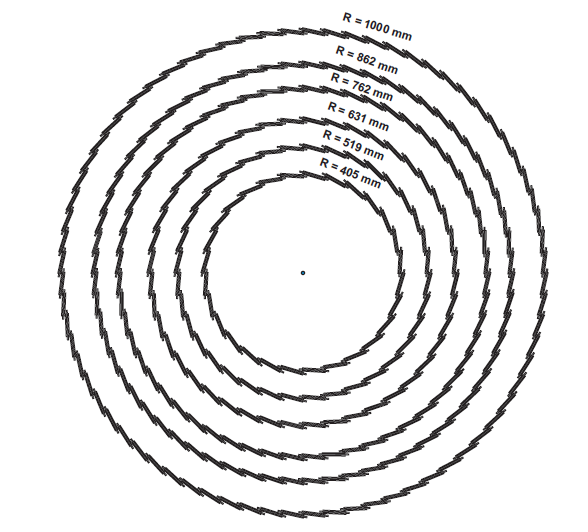}
\caption{Drawing of the arrangement of the staves in barrels. The tilt
angle is $10^{\circ}$.}\label{fig:3-Barrel}
\end{center}
\end{figure}

The total number of modules is $12272$ for full length stave layers while the
stub layer will have $512$ modules. That gives a total of
$12784$ modules in the whole barrel.

Figure~\ref{fig:3-Barrel} shows
the arrangement of the staves in the six barrel layers in the axial plane, 
including the tilt angle, which is specified to be at least $10$~degrees.

The barrel staves are arranged 
in a way that each layer is completely hermetic 
for $1\GeV/c$~tracks.  The rotation direction is chosen to minimize the charge
spread among strips due to the Lorentz angle. 

The total number of 
staves in one cylinder has been chosen to be a multiple of four. This allows 
simplifying the services 
routing and the design of structure supports.

Table~\ref{tab:cap2-barrelstaves} shows the number of staves that compose each
barrel layer of the strip detector in the Phase-II 
Upgrade~\cite{ATLAS:1502664,Clark:1477118}, the radii where
they are located and the active silicon length. The layers numbering on this 
table do not represent the actual numbering of the tracker layers.

\begin{table}[!htbp]
\begin{center}
\begin{tabular}{l|c|c|c}
\toprule
Layer & Radius & Number of staves & Active length ($z>0$) \\ \hline
1 & $405\mm$ & $28$ & $1275\mm$ \\
2 & $519\mm$ & $36$ & $1275\mm$  \\
3 & $631\mm$ & $44$ & $1275\mm$  \\
4 & $762\mm$ & $56$ & $1275\mm$  \\
5 (stub) & $862\mm$ & $64$ stubs & $196\mm$  \\
6 & $1000\mm$ & $72$ & $1275\mm$  \\
\bottomrule
\end{tabular}
\caption{Number of staves or stubs in each barrel layer of the Strips Tracker
for the Upgrade.}\label{tab:cap2-barrelstaves}
\end{center}
\end{table}

On the endcaps, the basic element is called ``petal'', analogous to the stave,
with a wedge shape that allows a disc to be built with full cover and 
reduced overlap. The current geometry of the petal foresees a structure 
$63\cm$ high, with a minimum width of around $10\cm$ and a maximum width
of around $30\cm$. A drawing showing the geometry and elements of one side of 
the petal can be seen in Figure~\ref{fig:3-Petal}.

\begin{figure}[!htbp]
\begin{center}
\includegraphics[scale=0.18]{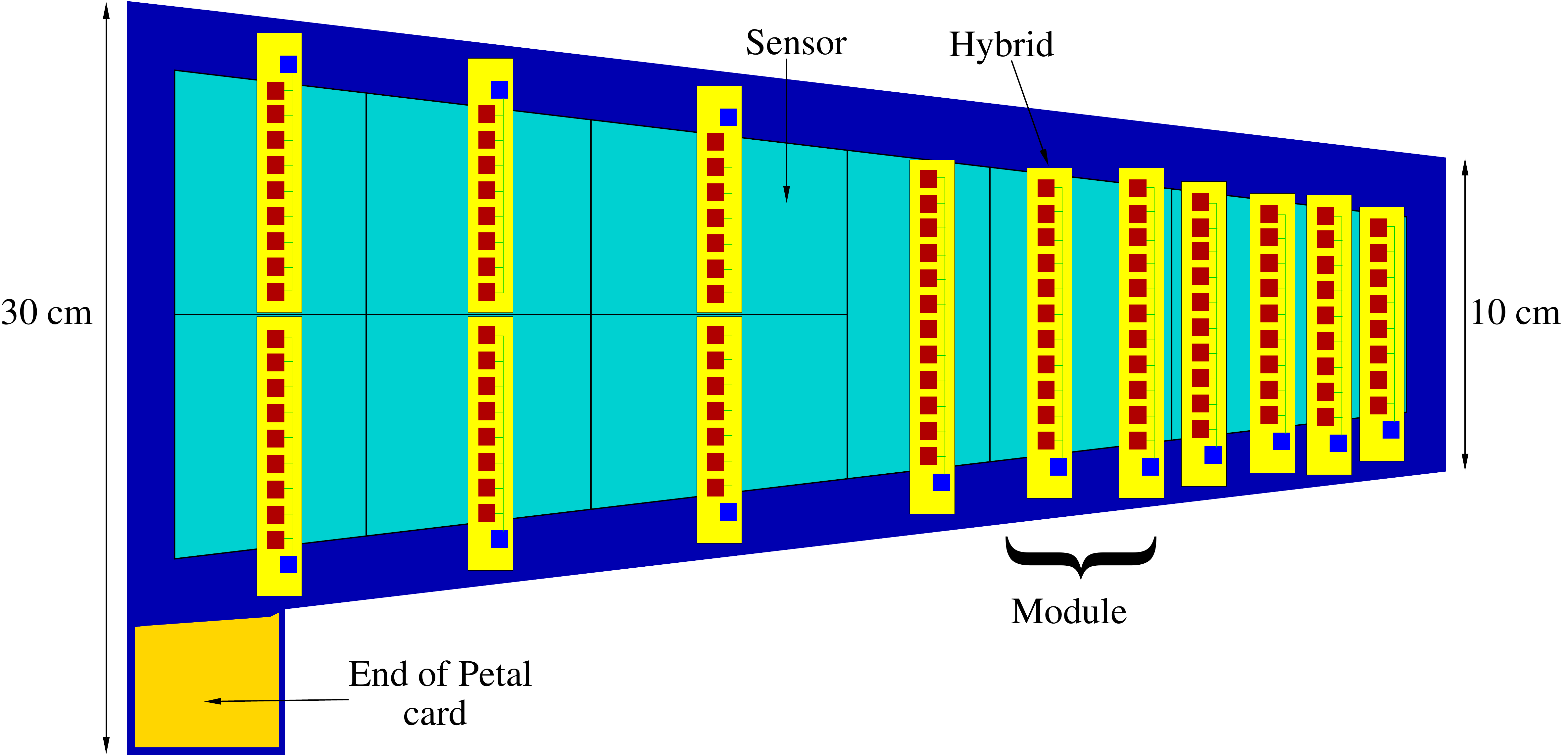}
\caption[Drawing of an endcap petal showing the 9 modules and the 
services.]{Drawing of an endcap petal showing the 9 modules and the 
services. The drawing does not reflect the actual device 
arrangement.}\label{fig:3-Petal}
\end{center}
\end{figure}

Each endcap disc is made from $32$ identical petals. The petals have
nine modules on each side, covering six rings, requiring a total of six
sensor types, with different geometries, and 14 hybrid variations. 

Since every petal has $18$ modules and there are seven discs on each endcap 
side, every disc has $576$ modules and the total module count for the endcaps 
is $8064$. The number is comparable to the barrel, with the added complexity
of having different types of modules. Both the high number of total modules
for the barrel and endcap structures and the variety of modules 
affect the production and quality
assurance planning~\cite{Blue:ModuleProduction}.

\begin{figure}[!htbp]
\begin{center}
\includegraphics[scale=0.5]{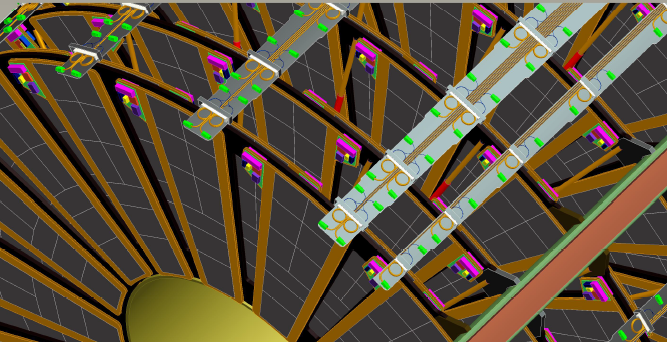}
\caption{The arrangement of petals in the 
endcaps.}\label{fig:3-Endcaplayouts}
\end{center}
\end{figure}

The petals are arranged in endcap discs that are hermetic for $1\GeV/c$,
with an overlap such that a petal covers an angle slightly greater than
$\pi/16$. The layout for the discs is shown in 
Figure~\ref{fig:3-Endcaplayouts}, with a ``castellated'' arrangement.

Table~\ref{tab:cap2-endcappetals} summarizes the 
geometrical characteristics of the 
endcap discs. The greater inner radius of the last disc, implemented by
removing the bottom module, is motivated by the integrated non-ionizing
radiation~\cite{Clark201397}.

\begin{table}[!htbp]
\begin{center}
\begin{tabular}{l|c|c|c}
\toprule
Disc & Inner radius & Outer radius & $z$ position \\ \hline
1 & $385\mm$ & $970$ & $1415\mm$ \\
2 & $385\mm$ & $970$ & $1582\mm$  \\
3 & $385\mm$ & $970$ & $1800\mm$  \\
4 & $385\mm$ & $970$ & $2040\mm$  \\
5 & $385\mm$ & $970$ & $2320\mm$  \\
6 & $385\mm$ & $970$ & $2620\mm$  \\
7 & $471\mm$ & $970$ & $3000\mm$  \\
\bottomrule
\end{tabular}
\caption[Active radii and $z$ positions of the endcap 
discs.]{Active radii and $z$ positions of the endcap 
discs~\cite{Clark201397}.}\label{tab:cap2-endcappetals}
\end{center}
\end{table}

Table~\ref{tab:3-Radiationlength} shows estimate values for the radiation
length of the various elements and the total sum of both a short-strips stave
and a petal. The numbers are based on the current prototyping with 
extrapolation to the use of lower mass components. 

For comparison, the current
SCT radiation lengths without the mechanical supports are $2.48\% X_0$ for the
barrel and $3.28\% X_0$ for the endcaps. This material reduction is one of the 
goals for the Phase II, in order to improve the resolution and performance of
the tracking, and is obtained thanks to the greater degree of 
mechanical support, power and services sharing in these new structures.

Although there is yet no material budget specified, the preliminary performance
simulation studies were done under the following assumptions for the detector 
materials 
and support structures:

\begin{itemize}
\item The inner pixel layers have a radiation length $< 1.5\%~X_0$ per layer.
\item The outer pixel layers have a radiation length $< 2.0\%~X_0$ per layer.
\item The inner short strips layers have a radiation length $< 2.5\%~X_0$ per
layer.
\item The outer long strips layers have a radiation length $< 2.0\%~X_0$ per
layer.
\end{itemize}

\begin{table}[!htbp]
\begin{center}
\begin{tabular}{c|c|c|c}
\toprule
\multicolumn{2}{c|}{Barrel} & \multicolumn{2}{|c}{Endcap} \\ \hline
Element & $\% X_0$ & Element & $\% X_0$ \\ \hline
Stave core & $0.55$ & Petal core & $0.47$ \\
Bus cable & $0.30$ & Bus cable  & $0.03$ \\
Modules & $1.07$ & Modules & $1.04$ \\
Module adhesive & $0.06$ & Module adhesive & $0.06$ \\ \hline
Total & $1.98$ & Total & $1.60$ \\
\bottomrule
\end{tabular}
\caption[Radiation length estimates for barrel stave and endcap petal. 
Power ASICs and EoS are not 
included.]{Radiation length estimates for barrel stave and endcap petal. 
Power ASICs and EoS are not 
included~\cite{ATLAS:1502664}.}\label{tab:3-Radiationlength}
\end{center}
\end{table}

In both the stave and petal cases, the basic unit is a module. For the barrel
staves, all modules in a stave are of one type, whereas for the endcaps, 
different rings
of the petal need different modules. Therefore, the petal construction and
testing involve many different kinds of sub-structures.

The ATLAS Strips Upgrade project comprises the design, construction and testing
of the silicon modules that will be used at the new detector. Single modules 
are mounted into the structures, the staves and petals. Therefore, not only 
individual modules have to be tested, but also the larger structures that 
include multiple modules.

\begin{figure}[!htb]
\begin{center}
\includegraphics[scale=0.7]{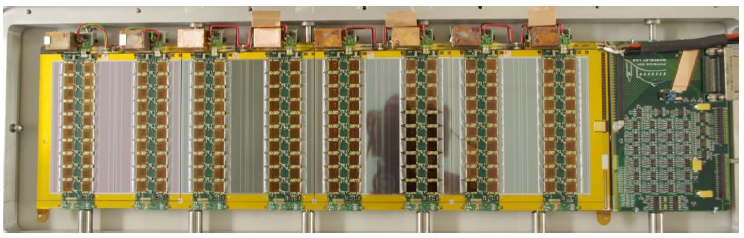}
\caption{Picture of a functional stavelet showing the 4 modules and the 
End of Stave card. }\label{fig:3-Stavelet}
\end{center}
\end{figure}

While the complete barrel staves and endcap petals will integrate 26 and 
18 modules, respectively, smaller structures have been prototyped as proof of
concept. In the case of the barrel, this structure is named ``stavelet'' and
is based on four single sided modules. 

\begin{figure}[!htb]
\begin{center}
\includegraphics[scale=0.4,angle=90]{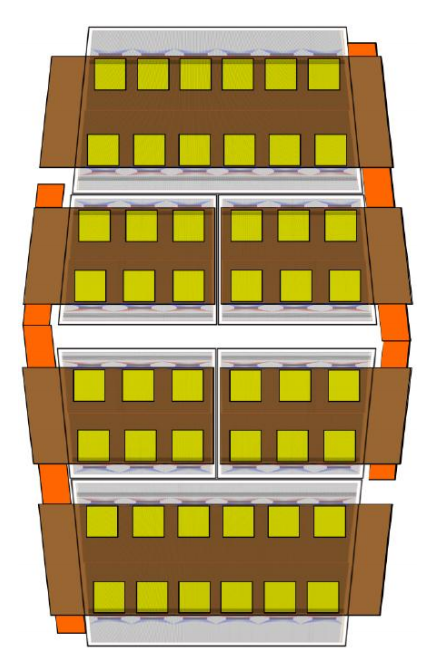}
\caption{Drawing of a petalet. The modules are unfolded to show both sides
of the petalet, the hybrids and services routing.}\label{fig:3-Petalet}
\end{center}
\end{figure}

For the endcap structures, the proof of concept structure is called the
``petalet''. It is based on the first and fifth rings, with three modules
on each side, for a total of six modules. 

Figure~\ref{fig:3-Stavelet} shows one of the actual stavelets that
was built at Rutherford Appleton Laboratory (RAL) and is currently at CERN,
undergoing tests. Some of the tests done on this and other stavelet
will be discussed in the following chapters of this thesis. 

One of the proposed petalet layouts is shown in
Figure~\ref{fig:3-Petalet}~\cite{LacastaPetalet}. The drawing shows an 
unfolded petalet, where
the six modules can be seen, as well as the services routing on the sides of
the modules. The first petalet prototypes are expected to be built during 
the first quarter of 2014.

These proofs of concept are aimed at providing insight with respect to the
feasibility of the design as well as to check the performance evolution as more
modules are added to the structure and the differences between powering 
schemes. This topic will be discussed in the next chapter.

The staves and petals concepts for barrel and endcaps of the strips tracker
upgrade are considered baseline. They are not, however, the only project that
has been working in the conception of the new strips tracker for ATLAS.

As an alternative to the stave concept, the so called ``supermodules'' have 
been considered. These supermodules are an evolution of the current
SCT design, trying to maintain some of its merits, such as mechanical and 
thermal stability, and true stereo space-point reconstruction. For further 
details, refer to~\cite{Clark201397}.

The next chapter discusses in detail the strips tracker upgrade, focused 
on the staves and petals structures and their constituents. The description
includes an overview
of the sensors, electronics, powering and the various testing structures for
both the barrel and the endcap.

\chapter{The ATLAS Silicon Strips Tracker Upgrade}

The ATLAS Phase-II Upgrade has been presented in the previous chapter together
with the Inner Tracker and some details of the strip detector. In this chapter
we will focus on the strips detector in greater detail, from the 
electronics and system points of view. 

\section{Silicon Sensors}

The proposed baseline sensors for the new strips detector are n-type implants
in p-type float-zone silicon substrate. The sensors have a
thickness of $320\pm 15\mum$, which could be reduced with additional 
costs~\cite{ATLAS:1502664}.

\begin{figure}[!ht]
 \begin{center}
  \subfigure[P-in-n detector]{
    \includegraphics[scale=0.5]{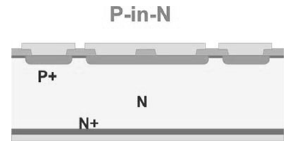}\label{fig:4-p_in_n}
  }
  \subfigure[N-in-n detector]{
    \includegraphics[scale=0.5]{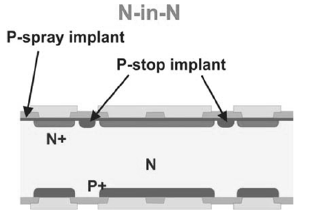}\label{fig:4-n_in_n}
  }
  \subfigure[N-in-p detector]{
    \includegraphics[scale=0.5]{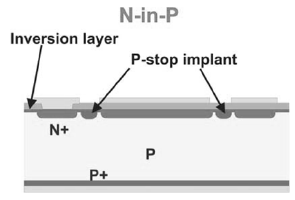}\label{fig:4-n_in_p}
  }
  \caption[Cross sections of the p-in-n, n-in-n and n-in-p detectors used in
  the studies on their radiation hardness.]{Cross sections of the p-in-n, 
  n-in-n and n-in-p detectors used in
  radiation hardness studies. The n-in-n detector shows P-spray 
  and P-stop implants and back side processing. The n-in-p detector shows 
  P-stop implants~\cite{1546445}.}\label{fig:4-cross_sections}
 \end{center}
\end{figure}

The choice of n-in-p technology was made after studying the 
radiation hardness and complexity of different types of 
combinations~\cite{1546445}:

\begin{itemize}
 \item P-type implants in n-type substrate. It is the usual choice for 
 radiation detectors in pixel and microstrips sensors because it
 is a simple technology due to the lower number of masks required for their
 processing, and therefore has lower cost. The fluence limit of this 
 technology is about $3\times 10^{14}~{\rm protons/cm^2}$, which is an order 
 of magnitude lower than the expected fluence on the strips tracker in the 
 HL-LHC. An example of this technology is shown in Figure~\ref{fig:4-p_in_n}.
 \item N-type implants in n-type substrate. These detectors are expected to
 withstand fluences of up to $10^{15}~{\rm protons/cm^2}$. However, this 
 technology is the most complex of the three: it requires ten mask levels and
 double sided processing. A cross section showing this
 type of technology with P-stop and P-spray implants can be seen in 
 Figure~\ref{fig:4-n_in_n}
 \item N-type implants in p-type substrate. This technology is more radiation
 hard, due to the fact that the bulk is already p-type, so no type 
 inversion occurs. On the negative side, these detectors are more complex
 than their p-type implants in n-type substrate counterpart.
 They need extra insulation, which achieved by different techniques:
 blank surface implants (named P-spray) or P-type junctions (named P-stop).
 A sample of the n-in-p technology is shown in Figure~\ref{fig:4-n_in_p}, where
 P-stop implants are included.
 
\end{itemize}

The fluence in the inner layers of the strips detector in the HL-LHC
is expected to be about $1.2\times 10^{15}$ $1\MeV$ neutron-equivalent for the
full operation period. For the outer layers the expectation is around 
$5\times 10^{14}$ $1\MeV~{\rm n_{eq}/cm^2}$. These numbers include a safety
factor of 2~\cite{Miyagawa:1516824}.

The choice of n-type implants in p-type substrate was made after considering
the fact that the n-type bulk becomes p-type after a $1\MeV$ neutron-equivalent
fluence of $10^{13}~{\rm cm^{-2}}$. This means that the SCT approach for the 
sensors is not suitable for the HL-LHC upgrade. Even considering the higher
radiation hardness of the n-in-n technology, it is a double sided process,
more complex than n-in-p and there are less foundries available. These 
difficulties lead to higher production costs and the n-in-n technology is
harder to handle and test~\cite{Affolder:1381518}.

\subsection{Barrel Sensors}

The barrel sensors have an area of $97.54\times 97.54\mm^2$ with $1280$ active
strips plus two guard strips
across the sensor, with a strip pitch of $74.5\mum$. The size of the
sensor was chosen so they can be laid out on $6~{\rm inch}$ ($150\mm$)
diameter wafers~\cite{Unno:2011zzb}.

The wafer also includes 24 miniature sensors, $10\times 10\mm^2$, surrounding 
the main sensor, which are used to study radiation hardness properties of the
different isolation techniques. The initial testing of the n-in-p structures
for the ATLAS Upgrade began with the ATLAS07
specification~\cite{ATLAS07}, with results on the radiation 
hardness~\cite{Dervan:2012zz} and charge collection efficiency comparison
of different substrates~\cite{Casse:ChargeCollection}.

There are two sensor
variations, as it was mentioned in the previous chapter, there will be short
strips and long strips, depending on the track density of the regions where the
sensors are located. The short strips have a length of $23.82\mm$, which
results in four rows of strips in the sensor. The long
strips are $47.755\mm$, therefore a sensor accommodates two rows of long 
strips~\cite{ATLAS:1502664}.

Complete details about the sensors specifications and manufacturing process 
can be found in the two specification documents that have been written so far.
These are the technical specifications for the ATLAS07~\cite{ATLAS07} and the 
ATLAS12A/ATLAS12M~\cite{ATLAS12} sensors.

A number of ATLAS07 sensors have been fabricated following these specifications
and have been tested under various conditions. The baseline material is 
float-zone (FZ) p-type silicon.
The read-out is
AC coupled and the biasing of the strips is implemented with poly-silicon 
resistors.

As mentioned above, the strip pitch in these sensors is $74.5\mum$, whereas 
the read-out implant strips are $16\mum$ wide highly doped 
n-type implants and the read-out strips are $22\mum$ aluminium depositions over
the n-implant strips~\cite{ATLAS12}.

Some of the specifications for pre-irradiated sensors in the ATLAS12
specification are summarized in Table~\ref{tab:4-ATLAS12}.

\begin{table}[!ht]
\begin{center}
 \begin{tabular}{l|c}
  \toprule
  Parameter & ATLAS12 specification \\ \hline
  Thickness & $310\pm 25\mum$ \\
  Crystal orientation & $<100>$ \\
  Outer dimension & $97.54\times 97.54\mm^2$ \\
  Strip segments & $4$ \\
  Strip segment length & $23.86\mm$ \\
  Strip pitch & $74.5\mum$ \\
  Strip implant width & $16\mum$ \\
  Strip read-out width & $22\mum$ \\
  Depletion voltage & $< 300~{\rm V}$ \\
  Maximum operating voltage & $600~{\rm V}$ \\
  Leakage current (normalized to $20^{\circ}{\rm C}$) & $< 2~{\rm \upmu A/cm^2}$ at $600~{\rm V}$ \\
  Maximum fluence & $1.2\times 10^{15}$ $1\MeV~{\rm n_{eq}/cm^2}$ \\
  \bottomrule
 \end{tabular}
\caption[Summary of ATLAS12 specifications 
pre-irradiation.]{Summary of ATLAS12 specifications 
pre-irradiation~\cite{ATLAS12}.}\label{tab:4-ATLAS12}
\end{center}
\end{table}

The layout of the wafers for the ATLAS12 specification is shown in 
Figure~\ref{fig:4-ATLAS12wafer}, which corresponds to the axial-stereo sensor,
two strips rows have an axial orientation, that is, parallel to the sensor
edge. The other two rows have a $40~{\rm mrad}$
stereo angle. The various P-stop isolation structures that have been 
implemented for 
testing are also shown, together with the Punch-Through Protection (PTP)
function~\cite{ATLAS12,Hara:1484223}.

\begin{figure}[!ht]
 \begin{center}
  \includegraphics[scale=0.5]{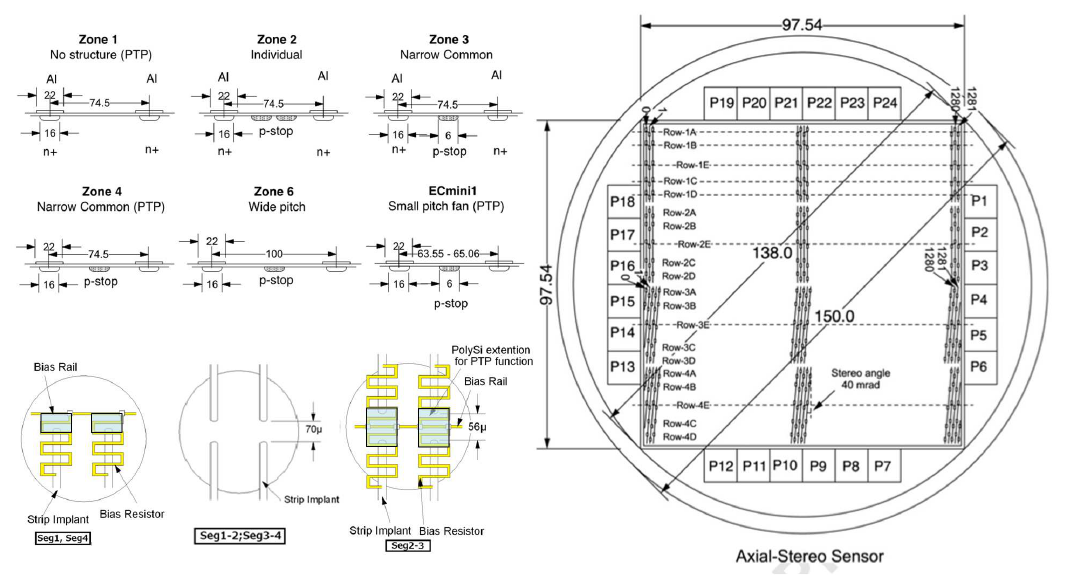}
  \caption[Wafer layout defined in ATLAS12 specifications for the barrel 
  short strips sensor prototypes.]{Wafer layout defined in ATLAS12 specifications for the barrel 
  short strips sensor prototypes~\cite{Hara:1484223}. On the left, top to
  bottom are the P-stop isolation  and the PTP structures. On the right, the
  wafer mask for an ATLAS12M sensor. Measures of the wafer
  drawing are in millimetres, measures of the PTP and isolation schemes are in
  micrometres.
  }\label{fig:4-ATLAS12wafer}
 \end{center}
\end{figure}

ATLAS12 describes different sensor types, being the most relevant the 
ATLAS12A, with all strips axial (A = axial), and ATLAS12M, with two rows of 
strips axial and the other two rows having a $40~{\rm mrad}$ stereo 
angle~\cite{ATLAS12} (M = mixed). Since ATLAS12A has been chosen as the 
baseline,
the stereo angle in the detector will be achieved by rotating the sensors.

This is the overall description of the barrel sensors. The exact mask is yet 
to be defined, including the Punch-Through Protection and
the isolation structures. 

\begin{figure}[!ht]
\begin{center}
\subfigure[Stereo strips on a sensor]{
\includegraphics[scale=0.4]{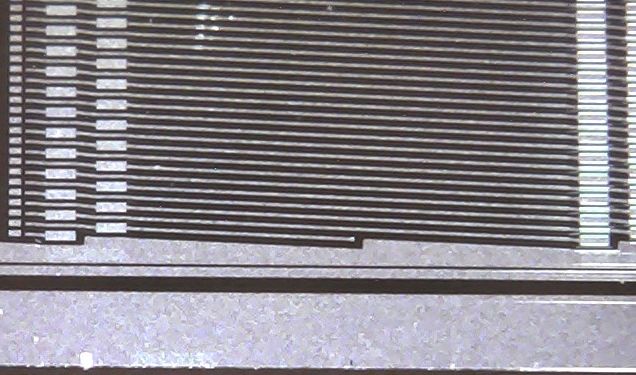}
\label{fig:4-stereo_strips}
} \quad
\subfigure[Axial strips on a sensor]{
\includegraphics[scale=0.4]{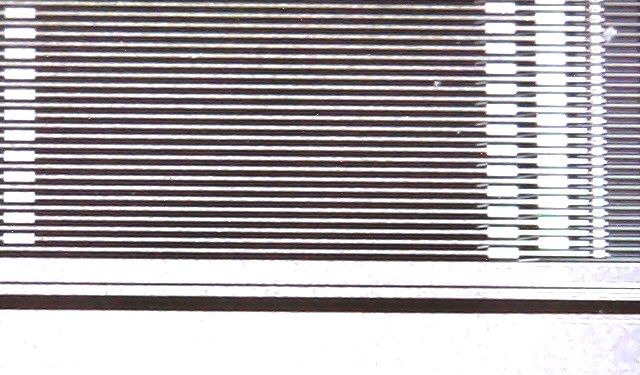}
\label{fig:4-non_stereo_strips}
}
\caption{Sensors on a barrel module: both stereo and axial strips on the
same sensor.}\label{fig:4-barrel_sensors}
\end{center}
\end{figure}

The batches that have been produced so far include higher quality 
(Float-Zone 1, FZ1) or Grade A sensors as well as lower quality (Float-Zone 2,
FZ2) or Grade B sensors. The main difference between them is that FZ1 sensors
have a leakage current around two orders of magnitude lower than FZ2 sensors. 
The isolation structure used on the main sensor is equivalent to the Zone 3
shown in Figure~\ref{fig:4-ATLAS12wafer}. An example of the ATLAS07 sensors
can be seen in Figure~\ref{fig:4-barrel_sensors}, which shows the two types
of strips: axial and stereo. 

More details about the batches of the ATLAS07 sensors can be found 
in~\cite{Unno:2011zzb}. The devices that have been tested at CERN for the
barrel structures, that is, modules and stavelets, all use ATLAS07
sensors. There are some differences between the ATLAS07 and ATLAS12 sensors,
mainly on the PTP structures and the existence of all axial strips sensors.

\subsection{Endcap Sensors}

The endcap sensors that will be used to build the petals need radial strips, 
pointing at 
the beam line, for accurate measurement of the $r\phi$ coordinate. To achieve
this, the sensors have a wedge shape and there need to be several sensor 
shapes~\cite{Hessey:1514636}. The size of the sensors need to be such
that all shapes can be fabricated on 6~inch wafers.

In the endcap petals, the stereo angle of $40~{\rm mrad}$ is achieved by
rotating the strips within the sensors by $20~{\rm mrad}$. Each side of the
petal points slightly away from the beam line, but the angle bisector of
the two sides points at the beam line.

The sensors are, like in the barrel, n-in-p type and the specifications are 
the same, except for the wedge shape and the different geometries.

The petal sensors will be divided in pairs of strips rows, due to the ABC130 
chip, which is designed to read-out two rows of strips. The inner ring
has shorter strips to accommodate the higher hit occupancy. This ring
has sensors with four pairs of rows and hence will hold four read-out hybrids.
The second ring has two pairs of strips and two read-out hybrids. The four
outer rings have one pair of strips and hence one read-out hybrid for each
module.

\begin{figure}[!ht]
 \begin{center}
  \subfigure[Wafer with a big petalet sensor.]{
    \includegraphics[scale=0.52]{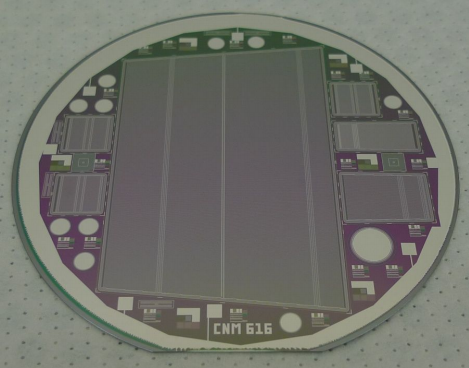}
    \label{fig:4-PetalSensorsCNM_big}
  }
  \subfigure[Wafer with small petalet sensors.]{
    \includegraphics[scale=0.5]{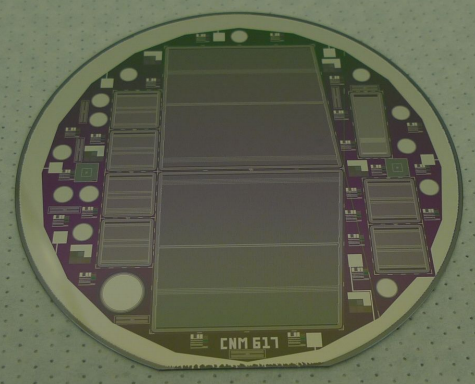}
    \label{fig:4-PetalSensorsCNM_small}
  }
  \caption[Pictures of wafers with endcap sensors produced by 
  CNM-IMB.]{Pictures of wafers with endcap sensors produced by 
  CNM-IMB~\cite{RD50Petalet}.}\label{fig:4-PetalSensorsCNM}
 \end{center}
\end{figure}

The strip pitch at the bond-pad region is made as close to the barrel sensors
strip pitch as possible, that is, $74.5\mum$. This requirement constrains the
number of chips on a hybrid for each of the rings~\cite{ATLAS:1502664}.

Some prototypes of the endcap petal sensors have been produced already. These  
are the three 
types that are needed for the petalet proof-of-concept and are fabricated by
Centro Nacional de Microelectrónica-Instituto de Microelectrónica de Barcelona
(CNM-IMB), a research centre owned by the Spanish National Research Council
(Consejo Superior de Investigaciones Científicas, CSIC). 
Figure~\ref{fig:4-PetalSensorsCNM} shows two wafers of the three
sensor types used in the petalet~\cite{RD50Petalet}.

The small sensors that have been fabricated by CNM-IMB have a depletion voltage 
of $68.2\pm 7.63~{\rm V}$. The big sensors have a depletion voltage of
$70.33\pm 9.45~{\rm V}$. These are the measurements performed by CNM before 
cutting the sensors from the wafers. The measurements have been repeated by the
institutes receiving the sensors after cutting and they are consistent with the
values obtained by CNM~\cite{PetaletModules}.

\begin{figure}[!hbt]
 \begin{center}
  \includegraphics[scale=0.75]{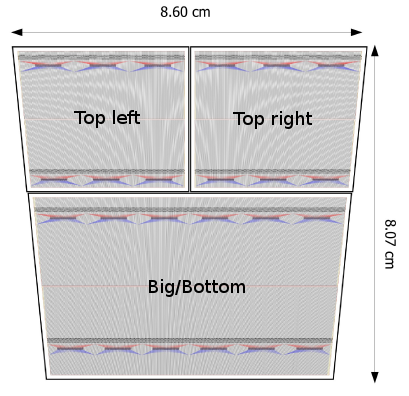}
  \caption{A drawing that shows the three different petalet  sensors, 
  including the size of the 
  petalet.}\label{fig:4-PetaletSensors}
 \end{center}
\end{figure}

The big sensors are also called ``bottom sensors'' and the small sensors are
the ``top right'' and ``top left'' sensors, depending on their position in the
petalet. Figure~\ref{fig:4-PetaletSensors} shows a drawing of the three sensors
of the petalet. All types of sensors in the petalet have one pair of rows. 
The petalet is designed
for the ABCn25 chip and each sensor will be read-out with one hybrid holding
two columns of chips, as in the barrel modules.

The petalet is conceived as the small petal that helps understanding the
whole model and the issues that are specific to the endcap region. This 
is done with a combination of the elements at the 
innermost radius and elements from the region where the petal elements are
split in two separate sensors.

A plan for a full sized petal with the ABCn25 has not been considered. The
ABC130 chip will be used to build the first full sized petal.

\section{Front-End and Read-Out Electronics}

The front-end electronics collect the charge on the strips and compare it with
the configured threshold. The read-out electronics then process all of the hit
strips in the event and 
send the signals outside the detector through the appropriate path, in a series
of stages. This signal
path is represented in Figure~\ref{fig:4-signalspath}.

\begin{figure}[!htb]
 \begin{center}
  \includegraphics[scale=0.55]{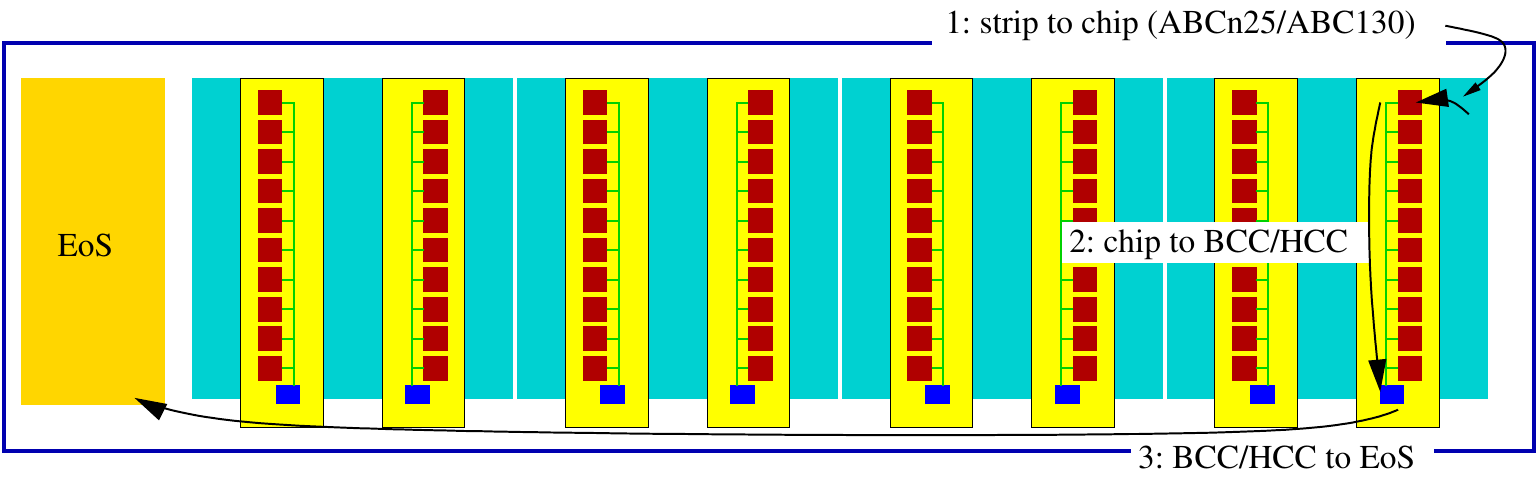}
  \caption{Path of the signals from the strip to the End of Stave 
  card.}\label{fig:4-signalspath}
 \end{center}
\end{figure}

First, the collected charge on the strip is processed at the chip, which has
a link with another chip that aggregates the signals from the front-end chips
and this other chip has, in turn, another link with the read-out electronics.

\subsection{The ABCn25 Chip}

A front-end electronics chip for the strips detector have been prototyped on 
$250~{\rm nm}$ technology, the chip is named ABCn250 or ABCn25, with
128 channels.
It is a binary read-out chip that only provides hit information and no charge
information. A drawing of the block diagram for this chip is shown in
Figure~\ref{fig:4-ABCN25diag}. It is very similar to the present SCT chip as
the functionality required is the same~\cite{Campabadal:994402}.

\begin{figure}[!ht]
 \begin{center}
 \subfigure[Block diagram of the ABCn25~\cite{Dabrowski:1273530}.]{
  \includegraphics[scale=0.4]{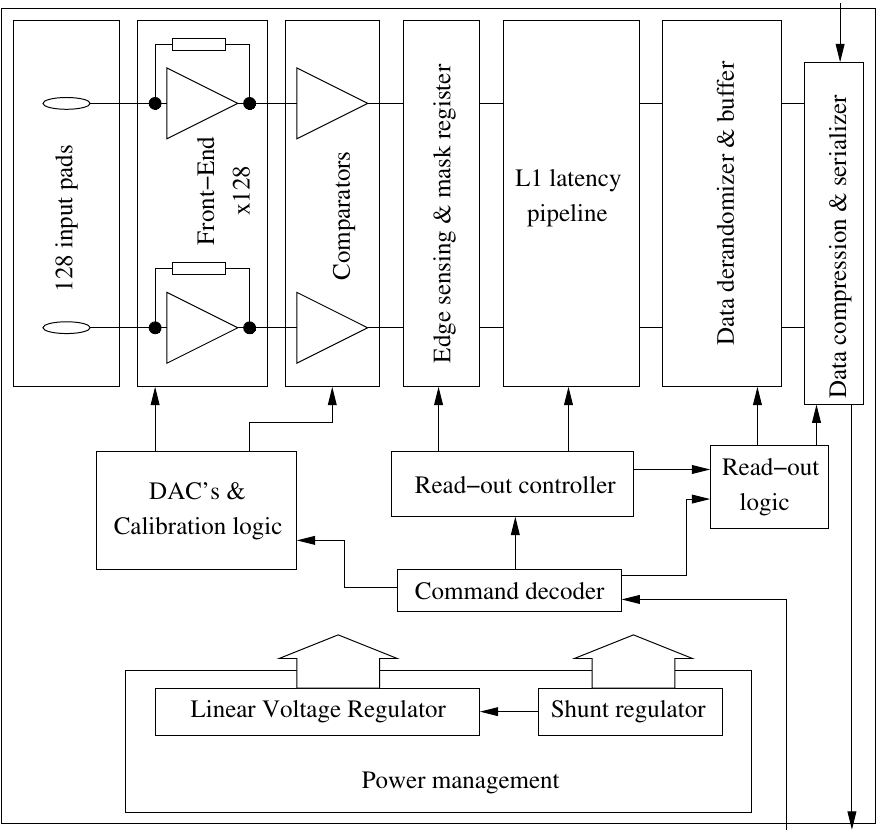}\label{fig:4-ABCN25diag}
  }
  \subfigure[Picture of an ABCn25 chip.]{
  \includegraphics[scale=0.23]{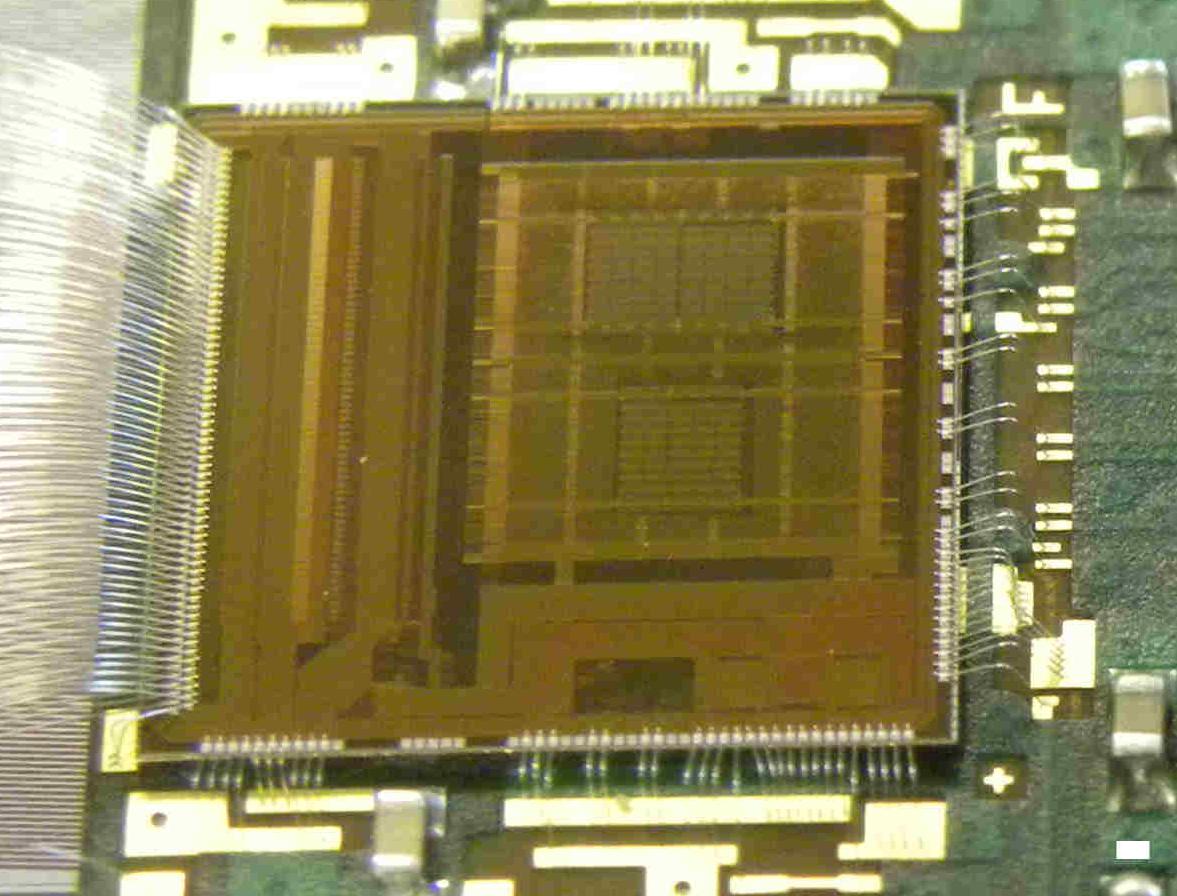}\label{fig:4-ABCN25pic}
  }
  \caption[Functional blocks diagram and picture of an ABCn25 chip.]{Functional 
  blocks diagram and picture of an ABCn25 chip. The 
  picture shows the 128 input pads to the left of the chip, the rest of 
  the bond-pads are for addressing and data reception and transmission.}
 \end{center}
\end{figure}

A picture of one of the chips mounted on a hybrid is shown in 
Figure~\ref{fig:4-ABCN25pic}. The wire-bonds to the left of the chip connect 
the 128 strips to the ASIC inputs. They are divided in two
sets of 64 channels, one with short bonds and the most visible ones are the
long bonds.

The chip is implemented with IBM $250~{\rm nm}$ CMOS 
technology. The design of the final read-out chip is foreseen to be 
based on $130~{\rm nm}$ technology. Some of the relevant features of the 
ABCn25 chip are~\cite{Dabrowski:1273530}:

\begin{itemize}
 \item Analogue gain of $100~{\rm mV/fC}$ with an integral non linearity lower
 than $3\%$ for input charges in the range $(-10, +10)~{\rm fC}$.
 \item Equivalent Noise Charge lower than $800~e^{-}$ for a detector 
 capacitance up to $5~{\rm pF}$.
 \item Each channel has an internal calibration capacitor at their input to 
 allow the injection of test charges up to $16~{\rm fC}$ with 
 $0.0625~{\rm fC/step}$, using an 8-bit calibration DAC.
\end{itemize}

There were technical and economical reasons to do an initial implementation of
the ABCn chip in $250~{\rm nm}$ technology instead of the final $130~{\rm nm}$.
The designers were able to re-use some of the functional blocks that 
were designed for the SCT, so the design could be finished in a
relatively short time, saving initial chip design costs.

The power consumption of the ABCn25 chip is split between the digital and the
analogue part. The analogue function requires $40~{\rm mA}$ while the digital
blocks require $90~{\rm mA}$ at $V_{dd} = 2.5~{\rm V}$~\cite{Affolder:1516555}.

\subsection{The ABC130 Chip}

The design of the ABC130 was finished in 2013~\cite{Affolder:1516555} and the 
first produced units arrived by the end of 2013. The detector prototypes 
that have been tested within the scope of this thesis are based on the ABCn25 
chip.

One of the reasons to embrace the $130~{\rm nm}$ technology for the ATLAS 
Upgrade binary read-out chip is the additional 
radiation hardness with respect to $250~{\rm nm}$ 
technology~\cite{Gonella2007750}.

The main difference from the ABCn25 is that the ABC130 has 256 read-out 
channels on a single chip. One chip reads out two rows of strips on a sensor,
instead of just one.

\begin{figure}[!hb]
 \begin{center}
  \includegraphics[scale=0.55]{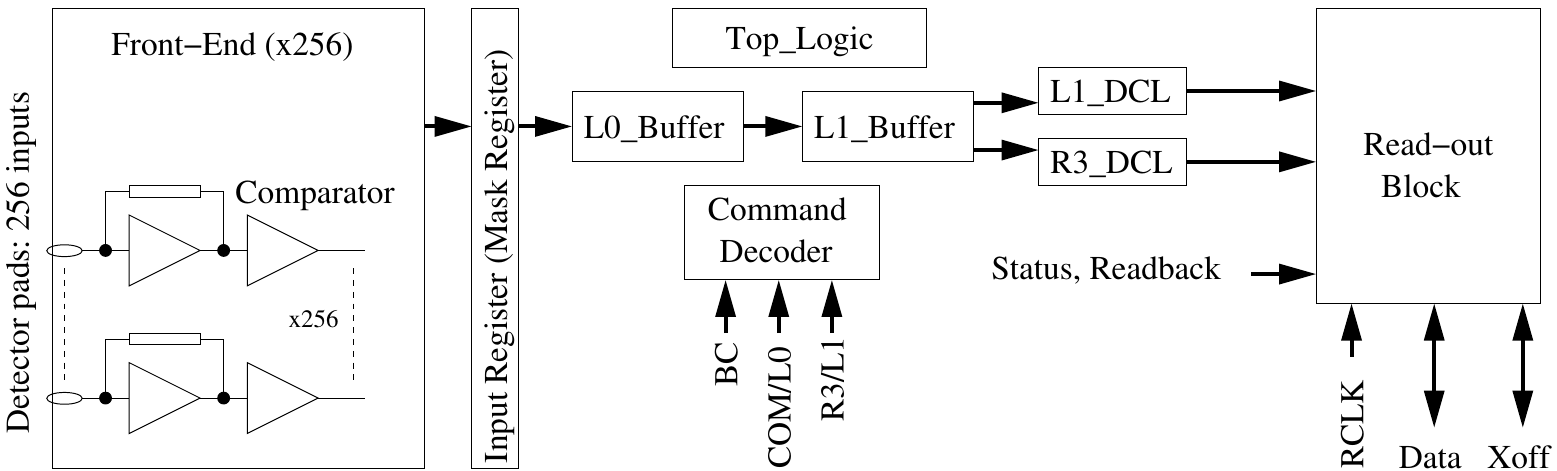}
  \caption[Block diagram of the ABC130.]{Block diagram of the ABC130~\cite{ABC130}.}\label{fig:4-ABC130diag}
 \end{center}
\end{figure}

Figure~\ref{fig:4-ABC130diag} shows a functional block diagram of the ABC130 
chip~\cite{ABC130}, where the blocks are different from the ABCn25 chip.

Some of the most important features that differ from the previous chips are
the following:

\begin{itemize}
 \item External supply voltage of $1.3~{\rm V}$, down from $2.5\V$.
 \item Data flow controlled by three trigger types: L0, R3 and L1
 \item Read-out clock up to $160~{\rm Mbps}$.
 \item Bi-directional read-out to allow the system to operate if one chip
 fails.
\end{itemize}

Performance optimization of the ABC130 is done for short strips. The 
post-irradiation noise for a load capacitance of $5~{\rm pF}$ is specified
to be below $650~e^{-}$. This load corresponds to the maximum capacitance
of a $47.755\mm$ long strip.

The gain at the discriminator input is specified to be $95~{\rm mV/fC}$ for
nominal bias current, with a linearity better than $15\%$ up to $8~{\rm fC}$.

For the calibration signals, the on-chip capacitor is $60~{\rm fF}$. The 8-bit
DAC provides an amplitude step of $0.586~{\rm mV}$ in the $0$-$150~{\rm mV}$
range, giving a charge step of $0.035~{\rm fC}$ in a charge range from $0$ to
$9~{\rm fC}$.

Regarding the power consumption of the chip, the estimation for the
analogue part is $160~{\rm \upmu W}$ per channel for short strips and 
$300~{\rm \upmu W}$ per channel for long strips. These numbers assume an 
analogue
voltage of $1.2~{\rm V}$ with $34~{\rm mA}$ drawn current for the 256 channels
in case of short strips. The current almost doubles for long 
strips~\cite{Affolder:1516555}.

\subsection{Hybrids}

For the ABCn25 based hybrids, part of the Stave09 programme, the hybrids are 
kapton flex circuits, $24\times 107.35\mm^2$ in size. The kapton circuits hold
the passive components and the chips to read-out the silicon strips.

These hybrids also include a Buffer Control Chip (BCC) that allows the 
multiplexing
of the signals to and from the chips on one hybrid to the read-out system.

The ABCn25 chips are connected in a chain, in a way that if a chip fails, the
rest of the chips in the chain are also lost. The chain direction can be 
changed by swapping the master and end chips.

The ABCn25 based hybrids for the barrel contain 20 chips each. This means that 
a fully populated hybrid requires a current of around $4~{\rm A}$ at 
$2.5~{\rm V}$~\cite{Affolder:1516555}. 

\begin{figure}[!ht]
 \begin{center}
  \includegraphics[scale=0.25]{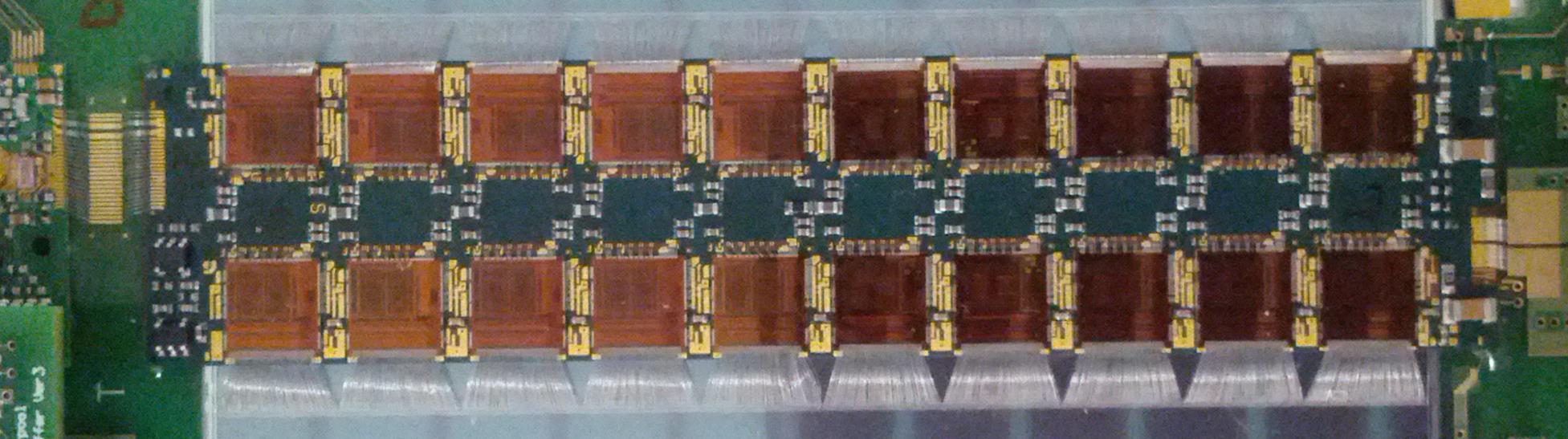}
  \caption{A hybrid with 20 ABCn25 chips, with the bonds to the sensor (top and
  bottom), to the
  powering (right) and to the BCC board (left) shown.}\label{fig:4-ABCn25_hybrid}
 \end{center}
\end{figure}

Figure~\ref{fig:4-ABCn25_hybrid} shows a picture of a barrel hybrid, with the 
20 chips wirebonded. Figure~\ref{fig:4-BCC} shows a BCC chip
on a module test frame with the wirebonds to the hybrid on the bottom part of
the picture.

\begin{figure}[!ht]
 \begin{center}
  \includegraphics[scale=0.35]{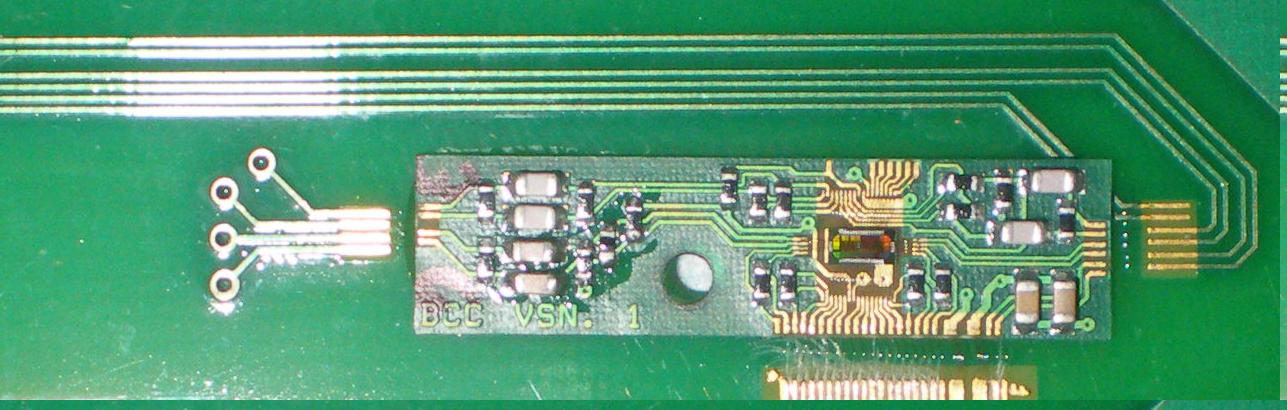}
  \caption{A BCC board.}\label{fig:4-BCC}
 \end{center}
\end{figure}

The hybrids for the ABC130 chips follow the same philosophy and are built using
kapton flex circuits but of smaller size, $16\times 97.54\mm^2$, just 
covering the length of a sensor. Also, the layout has been redesigned to
accommodate the different number of chips, 10 instead of 20, because each
chip reads out 256 strips.
In addition, instead of having a Buffer Control Chip
to aggregate and send the signals to and from the hybrid, the chip has been
redesigned in a $130\nm$ process and named Hybrid Control Chip (HCC). The BCC
is placed on a board separate from the hybrid and connected through wirebonds,
whereas the HCC is placed on the hybrid.

A limited level of redundancy is added in the read-out
chain at the hybrid level. The ABC130 chip 
allows data transmission in both directions, forming two possible data loops
towards the HCC. This means that, unlike what happens with an ABCn25 
hybrid, failure of one chip only means losing that one chip, instead of all
the chips in the chain from the faulty chip.

The barrel hybrids for the ABC130 hold 10 chips each. This hybrid requires, 
based on the current consumption of the 
chip, $1.51~{\rm W}$ in the case of short strips. There is a significant 
reduction power of the power consumption compared to the ABCn25 based hybrids.

\subsection{End of Stave/Petal Card}

The interface between the Hybrid Control Chips and the off-detector 
read-out systems is done
through the End of Stave (EoS) and End of Petal (EoP) cards. These contain a
controller, a DCS dedicated chip and interfaces for optical communication with
the off-detector system.

\begin{figure}[!ht]
\begin{center}
 \includegraphics[scale=0.23]{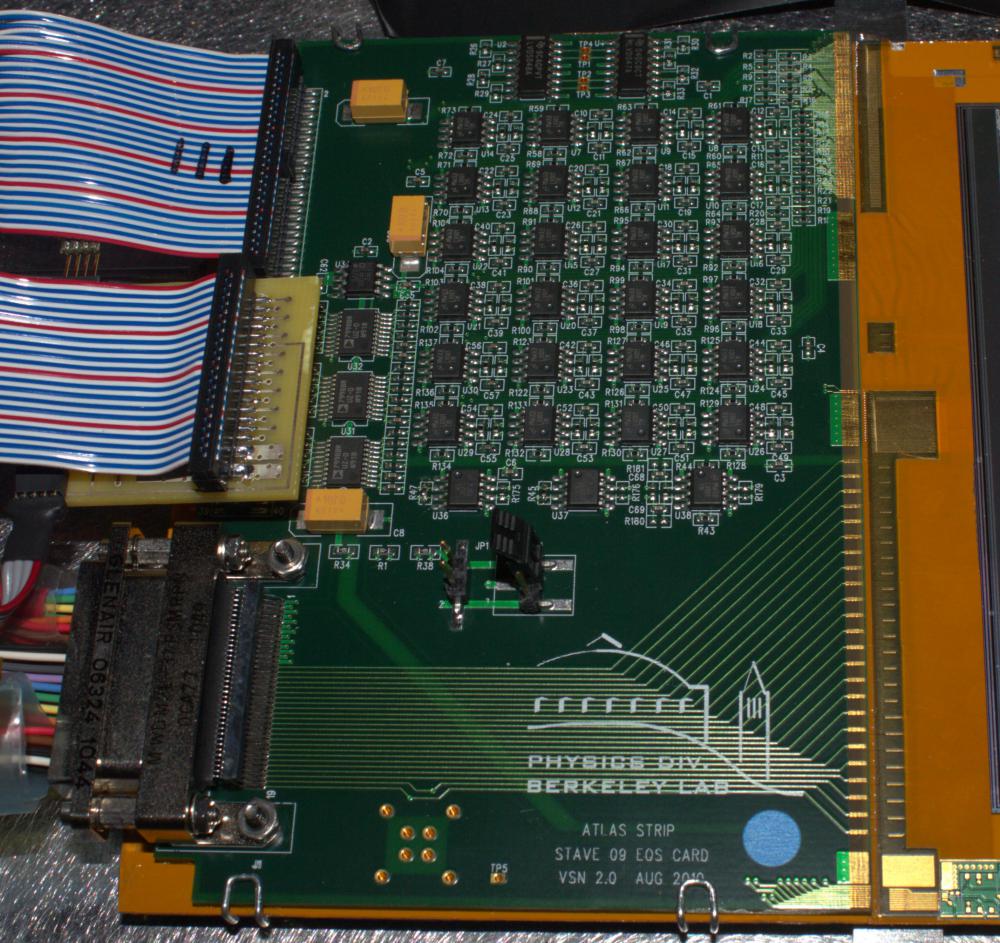}
\caption{End of Stave card from a four-module stavelet.}\label{fig:4-EoS}
\end{center}
\end{figure}

Figure~\ref{fig:4-EoS} shows the End of Stave card used on a four-module 
stavelet, with only electrical interfaces. This board is designed to support
up to 12 modules on a stave, by having the 48 streams that are needed to 
read-out the 24 BCCs (two streams per chip, one per column of chips). The two 
flat cables transmit the data to and from the BCCs, as well as the temperature
sensors read-out and the power for the LVDS buffers on End of Stave card. 
The bottom connector is 
used on this EoS card to connect the high voltage to bias the sensors. On the
right side the card is connected to the bus tape using wire-bonds.

The optical interface to be used in the final stave design
is implemented using the GigaBit Transceiver (GBT), 
a radiation tolerant chip that allows bi-directional optical transmission for
data read-out, trigger, timing, control and monitoring links. The point to
point link speed is specified to be $5~{\rm Gbps}$. Both multi-mode 
and single-mode transmission are being considered, using $850\nm$ and $1310\nm$ 
wavelengths, respectively.

\begin{figure}[!ht]
 \begin{center}
  \includegraphics[scale=0.5]{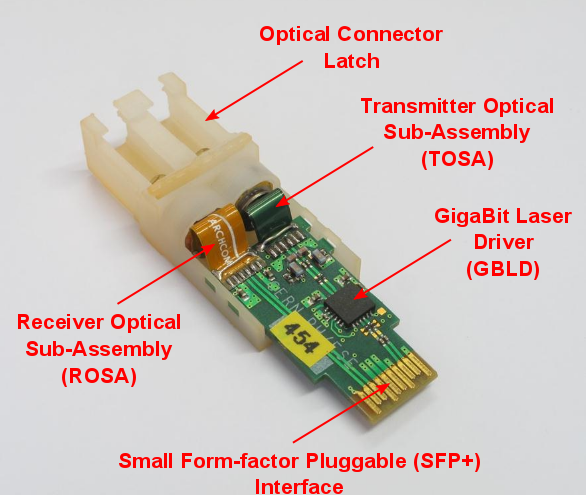}
  \caption[Picture of a VTRx showing the components of the transceiver
    package.]{Picture of a VTRx showing the components of the transceiver
    package~\cite{1748-0221-8-03-C03004}.}\label{fig:4-VTRx}
 \end{center}
\end{figure}

The GBT chip is part of the Versatile Link transceiver (VTRx), which is 
designed to use optical fibres with standard LC connectors on an SFP-like
package. The commercial Small-Form factor Pluggable (SFP)
packages have a metallic case which is, in principle, unsuitable for operation
in the HL-LHC environment. An option is to remove the case and design a custom 
connector latch with non metallic materials~\cite{1748-0221-8-03-C03004}. A 
picture of one of the transceiver prototypes is shown in 
Figure~\ref{fig:4-VTRx}.

\section{Powering Schemes}

The power delivery in the current silicon detectors at the LHC is made 
independently
for each detector element. In the case of the ATLAS SCT,
each module is powered with its own line, which means this sub-system uses
$4088$ independent power supplies, one for each module, what involves a set 
of cables in order to supply power to the hybrids, power to the read-out
and trigger, timing and control signals, and sensor 
bias~\cite{Phillips:1091485}. 

Since the power supplies have to be placed at a safe distance from the 
irradiated area, the power lines in the current ATLAS detector are in the 
order of $100\m$ long. An immediate consequence is that the cable resistance 
causes heating and waste of power in the cables. Between fifty and eighty 
percent of the low voltage power supplies output power is lost in the cables. 

\begin{figure}[!htb]
 \begin{center}
  \includegraphics[scale=0.8]{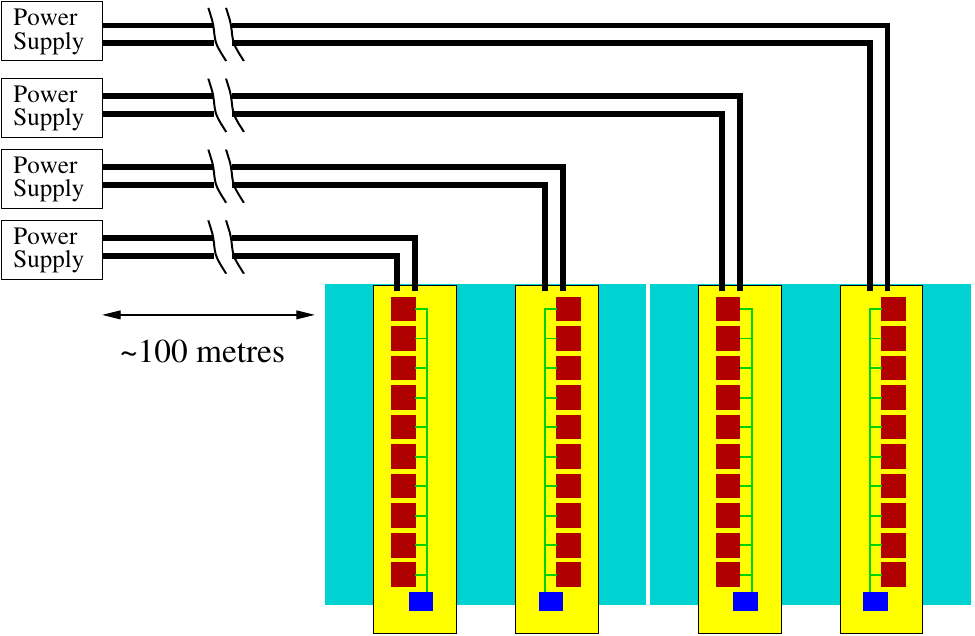}
  \caption{Independent powering of hybrids.}\label{fig:4-PoweringModules_Indep}
 \end{center}
\end{figure}

This approach is shown in 
Figure~\ref{fig:4-PoweringModules_Indep}, where each module has two hybrids,
independently powered by separate cables and power supplies.
This power distribution model is not suitable for the electronics in the
HL-LHC trackers. The power calculation for a short strips stave with 26 
hybrids holding 10 ABC130 chips each yields $78.5~{\rm W}$ per stave and 
$37~{\rm kW}$ for the whole tracker barrel~\cite{Affolder:1516555}. 

The petal consumes around $30~{\rm W}$ in its current design, leading to a 
power consumption of the endcaps of
around $13~{\rm kW}$ for all the $448$ petals.
The current SCT, with a consumption of $3.1~{\rm mW}$ per channel, has a total
power consumption of $19.5~{\rm kW}$, including both the barrel and the 
endcaps.

Considering this great increase in power consumption, due to the much higher 
number of read-out channels, the following problems have to be considered for the
high luminosity upgrade~\cite{Weber200844}:

\begin{itemize}
 \item More power lines would be needed in an independent powering scheme, 
 leading to an equivalent increase in material.
 \item That increased volume of cables also causes lack of space in the tracker
 volume.
 \item Very low power efficiency, more heating and wasted power.
\end{itemize}

The current space allocated for the services on the LHC experiments turned out
to be barely sufficient for the trackers. In an upgrade scenario, there is
absolutely no room to route more cables. Additionally, a requirement for the 
HL-LHC trackers is that they should not be more massive than the current ones, 
to avoid degrading the resolution.

Several alternative proposals for the power distribution have been made. The
solution is aimed at reducing the current on the long power distribution 
cables. In order to achieve this, the two techniques under study are:

\begin{itemize}
 \item Serial powering of modules is an innovative and elegant solution that 
 relies on a constant current provided to the hybrids.
 \item DC-DC conversion is a more traditional approach, in which power 
 transmission  takes place at a high constant voltage, with low 
 current, then a conversion to a lower voltage and higher 
 current is done local to the tracker.
\end{itemize}

A brief description of these two approaches follows.

\subsection{Serial Powering}

In the serial powering (SP) scheme, a constant current is fed to a series of 
elements. A local shunt regulator circuitry provides the voltage to each 
element. In this context, an element can be either a module or a hybrid.

The voltage across the chain of elements (e.g. modules) is $n$ times the 
element voltage and the reference potential of each module is different. These
different reference levels require an AC coupling or optical decoupling of the 
digital control and data signals~\cite{Weber200844}.

\begin{figure}[!hbt]
 \begin{center}
  \includegraphics[scale=0.8]{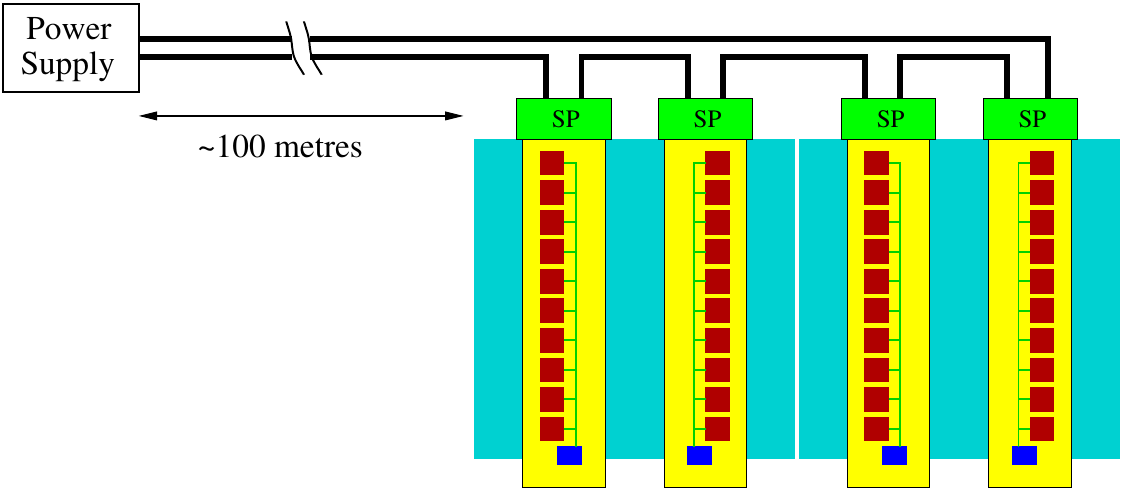}
  \caption{Serial powering of modules. In this case, each hybrid is at a 
  different voltage reference, so this scheme is called ``chain of
    hybrids''.}\label{fig:4-PoweringModules_SP}
 \end{center}
\end{figure}

Between 10 and 20 modules would be chained in series, leading to a
great reduction in the number of power cables, as such that less cables would 
be required in the HL-LHC trackers than in the current LHC trackers.

A sample of the serial powering technique is shown in 
Figure~\ref{fig:4-PoweringModules_SP},
with two modules, each with two hybrids, powered with constant current.

The voltage is regulated locally on each hybrid, by means of a shunt 
regulator. The current overhead to keep the right voltage on each hybrid can 
be low if all hybrids are electrically similar, drawing about the same
current. Otherwise, the hybrid with the highest consumption sets the current
draw on every hybrid of the chain, which can lead to efficiency losses when
power consumption is uneven.

\begin{figure}[!ht]
 \begin{center}
  \includegraphics[scale=0.8]{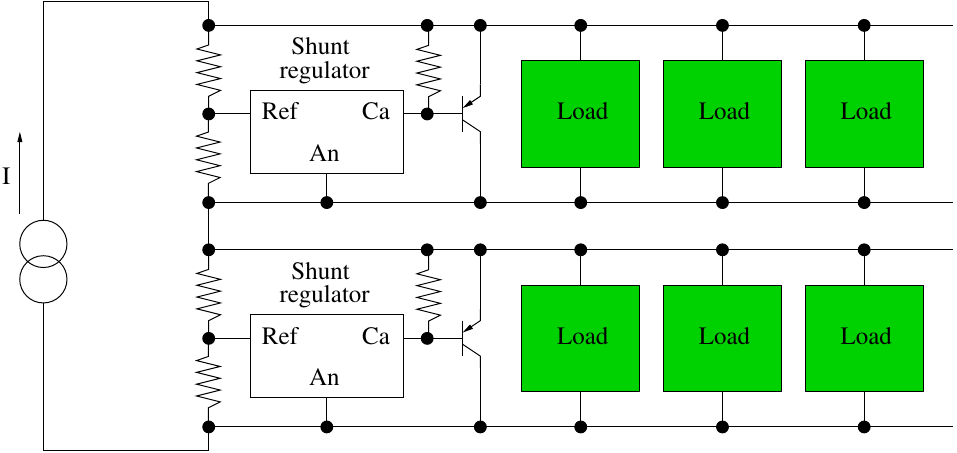}
  \caption{Serial powering with one shunt regulator and shunt transistor
  external to the load.}\label{fig:4-SP_ShuntRegs}
 \end{center}
\end{figure}

An example of such an implementation is shown in
Figure~\ref{fig:4-SP_ShuntRegs}, which depicts one of the alternatives to power
two modules. Other options are having one shunt regulator and one shunt
transistor per load unit (chip) or having one shunt regulator for a group of 
chips and one shunt transistor in each chip~\cite{Tic:1235825}.

A preliminary study of this powering scheme was performed using ATLAS SCT
barrel hybrids~\cite{Phillips:1158648}, with promising results. Currently, 
there is a wide community studying the performance of serial powering using
the Phase-II Upgrade modules and hybrids. Several devices implementing serial
powering have been constructed. 

There are single modules as well as stavelets,
the latter either chaining modules or hybrids. The chain of modules means that
both hybrids on a module have the same reference voltage, while the chain of
hybrids implementation forces all hybrids to have different reference 
voltages.

\begin{figure}[!ht]
 \begin{center}
  \includegraphics[scale=0.9]{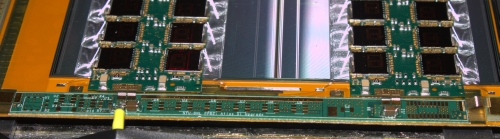}
  \caption{Serial powering board on a serially powered four-modules
  stavelet.}\label{fig:4-SPPboard}
 \end{center}
\end{figure}

Figure~\ref{fig:4-SPPboard} shows a serial powering board used in one of the
serially powered stavelets, assembled at RAL. The powering configuration of 
this stavelet is a chain of modules.

Some test results performed on serially powered devices, in the scope of 
this thesis, will be presented in chapter~\ref{chap:results}. The devices will
be described later in this chapter.

\subsection{DC-DC Conversion}\label{sec:4-dcdc}

The use of local DC-DC conversion is, \textit{a priori}, a more natural 
approach than serial 
powering. Both independent powering and parallel powering can be utilized with
DC-DC conversion, without completely losing the advantage of cable material 
reduction.

A popular DC-DC architecture is the buck converter, which uses an inductor as
the energy storage unit, switching transistors, a switch control unit and a 
filtering
capacitor. The efficiency of this architecture is high and it has the 
capability of delivering high current to the device~\cite{Weber200844}.

\begin{figure}[!htb]
 \begin{center}
  \includegraphics[scale=0.8]{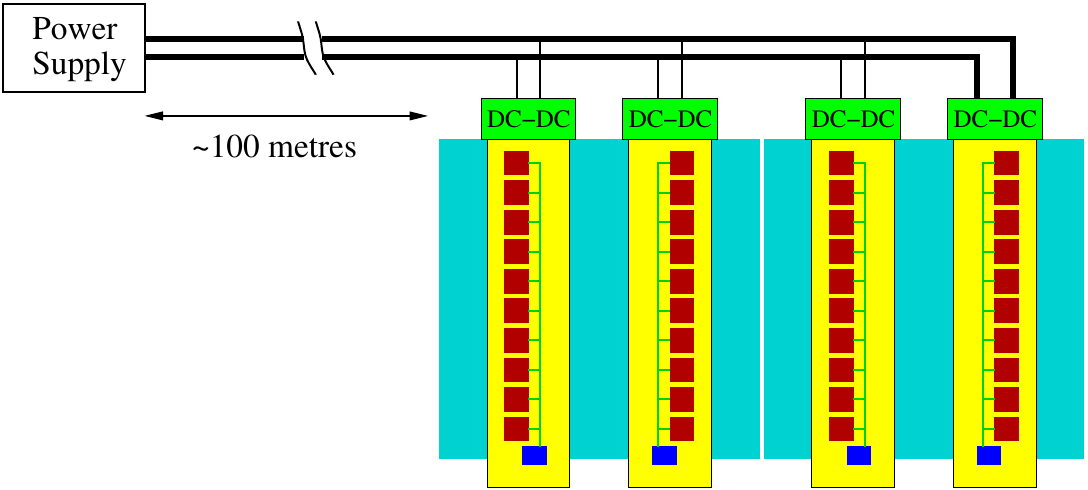}
  \caption{Powering of modules with DC-DC conversion. The drawing shows the 
  parallel powering scheme.}\label{fig:4-PoweringModules_DCDC}
 \end{center}
\end{figure}

A sample scheme of this approach is shown in 
Figure~\ref{fig:4-PoweringModules_DCDC}, where two modules, each with two 
hybrids, are powered by a higher voltage supply
that gets converted down at the hybrids.

One of the challenges for DC-DC converters in the HL-LHC trackers is to 
minimize the material, through miniaturization of the device. In addition, the
converters have to cope with an extremely high radiation environment and 
operate in a strong magnetic field. As a reminder, the ATLAS solenoid magnet 
is $2~{\rm T}$ and the CMS magnet is $4~{\rm T}$. 

This last detail limits the implementation of the inductor to an air core 
version, not allowing the use of ferromagnetic cores, in order to avoid 
saturation. As a result, the size
of the inductor increases for a target inductance value.

A low noise and low mass DC-DC converter has been developed at CERN and is now
the baseline for the DC-DC powering schemes in the Strips Tracker Phase-II 
Upgrade community~\cite{1748-0221-6-11-C11035}. Three versions of the 
converter modules that
have been designed are shown in Figure~\ref{fig:4-dcdc_converters}.

\begin{figure}[!ht]
 \begin{center}
 \subfigure[AMIS2 converter]{
  \includegraphics[scale=0.4]{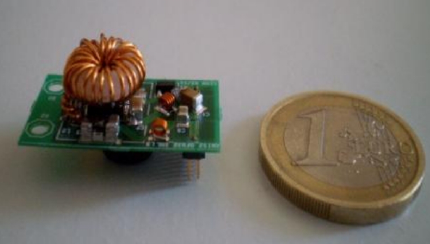}
 }
 \subfigure[SM01C converter]{
  \includegraphics[scale=0.5]{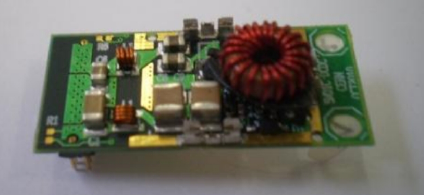}
 }
 \subfigure[STV10 converter]{
  \includegraphics[scale=0.63]{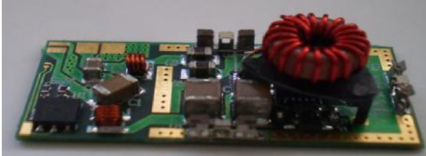}
 }
  \caption[Different versions of the DC-DC converter designed at CERN
    for the ATLAS and CMS trackers upgrade.]{Different versions of the DC-DC converter designed at CERN
    for the ATLAS and CMS trackers upgrade~\cite{1748-0221-6-11-C11035}.
  }\label{fig:4-dcdc_converters}
 \end{center}
\end{figure}

The AMIS2 converter~\cite{1748-0221-5-11-C11016} is designed for the 
$130\nm$ CMOS chips, providing up to $3~{\rm A}$ with an output voltage between
$1.2$ and $5\V$, and a peak 
efficiency of $80\%$. The input voltage range is $7$ to $10$ volts. This 
converter cannot deliver enough current to the ABCn25 based hybrids, as these 
require about $5~{\rm A}$ peak current.

\begin{figure}[!htb]
 \begin{center}
  \includegraphics[scale=0.65]{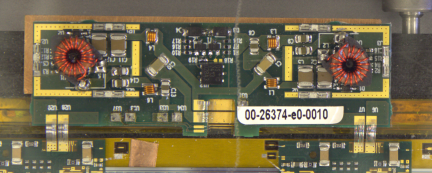}
  \caption{A tandem DC-DC converter with dual output.}\label{fig:4-DualDCDC}
 \end{center}
\end{figure}

The other two converters, SM01C and STV10, were designed to 
meet this requirement. The first, SM01C, is a pluggable
converter, whereas the STV10 is a connector-less, wirebondable 
module~\cite{1748-0221-6-11-C11035}. The STV10 is the power module used
in the construction of the stavelets and petalets. 

With respect to the noise performance, the AMIS2 is designed to be suitable
for the High Luminosity LHC tracker front-ends. The two designs of DC-DC 
  converters for ABCn25 chips have slightly higher noise than the AMIS2.

For stavelets with DC-DC converters, a common connection from the bus tape 
to the converters input is needed for the two converters of a module, in order 
to have good noise performance~\cite{DCDCstarpoint}. This requirement 
leads to the design of the so-called ``tandem'' converter, which consists of 
two DC-DC converters placed on the same board and with a common reference 
plane.

The tandem converter has a single power input and two outputs, placed at the
locations of the hybrids in a module. A re-manufacture was done by partially 
splitting
the ground plane and this converter has been used for the first full length
stave with DC-DC powering and using the ABCn25 chip. One of these converters is
shown in Figure~\ref{fig:4-DualDCDC}.

\begin{figure}[!htb]
\begin{center}
 \subfigure[DC-DC converters with their shields removed.]{
   \includegraphics[scale=1.0]{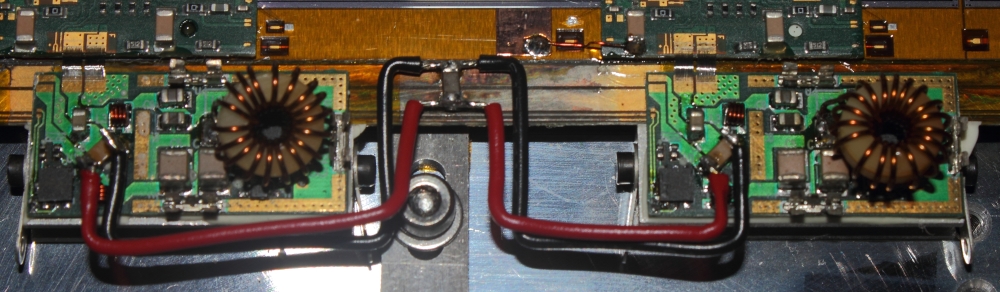}
 }
 
 \subfigure[DC-DC converters with the shields covering
   the coils.]{
   \includegraphics[scale=1.0]{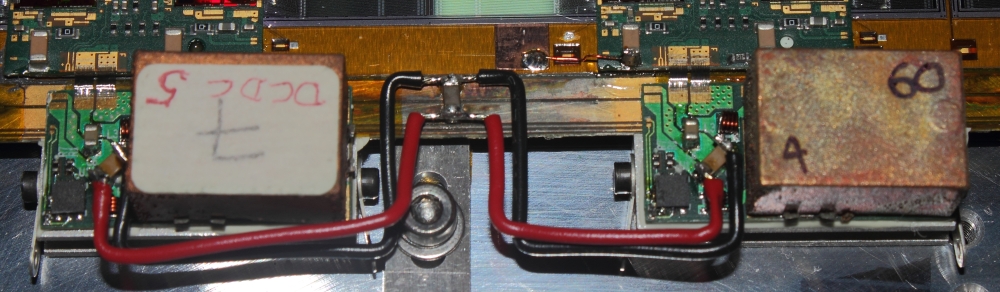}
 }
 \caption{DC-DC converters mounted on a 
 stavelet.}\label{fig:4-DCDCconvertersOnStavelet}
 \end{center}
\end{figure}

One of the difficulties in using DC-DC converters is the big inductor, which
generates interferences to the read-out electronics and forces the use of
shields to attenuate the emitted electromagnetic field. The shields are made of
a plastic box that is covered with a copper paint layer. This shield 
increases the amount of material that the DC-DC converter adds to the tracker.

In addition, due to the high power that has to be switched, the converter needs
to dissipate some of it, generating a lot of heat. This heat
needs to be extracted, increasing the cooling system requirements.

Figure~\ref{fig:4-DCDCconvertersOnStavelet} shows two converters on a 
stavelet, powering
one hybrid each. The first image shows the converter without the EMI shield
covering the big coil. The switching ASIC is located underneath the coil.

The star point implemented with wire for the two converters of one module can
be seen in the centre of the pictures.

\subsection{Advantages of DC-DC and SP}

Each of the two techniques presented here have their own advantages and
disadvantages. The two have clear advantages over independent powering. The 
different solutions are described next.

The main advantages of the serial powering approach with respect to DC-DC 
conversion are lower mass and potentially higher efficiency with increasing 
number of modules~\cite{Phillips:1158648}. The geometry of the DC-DC converters 
is also more difficult to accommodate into the tracker
volume due to the reduced clearances caused by the shield around the coil.

However, while this is true for the barrel staves, as was described in 
section~\ref{sec:3-stripsupgrade}, the endcap 
hybrids have different number of chips for different rings. Due to this
difference, the power requirements differ for each hybrid on the endcaps. As 
a result, the use of serial powering in the endcap would require adding power 
dumps in the hybrids with lower power consumption, leading to waste of 
power and a greatly reduced efficiency.

The consequence is that the endcap petals are forced to use DC-DC conversion
in order to obtain an adequate power efficiency.

In the case of the barrel, all hybrids in a stave hold the same number of 
chips. Therefore, there
are no efficiency losses unless there is a change in the power consumption of a
hybrid, such as a disconnected chip.

In addition, serially powered modules and stavelets tend to show more 
susceptibility to external noise and are harder to protect from common mode and
differential mode noise pickup.

Both DC-DC conversion and serial powering techniques lead to reduced
number of power supplies and rack space in comparison to an independent 
powering scheme. Also, as we have already discussed, the gain in efficiency 
leads to much less wasted power and cable material inside the detector.

With respect to a possible disadvantage of using either DC-DC conversion or 
serial powering, in both cases the number of lost modules in case of failure of
a power line is high. However, the number of connections that may
fail is greatly reduced, compared to a more robust, independent powering 
scheme. In addition, the savings of these two powering 
schemes can be exploited to engineer robust module 
connections~\cite{Weber2007844}.

\begin{table}[!ht]
 \begin{center}
  \begin{tabular}{p{5.1cm}|p{2.5cm}p{2.5cm}p{2.6cm}}
  \toprule
   \small Attribute & \small IP & \small SP & \small DC-DC \\
   \hline
   Power efficiency & $10-20\%$ & \small $60-80\%$ & $60-80\%$ \\
   Local regulator efficiency & \small N/A & $\approx 90\%$ & $ > 80\%$ \\
   Number of power cables & \small 4 per hybrid &  \small  $2n$ reduction& \small  $2n$ reduction \\
   Protection & Yes & Yes & Not known yet \\
   \bottomrule
  \end{tabular}
\caption{Comparison of some features of the three powering schemes: Independent
Powering (IP), Serial Powering (SP) and DC-DC 
conversion.}\label{tab:4-PoweringFeatures}
 \end{center}
\end{table}

Some features of the various powering schemes, namely independent powering,
serial powering and DC-DC conversion, are shown in 
Table~\ref{tab:4-PoweringFeatures}~\cite{Weber200844}. The protection features
are conceived in order to address the event of a connection failure or an 
over-current or over-voltage situation. 

In the independent powering case, the protection mechanism is the simplest that
can be conceived, that is separate cables for each hybrid. In the serial 
powering approach there are over-current protection mechanisms and redundant 
regulators protecting the devices.

\section{Sensor Biasing}\label{sec:4-HVmux}

The biasing of the sensors, in the case of the barrel, is done at the whole 
module level. This means that the four short strips rows are biased using the 
same high voltage line. When scaling the numbers to full staves and the barrel 
cylinders, this leads to over $12000$ high voltage (HV) lines, only for the 
strips tracker barrel.

The endcap discs hace a total of $8064$ sensors,
summing up to more than $20000$ high voltage lines for
the whole microstrips tracker.

Independent high voltage lines for each sensor would be the natural approach, 
as it permits disabling individual modules
(for instance, in case of malfunction) without disturbing the rest, as well as
individual leakage current measurement.

However, the amount of cable material needed for this is very high, to allow
for these two particular features. A means to reduce the number of cables is
being studied, by using a technique that is called ``high voltage 
multiplexing''.

In this approach, the high voltage lines for all the modules on a stave, or the 
modules on one stave side are multiplexed at the stave, feeding the staves with
only two high voltage cables,
reducing the total number of lines by a factor of $13$.

As indicated in Table~\ref{tab:4-ATLAS12}, the maximum operating voltage 
for the strips sensors is around $600~{\rm V}$. Therefore, the key factor in 
the high voltage multiplexing approach is to find a radiation hard switch rated 
above $500~{\rm V}$. In addition, a desirable feature is to allow measuring the 
leakage current on each sensor individually.

\begin{figure}[!ht]
 \begin{center}
  \includegraphics[scale=0.4]{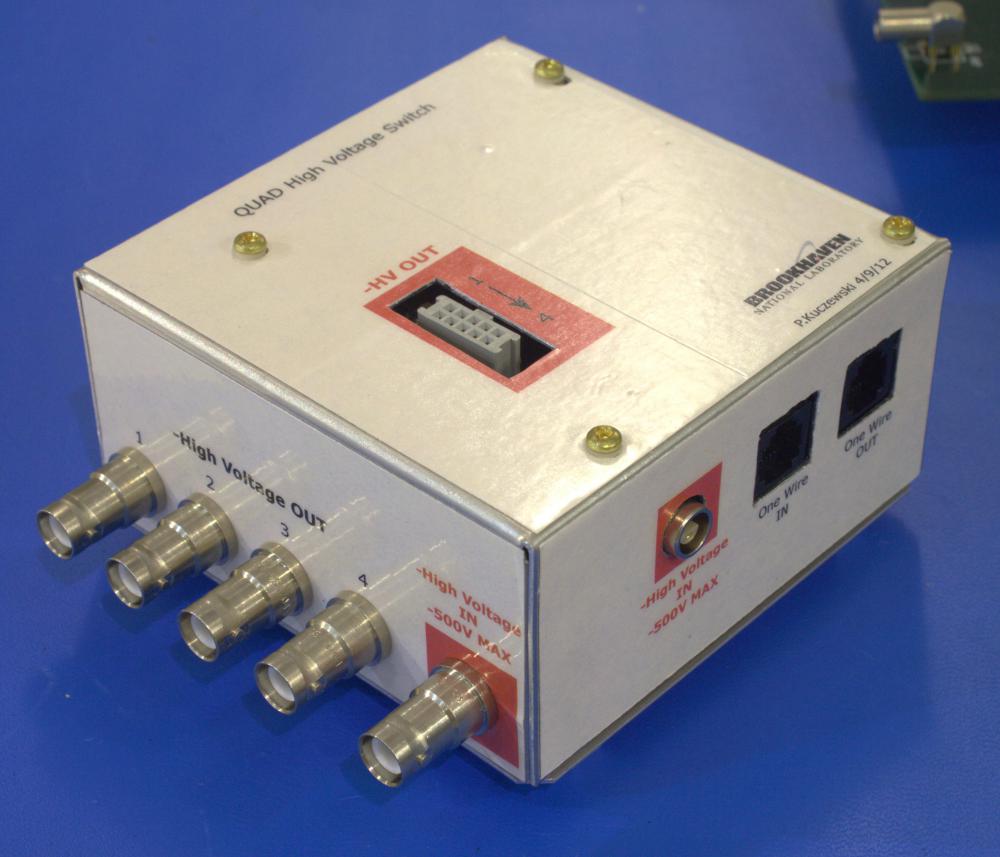}
  \caption{SiC based high voltage multiplexer developed at 
  BNL.}\label{fig:4-HVmux}
 \end{center}
\end{figure}

A four channel multiplexer based on silicon carbide (SiC) field-effect
transistors (FET) has been developed and tested on four-module
stavelets. 
There is no performance degradation of the modules and the only 
disadvantage is
the lack of individual leakage current sensing for this particular prototype.

The SiC transistors switch each of the four channels, which
allows the individual disconnection of HV channels. A picture of this
multiplexer is shown in Figure~\ref{fig:4-HVmux}. Results of the tests with
this device on a DC-DC powered and a serially powered stavelet are presented
in chapter 5.

The investigation on radiation hard, high voltage devices that can be used to 
multiplex up to 13 lines, corresponding to one side of a full length, $130\nm$ 
based stave, is ongoing. The existing MOSFET and bipolar transistors that
can operate above $500~{\rm V}$ are not as radiation hard as required. 
Materials such as silicon carbide (SiC) and gallium nitride 
(GaN) are the main candidates for implementing these 
multiplexers~\cite{ATLAS:1502664}. These materials have a wider band-gap what
gives them a greater intrinsic radiation hardness~\cite{SicRadiation}.

\section{The Stavelet Concept for the Tracker Barrel}

The stavelet proof-of-concept of the tracker stave is a multi-module structure,
holding four individual modules in its first incarnation. The initial stavelets
were single sided, others have been assembled with two 
sides~\cite{1748-0221-9-03-P03012}.
Both DC-DC conversion and serially powered modules and stavelets (with chain
of hybrids and chain of modules) have been built and tested. 

\subsection{Silicon Microstrip Modules}

The stavelets tested at CERN are single sided and consist of four individual 
silicon modules. Each module has a short strips silicon sensor, with integrated 
electronics for the read-out and control of the module. The sensors and the
individual electronic components have been described earlier in this chapter. 

\begin{figure}[!htb]
 \begin{center}
 \subfigure[A single module attached to its frame. The picture shows the two 
  hybrids, the sensor, serial powering line on the bottom and the BCC boards
  on the top of the module. The module is placed on an aluminium cooling 
  plate.]{
  \includegraphics[scale=0.95]{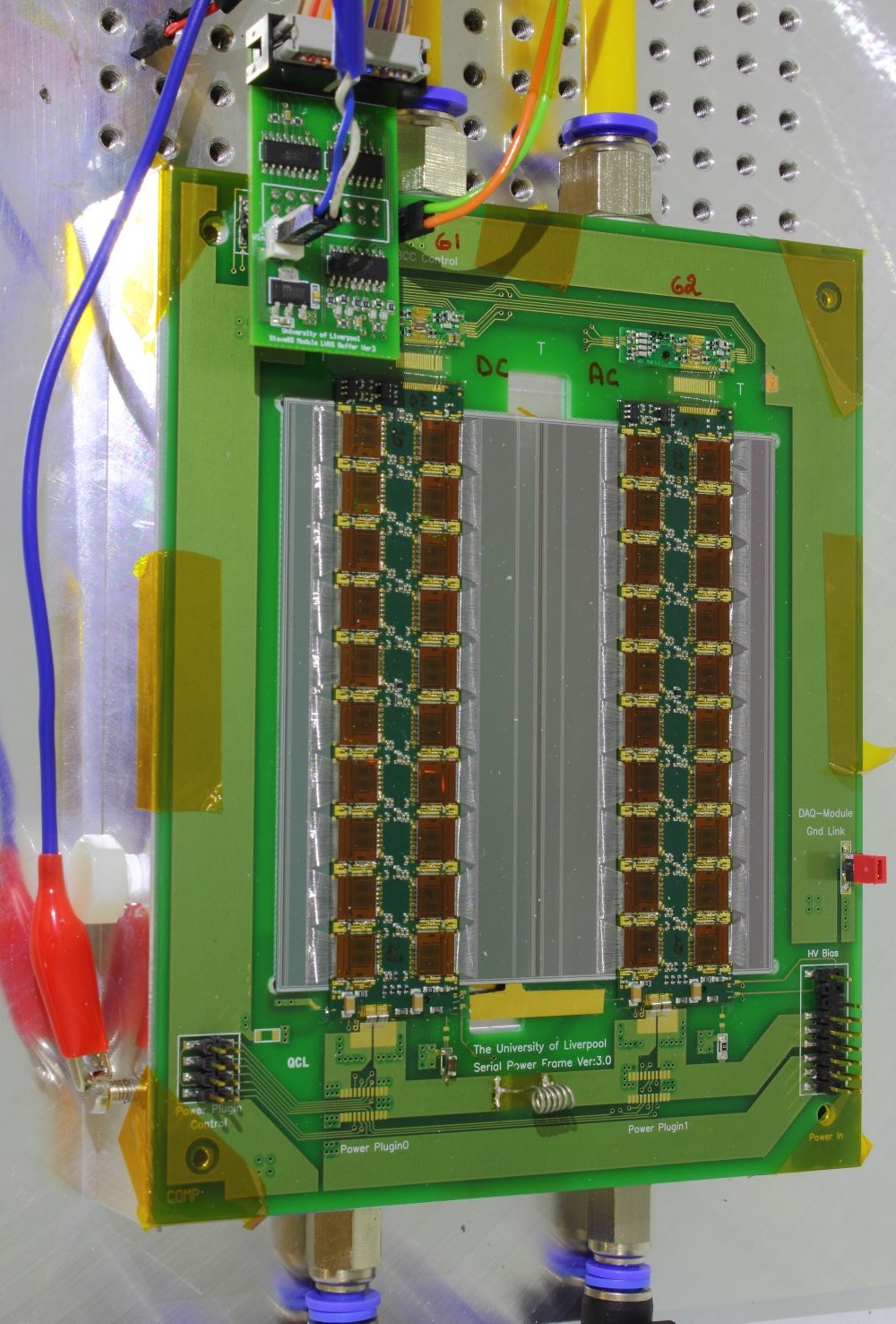}
  \label{fig:4-SPmodule}
  } \quad
  \subfigure[A single DC-DC module on the stavelet. The DC-DC converters
  are at the bottom, the BCC boards are on the top and the module is mounted
  on a bus tape that routes the powering and signal lines.]{
  \includegraphics[scale=0.95]{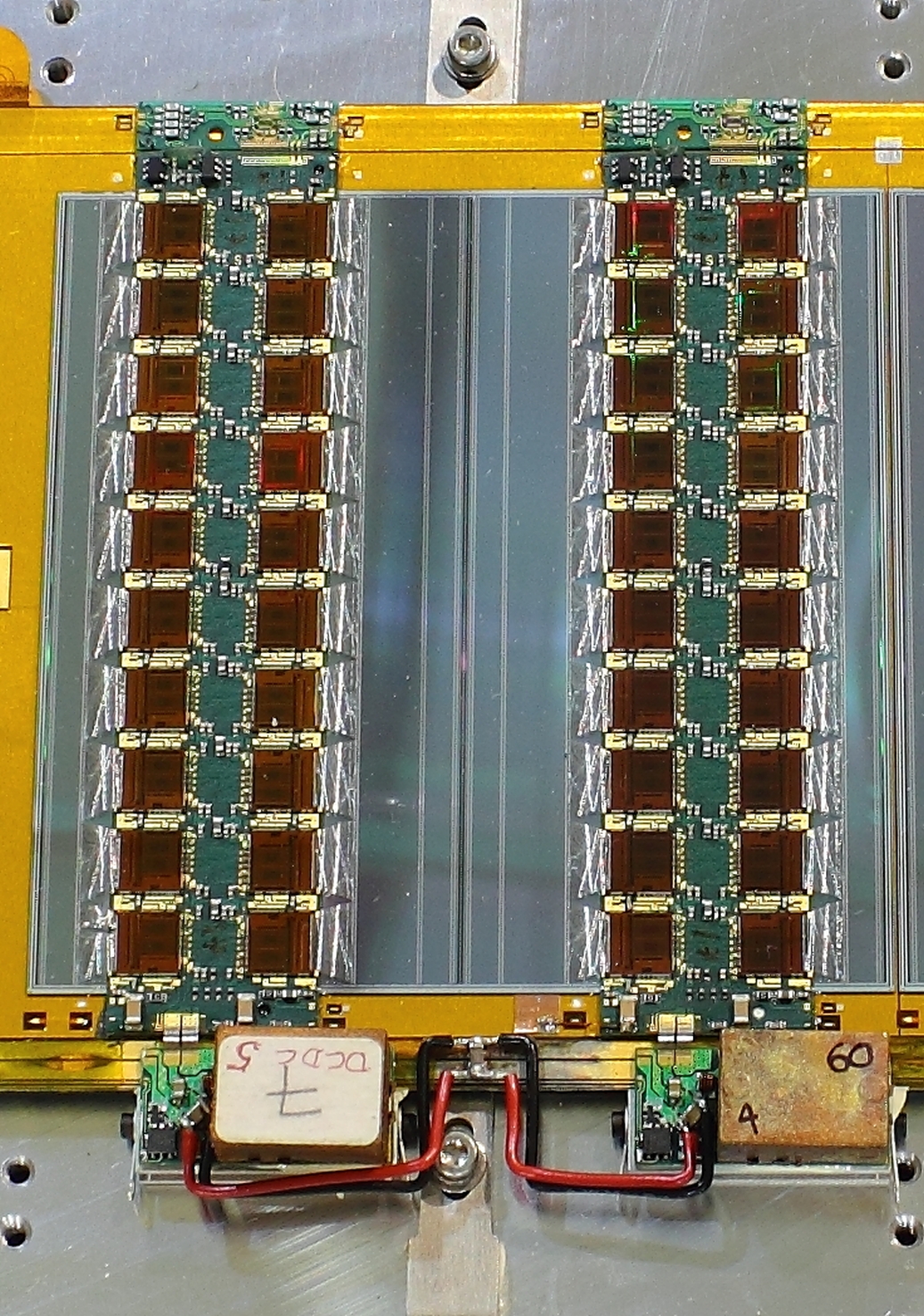}
  \label{fig:4-DCDCmodule}
    }
\caption{Single modules on a test frame (serially powered) and on a stavelet
(DC-DC powered).}\label{fig:4-SingleModules}
 \end{center}
\end{figure}

Figure~\ref{fig:4-SPmodule} shows a single module, mounted on a test frame that
interfaces with the power, sensor bias and DAQ system. The two hybrids with 
20 read-out ABCn25 chips are shown, the BCC boards are on the top of the 
picture, the powering and sensor bias come from the bottom right and the 
Low Voltage Differential Signalling (LVDS) come from the top left.

The powering scheme is implemented on the test frame, the picture shows a
serially powered module on such test frame. Also, the BCC is mounted on the 
test frame due to the lack of space on the hybrid. 

The picture shows the yellow cooling pipes on the top, the vacuum connectors on
the bottom, the LVDS buffer board at the top left side of the module and the
powering connector at the bottom right (without the board connected to the 
module). The module jig can be seen from the perspective on the left hand
side of the module.

Figure~\ref{fig:4-DCDCmodule} shows a DC-DC powered module mounted on a 
stavelet. The DC-DC converters power one hybrid each and they are connected in
a star configuration.

All the modules constructed to date use the ABCn25 ASICs, with 128 channels 
each. Each hybrid contains 20 chips that are arranged 
in two columns, and the read-out from the BCC is done for each column (called 
``chip column'').

The single modules on a test frame are usually tested using an aluminium block
to cool down the electronics, with a vacuum pump pulling down the module,
in order to have a good thermal contact between the cooling block and the 
module.

The connection to the read-out system, the HSIO board, is done by means of a
16-pin IDC ribbon cable to a Low-Voltage Differential Signal (LVDS) buffer 
board, that is in turn plugged to the module test frame. 

The low voltage and sensor bias power are connected via two
connectors (Molex for low voltage and LEMO for high voltage) that are plugged
into a PCB on the test frame.

Typically, the power for the hybrids is provided by a custom current 
source~\cite{1748-0221-6-01-C01019} in
the case of serially powered modules, which is capable of providing up to
$6~{\rm A}$. A serially powered module is usually powered with $5~{\rm A}$
and the voltage drop is around $6~{\rm V}$. Alternatively, a commercial DC
power supply may be used, set to limit the current around $5\A$.

To power DC-DC modules, a commercial DC power supply is normally
used, capable of providing $10~{\rm V}$ and around $3~{\rm A}$.

For the sensor bias, there are prototypes of the SCT high voltage cards 
available, that work on a VME crate. These cards are compatible with the 
requirements of the sensors. At CERN, this is 
typically used with stavelets. When testing single modules, a commercial high 
voltage unit is normally used.

\subsection{Multi-module Structures}

Several multi-module prototypes of these short strips modules have been 
designed and constructed, consisting of 4 modules in a single sided structure. 
These prototypes were conceived in order to assess the potential problems and
challenges that appear when multiple modules are powered and read-out
together.

Such multi-module structure is called ``stavelet'' in the case of barrel
modules, and is currently being used as the proof of concept of the 
stave design. 

The stavelet consists of a low mass carbon composite core, that provides
mechanical support and cooling, the latter through the embedded 
aluminium cooling pipes.

\begin{figure}[!htb]
 \begin{center}
 \subfigure[DC-DC powered stavelet.]{
   \includegraphics[scale=0.5]{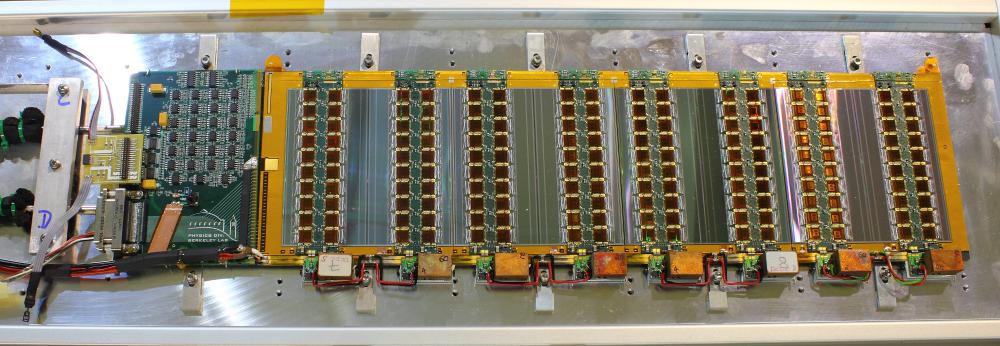}
   \label{fig:4-DCDCstavelet}
 }

 \subfigure[Serially powered stavelet.]{
   \includegraphics[scale=0.5]{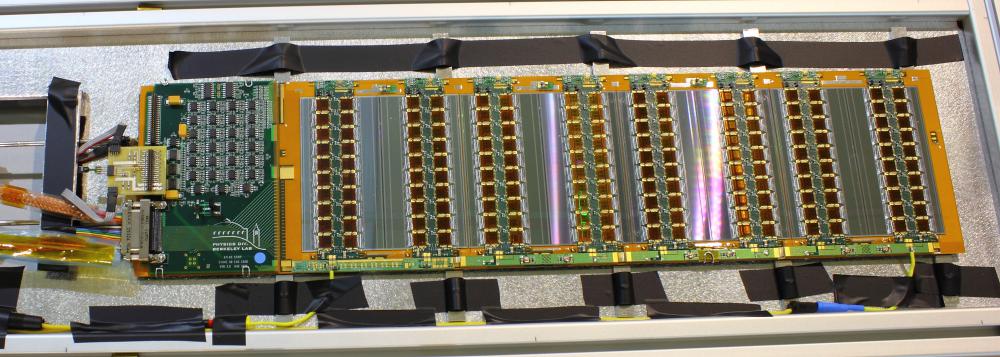}
   \label{fig:4-SPstavelet}
 }
  \caption[The two stavelets built at RAL and tested at CERN.]{The 
  two stavelets built at RAL and tested at CERN. Both pictures 
  show the four modules, the end of stave card to the left, as well as the 
  cooling pipes. The powering is placed at the bottom side of the 
  modules.}\label{fig:4-stavelets}
 \end{center}
\end{figure}

The modules are mounted on the stavelet using precision mechanical tools and
attached to the core with glue~\cite{1748-0221-9-03-P03012}.
The structure of the stavelet can be seen in both pictures of 
Figure~\ref{fig:4-stavelets}, with the 4 
sensors and the EoS card. Each of the sensors has the hybrids on top of it, 
holding the 20 ABCn25 chips. 

Figure~\ref{fig:4-DCDCstavelet} shows a DC-DC powered stavelet, and the
DC-DC converters with the EMC shield covering the coils can be seen at the 
bottom of the 
picture. The left side of the picture shows the End of Stave card and cooling 
pipes. The stavelet is mounted on an aluminium plate with a Bosch profile
attached for transportation. Figure~\ref{fig:4-SPstavelet} 
shows the serially powered stavelet in a similar manner. Instead of DC-DC
converters, the bottom side of the picture shows the serial power regulation.

Power and LVDS control signals are routed along the stavelet on a bus tape 
that has a copper trace layer under an aluminium 
shield~\cite{1748-0221-7-02-C02028}. The high voltage lines are routed 
independently for each module. Other stavelet implementations use
a shieldless tape~\cite{1748-0221-9-03-P03012}.

The tapes for the DC-DC and the SP stavelets are slightly different, as in
the SP stavelet it has a segmented serial power chain along one of the edges.
In the DC-DC stavelet, that edge has simple parallel lines. 

In addition, the width of the DC-DC stavelet is increased in order to add
mechanical support for the DC-DC converters. The increase is achieved by means
of aluminium pieces glued to the stavelet edge, what also adds a cooling path
to the converters~\cite{1748-0221-7-02-C02028}.

Other features are equal between the two stavelets, such as the AC-coupled
interface of the BCC board with the End of Stave card, which is unnecessary 
for the DC-DC stavelet. The EoS card is the same design for both. However, in
the DC-DC stavelet, the power traces cannot cope with the current needed. 
Therefore, they have been bypassed using thick 
copper~\cite{1748-0221-7-02-C02028}.

The DC-DC stavelet is typically powered using a commercial DC power
supply. The voltage is normally set so the voltage sense at the converters
inputs is very close to $10~{\rm V}$, accounting for the losses in the cables
and the traces. The current drawn when all eight hybrids are powered is around
$10~{\rm A}$.

The Serially Powered stavelet, as it follows the chain of modules scheme, is
powered by a commercial DC power supply, set to work in current limit mode, 
with the current limit set between $9~{\rm A}$ and $10~{\rm A}$. The typical 
current setting is $9.5~{\rm A}$, with a typical voltage drop around 
$12~{\rm V}$.

Sensors on a stavelet are normally biased using SCT high voltage cards as 
described in the previous section. These cards can deliver up to $500~{\rm V}$. 
They require a low pass filter in order to reduce noise on the lines. Also, a 
high voltage multiplexer like the one described in Section~\ref{sec:4-HVmux} 
is typically used in the tests.

The following chapter describes the tests and results performed on various
devices that were tested at CERN, in a facility in Building 180. These devices
are a serially powered single module, a serially powered stavelet
and a DC-DC powered stavelet. A DC-DC powered full length stave with 12 modules 
on one side was also tested at Rutherford Appleton Laboratory (RAL).

\chapter{Results and Discussion of the Structures for the Silicon Tracker 
Upgrade}\label{chap:results}

\section{Tests of Modules and Stavelets }

\subsection{Infrastructure}

The test facility at CERN is located in Building 180 (B180), which was 
initially put in place in 2010. After summer 2011, it has been regularly 
used for the Strips Tracker Upgrade projects: staves and supermodules. 

This facility includes a
clean room that is used for regular testing and assembly of strips and pixels
projects, and a bunker area outside the clean room. The clean room is shown in
Figure~\ref{fig:5-infrastructure}.

\begin{figure}[!htbp]
\begin{center}
\includegraphics[scale=0.72]{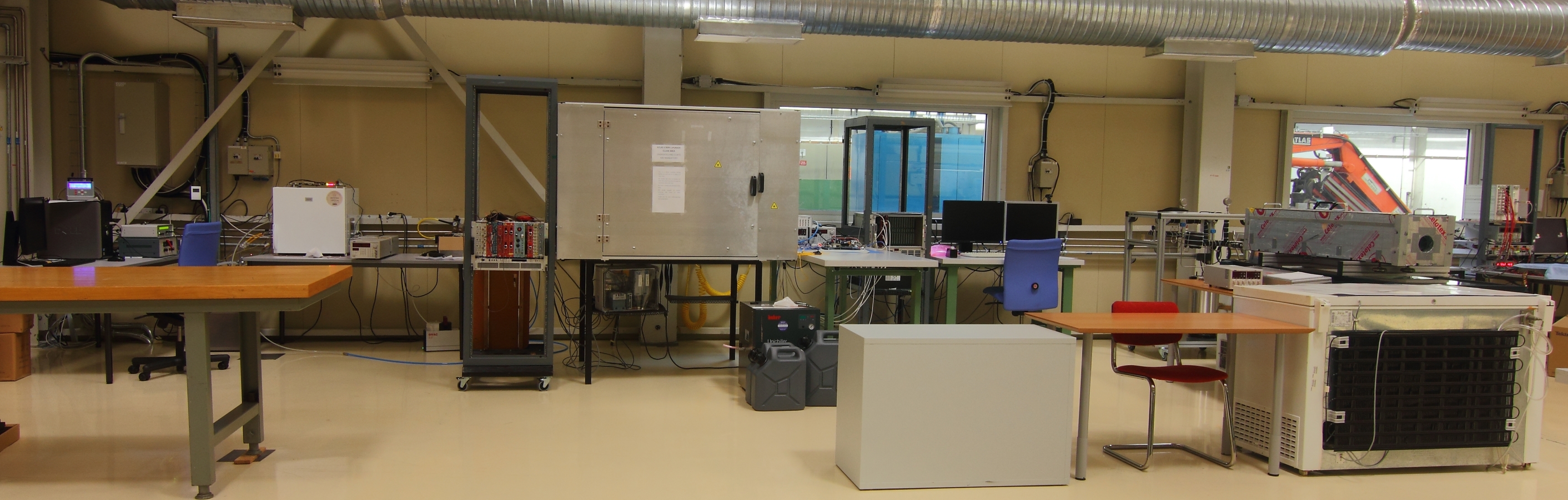}
\caption{The clean room in Building 180 at CERN.}\label{fig:5-infrastructure}
\end{center}
\end{figure}

The clean room is equipped with compressed air lines.
A gowning room is placed before the main entrance and there is an equipment 
entry area that is used to bring in and take out big equipment and furniture.

\subsection{Data Acquisition System for System Tests}

The Data Acquisition (DAQ) system used for the tests of modules and stavelets
consists of the read-out board, the firmware running on it, and the software
running on a computer that controls the whole system.

\subsubsection{The HSIO Board}

The ATLAS Strips Upgrade project is currently using an updated version of 
SCTDAQ software 
for detector hardware development. The legacy VME based system used in the 
current SCT cannot scale to 
read-out larger objects like stavelets. For this reason, the next generation 
of read-out hardware/firmware is being 
developed using an FPGA based board called the HSIO (High-Speed I/O) board,
designed at SLAC. A picture 
of the HSIO board is shown in Figure~\ref{fig:4-HSIO}.

\begin{figure}[!ht]
 \begin{center}
  \includegraphics[scale=0.45]{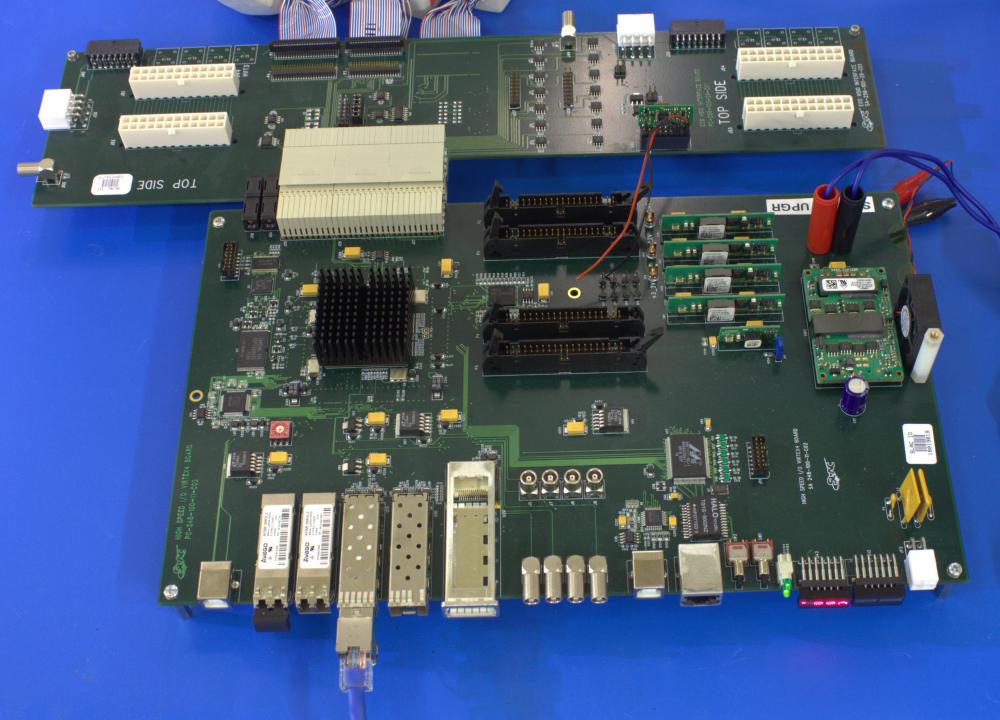}
  \caption{HSIO board used in the Strips Upgrade DAQ system.}\label{fig:4-HSIO}
 \end{center}
\end{figure}

The HSIO is a stand-alone unit with a Virtex-4 FPGA and many interface options. 
It has a variety of standard network connections, such as copper and optical
Gigabit Ethernet, making Ethernet the preferred communications protocol. 

The HSIO board is 
connected to an interface board that allows the connection of 
single sided modules and stavelets.  

In order to communicate with a computer, the currently supported interfaces 
are the $100~{\rm Mbps}$ Ethernet connection as well as the $1~{\rm Gbps}$
copper connections through an SFP connector that is plugged into the HSIO
board. The firmware version used for the tests described in the next chapter
use a direct communication over Ethernet, not relying on any higher
layer protocols.

\subsubsection{SCTDAQ Software for the Upgrade}

The SCTDAQ software used in the prototype tests for the Upgrade is based on the
SCTDAQ software that was used for the system tests of the current SCT. It has 
been modified to support the new HSIO system and to handle multiple modules.

Figure~\ref{fig:4-BurstData} shows the Burst Data window of the SCTDAQ software
as seen with up to four visible modules. The drawing on the top depicts the 
structure of a four module stavelet, which permits the visual identification of
the conditions of each module: bias voltage and leakage current, temperatures
of the hybrids, low voltage power status and temperature/humidity sensors at
the pipes.

The four pads on the bottom show the hit occupancy for every strip on the 
module, with one pad for each hybrid column (1280 strips per pad). There is
a tab for each module configured in the software. 

\begin{figure}[!htbp]
\begin{center}
\includegraphics[scale=0.7]{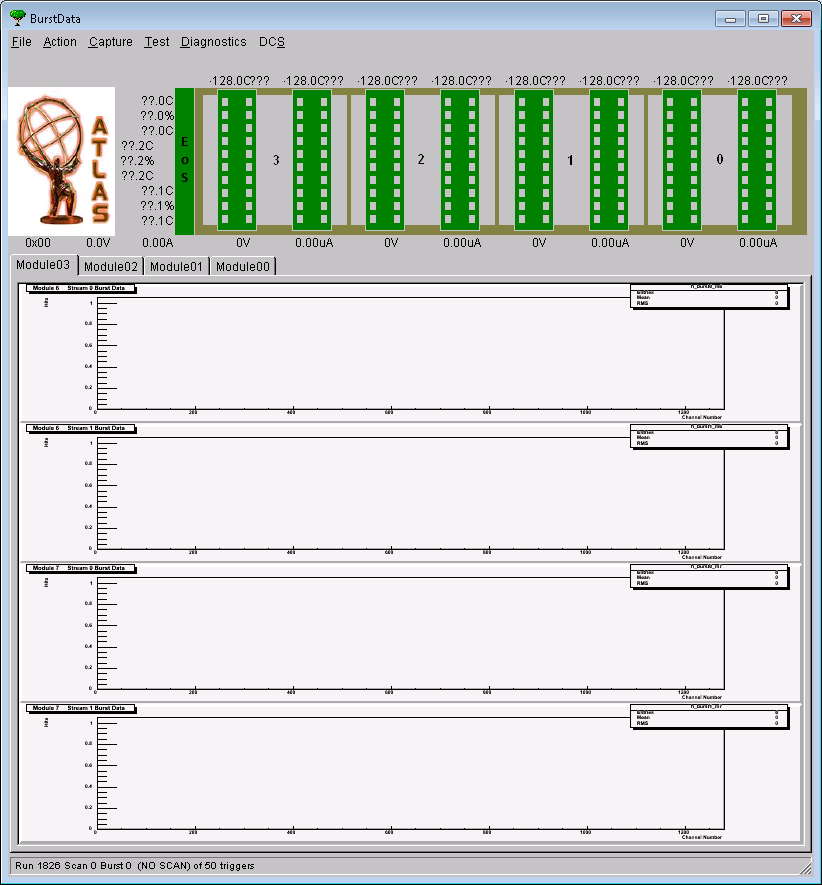}
\caption{The Burst Data window of the new SCTDAQ 
software.}\label{fig:4-BurstData}
\end{center}
\end{figure}

The software runs on either Windows or Linux computers, in a ROOT 
environment~\cite{ROOTfw}. It consists of compiled libraries supporting the
various interfaces with the devices and a set of ROOT macros that use these
libraries to implement the tests. The supported devices range from VME crates
to commercial power supplies. 

Configuration files and ROOT macros are used to control the power supplies or 
VME crates in use, modules connected to the HSIO and parameters of the modules.
Results of the tests are saved to raw ROOT files, which are usually processed
and the final output are text files and, for some tests, plots saved in 
PostScript or PDF files. The environmental variables, such as hybrids 
temperatures and sensors leakage currents are also recorded.

The ROOT macros enable flexible configuration and setup of tests. Automation
is possible and frequently used for long term tests, scans and varying 
parameters of the tests. Interlocks are used to power off the hybrids in case
of a cooling failure, thus preventing the destruction of the devices when 
running unattended tests.

It is possible to include support for additional devices by means of a National 
Instruments 
VISA driver that is included in the software, for equipment supporting this 
system. Other devices that require a specific driver may also be supported
with an extra effort.

\subsection{Electronic Equipment}

The testing facility is equipped with the necessary material to run
tests on upgrade modules and staves, and to do minor repairs, modifications
and adjustments.

Two computers control the two HSIO systems that are available. One of the
HSIO boards is used with single modules, either in the clean room
or in the radiation zone. The second HSIO is normally used to test stavelets. 

Several commercial and non-commercial power supplies are available in the room.
A Keithley 2410 source meter is used to bias the sensors, controlled through
its GPIB or RS-232 interfaces. A variety of TTi power supplies are used to 
power the HSIO and the modules or stavelets, also controlled through a GPIB
interface. A custom current source used to power hybrids on the chain of
hybrids configuration of a serially powered module or 
stavelet~\cite{Villani2011238} can be controlled through USB. Last, a VME
crate holding SCTHV boards which are the pre-production series for
the SCT high voltage have four channels and are used to bias the sensors on
stavelets~\cite{SCTHV}.

The chance of damaging the modules is minimized
by means of an interlock system developed at the University of 
Cambridge~\cite{CambridgeInterlock} that monitors the temperature of the 
devices. In case of failure of the cooling plant, the power to the modules is 
interrupted. This way, unattended, long term tests can be safely performed.

An optical table, assembled in the University of Freiburg, is used to hold
an X-Y motion stage, covered by a light-tight enclosure built at CERN. 
Inside the enclosure, modules and stavelets can be placed to be tested in a
dark environment. The motion stage can be used to map the electromagnetic 
field of a DC-DC stavelet, inject electromagnetic noise on a stavelet at 
different positions and inject charge on the strips with a laser.

\subsubsection{Single module DAQ stand}

This setup uses an HSIO board, powered by a linear power supply set to $12\V$,
and a Huber chiller with a $20\%$ glycol-water mix. A Julabo chiller has also 
been used to test single modules. This chiller uses an oil based coolant 
which can be cooled below $-20\degC$. Low temperature tests down to $-20\degC$
have been performed on a single module.

A vacuum pump is used to hold the module in the aluminium block, in order to 
have a good thermal contact between the module and the coolant. In addition, an
interlock monitors NTC temperature sensors on the hybrids. If the 
temperature goes
above the set threshold (usually around $40\degC$), the hybrid power is turned 
off. This may happen in the event of a cooling failure or a vacuum problem that
causes the module to lose thermal contact with the cooling block.

The single modules are tested inside a small freezer that is used to keep 
the light from falling on the biased silicon sensor.

A compressed air line is available and can be connected to the freezer in order
to keep the relative humidity in the freezer low, thus avoiding condensation 
when 
testing the module at low temperatures.

\subsubsection{Stavelets DAQ stand}

This setup is similar to the single module DAQ system. However, some pieces
are slightly different:

\begin{itemize}
\item A VME crate with two pre-production SCTHV cards controlled from the
computer, via a PCI-VME card. These cards have four output channels each,
which permit the biasing of up to eight modules without multiplexing.
\item There is no extra power supply to power any LVDS buffer board. The 
stavelet has the LVDS buffers on the End of Stave card, powered through the
HSIO.
\item Dual channel power supply to provide power to two stavelets at the
same time.
\item The interlock monitors the inlet and outlet pipes of the stavelets. Small
increases in the temperature of either is interpreted as a cooling failure.
\end{itemize}

\subsection{Cooling Plant}

\begin{figure}[!htb]
\begin{center}
\subfigure[Old Huber chiller]{
  \includegraphics[scale=0.9]{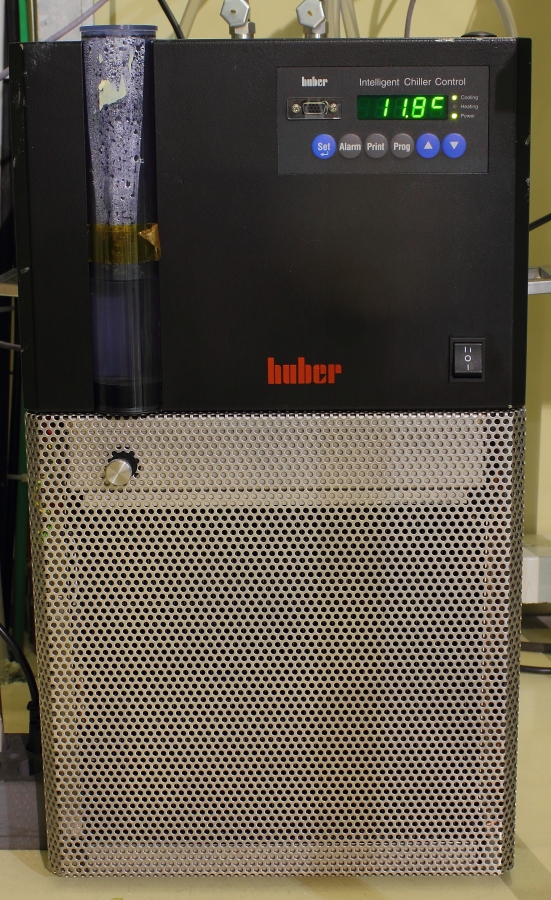}
}
\subfigure[Huber Unichiller UC015]{
  \includegraphics[scale=0.9]{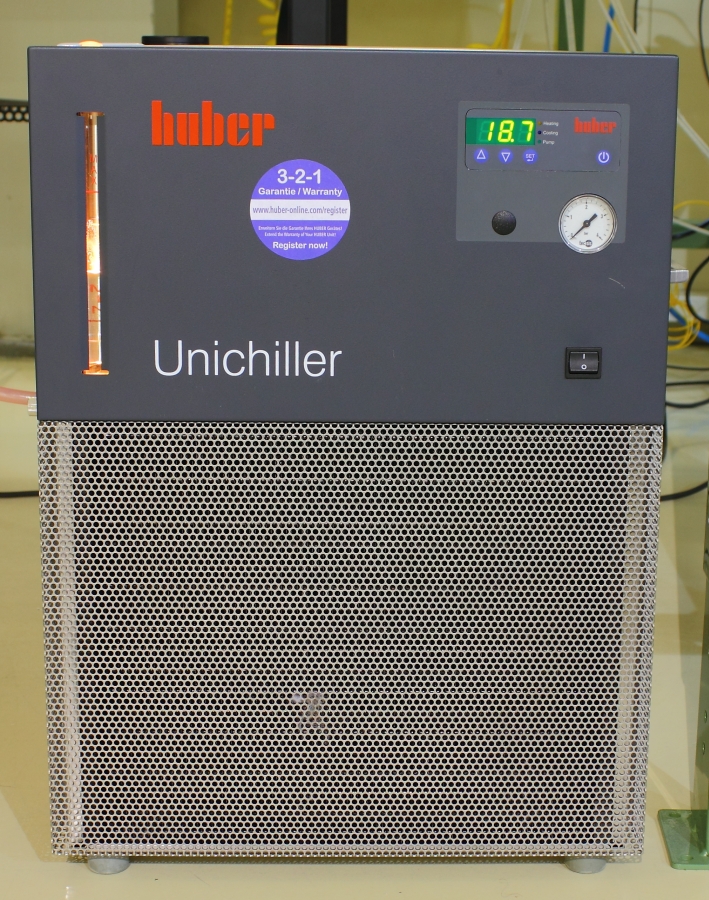}
}
\subfigure[Julabo FP50 chiller]{
  \includegraphics[scale=0.9]{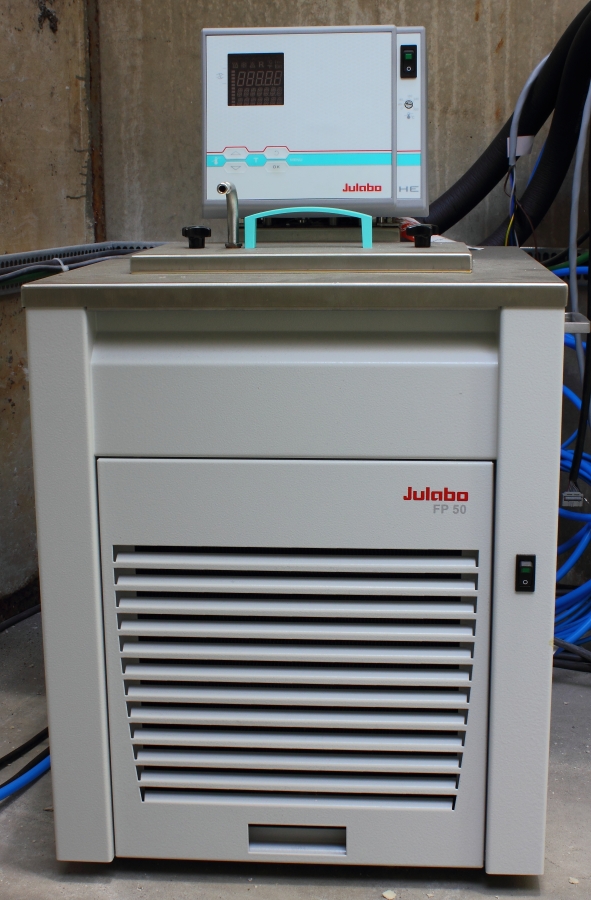}
}
\caption{Chillers used in the B180 test facility.}\label{fig:5-chillers}
\end{center}
\end{figure}

There are three chillers at the testing facility. Two of them are inside 
the clean room, they both are Huber chillers with a water-glycol
mix at a $20\%$ glycol concentration. The single module setup uses an old
version of the Unichiller 007-MPC and the stavelets setup has a Unichiller
UC015.

The third chiller is located in the radiation bunker, it is a Julabo FP50
running with an oil-based coolant. It has been operated
at $-20\degC$ with a
single module in the clean room. The three chillers are shown in 
Figure~\ref{fig:5-chillers}.

The typical temperature setting of the chillers in the clean room is $12\degC$.
The temperature and humidity are
monitored using SHT71 sensors in order to avoid going below the dew point, 
thus avoiding 
condensation. 

Lower temperature settings are used for some tests and, for 
these tests, dry air is flushed into the modules or stavelets enclosures to 
prevent condensation.

\subsection{Performance Tests and Calibration}

This section describes the typical performance tests and calibration functions
that are performed on the modules and stavelets. The tests are typical for 
describing the silicon microstrips devices performance, and they are the same
as the ones performed on the SCT modules. 

The calibration functions are normally performed before the tests, in order
to have adequate timing and response variability for all the channels.

\subsubsection{Strobe Delay}

The Strobe Delay (SD) test is one of the calibrations that have to be done 
before testing the modules. It varies the phase of the charge injection 
relative to the
Level 1 Accept (L1A) command. An optimal setting of the SD for each 
chip is important for the accuracy of the threshold calibration. The test 
consists in scanning the timing while injecting a charge. To achieve $100\%$ 
hit efficiency when the timing is correct, while at the same time minimizing
noise, the injected charge is $4~{\rm fC}$
and the threshold is set to $2~{\rm fC}$~\cite{Bilge07}. 

The delay of the strobe (with respect to the clock phase) can 
be tuned 
for individual chips within a range of around $50~{\rm ns}$, controlled by a 
6-bit SD register (1 DAC step $\approx 0.8~{\rm ns}$). Other SD
ranges are possible, for which 1 DAC step is equal to $1$ or $1.3~{\rm ns}$ 
respectively. The SD is actually recorded in an 8-bit register but 
only the first 6 bits of this are used, resulting in a delay range of 0 to 
63~\cite{TwikiSCTDAQDown}.

\begin{figure}[!htbp]
 \begin{center}
  \includegraphics[scale=0.6]{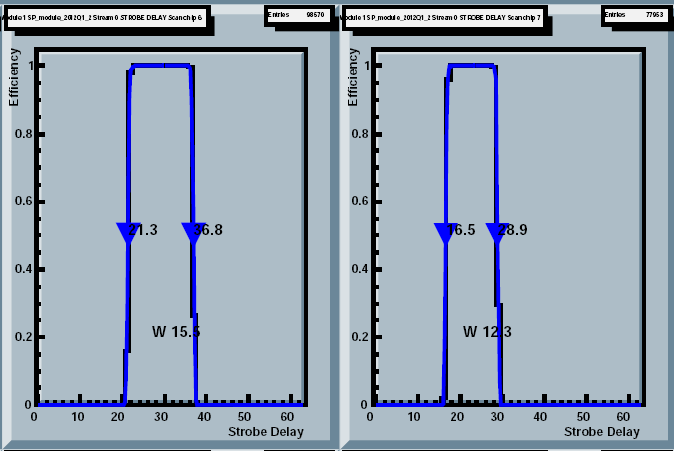}
  \caption{Strobe Delay data for the fit on two adjacent chips.}\label{fig:strobedelay_chip}
 \end{center}
\end{figure}

The SD test does a scan of the SD setting, from 0 to 63,
sending 200 triggers at each SD value. The resulting average hit efficiency 
for all channels of a single chip as a function
 of the SD setting then looks as shown in 
Figure~\ref{fig:strobedelay_chip}, where the efficiency for two adjacent chips 
is shown. The resulting SD settings are calculated 
taking the SD value that is at the $25\%$ of the working range, defined as the 
interval between rise and fall times of the curve

\subsubsection{Trim Range}

The channel trims are always set to minimize the variations of the response
between channels. An untrimmed hybrid is spotted when the offset shows big 
variations from chip to chip and within the same chip as well. 

The trim range scan injects a $1~{\rm fC}$ charge for all events and does 
threshold scans for different trim DAC settings. The test ultimately generates
optimal Trim Range settings that can be applied to each of the channels.

The analysis involves determining how many channels are trimmable for each 
trim range setting. Then the trim range setting that maximizes the number of 
trimmable channels is chosen~\cite{Bilge07}.

\subsubsection{Three Point Gain and Response Curve}

The 3-point gain test consists in performing threshold scans at three different
injected charges \cite{Bilge07}. There are two typical 3-point gain scans, one 
with a central injected charge of $1~{\rm fC}$ and another one that injects 
charges around $2~{\rm fC}$. 
They correspond to a fairly linear region of the response curve and the 
injected charges are $\left\{ 0.52, 1.0, 1.48\right\}~{\rm fC}$ and 
$\left\{ 1.5, 2.0, 2.5\right\}~{\rm fC}$, respectively. 

The purpose of this test is to verify the analogue performance of the modules.
The performance is measured with respect to the Equivalent Noise Charge 
($e^{-}{\rm ENC}$), 
which is measured in electrons ($e^-$). This$\ENC$ value is the result of 
dividing the output 
noise by the gain and converting the input noise in ${\rm fC}$ to electrons 
($1~{\rm fC} \approx 6242~e^-$). 

The output noise, which is measured in ${\rm mV}$, is the 
variance of the distribution of the threshold for a particular injected charge.
The gain, measured in $~{\rm mV}/{\rm fC}$, is calculated from a linear 
fit of the three threshold scan points~\cite{Bilge07}.

The extrapolated offset is obtained from the linear fit and its spread 
is an indicator of the trimming uniformity.
 
\begin{figure}[!htbp]
 \begin{center}
  \includegraphics[scale=0.5]{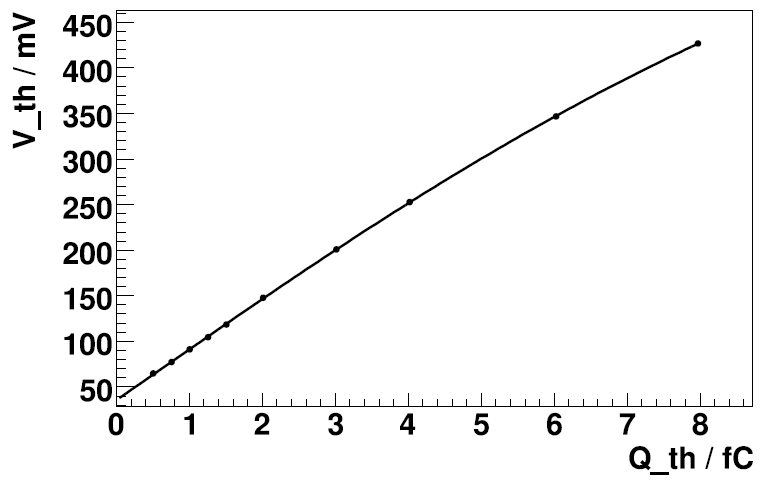}
  \caption{Quadratic fit to a response curve.}\label{fig:4-responsecurve}
 \end{center}
\end{figure}

The response curve is an extension of the three point gain test to $10$ scan 
points, where the injected charges are set to 
$0.5, 0.75, 1.0, 1.25, 1.5, 2, 3, 4, 6$ and $8~{\rm fC}$. In the current 
SCT, this test is used to get a more precise measurement of the gain and
offsets that can be used to update the configuration~\cite{Bilge07}. A 
quadratic fit to a response curve can be seen in 
Figure~\ref{fig:4-responsecurve}. This plot shows the non-linearity of the 
pre-amplifier in the chips as the input signal increases.

\subsubsection{Double Trigger Noise}

The goal of the Double Trigger Noise (DTN) test is to assess the module 
susceptibility to electrical and optical pickup during the 
read-out~\cite{Bilge07}. 

The test is performed by sending two L1A triggers 
to the modules, separated by a specified number of $40{\rm MHz}$ clock periods. 
The first 
event of each pair is discarded, leaving the second event to be recorded. The 
spacing of the two triggers is varied from 120 to 160 clock periods, to 
determine the occupancy of the modules at various points during the read-out 
cycle. 

A trigger spacing of 132 is of particular interest as this is equal to 
the depth of the read-out pipeline: the second event records the occupancy of 
the module as the read-out cycle of the first event 
starts~\cite{Phillips:2007db}.

The typical DTN test is done with three 
different thresholds: $0.5~{\rm fC}$, $0.75~{\rm fC}$ and $1~{\rm fC}$. Usually 
healthy modules and DAQ setups show no hits for $0.75~{\rm fC}$ and
$1~{\rm fC}$ thresholds and a rather low number of hits for the $0.5~{\rm fC}$
threshold.

\subsection{Modules and Stavelets}

Currently, there are three devices available for testing in the B180 facility:
a serially powered single module, a DC-DC powered stavelet and a serially 
powered stavelet.

The serially powered (SP) module arrived at the beginning of 2012 from the
University of Liverpool.

The DC-DC powered stavelet arrived at the beginning of 
2012 from the Rutherford Appleton Laboratory (RAL), where it was assembled.
The serially powered stavelet arrived at the beginning of 2013 from
RAL.

A DC-DC powered stave with twelve modules on one side was assembled at RAL and 
has been tested there in the context of this thesis work.

The results presented in this chapter refer to these four devices.

\begin{figure}[!htb]
\begin{center}
\includegraphics[scale=0.8]{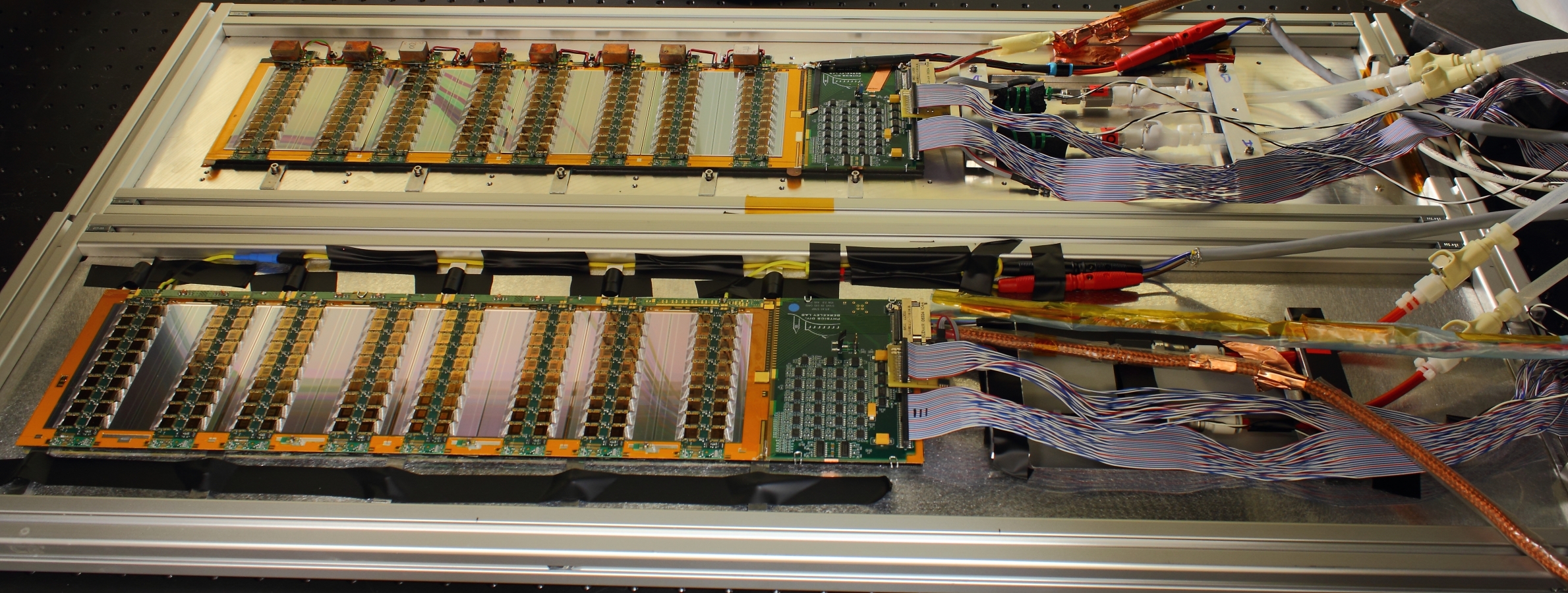}
\caption{DC-DC stavelet (top) and SP stavelet 
(bottom).}\label{fig:5-stavelets_in_enclosure}
\end{center}
\end{figure}

A picture of the serially powered module can be seen in 
Figure~\ref{fig:4-SPmodule}, and pictures of the two stavelets are shown in 
Figure~\ref{fig:4-stavelets}. Both stavelets are shown together
in Figure~\ref{fig:5-stavelets_in_enclosure}, with the four modules on the left
side of the picture, the End of Stave card around the centre of the picture and
the services (powering, cooling and bias) on the right side.

All three have ABCn25 based hybrids using BCC boards. The sensors are from
the ATLAS07 production batches, which means they all have two rows of axial 
strips and two rows of strips with a stereo angle of $40\mrad$. The SP module 
has a FZ2 sensor, while the two
stavelets have a mixture of FZ1 and FZ2 sensors.

The four modules mounted on the DC-DC stavelet come from the University of 
Liverpool. The characteristic parameters of the sensors are summarized in 
Table~\ref{tab:5-sensors_dcdc_stavelet}. The parameters were measured before 
the module assembly, in a clean room at $22\degC$. Both the leakage current
and the capacitance are measured for the whole sensors, that is, for
$5120$ strips. The sensor of module LIV-M14, a FZ1 
sensor, has insufficient oxide in the guard rings, leading to transients in the
leakage current above $300\V$ bias voltage~\cite{ArgosAUW12}.

\begin{table}[!htb]
\begin{center}
\begin{tabular}{l|ccccc}
\toprule
 Module & Type & Isolation & $V_{\rm dep}$ & $C@250\V$ & $I_{\rm leak}@250\V$ \\
\hline
0 (LIV-M16) & FZ2 & P-stop, $10^{13}\cm^{-3}$ & $145\V$ & $2.63\nF$ & $10.0\uA$ \\ 
1 (LIV-M14) & FZ1 & P-stop, $4\times 10^{12}\cm^{-3}$ & $240\V$ & $2.95\nF$ & $0.24\uA$ \\ 
2 (LIV-M15) & FZ2 & P-stop, $10^{13}\cm^{-3}$ & $130\V$ & $2.63\nF$ & $9.29\uA$ \\ 
3 (LIV-M17) & FZ2 & P-spray, $2\times 10^{13}\cm^{-3}$ & $130\V$ & $2.63\nF$ & $6.62\uA$ \\ 
\bottomrule
\end{tabular}
\caption[Characteristics of the sensors on the DC-DC
stavelet.]{Characteristics of the sensors on the DC-DC
stavelet: wafer type, isolation structures 
of the strips and concentration dose, 
depletion voltage, capacitance with $250\V$ bias and leakage current with
$250\V$ bias.}\label{tab:5-sensors_dcdc_stavelet}
\end{center}
\end{table}

The modules used to assemble the serially powered stavelet
came from both the University of Liverpool and the University of Cambridge. The
summary of the parameters for these sensors is shown in 
Table~\ref{tab:5-sensors_sp_stavelet}.

\begin{table}[!htb]
\begin{center}
\begin{tabular}{l|ccccc}
\toprule
 Module & Type & Isolation & $V_{\rm dep}$ & $C@250\V$ & $I_{\rm leak}@250\V$ \\
\hline
0 (CAM-M2) & FZ1 & P-stop,   $4\times 10^{12}\cm^{-3}$ & $202\V$ & unknown & $0.31\uA$ \\ 
1 (LIV-M21) & FZ2 & P-spray, $2\times 10^{12}\cm^{-3}$ & $140\V$ & $2.61\nF$ & $6.63\uA$ \\ 
2 (LIV-M12) & FZ2 & P-spray, $2\times 10^{12}\cm^{-3}$ & $110\V$ & $2.63\nF$ & $16\uA$ \\ 
3 (CAM-M4) & FZ1 & P-spray, $4\times 10^{12}\cm^{-3}$ & $202\V$ & unknown & $0.78\uA$ \\ 
\bottomrule
\end{tabular}
\caption[Characteristics of the sensors on the serially
powered stavelet.]{Characteristics of the sensors on the serially
powered stavelet: wafer type, isolation structures 
of the strips and concentration dose, 
depletion voltage, capacitance with $250\V$ bias and leakage current with
$250\V$ bias.}\label{tab:5-sensors_sp_stavelet}
\end{center}
\end{table}

\subsection{Tests Description}

The performance tests were done under controlled and measured conditions.
Various conditions were applied to the modules and stavelets, such as different
coolant temperatures,  sensor bias,  hybrid power parameters and
external electromagnetic interference.

For some of the tests, multiple measurements were done at each setting, in 
order to ensure temporal stability of the tests conditions. In these cases, 
the average
is calculated as well as the deviation, represented as error bars in the plots.
There usually is a small variation of the measured noise values between 
consecutive tests under the same conditions.

The typical performance test is the three point gain test, centred at $1\fC$.
It is used for parameter scans and long term stability tests, through the
automation allowed by the SCTDAQ software. For some of the tests, a Double 
Trigger Noise (DTN) test is also run. Both tests were described in the previous
chapter. 

Every time a noise measurement is started for a particular test, a
SD is run in order to always have the optimal timing between the 
charge injection and the read-out. This is particularly important when the
coolant temperature is changed, because the optimal SD settings are 
dependent on the temperature.

\section{Noise Model}\label{sec:5-noisemodel}

A noise model for the SCT modules, using the ABCD3TA chip, had already been 
developed~\cite{Burdin:1237817,SCTnoiseThesis}. The main difference between 
the ABCD3TA and
the ABCn25 chips is the input to the front-end chip, which changes from a 
bipolar transistor to a MOSFET. The noise models differ for both.

Using the EKV model for the MOSFET transistors, another noise model was 
developed for the first precursor of the ABCn25, which uses the same chip
architecture and a similar fabrication process. This model, presented 
in~\cite{1589268}, is applied to the current ABCn25 chip
which is used in the modules and stavelets tested.

\begin{equation}
 {\rm ENC_{Id}} = \frac{e^3}{12\sqrt{3}}\sqrt{\frac{4k_B T_{\rm chip} 
 n \gamma}{g_m}} \frac{C}{\sqrt{t_{\rm peak}}}\frac{1}{q}
\end{equation}

The Gate-Induced Current (GIC) noise component is

\begin{equation}
 {\rm ENC_{GIC}} = \frac{e^3\sqrt{2}}{9\sqrt{15}}\sqrt{\frac{k_B T_{\rm chip} 
 \gamma}{n g_m}} \frac{C_{\rm OX}}{\sqrt{t_{\rm peak}}}\frac{1}{q}
\end{equation}

The noise from the feedback transistor is

\begin{equation}
 {\rm ENC_{If}} = \frac{e^3\sqrt{2}}{9\sqrt{3}}\sqrt{\frac{K_a 
 }{WL}} \frac{C}{\sqrt{C_{\rm OXU}}}\frac{1}{q}
\end{equation}

The correlation term is

\begin{equation}
 {\rm ENC_{corr}} = \frac{e^3}{18}\sqrt{\frac{k_B T_{\rm chip} 
 \gamma}{g_m}} \frac{\sqrt{C_{\rm OX}C}}{\sqrt{t_{\rm peak}}}\frac{1}{q}
\end{equation}

The flicker noise due to each feedback transistor is

\begin{equation}
 {\rm ENC_{thermal}} = \frac{e^3}{18}\sqrt{\frac{5}{3}}
 \sqrt{k_B T_{\rm chip}n\gamma g_m^{f} t_{\rm peak}}\frac{1}{q}
\end{equation}

Last, the noise due to the leakage current of the sensor is

\begin{equation}
 {\rm ENC_{leak}} = \frac{e^3}{18}\sqrt{\frac{5}{3}}
 \sqrt{k_B T_{\rm chip}n\gamma g_m^{f} t_{\rm peak}}\frac{1}{q}
\end{equation}

And the total noise is the sum in quadrature of those terms:

\begin{equation}
 {\rm ENC^2 = ENC_{Id}^2 + ENC_{GIC}^2 +
 ENC_{If}^2 + ENC_{corr}^2 + ENC_{thermal}^2
   + ENC_{\rm leak}^2 }
\end{equation}

Where:

\begin{itemize}
 \item $e$ is the Napier constant.
 \item $k_B$ is the Boltzman constant.
 \item $T_{\rm chip}$ is the chip temperature.
 \item $W$ and $L$ are the width and length of the transistor channel, 
 $W = 320\mum$ and $L = 0.5\mum$.
 \item $n = 1.45$ is the slope factor, technology dependant.
 \item $\gamma$ is a parameter of the MOSFET that depends on the working mode.
 Typically $\gamma = 2/3$ for the working range.
 \item $C$ is the total capacitance at the input of the pre-amplifier.
 \item $q$ is the electron charge.
 \item $g_m$ is the transconductance of the input transistor, and $g_m^f$ is 
 the transconductance of the transistor used as feedback resistor.
 \item $t_{\rm peak}$ is the peaking time, which for the ABCn25 is $22\ns$.
 \item $C_{\rm OX}$ is the gate oxide capacitance:
 $$C_{\rm OX} = WL\frac{\epsilon_r \epsilon_0}{t_{\rm OX}}$$
 With $t_{\rm OX}$ the effective gate oxide thickness and
 $$C_{\rm OXU} = \frac{\epsilon_r \epsilon_0}{t_{\rm OX}}$$
 is the gate oxide capacitance per unit area.
 \item $K_a$ is a technology dependant parameter, equal to $6\times 10^{-27}$ 
 for NMOS and $10^{-27}~{\rm C/m}^2$ for PMOS.
\end{itemize}

Using the following values for the parameters of the final design of the chip:

\begin{center}
\begin{tabular}{c|c}
\toprule
 Parameter & Value \\ \hline
 $t_{\rm OX}$ & $6\nm$ \\
 $I_d^f$ & $0.8\uA$ \\
 $I_d$ & $140\uA$ (nominal)\\
 $C_a$ & $0.1\pF$ \\
 $C_f$ & $0.14\pF$ \\
 $C_{str}$ & $1.5\pF$ \\
 \bottomrule
\end{tabular}
\end{center}

The total capacitance is:

$$C = C_a + C_f + C_{str} + C_{\rm sensor}$$

Where the sensor capacitance depends mostly on the strip length. It is in the 
order of $2.6\pF$ for short strips ($2.36\cm$) and $5\pF$ for long strips 
($4.78\cm$), in fully depleted sensors.

\begin{figure}[!htbp]
 \begin{center}
  \includegraphics[scale=0.5]{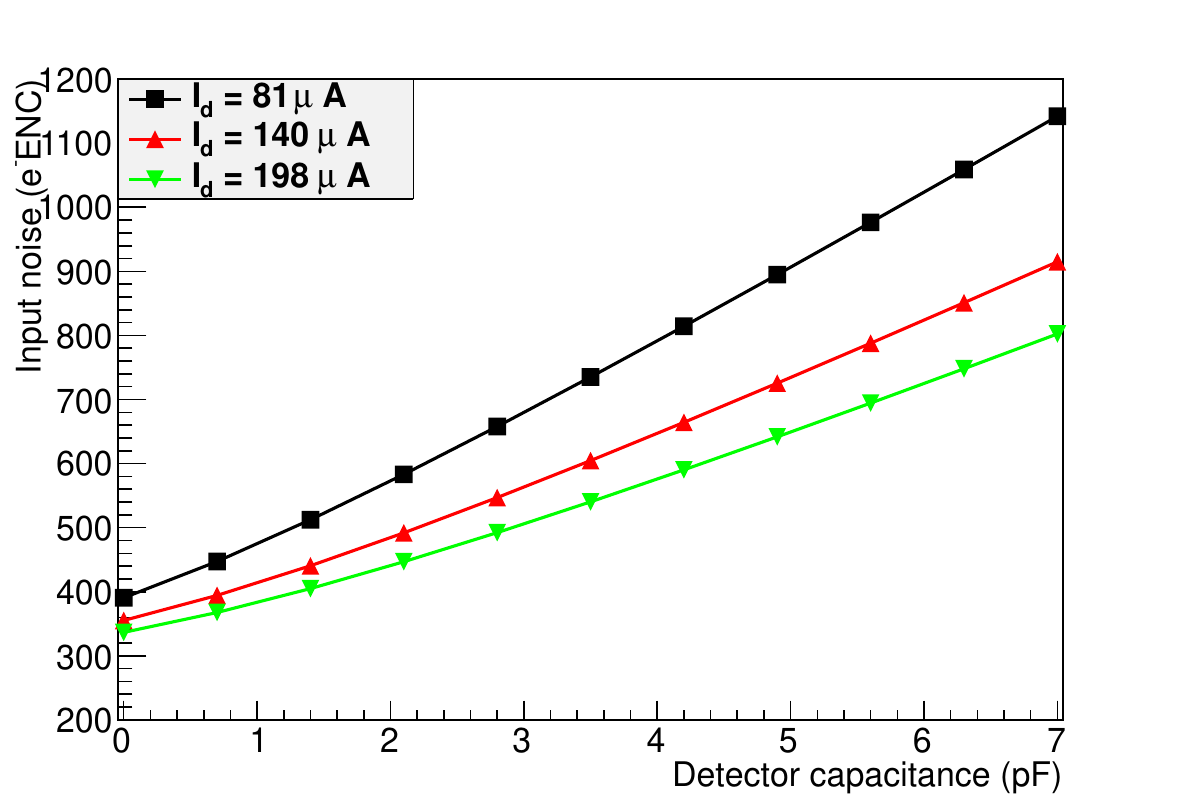}
  \caption{Result of the noise model for the ABCn25 chip with varying input 
  capacitance.}\label{fig:5-ABCn25model}
 \end{center}
\end{figure}

Figure~\ref{fig:5-ABCn25model} shows the noise evolution with detector 
capacitance, as calculated from the model, with three different bias currents 
for the pre-amplifier stage. This is the same result as the one shown 
in~\cite{Dabrowski:1273530}, confirming the validity of the model above.

\begin{figure}[!htbp]
 \begin{center}
  \includegraphics[scale=0.55]{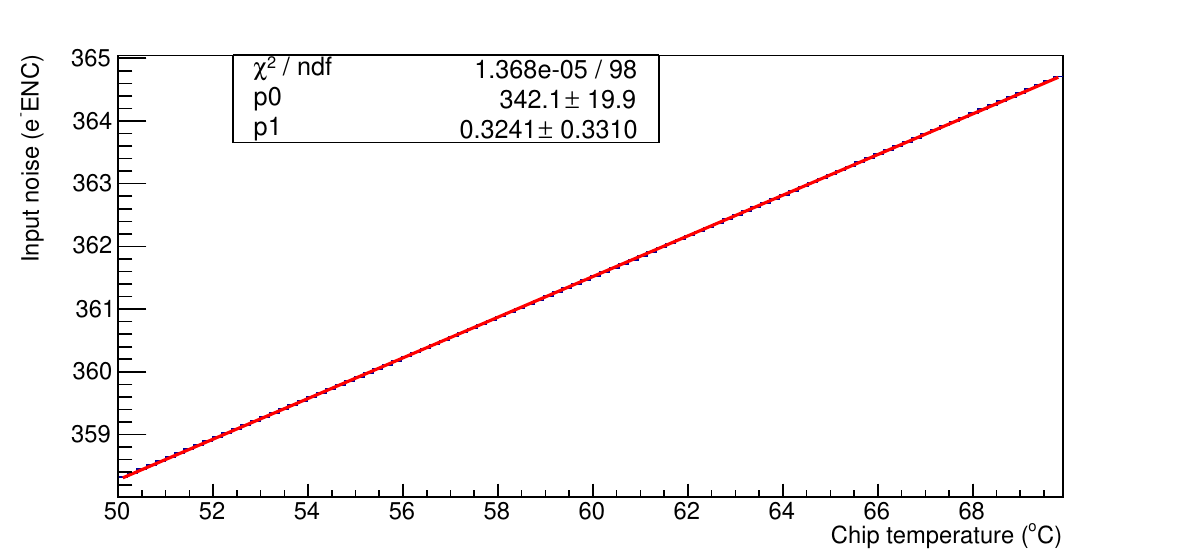}
  \caption{Result of the noise model for the ABCn25 chip with varying 
  temperature for unbonded channels.}\label{fig:5-ABCn25model_unbonded}
 \end{center}
\end{figure}

The model was also used with varying temperature values, to extract the
temperature dependence of the noise. This dependence is highly dependant on the
values of the parameters used, such as the bias current, which is nominally
$140\uA$ and this is the value used for the model.

Figure~\ref{fig:5-ABCn25model_unbonded} shows the noise plot with changing
temperature for unbonded channels, together with the fit result. The variation
of the noise with temperature is

\begin{equation}
\overline{\Delta {\rm ENC}/\Delta T}_{\rm unbonded} = 0.3241\pm 0.3310~e^{-}{\rm ENC/K} 
\end{equation}

\begin{figure}[!htbp]
 \begin{center}
  \includegraphics[scale=0.55]{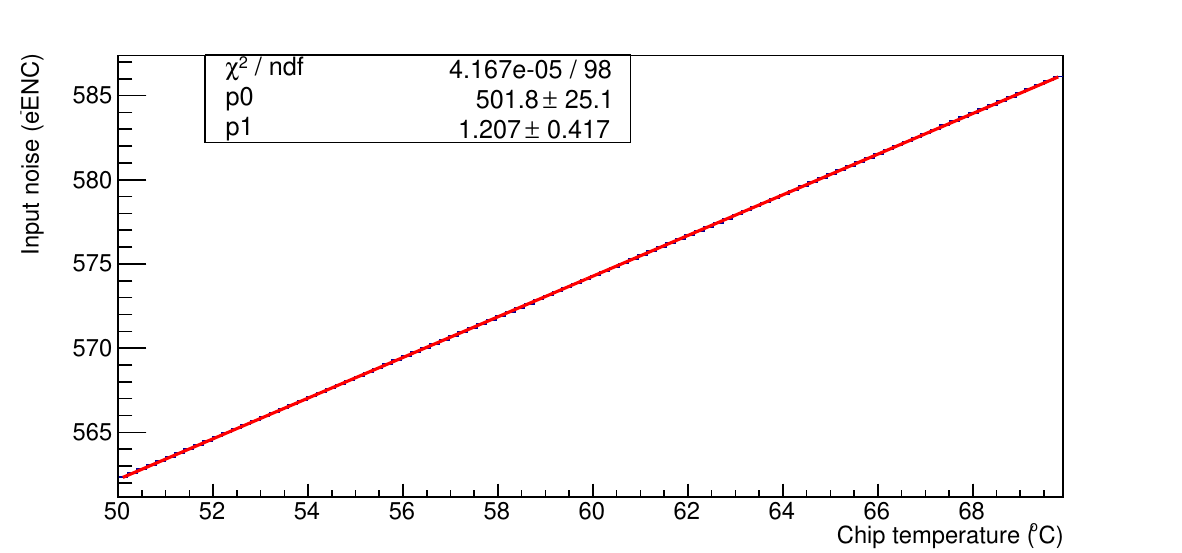}
  \caption{Result of the noise model for the ABCn25 chip with varying 
  temperature for short strips ($2.36\cm$).}\label{fig:5-ABCn25model_bonded}
 \end{center}
\end{figure}

Figure~\ref{fig:5-ABCn25model_bonded} shows the same plot for short strips,
$2.36\cm$ long, with a total capacitance around $2.82\pF$. The result, in this
case, is

\begin{equation}
\overline{\Delta {\rm ENC}/\Delta T}_{\rm short strip} = 1.207\pm 0.417~e^{-}{\rm ENC/K} 
\end{equation}

The capacitance value of $2.82\pF$ was chosen because it is the measured value
for the serially powered module that has been tested. A comparison between the 
model and experimental data is shown in section~\ref{sec:5-varyingtemperature}.

\section{Single Module Performance Tests}\label{sec:5-spmodule}

The single module tested at CERN uses serial powering in a chain of hybrids
configuration. It has a FZ2 sensor
with a leakage current in the order of $50\uA$ at $200\V$ and $20\degC$. The
module was assembled at the University of Liverpool and arrived at CERN in
January 2012.

The first measurements performed on this module were three point
gain tests to have the reference noise for this module. The chiller temperature
was set to $12\degC$, the serial powering current was set to $5\A$ and the 
sensor bias to
$200\V$.

\begin{figure}[!htbp]
 \begin{center}
  \includegraphics[scale=0.75]{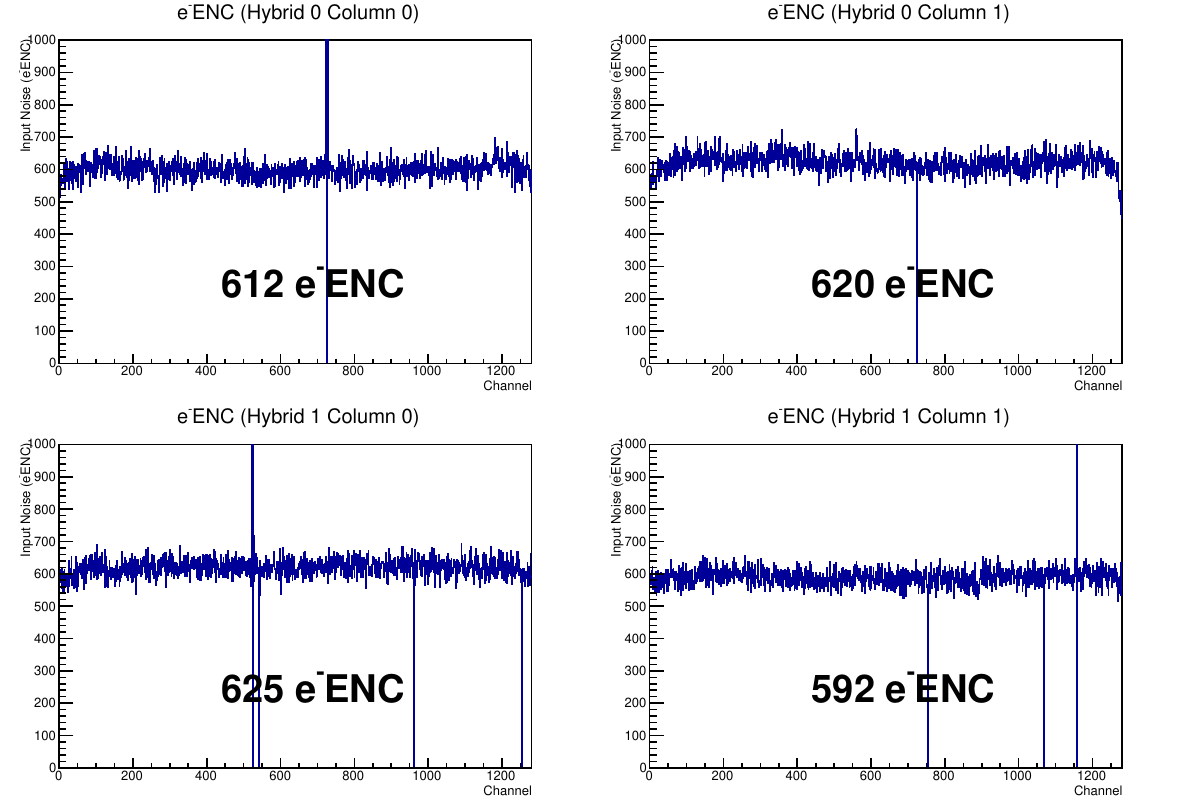}
  \caption{Reference noise ($e^{-}{\rm ENC}$) for the serially powered 
  module.}\label{fig:5-SPmoduleENCref}
 \end{center}
\end{figure}

\begin{figure}[!htbp]
 \begin{center}
  \includegraphics[scale=0.75]{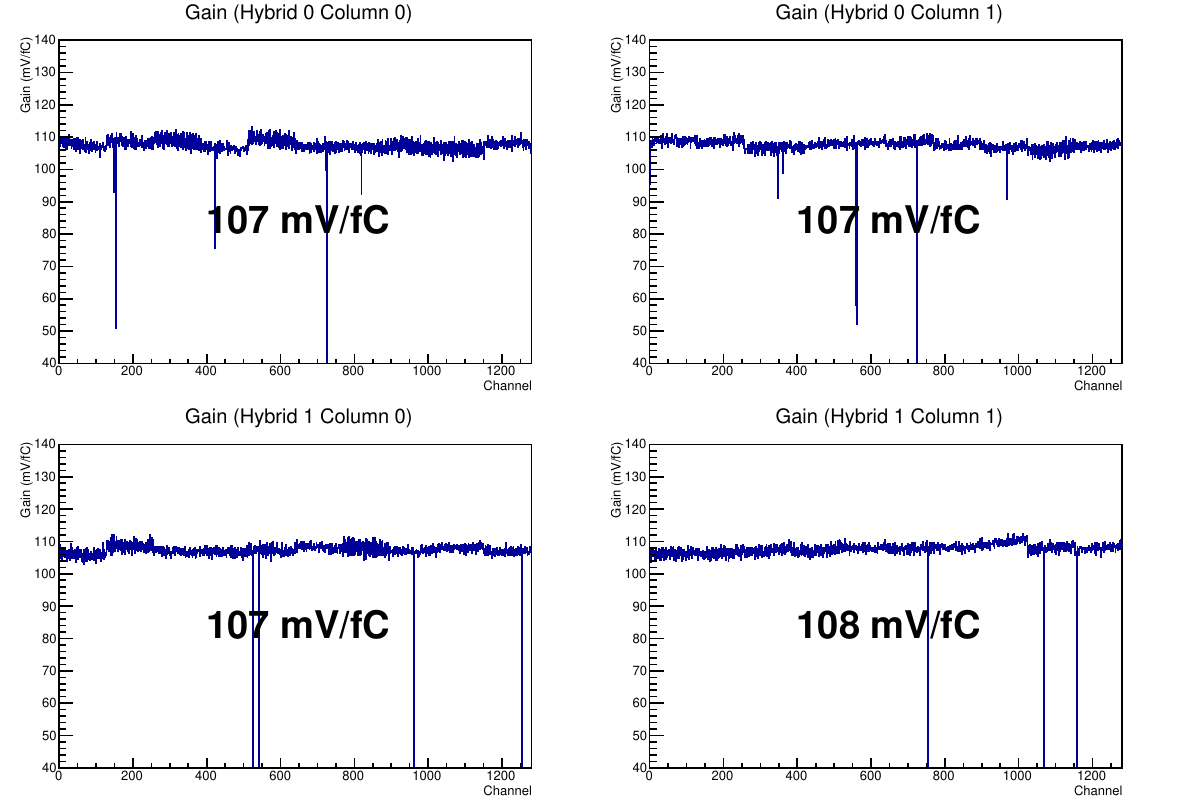}
  \caption{Reference gain (${\rm mV/fC}$) for the serially powered 
  module.}\label{fig:5-SPmoduleGainref}
 \end{center}
\end{figure}

Figure~\ref{fig:5-SPmoduleENCref} shows the input noise, measured in
electrons, for the channels of each column of chips of the module. 
Figure~\ref{fig:5-SPmoduleGainref} shows the gain (measured in$\mV$ per$\fC$) 
plot for all the channels. 

The values shown on each plot are the average values for each column. All
results in the next sections are presented as average values for
the chips, columns or whole modules.

Some of the channels show high noise or very low noise. Usually noise around
$400\ENC$ means the channel is unbonded and zero noise means the channel
is masked or dead on the chip. Those kinds of defects are normally masked when 
the trim range test is performed.

The results presented in the following sections are all with the correct trim 
range settings for the single module and the two stavelets.

\subsection{Varying Temperature}\label{sec:5-varyingtemperature}

This test is conceived as a means of extracting the variation of the module 
noise with temperature. The chiller is set to different temperature settings, 
then the noise is measured several times with three point gain tests. By using 
at
least three different temperature settings and fitting to a linear function, it 
is possible to extract the noise change rate with temperature.

The temperature of the coolant, $T_{\rm chiller}$, was set at various values, 
the temperature of one of the hybrids was monitored and noise measurements were 
repeated. The sensor 
bias was set to $150\V$ for all the noise measurements.

The temperature settings and measured hybrid temperature correspondence are 
shown in Table~\ref{tab:5-DNDT_temperatures} (temperatures are in Celsius). The 
temperature 
difference between both is not constant due to the effect of the dry air 
flowing inside the module enclosure and the fact that the coolant is heated 
slightly in the pipe due to the temperature in the room.

\begin{table}[!htb]
\begin{center}
 \begin{tabular}{c|cccccc}
 \toprule
  $T_{\rm chiller}[^{\circ}{\rm C}]$ &  $-6$ & $0$ & $6$ & $12$ & $18$ & $24$ \\ \hline
  $T_{\rm hybrid}[^{\circ}{\rm C}]$ & $11.6$  & $16.0$  & $19.7$ & $25.9$ & $30.8$ & $35.7$  \\
  \bottomrule
 \end{tabular}
\caption{Coolant and hybrid temperatures in the serially 
powered module.}\label{tab:5-DNDT_temperatures}
\end{center}
\end{table} 

The temperature of the room during these tests was constant at about
$24^{\circ}{\rm C}$, and the relative 
humidity is around $40\%$. The humidity in the module enclosure was 
kept low with a dry air flow, which is at room temperature. As a consequence,
the temperature 
inside the enclosure does not entirely follow the temperature variations of the 
coolant. 

\begin{figure}[!htbp]
\begin{center}
\includegraphics[scale=0.7]{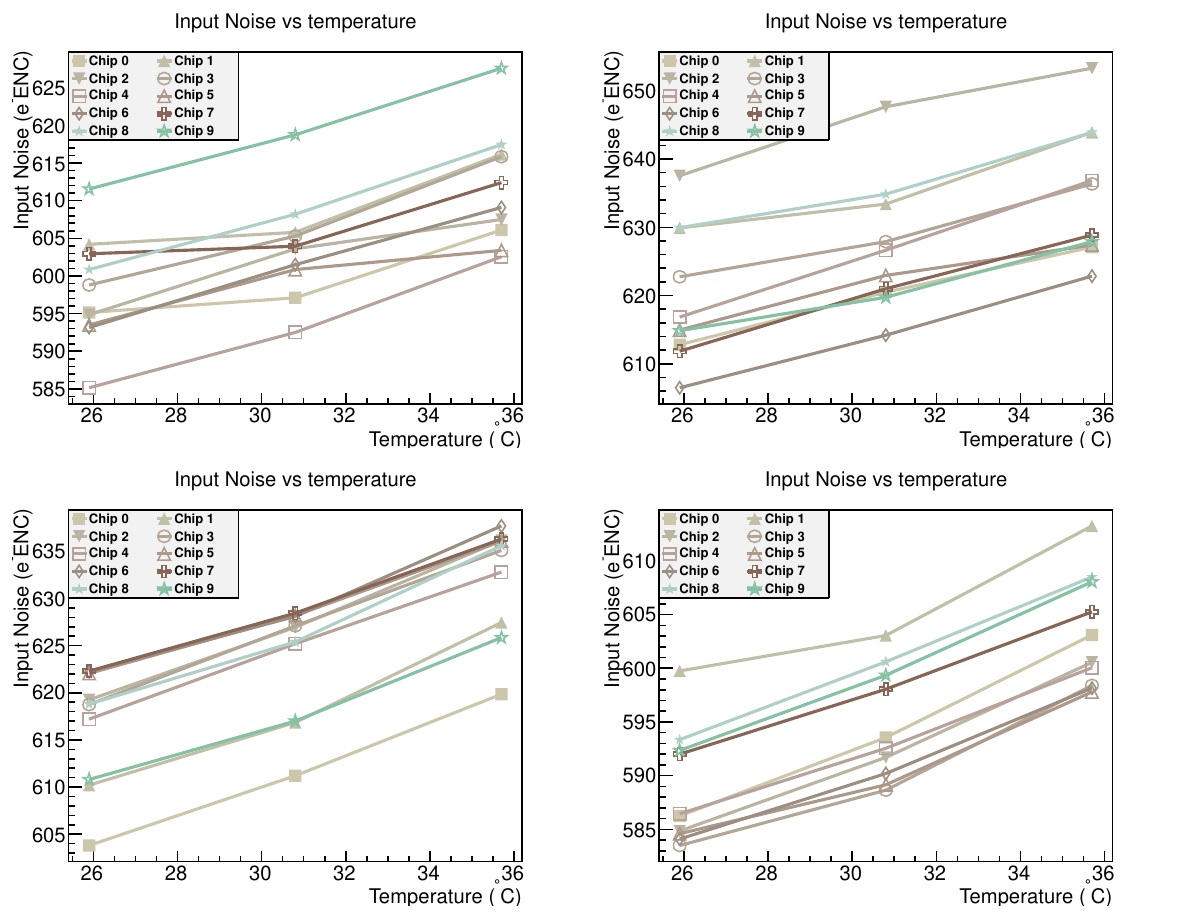}
\caption{Noise variation with temperature on the
SP module.}\label{fig:5-spmodule_noise_vs_T}
\end{center}
\end{figure}

\begin{figure}[!htbp]
\begin{center}
\includegraphics[scale=0.7]{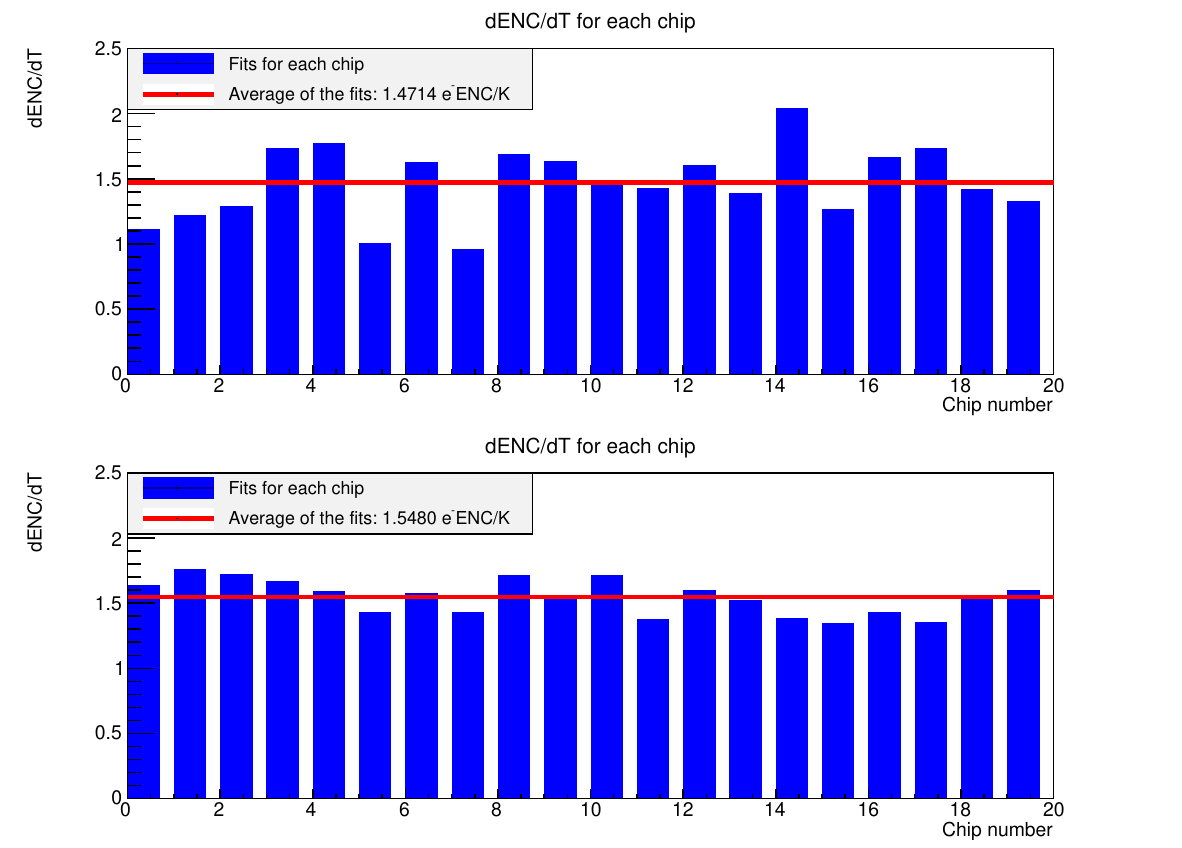}
\caption{Fit of the noise variation with temperature for all 
chips.}\label{fig:5-spmodule_DNDTfit}
\end{center}
\end{figure}

Measurements of the temperature inside the freezer and on the cooling 
pipes at the enclosure boundaries confirm that the dry air flow increases the
temperature of the air inside the freezer, causing a different
noise change with temperature for lower temperature settings than expected.
In this case, convection has a greater effect in the heat transfer when the 
temperature difference between the air and the coolant is so high.

This has the effect of the temperature measured on the hybrids not
having the same relation with the temperature on the chips.
This effect is presented in more detail below, where the very low temperature
tests are described.

Because of that effect, the lower temperature settings for this test did not 
entirely follow the trend of the upper part of the interval, in terms of noise. 
Therefore, the lower temperature settings were discarded for the study and
the only settings that were used are 
$T_{\rm chiller} = \left\lbrace 12, 18, 24\right\rbrace\degC$.

Figure~\ref{fig:5-spmodule_noise_vs_T} shows the noise behaviour with the 
measured hybrid temperature. Each subplot corresponds to a chip column and each
line shows the noise evolution on each chip.

The slope is very similar for most of the chips and for both columns on the 
same 
hybrid. Possible differences can be attributed to small temperature variations
among chips and to the noise estimation for each individual channel.

Figure~\ref{fig:5-spmodule_DNDTfit} shows the slopes for
each of the 20 chips in both hybrids. 

The average values and spreads for both hybrids are:

\begin{equation}
\overline{\Delta {\rm ENC}/\Delta T}_{\rm Hybrid~0} = 1.4714\pm 0.28~e^{-}{\rm ENC/K} 
\end{equation}

\begin{equation}
\overline{\Delta {\rm ENC}/\Delta T}_{\rm Hybrid~1} = 1.5480\pm 0.13~e^{-}{\rm ENC/K} 
\end{equation}

\begin{table}[!htb]
 \begin{center}
  \begin{tabular}{ccc}
   Module type & Model & Measured \\ \hline
   Outers & $4.06\pm 0.18\ e^{-}{\rm ENC/K}$ & $5.77\pm 0.27~e^{-}{\rm ENC/K}$ \\
   Middles & $3.89\pm 0.17\ e^{-}{\rm ENC/K}$ & $4.85\pm 0.28~e^{-}{\rm ENC/K}$ \\
   Inners & $2.84\pm 0.10\ e^{-}{\rm ENC/K}$ & $3.86\pm 0.21~e^{-}{\rm ENC/K}$ 
  \end{tabular}
  \caption{Predicted and measured $\Delta {\rm ENC}/\Delta T$ for SCT Endcap 
  modules.}\label{tab:5-sctendcaps_dndt}
 \end{center}
\end{table}

As a comparison, 
the temperature dependence predicted and measured for the current SCT modules 
in~\cite{Burdin:1237817} are shown in Table~\ref{tab:5-sctendcaps_dndt} (only 
endcaps). The results of the tests performed on the upgrade module show a 
temperature dependence of around 
$1.5~e^{-}{\rm ENC/K}$. In comparison, the dependence of the current SCT 
endcap modules was measured between $3.86$ and $5.77~e^{-}{\rm ENC/K}$.

\begin{figure}[!htb]
\begin{center}
\subfigure[Noise variation with temperature.]{
\includegraphics[scale=0.37]{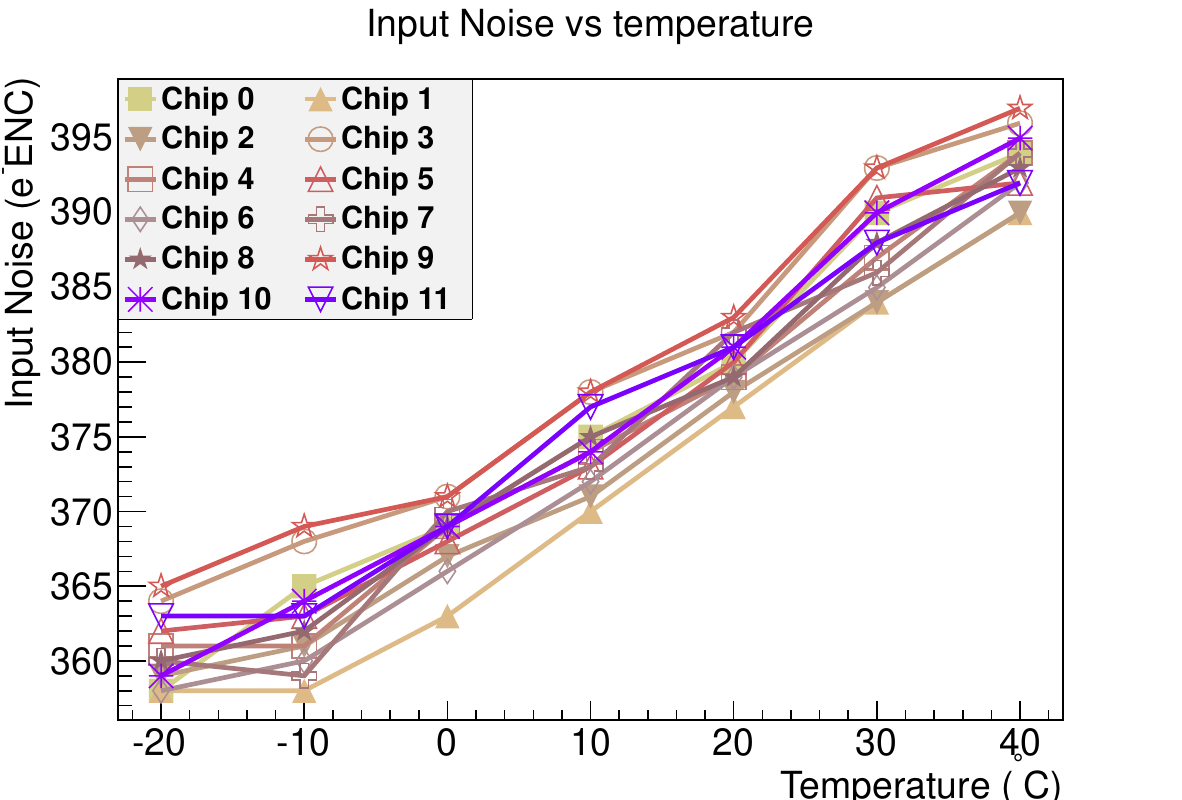}\label{fig:5-UnbondedHybrid_dENCdT_plots}
}
\subfigure[Linear fits of $\Delta {\rm ENC}/\Delta T$.]{
\includegraphics[scale=0.37]{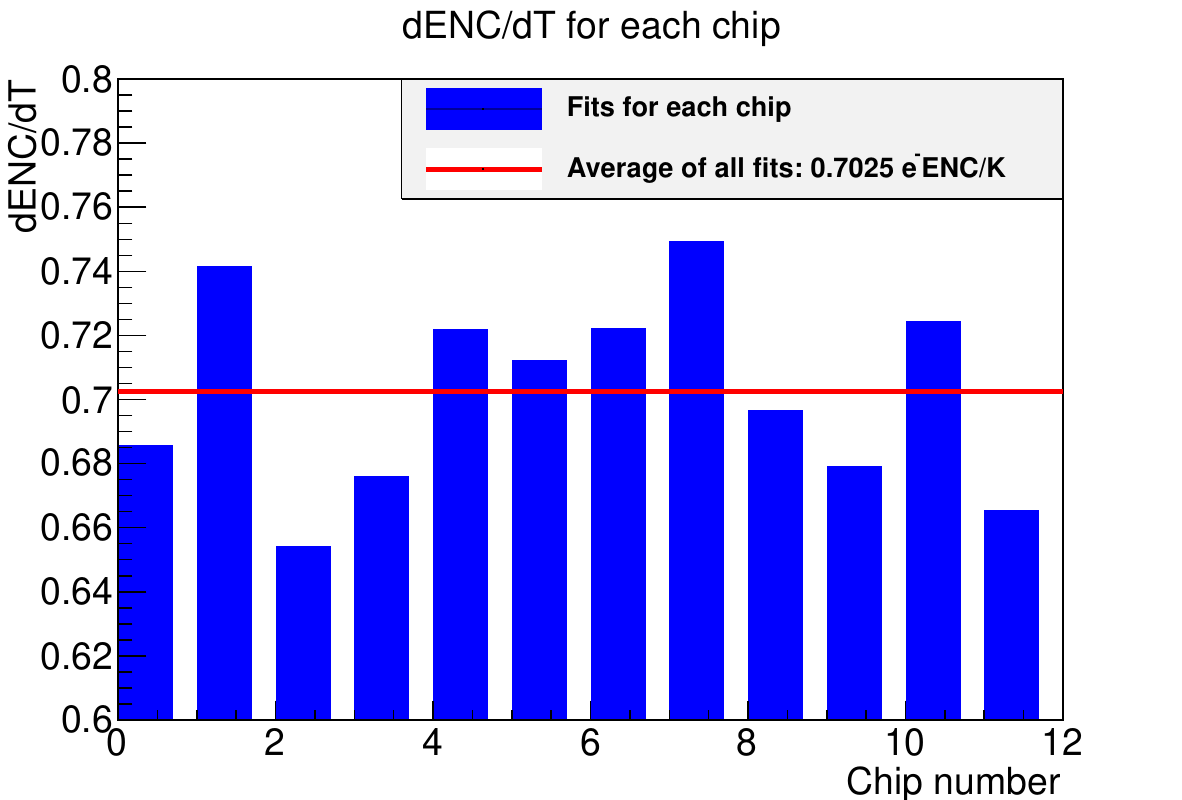}\label{fig:5-UnbondedHybrid_dENCdT_fit}
}
\caption{Noise variation with temperature for an unbonded 
hybrid.}\label{fig:5-UnbondedHybrid_dENCdT}
\end{center}
\end{figure}

The noise variation with temperature of a hybrid that was not bonded to a 
sensor was also studied in the University of 
Freiburg~\cite{FreiburgPetalet20130626}. The hybrid tested was one of the first
Petalet hybrids, with 12 ABCn25 chips on it (2 columns with 6 chips each).

Figure~\ref{fig:5-UnbondedHybrid_dENCdT_plots}
shows the result of this test and Figure~\ref{fig:5-UnbondedHybrid_dENCdT_fit}
shows the result of the linear fits for each chip and the average of all the
fits for the unbonded hybrid.

The first temperature point ($-20\degC$) was ignored for the fit, as it does
not entirely follow the expected linear trend of noise variation with 
temperature.

The linear fit for the rest of the data sets the chip contribution to the noise
variation with temperature at 

\begin{equation}
 \overline{\Delta {\rm ENC}/\Delta T}_{\rm Unbonded~Hybrid} 
 = 0.7025\pm 0.0306~e^{-}{\rm ENC/K} 
\end{equation}

This means that about half of the noise increase with temperature on a module 
is due to the chips, when no sensor is bonded to them. 

The model presented in section~\ref{sec:5-noisemodel} gave a much lower
value of the $\Delta {\rm ENC}/\Delta T$ for the unbonded hybrid, less than
half of the experimental result. However, the error on the model was quite 
large. For the bonded chips, the model gave a value approximately $25\%$ lower 
than the experimental result.

The results obtained for the model used the parameters values shown in 
section~\ref{sec:5-noisemodel}. Small variations in some of these values may 
lead to relevant changes in the results. For instance, the bias current of the
front-end circuitry is nominally $140\uA$, but it has a great influence in the
evolution of the noise with temperature. 

The differences were also observed between the SCT noise model and the measured
results, as shown in Table~\ref{tab:5-sctendcaps_dndt}.

\subsection{High Voltage Scans}

The change in sensor bias voltage effects the depletion depth of the p-n 
junction, when it is lower than the full depletion voltage. Noise decreases
as the bias voltage is increased. 

The leakage current increases and usually
reaches a semi-plateau after full depletion. If the bias voltage is further
increased, the sensor can reach breakdown and the leakage current increases 
again dramatically. 

The sensors used to build the modules for the upgrade programme are specified 
to withstand up to $1000\V$. 
  The voltage on the modules is 
limited by the hybrid circuitry to $500\V$, but the sensors are usually 
measured up to $1000\V$ before assembling the modules.

\begin{figure}[!htb]
\begin{center}
\includegraphics[scale=0.65]{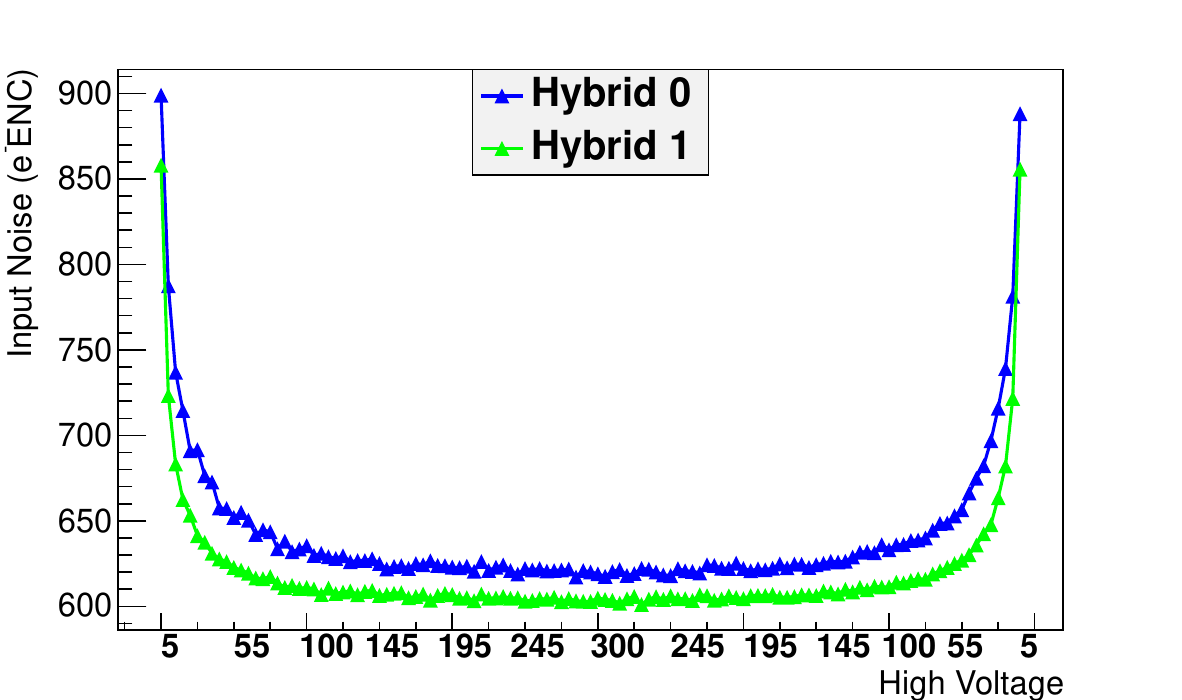}
\caption{Sensor bias voltage scan on the SP
module up to $300\V$.}\label{fig:5-spmodule_hvscan_updown}
\end{center}
\end{figure}

Tests with varying bias voltage were run with the serially powered
module. First the voltage was ramped up and then it was ramped down, measuring
the noise at every bias level.

\begin{figure}[!htb]
\begin{center}
\includegraphics[scale=0.65]{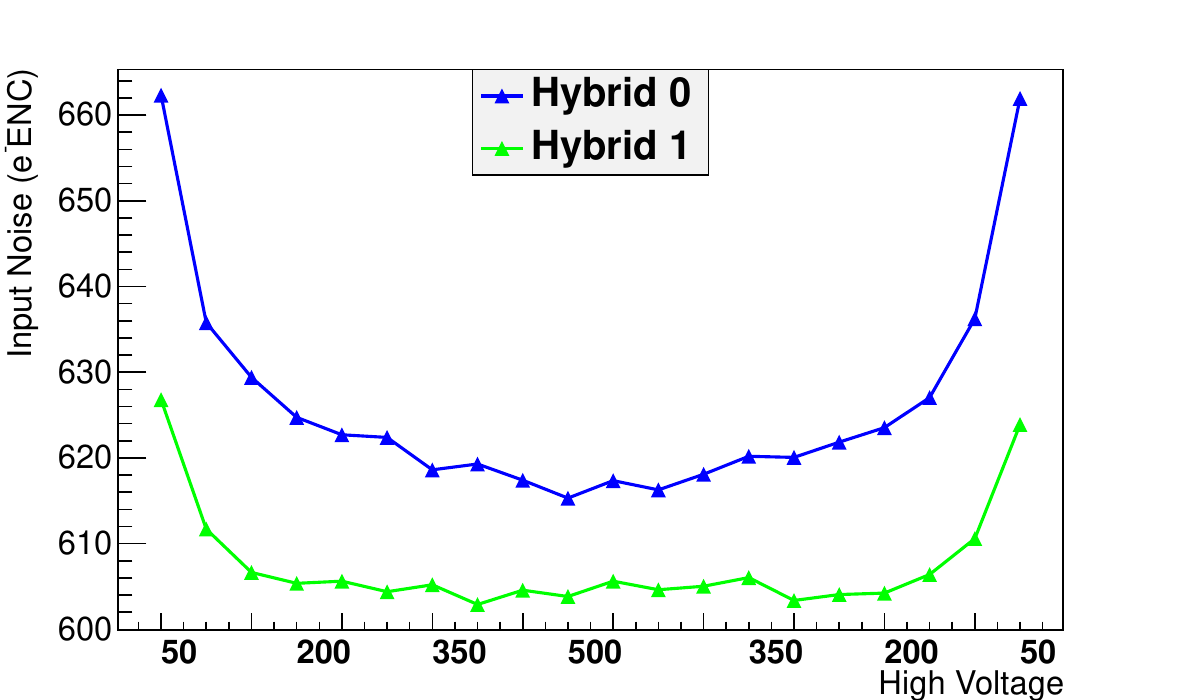}
\caption{Sensor bias voltage scan on the SP 
module up to $500\V$.}\label{fig:5-spmodule_hvscan_updown_500V}
\end{center}
\end{figure}

Figure~\ref{fig:5-spmodule_hvscan_updown} shows the results of a high voltage 
scan, first ramping up from $5\V$ to $300\V$ and then down from $300\V$ to 
$5\V$, in steps of $5\V$. The results are shown averaged for the two hybrids
in the module. The noise was calculated using three point
gain scans centred at $1\fC$. The coolant temperature was set to $12\degC$.

An additional test with the maximum bias voltage supported by the hybrid was done, 
ramping up to
$500\V$, from $50$ and with $50\V$ steps. 
Figure~\ref{fig:5-spmodule_hvscan_updown_500V} shows the results for this 
test.

There have been reports of higher noise for the second time the bias voltage 
is set to some value after ramping up to higher 
voltages~\cite{SvenDiplomaThesis}. This  module does not show that 
behaviour, while the temperature was very stable throughout the whole test.

\subsection{Low Temperature Tests}

In order to test if the electronic components are capable of operating at very
low temperatures, some tests were run below $-10\degC$. In order to avoid 
condensation and ice, dry air was flushed into the freezer where the module is
tested. The pipes of the chiller were covered with insulating foam
to avoid ice on them. The sensor was biased at $200\V$.

\begin{table}[!hbt]
 \begin{center}
  \begin{tabular}{c|c|c|c|c|c}
  \toprule
   $T_{\rm chiller}[^{\circ}{\rm C}]$ & $T_{\rm pipe}$ & $T_{\rm hybrid}$ & $T_{\rm freezer}$ & 
   $T_{\rm hybrid} - T_{\rm pipe}$ & $T_{\rm freezer} - T_{\rm pipe}$\\
   \hline
   $12$ & $13.4$ & $25.9$ & $18.1$ & $12.5$  & $4.7$ \\
   $-6$ & $-1.5$ & $12.1$ & $14.5$ & $13.6$  & $16.0$ \\
   $-12$ & $-5.8$ & $7.6$ & $10.4$ & $13.4$  & $16.2$ \\
   $-20$ & $-11.7$ & $2.5$ & $8.5$ & $14.2$  & $20.2$ \\
 \bottomrule
  \end{tabular}
\caption[Temperatures measured during low temperature testing]{Temperatures 
measured during low temperature testing: chiller setting, 
temperature at the input pipe to the freezer, temperature on the hybrid, 
temperature inside the freezer enclosure and temperature differences between 
the pipe and the hybrid as well as the freezer.
}\label{tab:5-temperatures}
 \end{center}
\end{table}

The disadvantage of using a normal dry air supply is the temperature of the 
air, which is typically around $24\degC$, which creates a temperature difference
between the sensor and the air. This does not affect the low temperature 
behaviour of the electronics, but it influences the measurement of noise with
varying temperature.

Table~\ref{tab:5-temperatures} shows the different temperatures measured for 
various chiller settings. The temperature was measured on the inlet pipe, next
to the freezer using a commercial NTC. The temperature on the hybrid is 
measured with an NTC that is part of the hybrid circuitry, with the hybrid
powered on. The temperature 
inside the freezer was measured using a SHT71 temperature and humidity sensor.

The temperature difference between the pipe and the hybrid is quite constant
while the temperature difference between the air in the freezer and the pipe
is greater for lower temperatures. As it was stated above, this is caused by
the dry air that is flushed into the freezer at room temperature.

The noise results for $12\degC$ on the chiller were shown in 
Figure~\ref{fig:5-SPmoduleENCref}. Results with lower temperature settings 
are shown next.

\begin{figure}[!htbp]
\begin{center}
\includegraphics[scale=0.75]{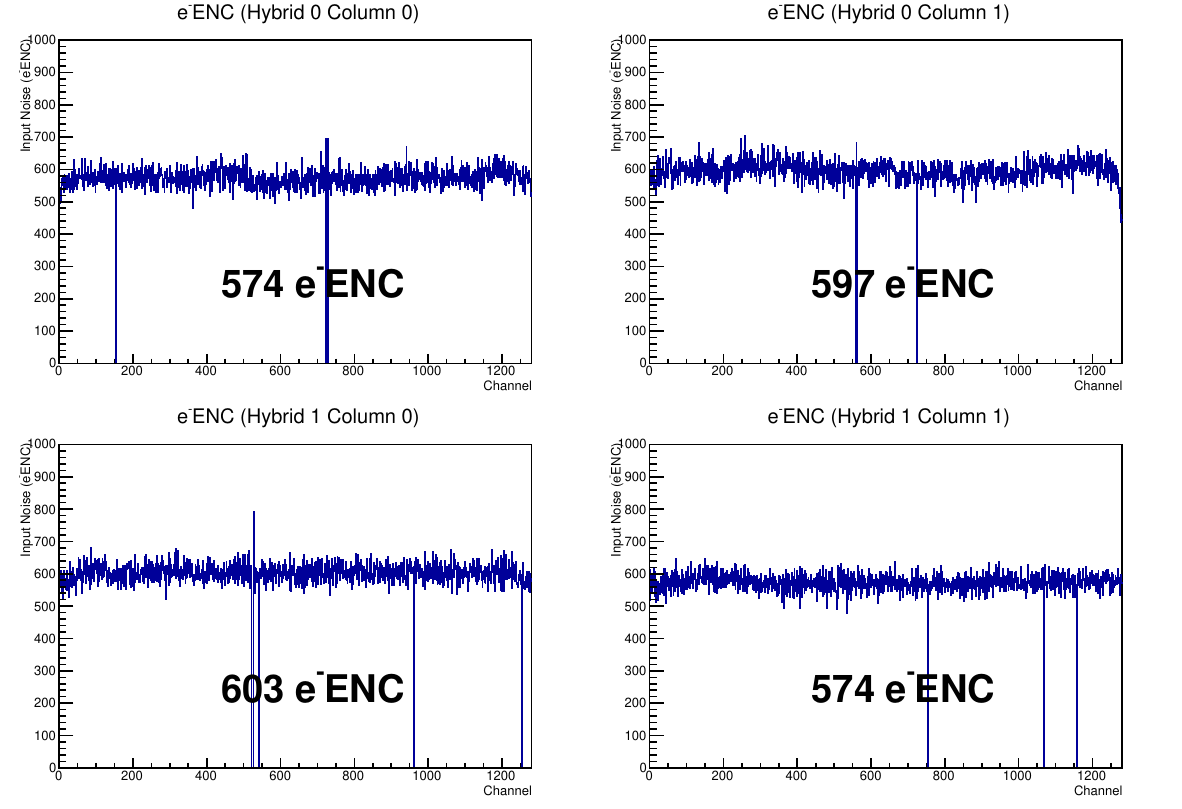}
\caption{Input noise ($e^{-}{\rm ENC}$) with chiller temperature set to $-6\degC$, three
point gain scans centred at $1\fC$.}\label{fig:5-SPmodule_m06deg_1fC}
\end{center}
\end{figure}

Figure~\ref{fig:5-SPmodule_m06deg_1fC} shows the input noise for a temperature
setting of $-6\degC$. The temperature on the hybrid is shown in
Table~\ref{tab:5-temperatures}.

Figure~\ref{fig:5-SPmodule_m20deg_1fC} shows the input noise when the
temperature was set to $-20\degC$. All columns show lower noise than the 
previous plot and all the channels were working correctly, except for those
that were not previous to this test. There are some channels in hybrid 0,
column 1, between channel numbers 589 and 593 that show up dead in 
Figure~\ref{fig:5-SPmodule_m06deg_1fC}, but look fine in 
Figure~\ref{fig:5-SPmodule_m20deg_1fC}. After reviewing the noise history
of those channels, it was found out that they are inefficient in the chip
and sometimes are masked by the software but were not during the test
at $-20\degC$. Despite the inefficiency, the input noise result is consistent
with the overall column noise, making them appear as normal healthy channels.

The temperature difference of the hybrids between the $-6\degC$ and the 
$-20\degC$ settings is $9.6\degC$. Assuming a $1.55~e^{-}{\rm ENC/K}$ variation, as
measured above, a noise difference of $14.88\ENC$ is expected. The 
variation observed lies between $16$ and $25\ENC$, which is consistent
with the expectation.

The conclusion of this test is that the system works correctly when operating
at much lower temperatures than the typical testing temperature of $12\degC$. 
The hybrid has been 
operated at a temperature below $3\degC$ and the LVDS drivers at a temperature
below $9\degC$.
 
\begin{figure}[!htbp]
\begin{center}
\includegraphics[scale=0.75]{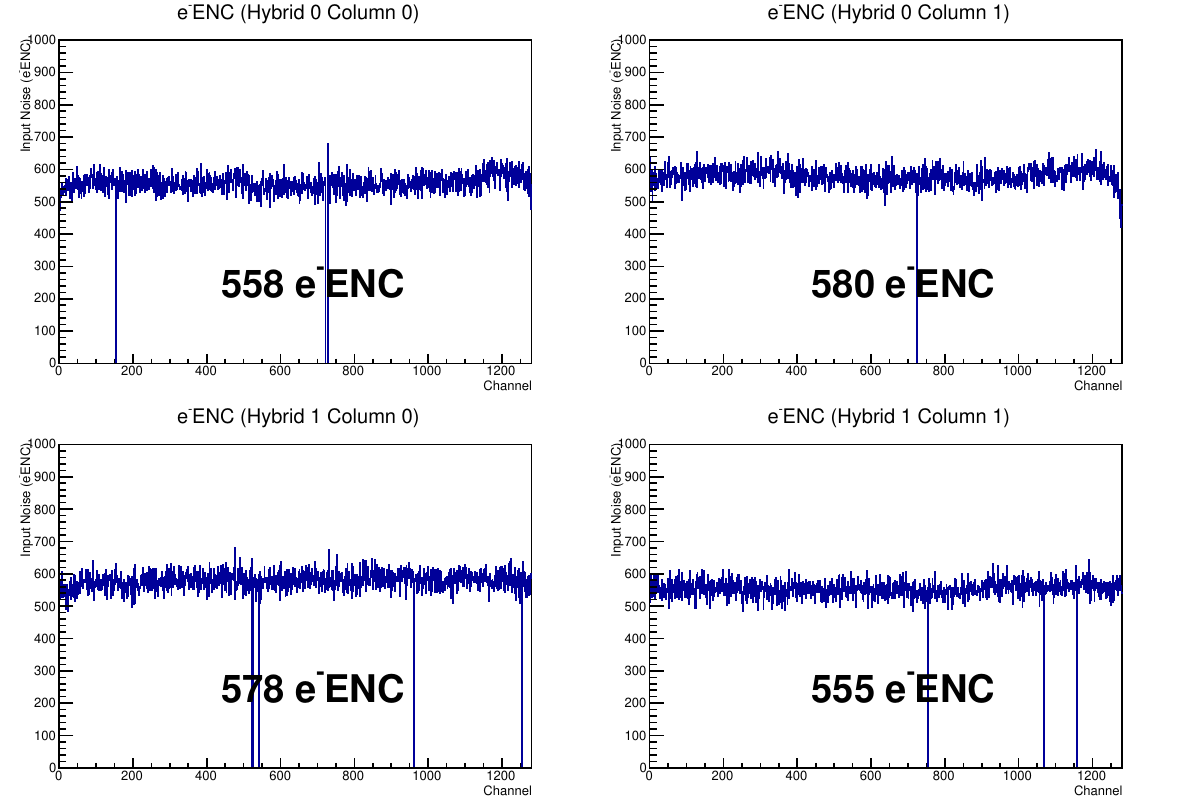}
\caption{Input noise ($e^{-}{\rm ENC}$) with chiller temperature set to $-20\degC$, three
point gain scans centred at $1\fC$.}\label{fig:5-SPmodule_m20deg_1fC}
\end{center}
\end{figure}

\section{Performance Tests of Four Module Stavelets}

\subsection{High Voltage Scans}

The behaviour of the modules when changing the bias voltage in the sensors
was also studied for the four module stavelets. The different types of sensors
used to build the stavelets present small differences in their behaviour when
changing the sensors bias.

\begin{figure}[!htb]
\begin{center}
\includegraphics[scale=0.8,trim=7 0 20 0,clip=true]{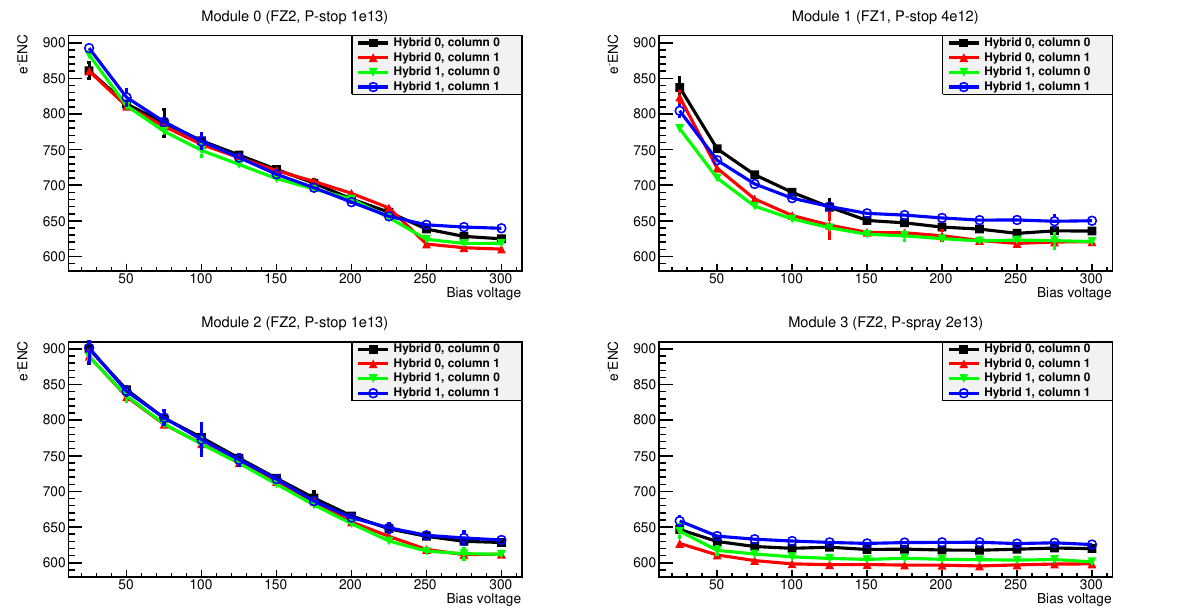}
\caption{DC-DC stavelet input noise ($e^{-}{\rm ENC}$) with sensor 
bias.}\label{fig:5-DCDC_ENC_vs_HV}
\end{center}
\end{figure}

Figure~\ref{fig:5-DCDC_ENC_vs_HV} shows the noise results with varying bias
voltage of the sensors. The scan consisted in running three point gain tests
centred at $1\fC$, and the chiller temperature was set to $12\degC$. 
The voltage
set to power the DC-DC converters was calibrated to measure $10\V$ in the sense
wires. The voltage on the power supply was $10.3\V$. At the time of these tests
and with the wires used, the voltage drop up to the DC-DC converters was around
$300\mV$.

The bias voltage was modified from $25\V$ to $300\V$, in $25\V$ steps. Two
three point gain scans were used to measured the noise, the points show the 
average value. The errors bars are the calculated deviations at each point. 

The noise results show a noise comparable to the estimated values for the
fully depleted sensors. The isolation structure for module 3 (P-spray) seems
to give the sensor a very low noise from very low bias. Also, the stronger 
P-stop
isolation on modules 0 and 2 seems to lower the noise with voltage later than
the weaker P-stop in module 1.

\begin{figure}[!tb]
\begin{center}
\includegraphics[scale=0.8,trim=7 0 20 0,clip=true]{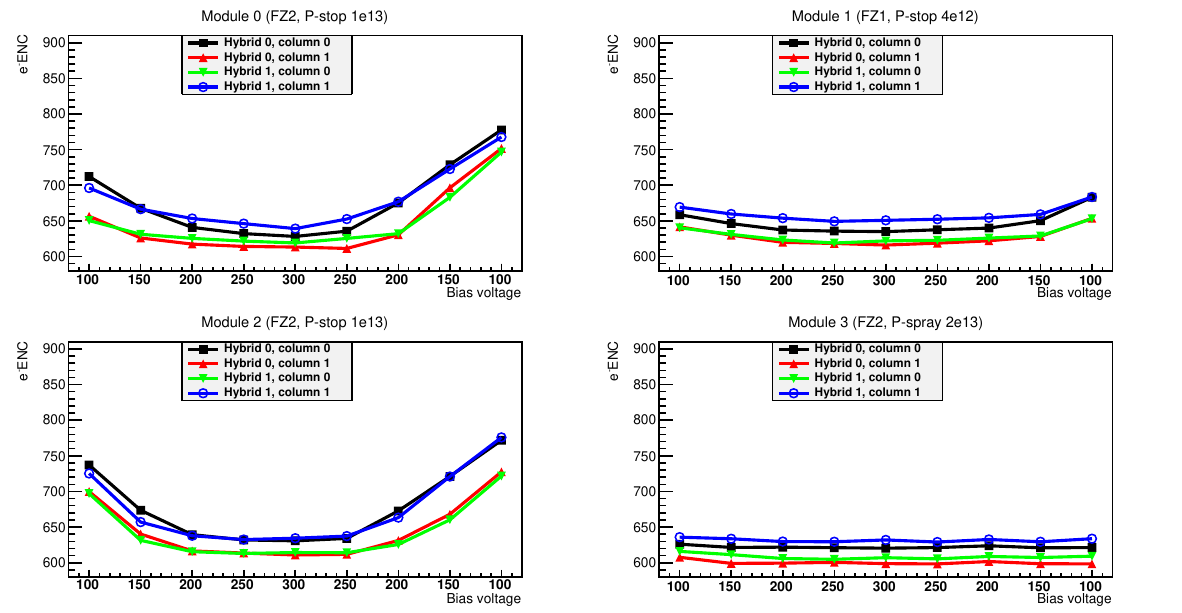}
\caption{DC-DC stavelet input noise ($e^{-}{\rm ENC}$) with sensor bias, first ramping up 
to $300\V$ and then ramping down.}\label{fig:5-DCDC_ENC_vs_HV_UpDown}
\end{center}
\end{figure}

Figure~\ref{fig:5-DCDC_ENC_vs_HV_UpDown} shows the results 
of a test similar to that shown in Figure~\ref{fig:5-spmodule_hvscan_updown},
with a ramp up of the sensor bias voltage first and then ramp down. For modules
0, 1 and 2, the effect described in the single module section is now present:
the noise for the same values of bias voltage is higher after passing through
a higher voltage.

The plots show the timeline of the whole test for each chip column,
one module for each graph~\cite{ArgosAUW12}. The same high voltage scans 
were done with the serially powered stavelet.

\begin{figure}[!htb]
\begin{center}
\includegraphics[scale=0.8,trim=7 0 20 0,clip=true]{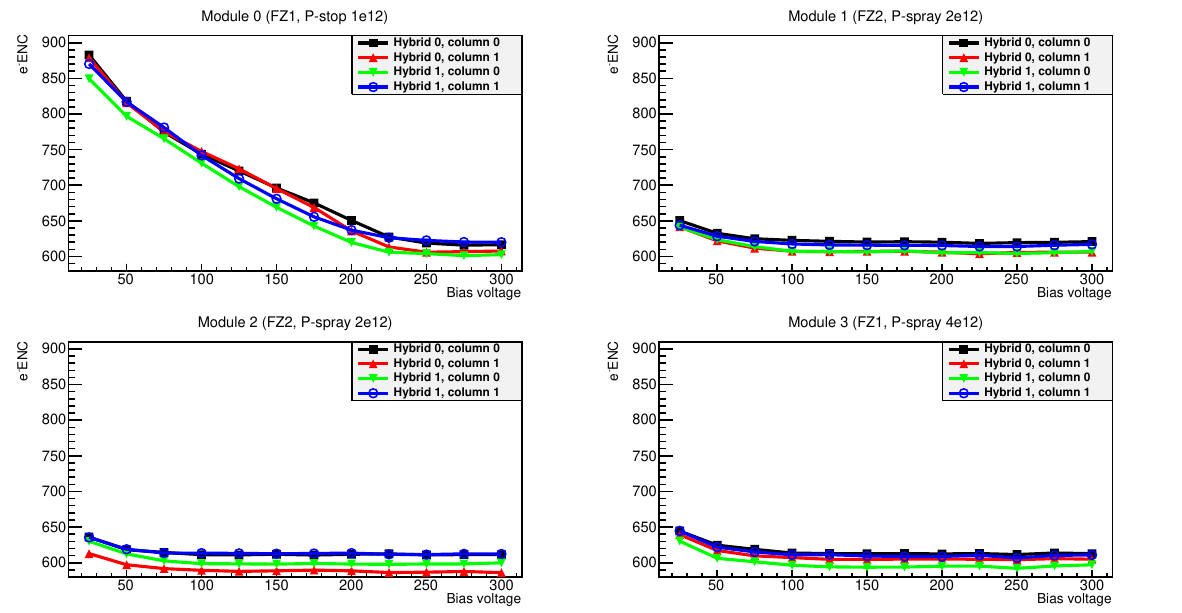}
\caption{SP stavelet input noise ($e^{-}{\rm ENC}$) with sensor 
bias.}\label{fig:5-SPstavelet_ENC_vs_HV}
\end{center}
\end{figure}

Figure~\ref{fig:5-SPstavelet_ENC_vs_HV} shows the noise evolution of the 
serially powered stavelet with the bias voltage. Like in the DC-DC stavelet,
the noise on modules with P-spray sensors starts at a much lower value with 
low bias voltage than the one P-stop sensor.

\begin{figure}[!htb]
\begin{center}
\includegraphics[scale=0.8,trim=7 0 20 0,clip=true]{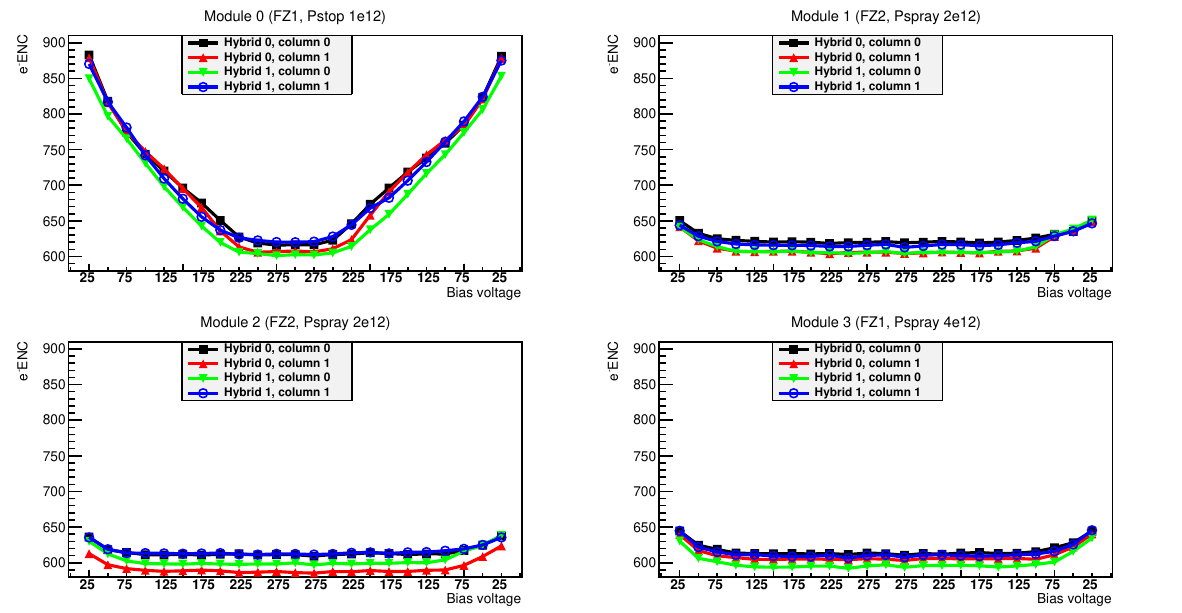}
\caption{SP stavelet input noise ($e^{-}{\rm ENC}$) with sensor bias, first ramping up 
to $300\V$ and then ramping down.}\label{fig:5-SPstavelet_ENC_vs_HV_UpDown}
\end{center}
\end{figure}

The tests were repeated with the SP stavelet, ramping up the bias voltage
and also ramping down. The result is shown in 
Figure~\ref{fig:5-SPstavelet_ENC_vs_HV_UpDown}. Some modules present the same
behaviour as modules 0 and 2 on the DC-DC stavelet, where the noise is higher 
after biasing at a lower voltage than a previous setting while ramping down. 

Module 0 has a P-stop isolation structure, but with a lower dose than
modules 0, 1 and 2 on the DC-DC stavelet. The noise increase exists but it
is less than for the DC-DC modules. Modules 1, 2 and 3
are P-spray with different doses. Module 3 is the one with highest dose and it
does not show a clear noise increase when going down in voltage, whereas 
modules 1 and 2 show
a small noise increase in most of the columns..

\subsection{Hybrid Powering Variations}

Both power distribution mechanisms, serial powering and DC-DC, admit a certain
range of input current and voltage. A study on how these variations affect the 
noise was done with both stavelets.

On the one side, the efficiency of the DC-DC converters varies with the input
voltage. The converters are operational for $V_{in}$ between $7$ and $15\V$.
The tests on the stavelet were performed in the range from $10.5$ to $14.5\V$.

The ``sweet spot'' for the DC-DC converters is $V_{in} = 10\V$, voltage at 
which the conducted noise is the lowest. Due to the high
current, this figure is normally quoted as the voltage at the sense wires that
are directly connected to the converters and not the voltage set at the power
supply. 

There is a typical voltage drop around $500\mV$ on the setup at CERN at the 
time of the hybrid powering tests. In order to have a voltage around $10\V$ 
at the converters inputs, the
output of the power supply is set to $10.5\V$. The voltages at the input of the
converters were measured for power supply settings of $10.5\V$ and $14.5\V$. 
These are shown in Table~\ref{tab:5-DCDCvoltages}. The drop on the bus tape is
$220\mV$ for higher currents and $170\mV$ for lower currents. The current shown
is the current drawn from the power supply after configuring the chips, without
any tests running. 

\begin{table}[!htb]
 \begin{center}
  \begin{tabular}{c c|c c c c}
  \toprule
   $V_{\rm PSU}[{\rm V}]$ & $I_{\rm PSU}[{\rm A}]$ & $V_{\rm M3}[{\rm V}]$ & $V_{\rm M2}[{\rm V}]$ & $V_{\rm M1}[{\rm V}]$ & $V_{\rm M0}[{\rm V}]$ \\ \hline
   $10.5$        &        $9.96$ &      $9.96$ &       $9.85$ &       $9.78$ &       $9.74$ \\
   $14.5$        &        $7.41$ &     $14.04$ &      $13.97$ &      $13.91$ &      $13.87$ \\
  \bottomrule
   \end{tabular}
\caption{Voltages measured at the DC-DC converters inputs, as well as the output
current of the power supply.}\label{tab:5-DCDCvoltages}
 \end{center}
\end{table}

In the DC-DC stavelet tests, the voltage is referenced to the approximate 
measurement on the sense wires, always accounting for a voltage drop on the
wires around $500\mV$.

The output voltage of the serial power regulator varies
slightly with the input current. The serially powered stavelet is typically run
with an input current of $9.5\A$. The stavelet has been tested also with input 
currents of $9\A$ and $10\A$. An increase in the current also increases the 
voltage across the hybrid, resulting in a higher hybrid temperature.

The different low voltage powering tests were performed with a sensor bias 
voltage of $250\V$ and a chiller temperature setting of $12\degC$. The DC-DC 
stavelet tests consisted in varying the input voltage to the converters from 
$10\V$ to $14\V$, in steps of $1\V$. Ten three point gain tests centred at
$1\fC$ were performed at each low voltage level. 

\begin{figure}[!htb]
\begin{center}
\includegraphics[scale=0.8,trim=7 0 20 0,clip=true]{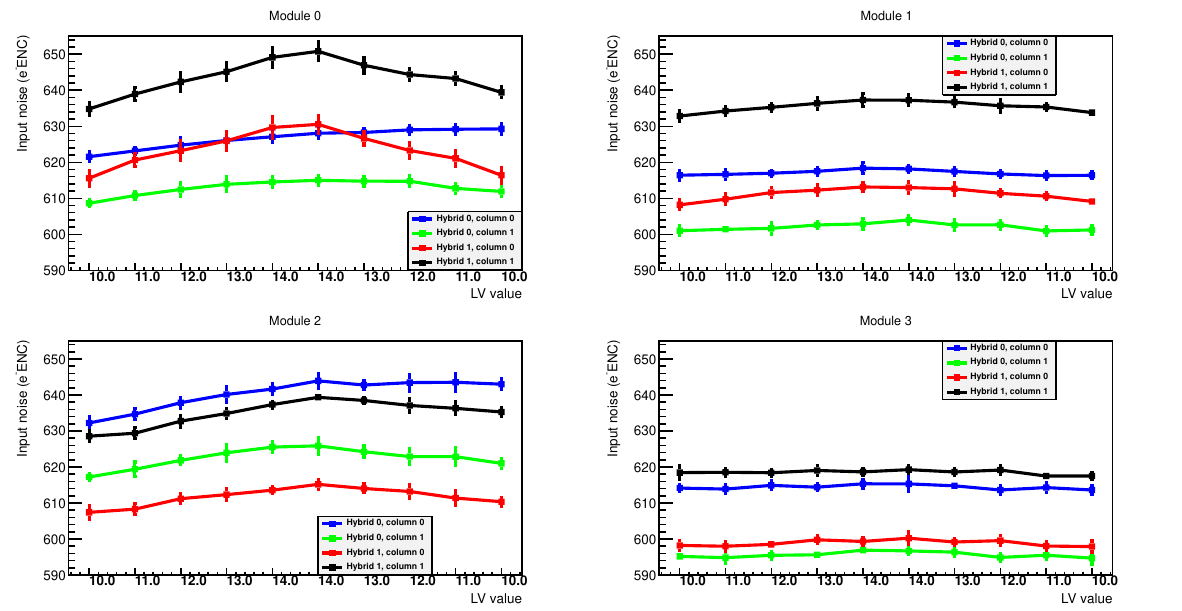}
\caption{DC-DC stavelet input noise ($e^{-}{\rm ENC}$) with low voltage value. Low voltage 
from $10\V$ to $14\V$, up and down, $1\V$ 
steps.}\label{fig:5-DCDC_LV_noise_HVfilter}
\end{center}
\end{figure}

Figure~\ref{fig:5-DCDC_LV_noise_HVfilter} shows the noise evolution of the 
DC-DC stavelet when the low voltage value is changed between $10$ and $14\V$.
Each line represents one chip column.  The points on the plots show the average 
value of the noise and the error bars represent the deviation from the average. 
The noise evolution is uneven for the four modules. It is essentially due to a 
higher noise on the converters when they operate at higher input voltage. 
However, modules 1 and 3 seem to be immune to this effect.

The noise trend is very clear on the second hybrid of module 0, whereas the 
first hybrid shows a mixed behaviour, with one of the columns always increasing
noise. The behaviour of module 2 is also mixed, with some increase up to $14\V$
supply voltage, but for one of the hybrids the noise does not decrease when the
voltage is reduced back to $10\V$.

\begin{figure}[!htb]
\begin{center}
\includegraphics[scale=0.8,trim=7 0 20 0,clip=true]{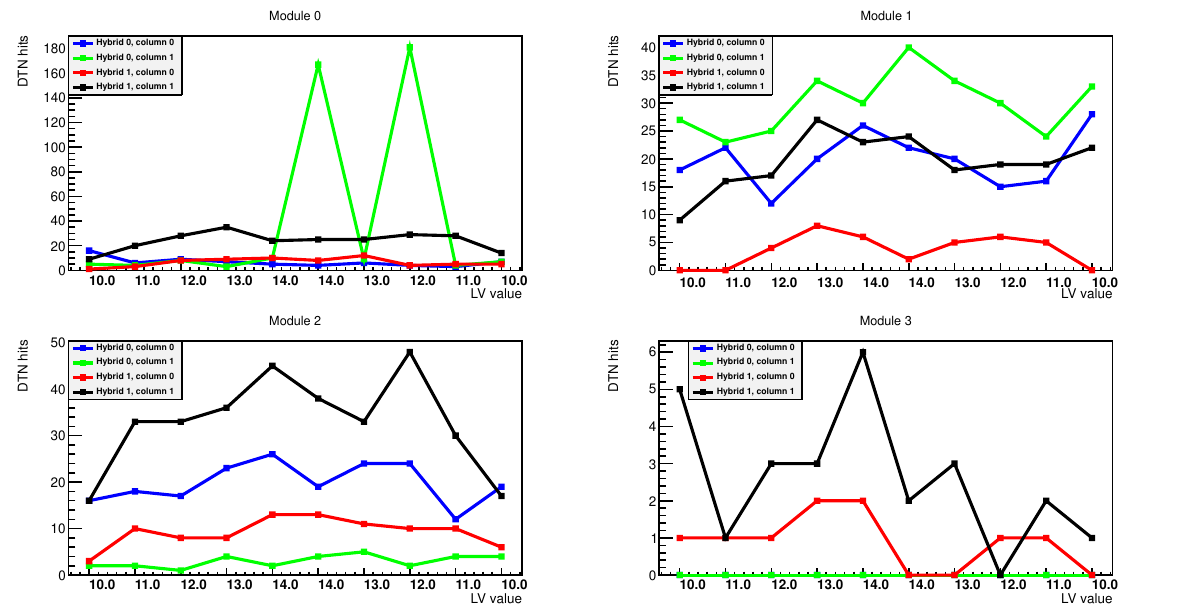}
\caption{DC-DC stavelet Double Trigger Noise variation with low voltage value 
for $0.5\fC$ threshold.}\label{fig:5-DCDC_LV_DTN_HVfilter}
\end{center}
\end{figure}

Some of the shields leave a small gap between the shield and the converter 
board. However, 
additional tests with a bigger gap on module 3 show it does not cause
the additional noise when increasing the hybrid powering voltage.

The temperatures of the hybrids were stable during the whole test, with 
variations around $0.1\degC$ for the hybrids closest to the EoS card and
smaller than $0.5\degC$ for the hybrids further away from the EoS card.

\begin{figure}[!htb]
\begin{center}
\includegraphics[scale=0.8,trim=7 0 20 0,clip=true]{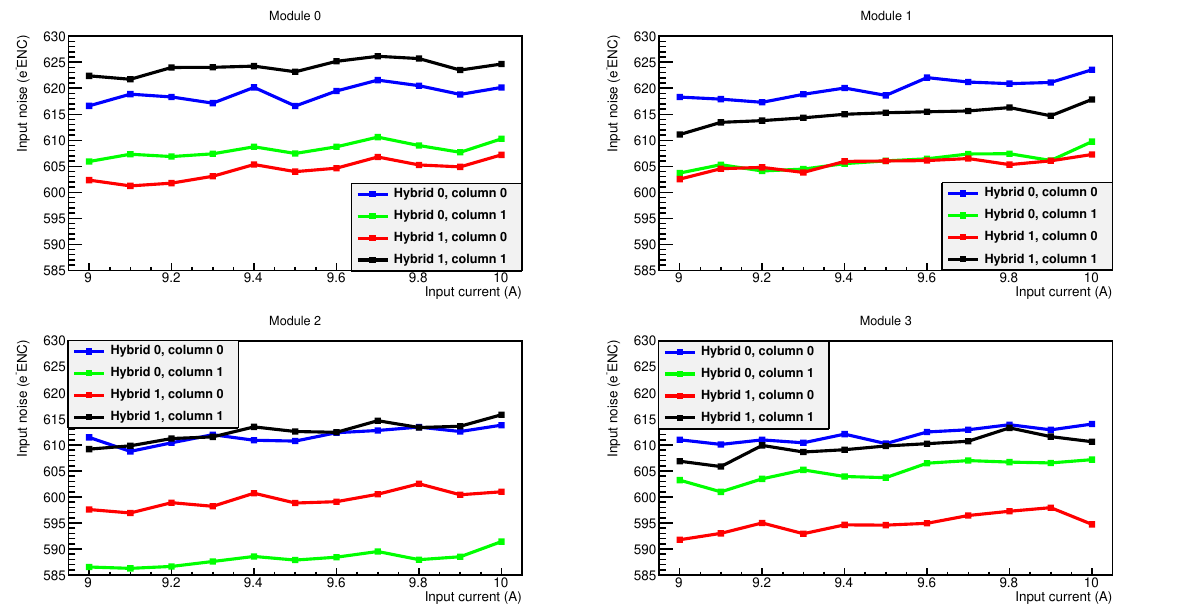}
\caption{SP stavelet input noise ($e^{-}{\rm ENC}$) with low voltage value. Input current 
varied between $9$ and 
$10\A$ in $0.1\A$ steps.}\label{fig:5-SP_LV_noise}
\end{center}
\end{figure}

In addition, Figure~\ref{fig:5-DCDC_LV_DTN_HVfilter} shows the Double Trigger 
Noise variation with the low voltage setting for the $0.5\fC$ threshold. The
plots show a very random variation of the number of hits with different
measurements, leading to the conclusion that the variation of the voltage
input to the DC-DC converters does not affect the noise pick-up on the 
stavelet.

The equivalent test on the serially powered stavelet was done varying the input
current from $9\A$ to $10\A$ and running three point gain tests. The results
are shown in Figure~\ref{fig:5-SP_LV_noise}, with the expected noise increase
when increasing the current, due to the higher temperature on the hybrids. 

One three point gain test was done at every current value. The hybrids
temperature increased between $2.4$ and $2.9\degC$, from the minimum current
to the maximum current setting. This variation would account for a noise 
difference between $3.7$ and $4.5\ENC$, which is observed in the results.

The same Double Trigger Noise measurement as with the DC-DC stavelet was done
for the serially powered stavelet when changing the input current. 
Figure~\ref{fig:5-SP_LV_DTN} shows the results, the first module has one 
column that is out of scale (hybrid 1, column 1). This is because it has a very 
large number of hits, 
greater than 10000, but with the same behaviour as all the others. As before, 
this is the Double Trigger Noise test with a threshold of $0.5\fC$ that allows
seeing actual variations of the pick-up. The trend shown in the plots
is random and it cannot be attributed to any additional noise pick-up caused
by the variations of the input current.

\begin{figure}[!htb]
\begin{center}
\includegraphics[scale=0.8,trim=7 0 20 0,clip=true]{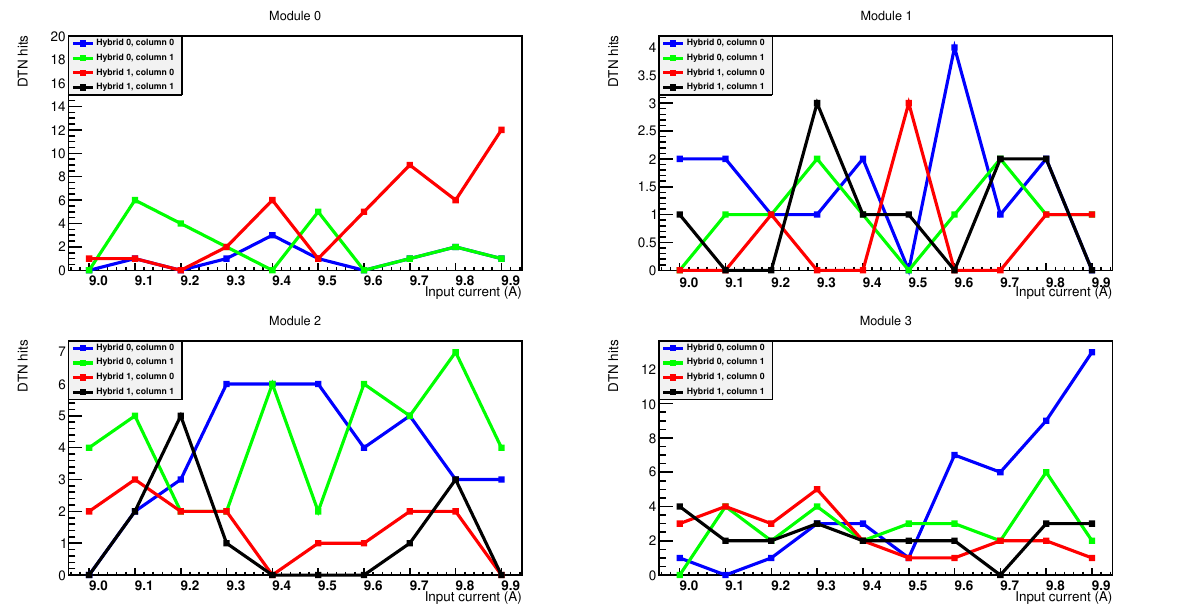}
\caption{SP stavelet Double Trigger Noise variation with low voltage value for 
$0.5\fC$ threshold.}\label{fig:5-SP_LV_DTN}
\end{center}
\end{figure}

\subsection{High Voltage Multiplexing}

The high voltage multiplexer presented in section~\ref{sec:4-HVmux} is also
available for testing. A comparison of high voltage scans with the multiplexer
and using the default single high voltage lines follows.

\begin{figure}[!htbp]
\begin{center}
\includegraphics[scale=0.75]{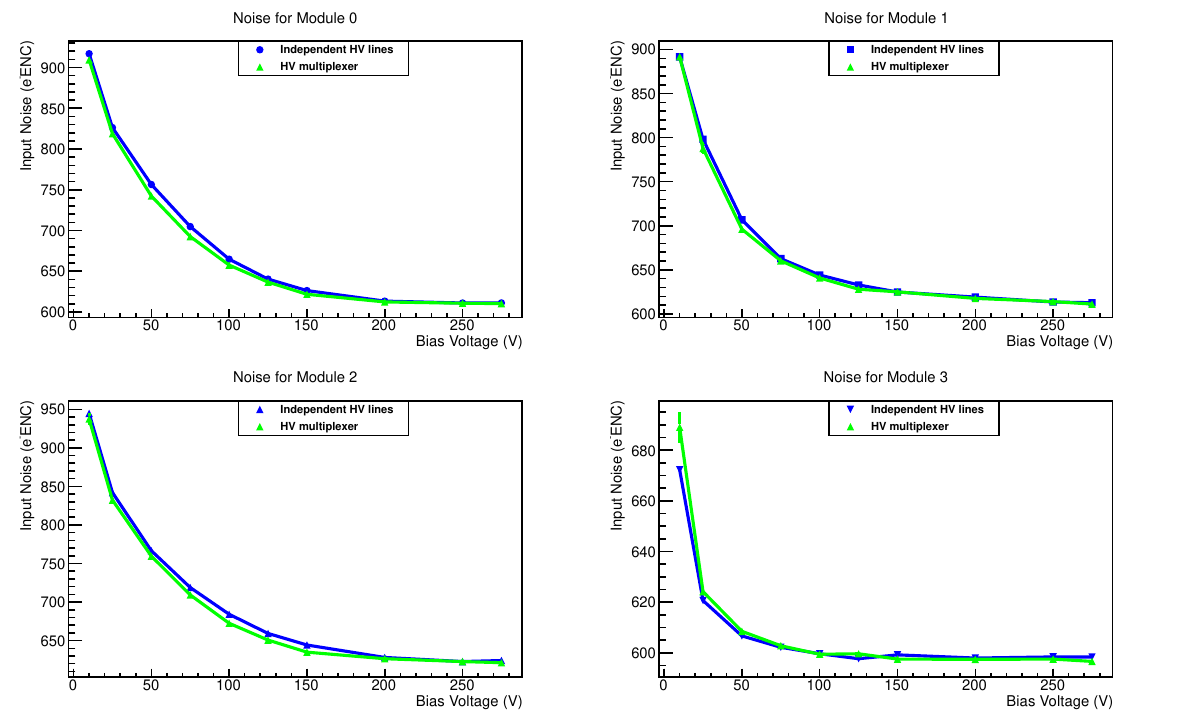}
\caption{Noise comparison of high voltage scans on the DC-DC stavelet, using 
the high voltage multiplexer and using independent biasing, with the chiller
temperature set to $6\degC$.}\label{fig:5-HVmux_DCDC_6deg}
\end{center}
\end{figure}

\begin{figure}[!htbp]
\begin{center}
\includegraphics[scale=0.75]{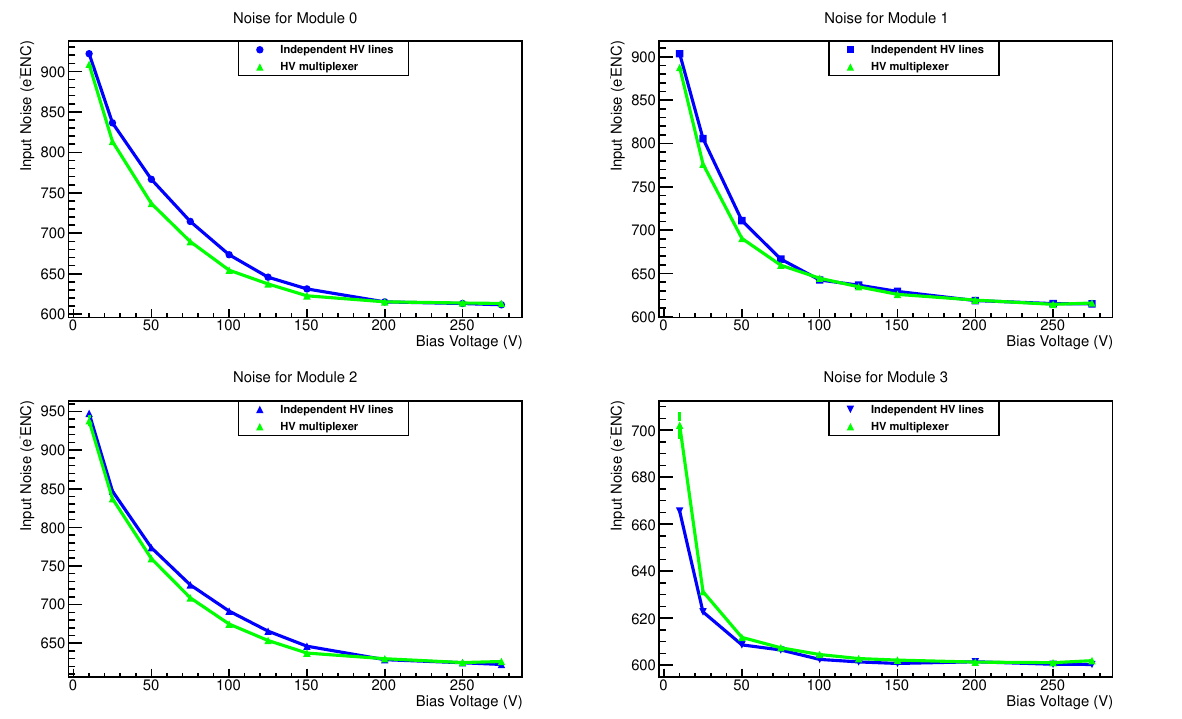}
\caption{Noise comparison of high voltage scans on the DC-DC stavelet, using 
the high voltage multiplexer and using independent biasing, with the chiller
temperature set to $9\degC$.}\label{fig:5-HVmux_DCDC_9deg}
\end{center}
\end{figure}

\begin{figure}[!htbp]
\begin{center}
\includegraphics[scale=0.75]{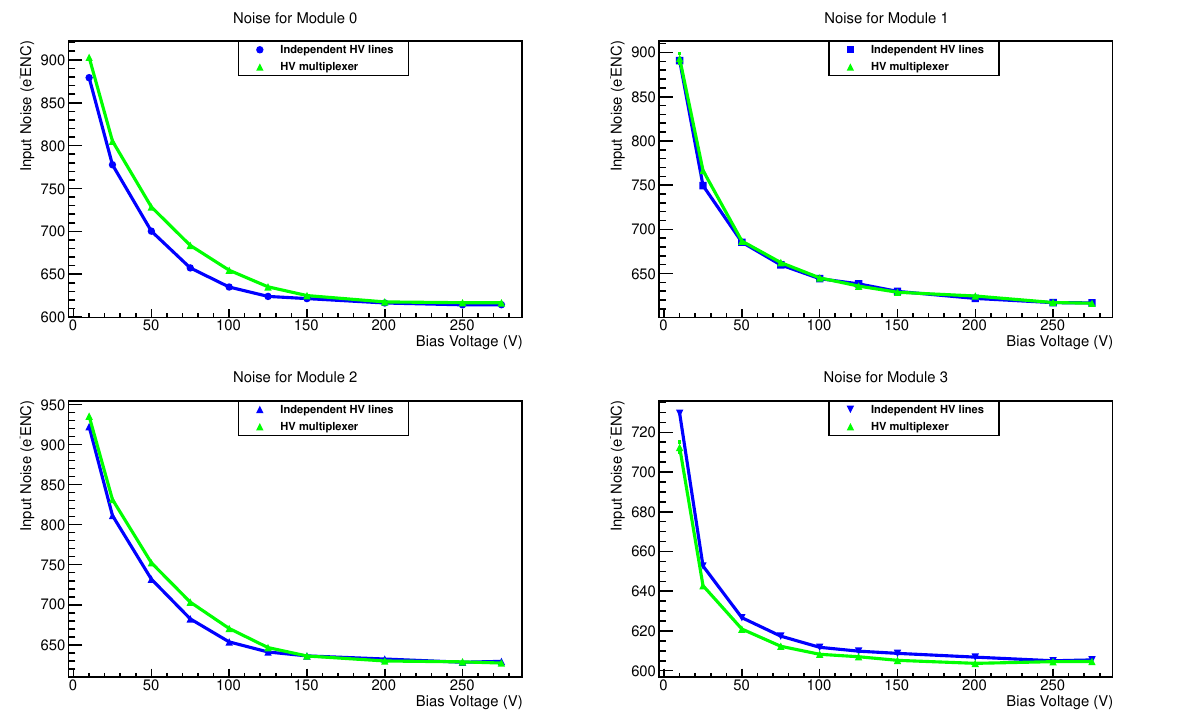}
\caption{Noise comparison of high voltage scans on the DC-DC stavelet, using 
the high voltage multiplexer and using independent biasing, with the chiller
temperature set to $12\degC$.}\label{fig:5-HVmux_DCDC_12deg}
\end{center}
\end{figure}

\begin{figure}[!htbp]
\begin{center}
\includegraphics[scale=0.75]{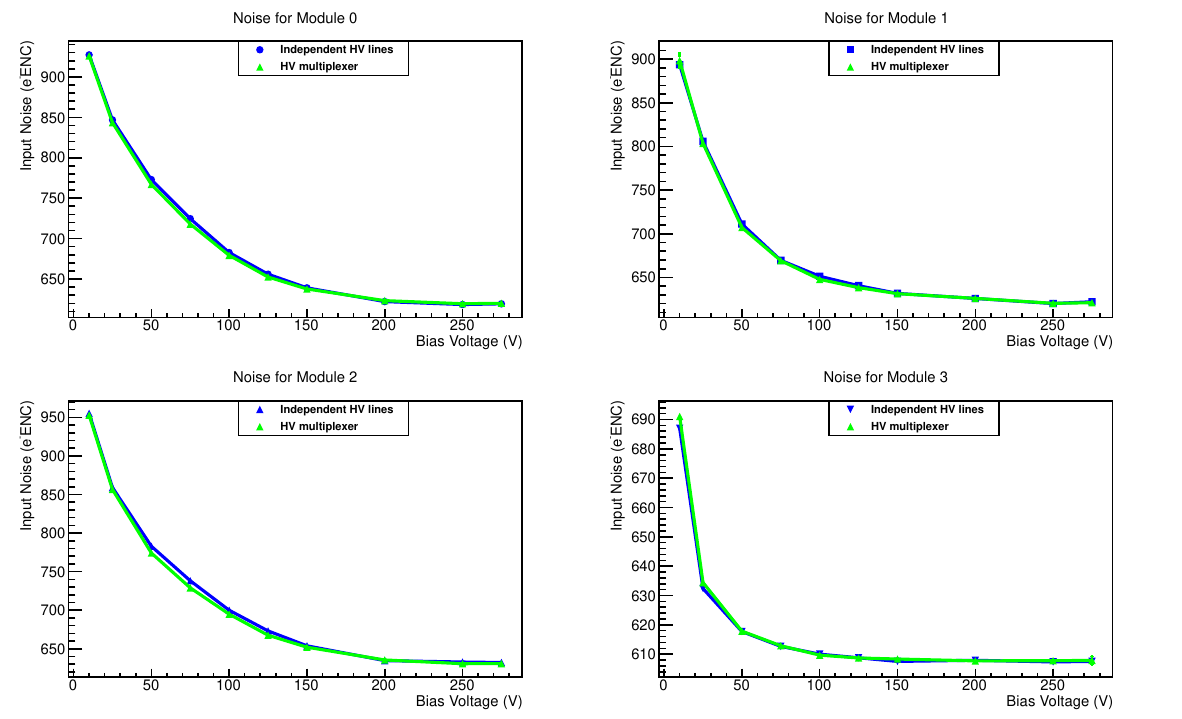}
\caption{Noise comparison of high voltage scans on the DC-DC stavelet, using 
the high voltage multiplexer and using independent biasing, with the chiller
temperature set to $15\degC$.}\label{fig:5-HVmux_DCDC_15deg}
\end{center}
\end{figure}

\begin{figure}[!ht]
\begin{center}
\includegraphics[scale=0.75]{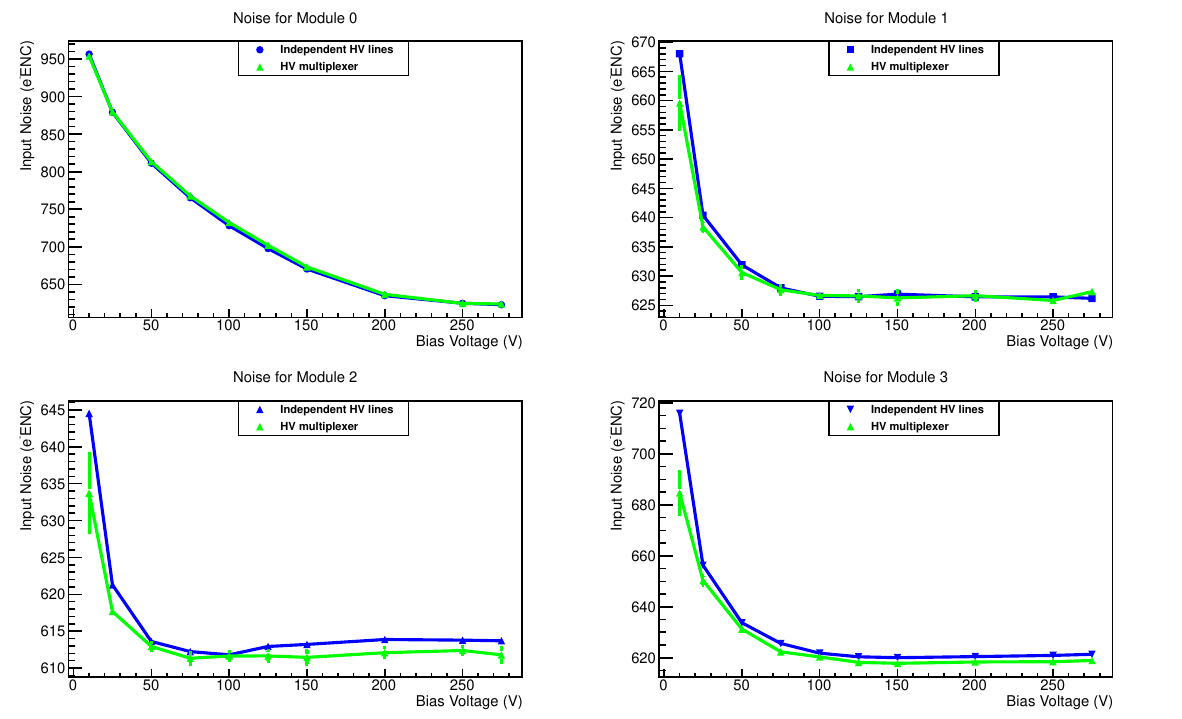}
\caption{Noise comparison of high voltage scans on the serially powered
stavelet, using 
the high voltage multiplexer and using independent 
biasing.}\label{fig:5-HVmux_SP}
\end{center}
\end{figure}

Figures~\ref{fig:5-HVmux_DCDC_6deg},~\ref{fig:5-HVmux_DCDC_9deg},~\ref{fig:5-HVmux_DCDC_12deg} 
and~\ref{fig:5-HVmux_DCDC_15deg} show the results of the high voltage scans on
the DC-DC stavelet with different temperature settings.

The bias voltages for both configurations and every temperature setting were
chosen from the ranges with sensor over-depletion and under-depletion 
voltages. These values are
$V_{\rm bias} = \left\lbrace 10, 25, 50, 75, 100, 125, 150, 200, 250, 
275\right\rbrace$.

The values shown are the average for every module. The error bars represent the
noise difference among the three measurements for each bias voltage value. They 
are not visible for most values due to the small noise differences at each bias 
voltage setting, being consecutive measurements.

The tests consisted in running high voltage scans, with three point gain tests
at each bias voltage, and three repetitions each. The chiller temperature was 
set to different values to check that the behaviour was consistent for all of
them. The low voltage powering of the hybrids was set to $10.5\V$, measuring 
$10\V$ on the sense wires of the converters.

The results 
show that multiplexing the high voltage lines has no impact on the noise of
the stavelet modules, regardless of the powering scheme used on the stavelets 
and the temperature setting of the cooling. The small differences are not 
relevant as they are of different signs for various tests and disappear when
the sensor is over-depleted.

Figure~\ref{fig:5-HVmux_SP} show the same comparison
for the serially powered stavelet. The results reflect a noise variation under
$5\ENC$ for over-depleted sensors. Only one test was done with this stavelet,
at $12\degC$ chiller setting and the hybrids temperatures were equal during 
both tests within $0.3\degC$.

\subsection{Electromagnetic Interference}

\subsubsection{Description of the Electromagnetic Aggressor}

A study of the electromagnetic interference was done with the DC-DC stavelet 
and a VersatileLink prototype. The VersatileLink project is described 
in~\cite{1748-0221-8-03-C03004}. It is a joint ATLAS, CMS and CERN project to 
develop a high speed fibre-optic transceiver suitable for use in high 
radiation environments, such as the Phase II Upgrade of the 
LHC. The device connects to two fibres using a modified low-mass SFP+ package 
with standard LC fibre connectors and couples to qualified, commercially 
available, radiation hard optical fibres.

High speed signals have the potential to broadcast by capacitive and inductive
coupling or by electromagnetic transmission to other parts of the detector. 
We suspected that the silicon strip read-out chips, being of limited
bandwidth, would be insensitive to signals transmitting at $4.8~{\rm Gbps}$, 
but this was the first test of this type performed with the VersatileLink
prototype. 

The situation is more complex than just a high frequency emission, due to the 
fact that typical high speed multiplexers frame the data they send in bigger 
groups of bits called frames, that 
repeat at a lower frequency, in this case $40~{\rm MHz}$. The current plan for 
the transmitter includes protocols that randomize the data in a given 
frame, meaning that even with data packets which are quite uniform, the 
actual bits sent by the line driver to the VL will have a more random 
appearance and put more signal power into higher 
frequencies~\cite{1748-0221-5-11-C11022}.

\subsubsection{Test Description}

The tests performed consisted in measuring
the input noise using three point gain tests centred at $1\fC$, and Double 
Trigger Noise 
tests, putting the VersatileLink prototype at different locations
with respect to the DC-DC stavelet. The locations were chosen as
both potential locations for the final device (near the EoS card) and also
locations with potentially high noise injection on the 
stavelet~\cite{SvenDiplomaThesis,ArgosVL}.

\begin{figure}[!htb]
\begin{center}
    \includegraphics{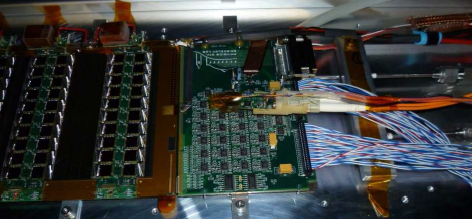}
    \caption{Position of the VersatileLink on top of the EoS
    card for one of the tests.}\label{fig:5-vl_position}
\end{center}
\end{figure}

An example of one of such locations is shown in Figure~\ref{fig:5-vl_position},
with the VersatileLink prototype lying on top of the EoS card. This position
is close to the electrical signals coming from the modules, as well as the
place at which the VersatileLink device would be located.

The cables used to power the VersatileLink were kept parallel to the cables
used to power and read-out the stavelet, in a similar arrangement that would be
a realistic scenario, but with data cables that would not be present. 
This way, the testing conditions are worse than they would be in the actual
detector.

The VersatileLink prototype is connected to a Bit Error Rate Tester (BERT),
which generates a Pseudo-Random Bit Stream (PRBS), simulating the data transfer
to and from a stave, at a rate of $4.8~{\rm Gbps}$.

The parameters of the tests were:

\begin{itemize}
\item Sensor bias: $250\V$.
\item Low voltage: $10.3\V$ ($10\V$ on the sense wires).
\item $T_{\rm chiller} = 12\degC$.
\end{itemize}

There was no light leak test performed with this prototype, due to the 
working wavelength of $1310\nm$, to which silicon detectors are insensitive.
This case is different to the current SCT modules that use $850\nm$ lasers,
which can generate signals on the silicon strips~\cite{Phillips:2007db}.

\subsubsection{Tests Results}

\begin{figure}[!htb]
\begin{center}
    \includegraphics[scale=0.5]{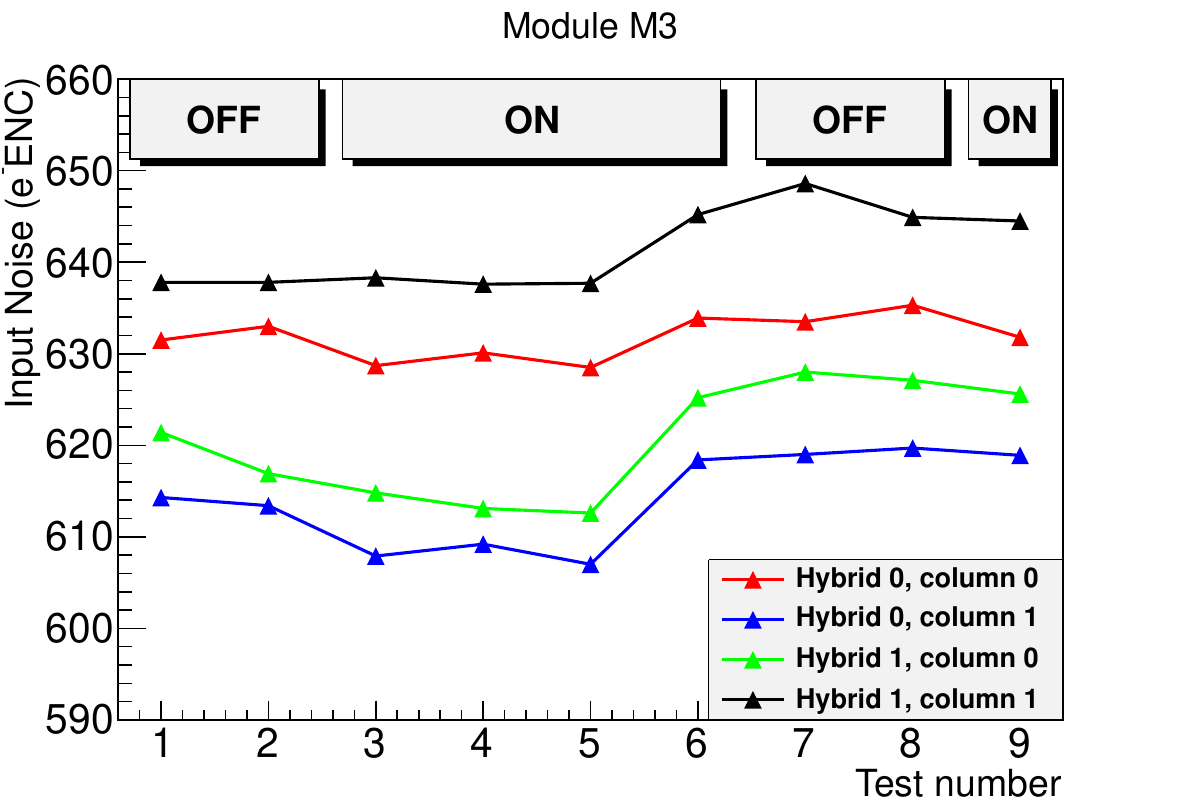}
    \caption{Noise measurements for module M3 with the interaction of the
    VersatileLink.}\label{fig:5-Versatile_M3}
\end{center}
\end{figure}

Figure~\ref{fig:5-Versatile_M3} shows the noise results for module M3 of the
stavelet. The values are the average for each chip column (10 chips). The plot
also shows the status (ON/OFF) of the VersatileLink prototype for each test.

The noise reference was established in tests number 1 and 2,  and the small
noise increase observed is consistent with the temperature increase also
observed on the stavelet during the tests, as shown in 
Figure~\ref{fig:5-Versatile_Temperatures}. The temperature increase was due to
the only available chiller at the time of the tests, which was not working
properly and could not sustain the cooling power for prolongued periods.

The test involved placing the
prototype in different positions and orientations, as well as turning it
off before the last measurement. The fact that it was turned off during
tests number 7 and 8 confirms the noise increase was not caused by the 
VersatileLink.

\begin{figure}[!htb]
\begin{center}
    \includegraphics[scale=0.5]{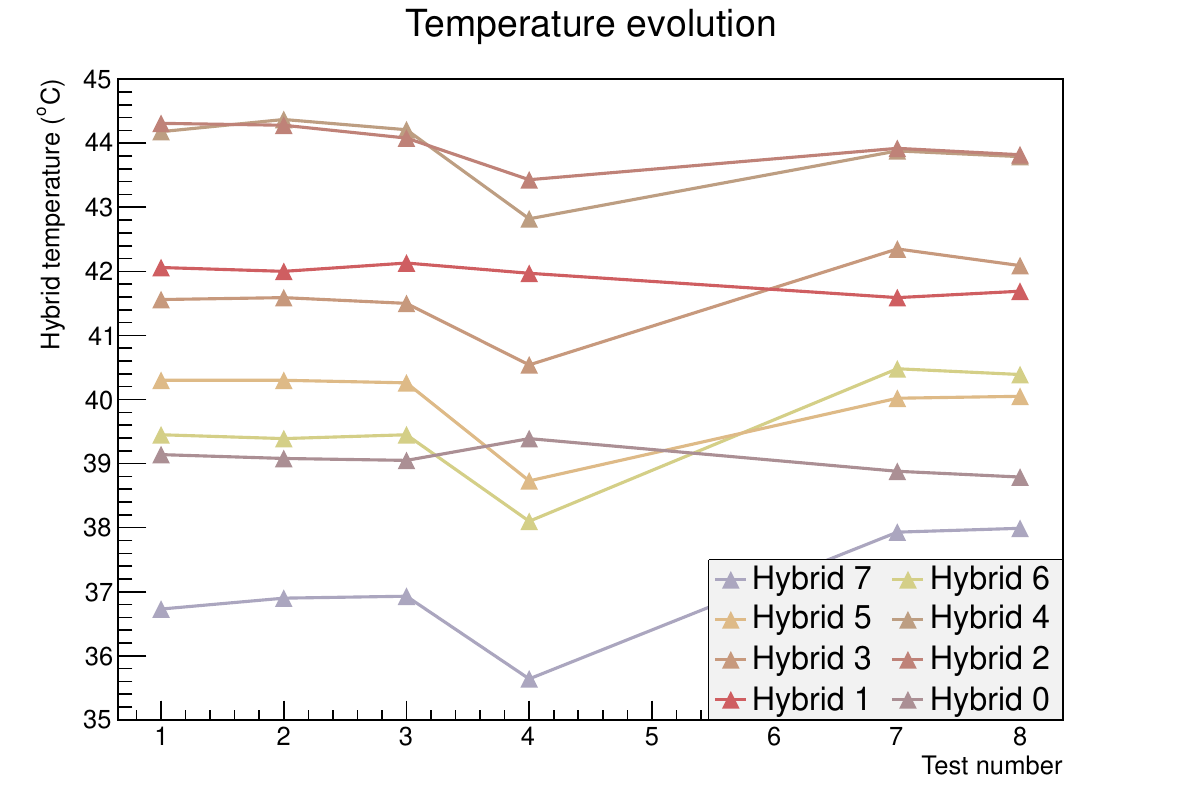}
    \caption{Temperatures measured on the stavelet during the noise 
    tests with the interaction of the
    VersatileLink.}\label{fig:5-Versatile_Temperatures}
\end{center}
\end{figure}

The tests were done in the following sequence:

\begin{itemize}
\item Tests 1 and 2: VersatileLink prototype turned off. Reference noise 
measurements.
\item Test 3: VersatileLink placed close to the EoS card, turned on. The power 
cables were routed along the stavelet.
\item Test 4: VersatileLink placed close to the EoS card, turned on. The power 
cables were folded back, following the same path as the optical fibres, which
is the realistic configuration.
\item Test 5: VersatileLink placed above the EoS card, turned on. 
\item Test 6: VersatileLink placed above a group of ABCn25 chips in module M3, 
turned on. 
\item Tests 7 and 8: VersatileLink turned off to get back a reference noise 
measurement. 
\item Test 9: VersatileLink placed above a group of ABCn25 chips in module M3,
turned on. 
\end{itemize}

\begin{figure}[!htb]
 \begin{center}
  \includegraphics[scale=0.75]{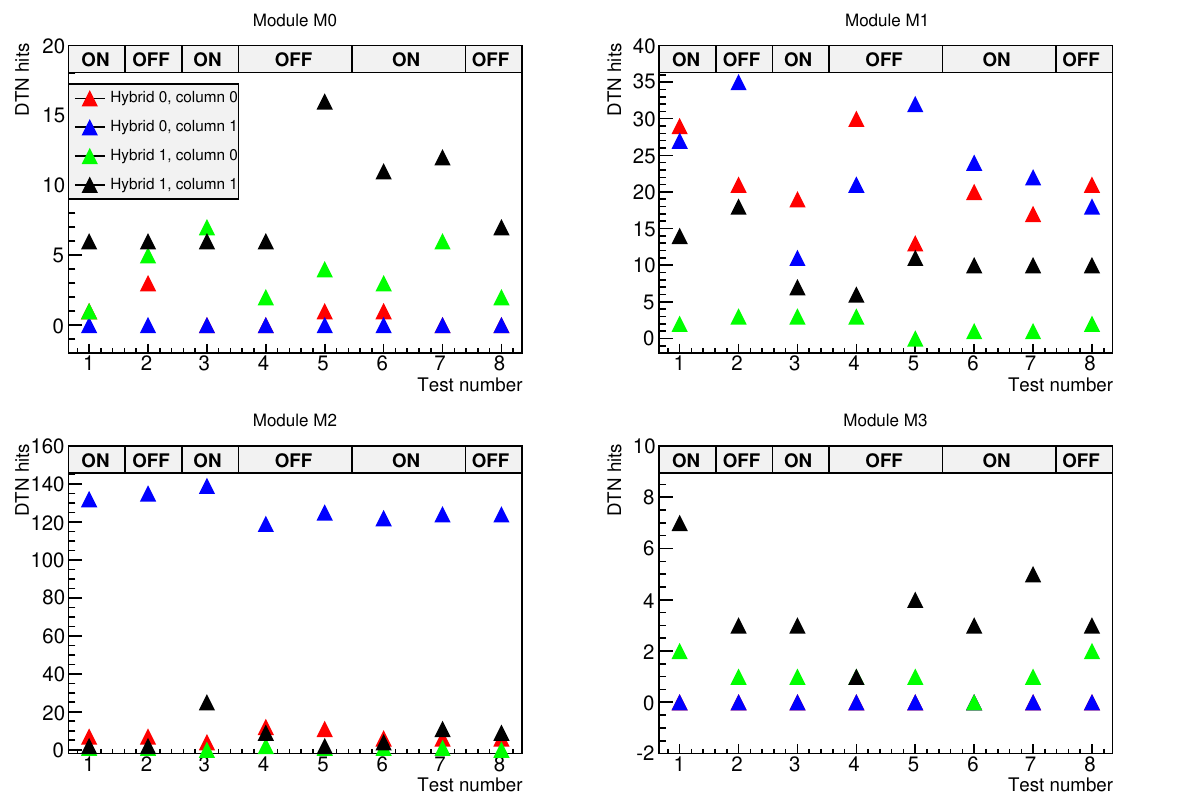}
 \end{center}
 \caption{Double Trigger Noise results for $0.5\fC$ threshold on the DC-DC 
 stavelet with the interaction of the     
 VersatileLink.}\label{fig:5-versatile_dtnplot}
\end{figure}

Figure~\ref{fig:5-versatile_dtnplot} shows the Double 
Trigger Noise results for another test performed. Only the $0.5\fC$ threshold 
is presented here, in order to be able to observe variations that might come 
unnoticed with the other higher thresholds. The variations are consistent with 
random noise and big variations from one test to another were not observed. 

Both $1\fC$ and $0.75\fC$ thresholds for all these Double Trigger Noise
tests resulted in zero hits for all columns (except for one that sometimes 
shows a few hits).

The tests labelling is different from the ENC tests. In the DTN case, the
tests sequence is the following:

\begin{enumerate}
\item VersatileLink on top of module M3, turned on.
\item VersatileLink on top of module M3, turned off.
\item VersatileLink on top of EoS card, turned on.
\item VersatileLink on top of EoS card, turned off.
\item VersatileLink on top of EoS card, turned off.
\item VersatileLink on top of EoS card, turned on.
\item VersatileLink on top of EoS card, rotated $90^{\circ}$, turned on.
\item VersatileLink on top of EoS card, turned off.
\end{enumerate}

One of the first prototype Versatile Link (VL) transceivers was used to test 
the ITk stavelet for immunity to the signals that the VL generates. 
Several tests were performed with different locations of the VL relative to 
the stavelet and the noise was compared to the cases where the VL was 
not powered on. All noise variations seen were consistent with temperature 
fluctuations of the system. There was no indication that the
VL contributed to increase the noise on the ABCn25 based system.

\subsection{Dual and Single Stavelet Tests}

The two stavelets, DC-DC and serially powered, have been tested separately and
together with the same DAQ, powering and cooling systems. This section presents
the results of the tests done with two stavelets running at the same time.

\begin{figure}[!htb]
 \begin{center}
  \subfigure[Noise comparison on the DC-DC stavelet alone and 
  together with the SP stavelet.]{
    \includegraphics[scale=0.65]{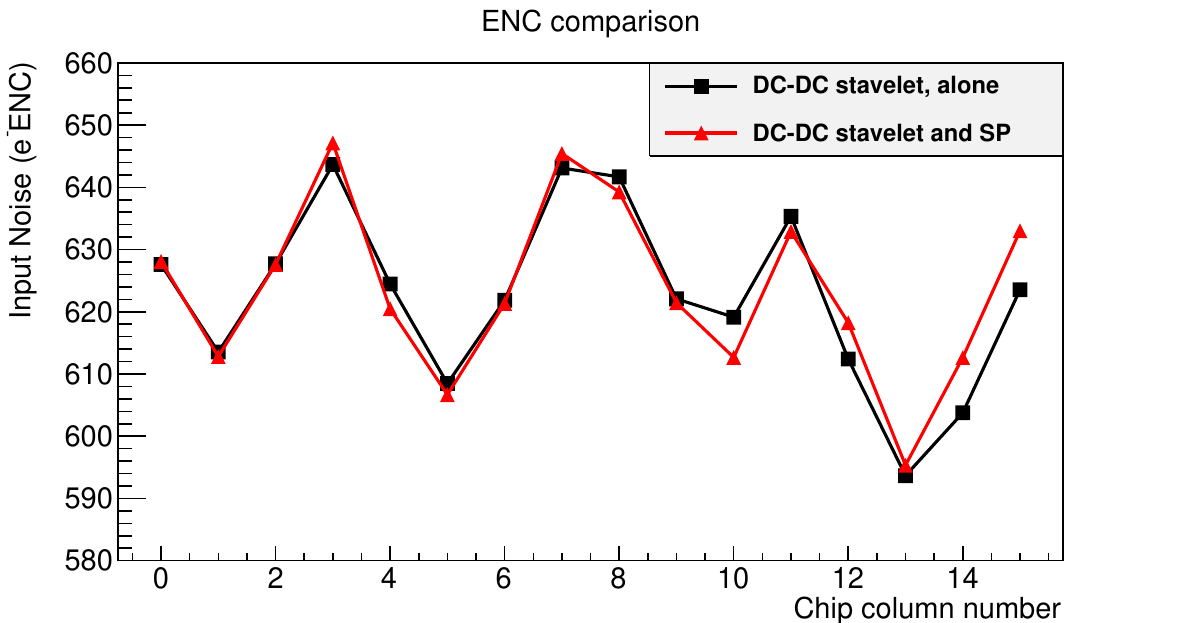}
  }
  \subfigure[Double Trigger Noise comparison on the DC-DC stavelet alone and 
  together with the SP stavelet.]{
    \includegraphics[scale=0.65]{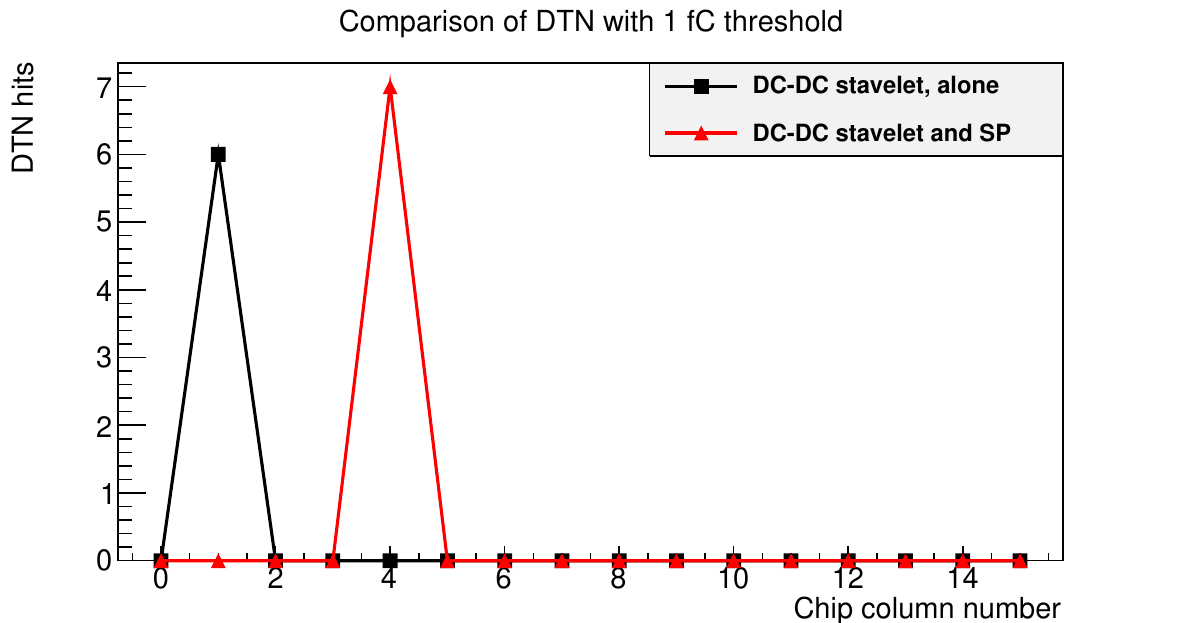}
  }
  \caption{Noise differences on the DC-DC stavelet when operated alone and when 
  operated with the SP stavelet at the same time.}\label{fig:5-TwoStavelets_DCDC}
 \end{center}
\end{figure}

The goal of this type of test is to assess if there is any kind of 
interference between two near-by stavelets. In particular, I was 
interested in studying the possible noise increase in one stavelet caused by
DC-DC converters placed close to the adjacent stavelet. Also, the common 
read-out using the same HSIO board and adjacent power lines, both hybrid power 
and sensor bias, could have an influence in the noise pick-up on the modules.

One of the stavelets is biased using independent high voltage lines, whereas 
the other is biased using the high voltage multiplexer. The results presented 
in Figures~\ref{fig:5-HVmux_DCDC_6deg} through~\ref{fig:5-HVmux_SP} show that
there is no difference between both biasing methods.

The cooling of the two stavelets when operating together was initially done by
connecting the serially powered stavelet in series with the DC-DC powered 
stavelet. The DC-DC stavelet was first in the cooling path, meaning it ran at
the same temperature as if it were operated alone. The SP
stavelet ran at higher temperature than it did when operated separately. 

Figure~\ref{fig:5-TwoStavelets_DCDC} shows the input noise results and the 
Double
Trigger Noise results for the DC-DC stavelet, when running separately and when
operated together with the serially powered stavelet. The chiller temperature
was set to $12\degC$ in both cases and the sensor bias was $250\V$. The 
stavelet sensor bias was done through independent high voltage lines.

The input noise results are presented per chip column, showing the average 
noise value 
for each group of 10 chips and the total number of hits for the Double Trigger
Noise plots.

\begin{figure}[!htb]
 \begin{center}
  \subfigure[Noise comparison on the serially powered stavelet alone and 
  together with the DC-DC stavelet.]{
    \includegraphics[scale=0.65]{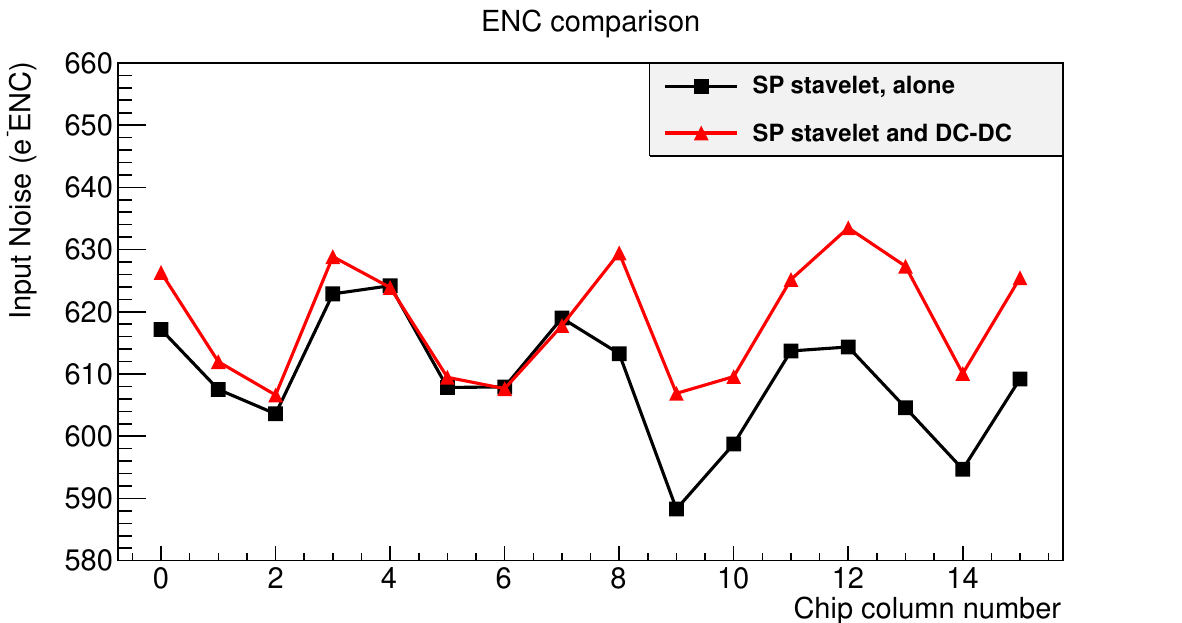}
  }
  \subfigure[Double Trigger Noise comparison on the serially powered stavelet alone and 
  together with the DC-DC stavelet.]{
    \includegraphics[scale=0.65]{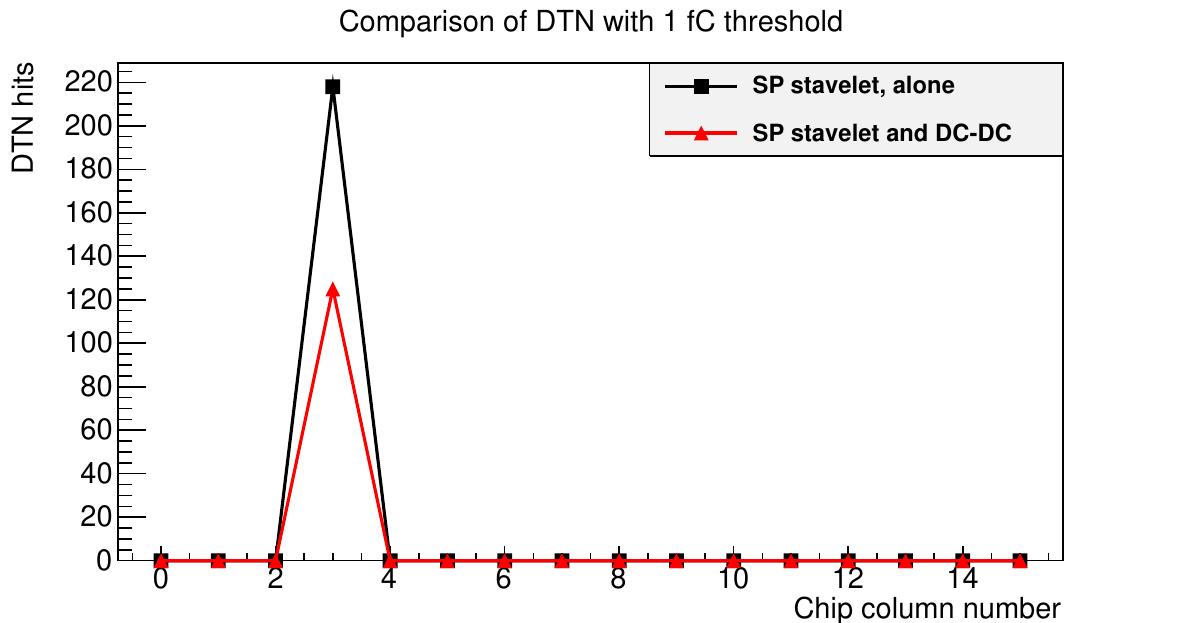}
  }
  \caption{Noise differences on the DC-DC stavelet when operated alone and when 
  operated with the SP stavelet at the same time.}\label{fig:5-TwoStavelets_SP}
 \end{center}
\end{figure}

The noise was measured by doing three point gain tests centred at $1\fC$. 
The plots show that there is no increase in the noise when operating both 
stavelets at the same time and the Double Trigger Noise is not affected either.

Figure~\ref{fig:5-TwoStavelets_SP} shows the input noise 
results and the Double Trigger Noise results for the serially powered stavelet,
again when running alone and when
operated together with the serially powered stavelet. The chiller temperature
was set to $12\degC$ in both cases and the sensor bias was $250\V$. The 
stavelet bias was done through the high voltage multiplexer in both cases. 
Results shown before in Figure~\ref{fig:5-HVmux_SP} confirm the noise is not
affected by the use of high voltage multiplexing.

The first tests resulted in a noise increase by around $70\ENC$ in some modules
of the serially powered stavelet, which was eliminated after using better 
twisted cables. The cables on the hybrids power supplies are twisted and using
a common mode choke in order to get rid of the common mode noise, which the 
serially powered stavelet is more affected by than the DC-DC stavelet.

\begin{figure}[!htb]
 \begin{center}
  \includegraphics[scale=0.65]{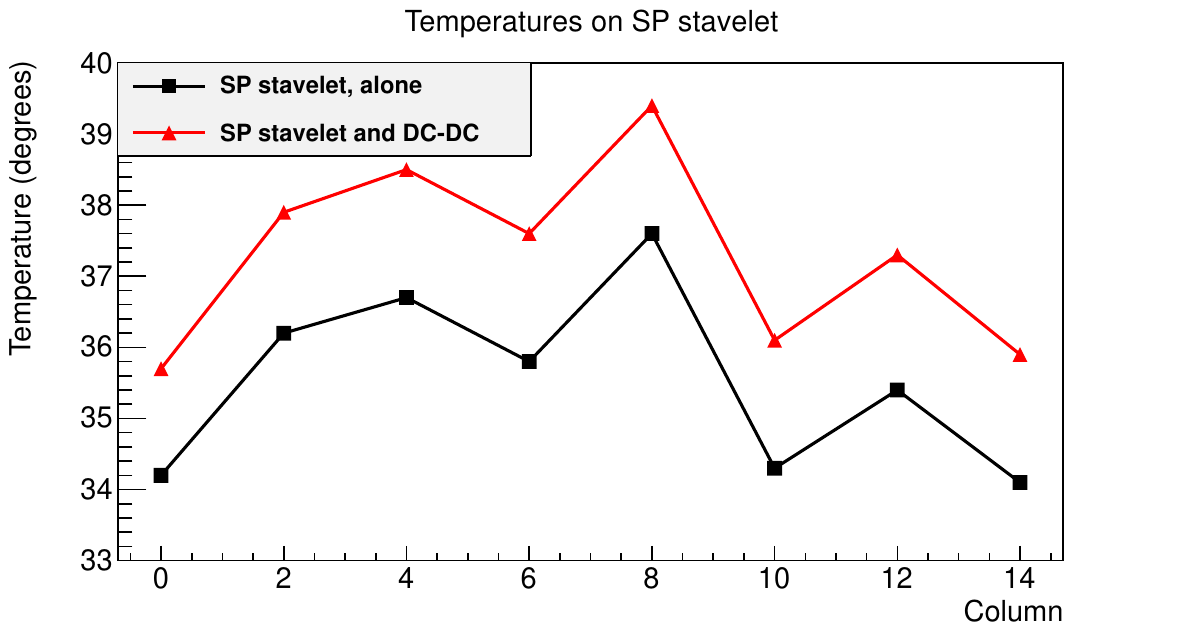}
  \caption{Temperatures recorded on the Serially Powered stavelet when operated
  alone and together with the DC-DC 
  stavelet.}\label{fig:5-TwoStavelets_Temperatures_SP}
 \end{center}
\end{figure}

Considering the temperature increase on the serially powered stavelet due to
the cooling configuration, these plots show that there is no increase in the 
noise caused by operation of both 
stavelets at the same time. The temperatures on each column in both cases are
shown in Figure~\ref{fig:5-TwoStavelets_Temperatures_SP}. There is a global 
increase of the hybrids temperatures between $1.5$ and $2\degC$.

As observed in the DC-DC stavelet, running both stavelets at the same time
does not have an effect on the Double Trigger Noise hits.

The Double Trigger Noise on the serially powered stavelet shows column 3 is
very noisy. However, all the other columns are clean at $1\fC$ threshold, as
expected. The noise on column 3 of this stavelet is related to the use of 
non-blind vias and the hits appear only at specific locations.

These results show that there is no negative influence originating in the
simultaneous read-out of the two stavelets.

\section{DC-DC Powered Stave}

The first full length stave was assembled at Rutherford Appleton 
Laboratory~(RAL) by the end of 2013. 
The Stave-250 is a 12-module stave with DC-DC powering. The denomination
``250'' comes from the use of the ABCn25 chip.

Initially, only one
of the sides has been populated with modules. After its finalization, a 
serially powered stave is started to be assembled. When complete, the decision
to populate the second side of the two staves will be taken based on their 
performance.

The results obtained with this stave are presented in this section.

\subsection{Assembly of the Stave}

The modules are glued on a $1.2$~metre long bus tape that is mounted on the
stave core. The bus tape routes the signals and power to the hybrids, as well 
as the sensors bias.

The core is  made of thermally conductive carbon foam in a honeycomb layout.
It provides support and cooling for the 24 modules that can be mounted on
the stave, one on each side. Titanium cooling pipes are embedded in the core,
providing the cooling to the whole stave volume.

All the modules used in the assembly of the stave are Grade A, with FZ1 
sensors. They come from multiple assembly sites:

\begin{itemize}
 \item University of Cambridge, United Kingdom (CAM).
 \item The University of Liverpool, United Kingdom (LIV).
 \item University of California, Santa Cruz, United States (UCSC).
 \item University of Freiburg, Germany (FREI).
\end{itemize}

The DC-DC converters used in the Stave-250 are a re-manufacture of the CERN 
tandem
converter, which has a dual output conceived to power two hybrids on one module
with a single board. The main motivation to build this kind of converter was
to have a star point for common referencing of the two hybrids on a module,
which for single converters is done with wire. By using a common board, the
star point is implemented together with the two converters on a single printed
circuit board.

The modification consists of a partially split ground plane, which originally 
was 
a solid ground plane for both converters and the splitting decreased the noise 
when tested on a module of a stavelet~\cite{TandemDCDC}.

In addition, the board includes a 1-Wire chip that is used to turn on and 
off each of the power outputs individually.
This converter was initially described in section~\ref{sec:4-dcdc}.

The modules are glued down on the stave bus tape using a vacuum jig that picks
up the module and the vacuum is released when the module is in place. This is
done to avoid tension on concave modules, so the module returns to its shape,
being pressed down with the vacuum cups.

SE4445 glue was used to glue the modules, curing overnight at room temperature.
After the modules, the BCC boards and the DC-DC converters are also glued down
and when cured, the wire-bonding is done.

The converters for the first three modules were stuck to the ceramics using
Araldite Rapid glue and then glued to the stave using double sided tape. This
procedure proved to provide poor thermal contact for the converters, and was
changed for the rest of the modules. The remaining nine converters used SE4445
and run cooler than the first three.

Assembly commenced from the end farthest from the End of Stave board. Initially
one module was glued down and, after checking the procedure was correct, two
more modules were mounted, together with the DC-DC converters and BCC boards.
Next, they were wire-bonded and the stave with three modules was tested.

After the first results, six more modules were assembled and the stave was
tested with nine modules. Last, the final three modules were mounted and the
results on a full single-sided stave were obtained.

\subsection{Setup at RAL}

The setup available at RAL is similar to the one at CERN. There are three TTi 
power supplies with dual outputs and a VME crate with three SCTHV 
preproduction boards, for a total of 12 high voltage channels. These boards 
are controlled through the SCTDAQ software in the same way as in the setup at 
CERN~\cite{SCTHV}.

Powering of the hybrids was split in three segments, with three different lines
on the bus tape, that can withstand the high total current needed for the
twelve modules. 
The total output power when all modules are powered on is around $300~{\rm W}$.
All the power supplies are controlled via USB using the SCTDAQ software.

A Neslab Thermoflex 1400 chiller with tap water is used to cool down the stave.
The temperature is set to $13\degC$, in order to avoid going below the dew 
point, which is typically around $10\degC$ in the lab at RAL.
A nitrogen supply is available to run the tests in dry conditions and to ensure
there is no condensation if the chiller is run at lower temperature.

In addition, there is a blow-off ${\rm CO_2}$ cooling system designed to cool
down the stave down to around $-30\degC$. It will be used in some of the tests
in the future.

\subsection{Results with Three Modules}

The first tests on the stave were done after three modules were placed and 
bonded on the stave. These three modules are:

\begin{itemize}
 \item Two modules from University of Cambridge: CAM-M5 (module 0) and CAM-M6 (module 1).
 \item One module from University of California, Santa Cruz: UCSC-M4 (module 2).
\end{itemize}

The first tests consisted in measuring the temperatures of the DC-DC converters
when powering up the hybrids. Initially, the coils were shielded with a copper
coated plastic case. This type of shield does not have a good heat transfer and
that caused a measured temperature on the coil of $140\degC$.

This high temperature could cause the destruction of the converter, so reducing
it was a requirement before proceeding with further tests. In a first 
iteration, the shield was replaced with a tinned copper one, but the 
temperature was still around $140\degC$. After filling the shield with thermal
pad, in contact with the coil, the temperature finally decreased to around
$60\degC$.

\begin{figure}[!htbp]
 \begin{center}
 \subfigure[Circuit diagram of the buffers at the EoS.]{
   \includegraphics[scale=0.55]{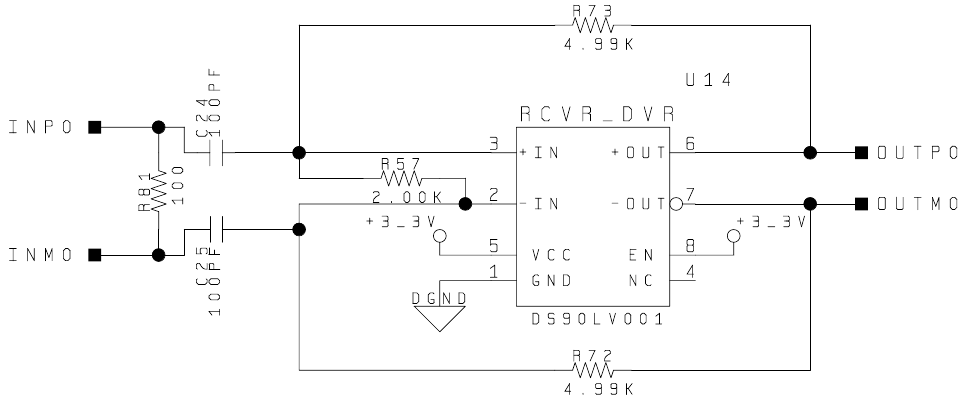}
   \label{fig:5-buffers}
 }
 
 \subfigure[Signal for hybrid 1 with $100~\Omega$ termination.]{
    \includegraphics[scale=0.35]{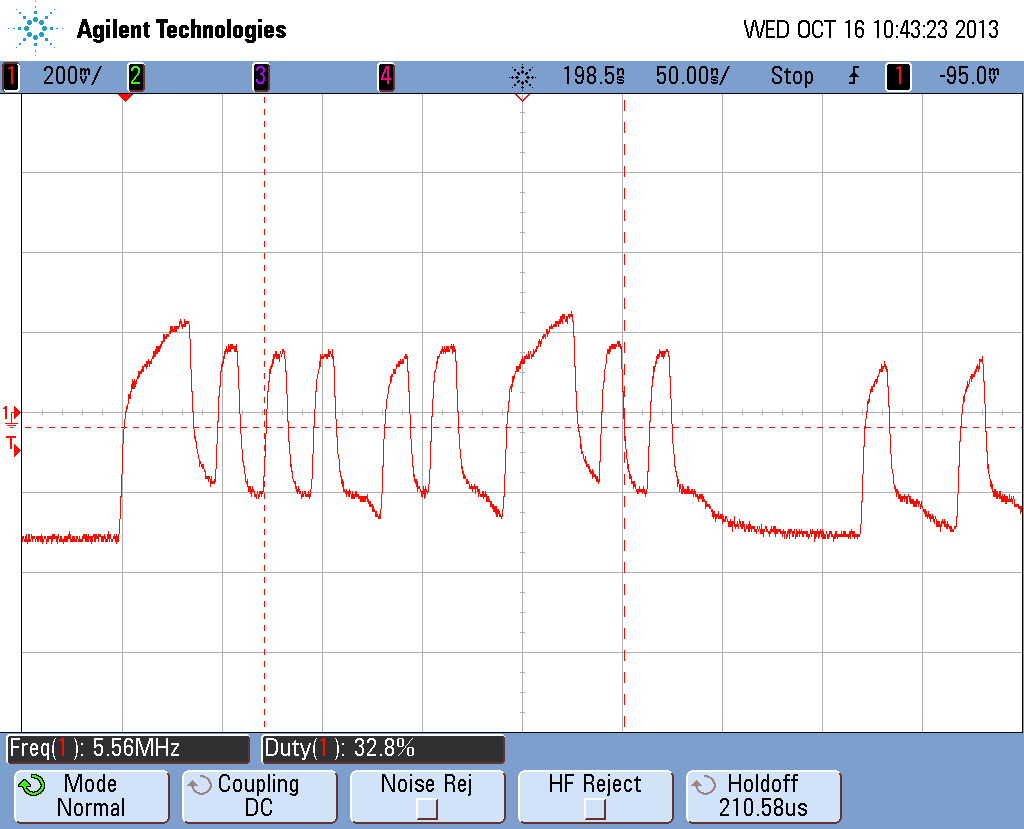}
    \label{fig:5-H1_rTerm100}
 }
 
 \subfigure[Signal for hybrid 1 with $56~\Omega$ termination.]{
    \includegraphics[scale=0.35]{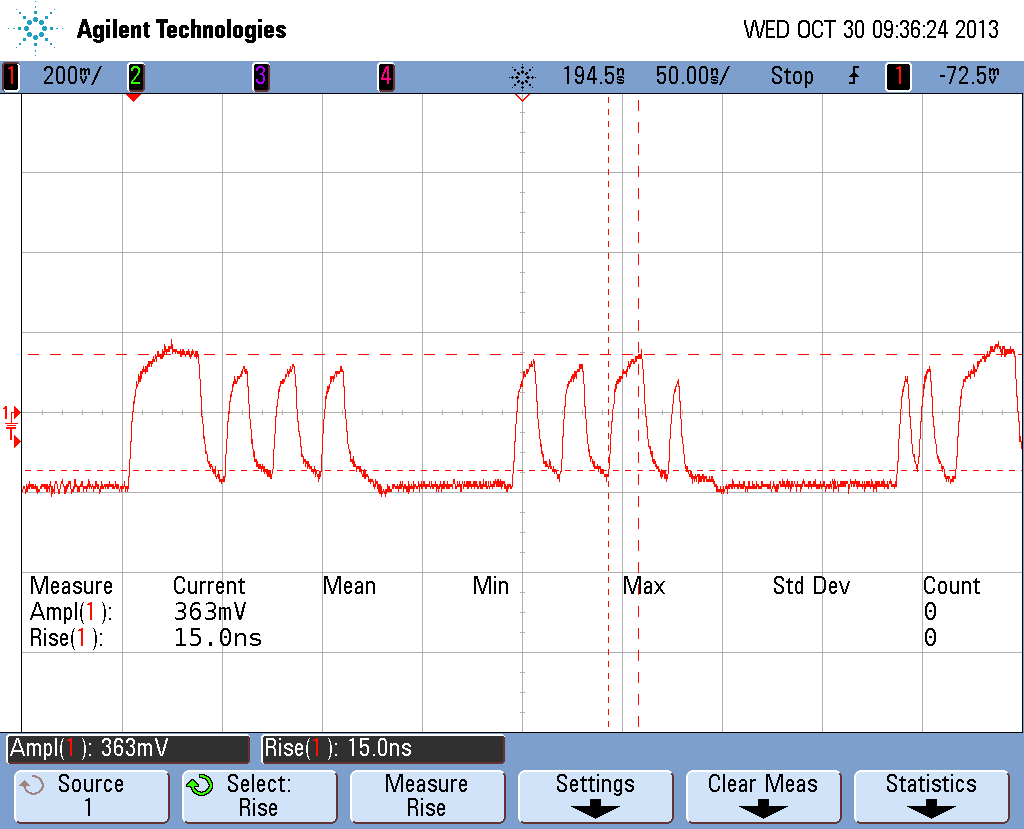}\label{fig:5-H1_rTerm56}
 }
  \caption{Signal integrity at the End of Stave buffers for different termination
  resistors.}\label{fig:5-stave_3modules_signals}
 \end{center}
\end{figure}

Initially, the communication with the modules was not stable and some streams
would have a lot of errors when trying to talk to the hybrids. Since the
stave bus tape is a longer one than the stavelets bus tape, a study of the
signals was done to assess the signal integrity. The critical point is on the
End of Stave card, right at the buffers for the BCC signals.

Figure~\ref{fig:5-stave_3modules_signals} show the analysis of the signals at
the termination resistor of the buffers (see Figure~\ref{fig:5-buffers}) 
on the End of Stave card. The $100~\Omega$ termination resistor that is 
typically used in LVDS signal transmission and used on the End of Stave card 
results in reflections that causes a badly formed signal, resulting in errors 
in the reception at the HSIO. 

The signal at the $100~\Omega$ termination resistor is 
shown in Figure~\ref{fig:5-H1_rTerm100}, showing that the fast signals cannot
reach their levels, as opposed to the signal at the $56~\Omega$ resistor, shown
in Figure~\ref{fig:5-H1_rTerm56}, with a lower signal amplitude but with well
defined levels.

The capacitive load difference of the long bus tape, combined with the limited
drive current of the BCC chips, are responsible for this signal integrity 
problem. However, the drawback of reducing the value of the termination 
resistors is a reduction in the signals amplitude. In some cases, the signal 
levels might be too low even though the reflections are eliminated.

\begin{figure}[!htbp]
 \begin{center}
 \includegraphics[scale=0.6]{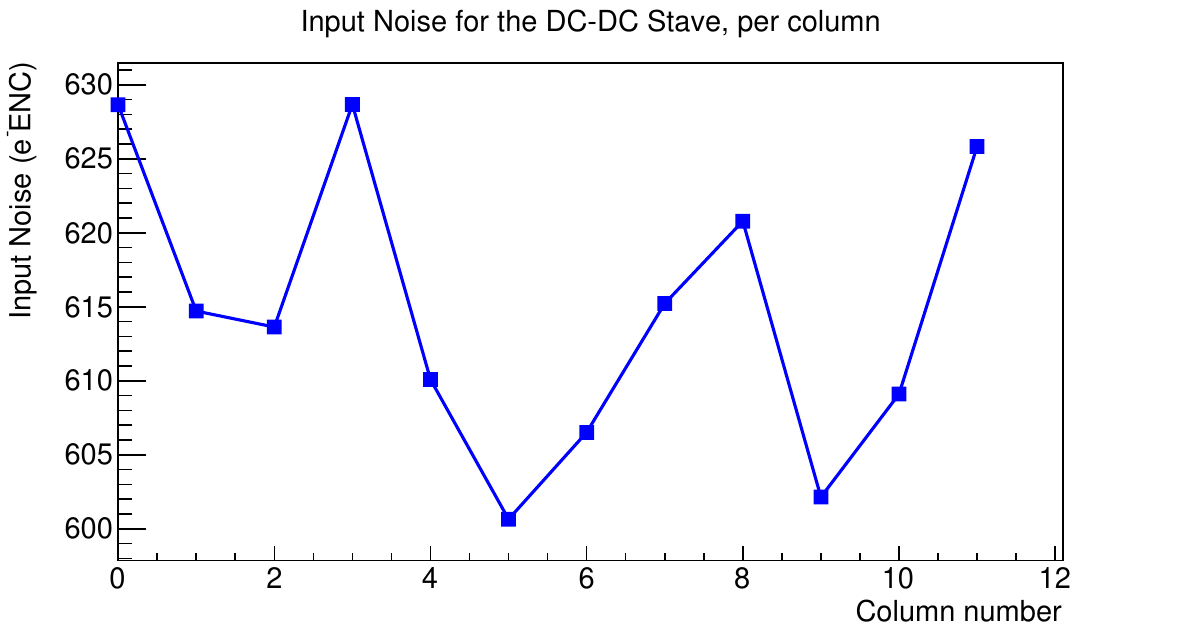}
  \caption{Input noise results of the DC-DC stave with the first three 
  modules.}\label{fig:5-stave_3modules_enc}
 \end{center}
\end{figure}

\begin{figure}[!ht]
 \begin{center}
\subfigure[DTN on the Stave 250 with three modules for high thresholds.]{
 \includegraphics[scale=0.6]{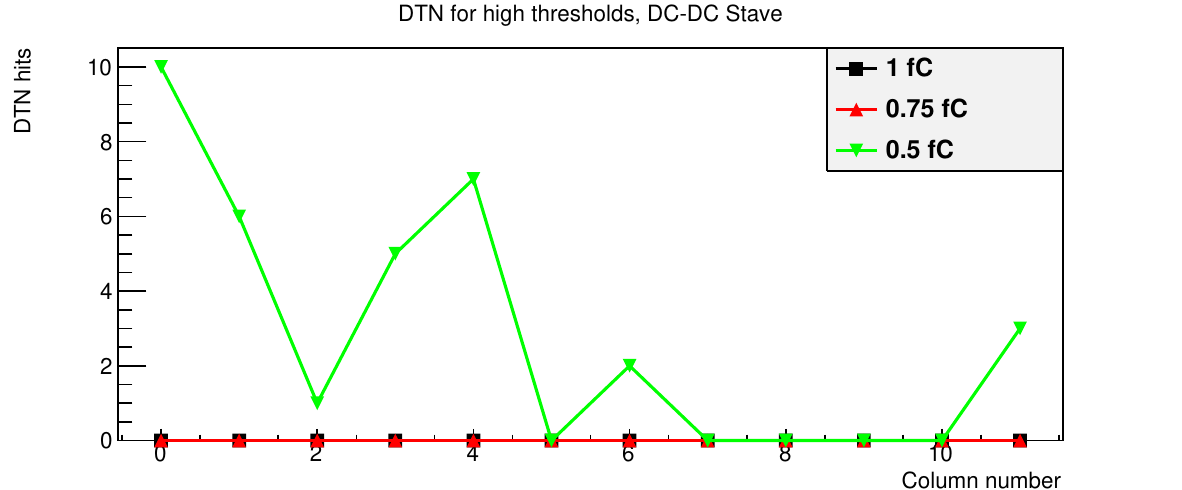}
}

\subfigure[DTN on the Stave 250 with three modules for low thresholds.]{
 \includegraphics[scale=0.6]{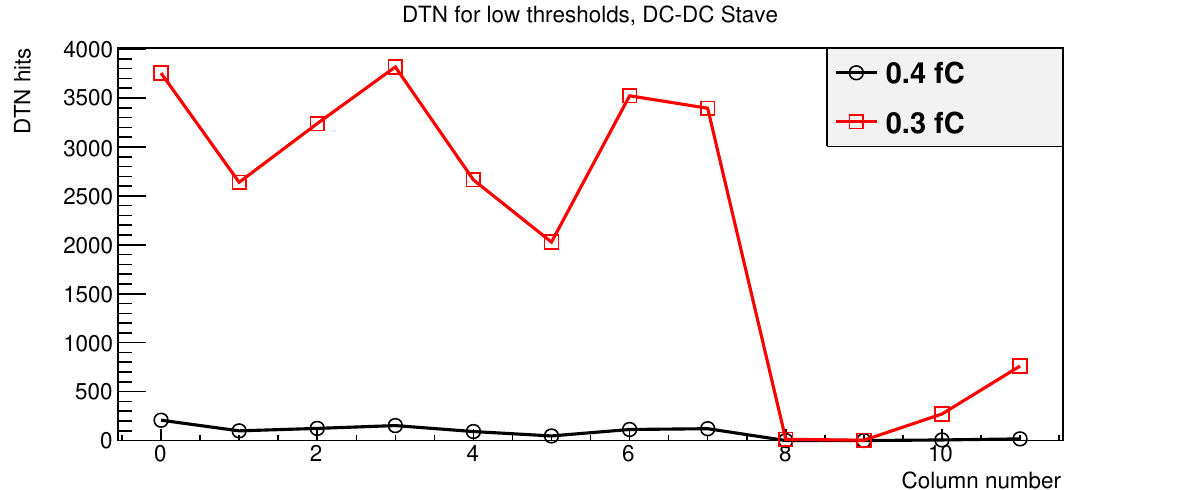}
}
  \caption{Double Trigger Noise of the DC-DC stave with three 
  modules.}\label{fig:5-stave_3modules_dtn}
 \end{center}
\end{figure}

The noise results for the stave with three modules are shown in 
Figure~\ref{fig:5-stave_3modules_enc}. The three modules have very good noise
results and the values are consistent with the results for the individual 
modules. The X axis represents the chip columns (3 modules, 6 hybrids, 2 
columns per hybrid results in 12 columns).
All columns have average noise values between $600$ and $630\ENC$.

In order to be able to run the input noise tests, the threshold scans had to be
started from $50\mV$, instead of the typical $0\mV$. Some hybrids were still
problematic from the signal integrity point of view and generating errors
at low thresholds. By starting at a higher threshold, these errors were 
eliminated and the initial threshold was set well below the $V_{t50}$ so the
noise measurements were not affected.

The Double Trigger Noise results are show in 
Figure~\ref{fig:5-stave_3modules_dtn}. The counts with $1$ and $0.75\fC$ 
thresholds are clean on all modules, while the count at $0.5\fC$ threshold
is not very high, being most of them due to some bad channels. The counts
at lower thresholds are as high as it would be expected.
The spikes appear in some
channels that show noisy behaviour at lower thresholds.

The bad channels are identified by looking at the Double Trigger Noise plots
generated by SCTDAQ, and spotting the channels that have hits at most of the
BCO delay settings. In some of the modules there is at least one bad channel
that could be masked to remove the Double Trigger Noise extra hits.

\subsection{Results with Nine Modules}

A picture of the stave with nine modules mounted is shown in 
Figure~\ref{fig:5-stave_9modules_picture}. The nine modules are mounted from 
the far end of the stave, shown at the left, and the End of Stave card is 
located at the right side of the picture.

\begin{figure}[!htb]
 \begin{center}
  \includegraphics[scale=0.55]{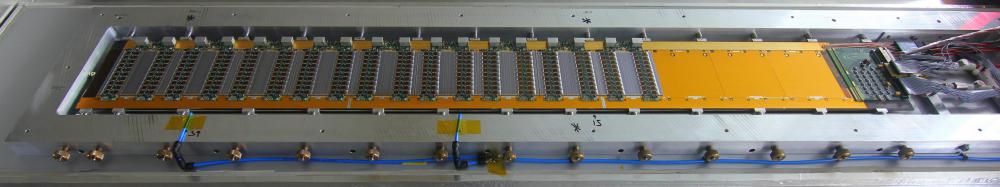}
  \caption{A picture of the DC-DC stave with nine 
  modules.}\label{fig:5-stave_9modules_picture}
 \end{center}
\end{figure}

The six modules mounted after the first three are the following:

\begin{itemize}
 \item Four from University of Cambridge: CAM-M9 as module 3, CAM-M10 as module 4, 
 CAM-M11 as module 5 and CAM-M3 as module 8.
 \item Two from The University of Liverpool: LIV-M31 as module 6 and LIV-M32 as
 module 7.
\end{itemize}

\begin{figure}[!htbp]
 \begin{center}
   \includegraphics[scale=0.6]{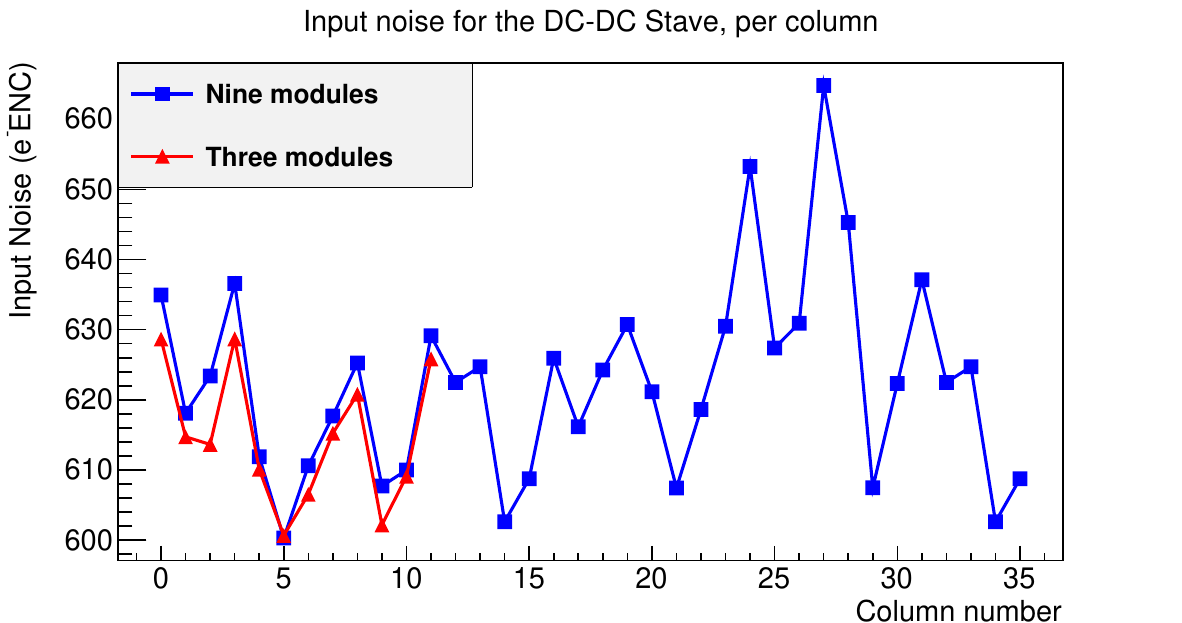}
  \caption{Input noise of the DC-DC stave with nine 
  modules.}\label{fig:5-stave_9modules_enc}
 \end{center}
\end{figure}

Figure~\ref{fig:5-stave_9modules_enc} shows the results for the 
input noise of the nine modules
on the stave. The noise values are consistent with the values measured on the
individual modules and they are all reasonable values. The results for the 
first three modules are included in the plot for comparison. They
do not show a great difference with respect to the values shown in 
Figure~\ref{fig:5-stave_3modules_enc}. The differences are consistent with the
temperature increase due to the next modules increased heat. The inner columns 
have a noise between
$600$ and $631\ENC$, while the outer columns show a noise between
$609$ and $665\ENC$.

The modules that present highest noise are LIV-M31 and LIV-M32, which are 
columns 24 to 31. These two modules have 
shield-less hybrids, which were designed without the bottom shield layer. 
However, the cause for the higher noise figures is not in the lack of shield,
but it is related to it. The flatness of the hybrids is not as good as other
shielded hybrids, which results in a different hybrid to sensors glue 
thickness, which effects the noise.

With the exception of those two modules, the noise on the stave with nine 
modules is between $600$ and $640\ENC$.

\begin{figure}[!htbp]
 \begin{center}
\subfigure[DTN on the Stave 250 with nine modules for high thresholds.]{
 \includegraphics[scale=0.6]{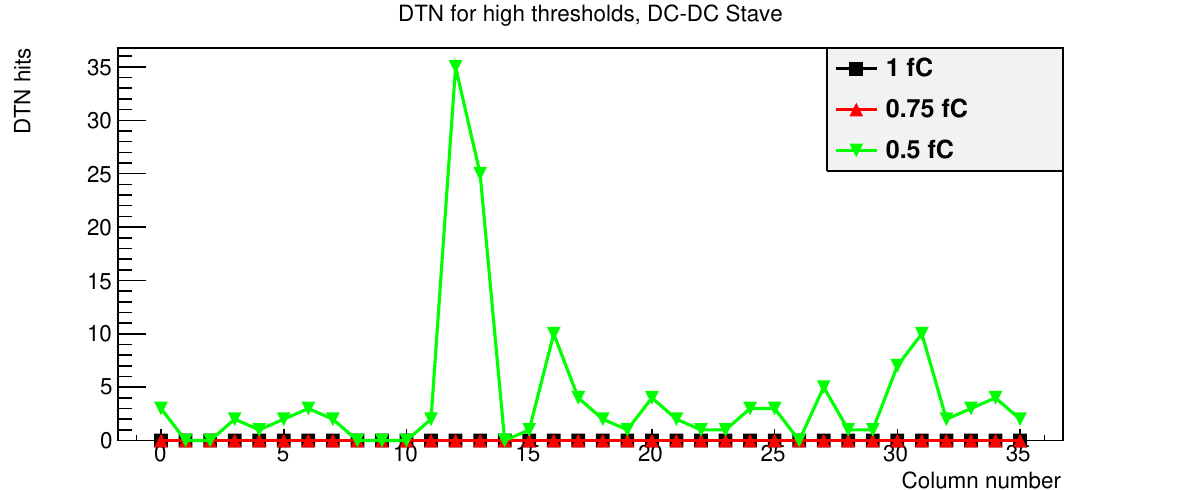}
}

\subfigure[DTN on the Stave 250 with nine modules for low thresholds.]{
 \includegraphics[scale=0.6]{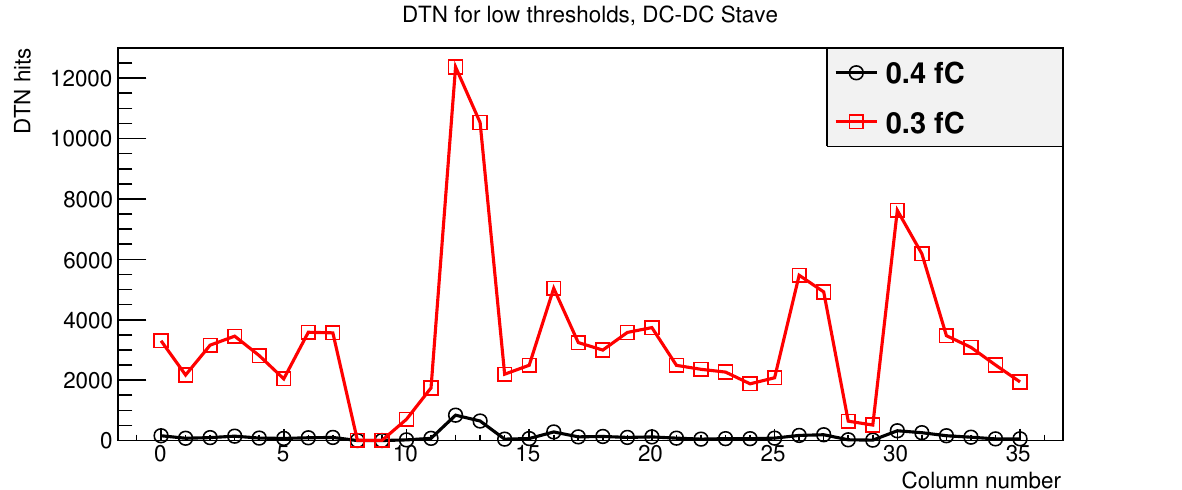}
}
  \caption{Double Trigger Noise of the DC-DC stave with nine 
  modules.}\label{fig:5-stave_9modules_dtn}
 \end{center}
\end{figure}

Figure~\ref{fig:5-stave_9modules_dtn} shows the Double 
Trigger Noise results on the stave with nine modules. As indicated in the 
results on three modules, the X axis represent the
chip columns (9 modules, 18 hybrids: 36 columns).
The higher threshold Double Trigger Noise results show no hits for the 
$1\fC$ and $0.75\fC$
thresholds, while some hits appear on the $0.5\fC$. The spikes appear in some
channels that show noisy behaviour at lower thresholds.

\begin{figure}[!htbp]
 \begin{center}
   \includegraphics[scale=0.6]{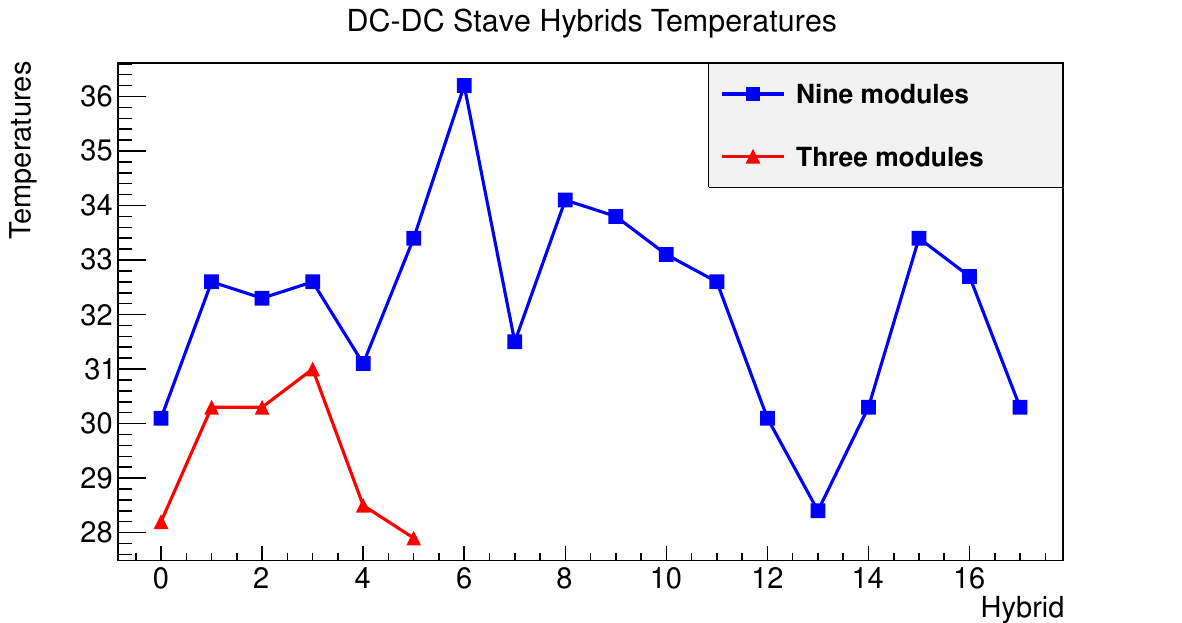}
  \caption{Hybrids temperatures on the DC-DC stave hybrids with nine 
  modules.}\label{fig:5-stave_temperatures}
 \end{center}
\end{figure}

These results show that the stave with nine modules worked very well, with good
noise results and very low noise pick-up. The temperatures plotted in 
Figure~\ref{fig:5-stave_temperatures}
were measured on the stave hybrids while being tested, both
with three modules and with nine modules. The first three modules show a 
temperature increase between $1.6$ and $5.5\degC$. These increased temperatures
give consistence to the small noise increase of the first three modules.

Results on the DC-DC Stave with nine modules were first presented at the ATLAS
Upgrade Week of November 2013~\cite{DCDCstave_AUW}. Additional results, such as
the DC-DC converters and the bus tape temperatures are shown in the talk.

\subsection{Results with Twelve Modules}

The last three modules that were mounted on the DC-DC stave were the following:

\begin{itemize}
 \item Three from University of Freiburg in Germany: FREI-M9 as module 9, 
 FREI-M11 as module 10 and FREI-M6 as module 11.
\end{itemize}

The same procedure for modules glueing as for the first nine was followed. The
DC-DC converters and the BCC boards were glued as for the previous six modules.
Testing conditions were the same as for the previous results.

\begin{figure}[!htb]
 \begin{center}
  \includegraphics[scale=0.58]{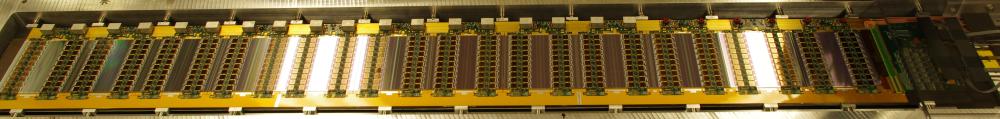}
  \caption{A picture of the DC-DC stave with all the modules on one side 
  modules.}\label{fig:5-stave_12modules_picture}
 \end{center}
\end{figure}

Figure~\ref{fig:5-stave_12modules_picture} shows a picture of one side of the 
complete stave, with the twelve modules, before bonding the DC-DC converters 
and BCC boards. The tests described next include repeating the same tests that 
were shown above for three and nine modules, as well as a low voltage scan and
a high voltage scan.

\begin{table}[!htb]
\begin{center}
\begin{tabular}{l|cccc}
\toprule
 Module & Isolation & $V_{\rm dep}$ & $C@250\V$ & $I_{\rm leak}@250\V$ \\
\hline
0 (CAM-M5) &  P-stop, $4\times 10^{12}\cm^{-3}$ & $190\V$ & $2.63\nF$ & $0.23\uA$ \\ 
1 (CAM-M6) &  P-stop, $4\times 10^{12}\cm^{-3}$ & $240\V$ & $2.95\nF$ & $0.27\uA$ \\ 
2 (UCSC-M4) &  P-stop, $4\times 10^{12}\cm^{-3}$ & unknown & unknown & $0.27\uA$ \\ 
3 (CAM-M9) &  P-stop, $4\times 10^{12}\cm^{-3}$ & $190\V$ & $3.28\nF$ & $0.27\uA$ \\ 
4 (CAM-M10) &  P-stop, $4\times 10^{12}\cm^{-3}$ & $200\V$ & $3.29\nF$ & $0.21\uA$ \\ 
5 (CAM-M11) &  P-stop, $4\times 10^{12}\cm^{-3}$ & $180\V$ & $3.28\nF$ & $0.21\uA$ \\ 
6 (LIV-M31) &  P-stop, $4\times 10^{12}\cm^{-3}$ & $250\V$ & $2.93\nF$ & $0.38\uA$ \\ 
7 (LIV-M32) &  P-stop, $4\times 10^{12}\cm^{-3}$ & $190\V$ & $2.92\nF$ & $0.17\uA$ \\ 
8 (CAM-M3) &  P-stop, $4\times 10^{12}\cm^{-3}$ & unknown & unknown & $0.25\uA$ \\ 
9 (FREI-M9) &  P-stop, $4\times 10^{12}\cm^{-3}$ & unknown & unknown & $0.20\uA$ \\ 
10 (FREI-M11) &  P-stop, $4\times 10^{12}\cm^{-3}$ & $250\V$ & unknown & $0.30\uA$ \\ 
11 (FREI-M6) &  P-stop, $4\times 10^{12}\cm^{-3}$ & $190\V$ & unknown & $0.20\uA$ \\ 
\bottomrule
\end{tabular}
\caption[Characteristics of the sensors on the DC-DC
stave.]{Characteristics of the sensors on the DC-DC
stave: isolation structures 
of the strips and concentration dose, 
depletion voltage, capacitance with $250\V$ bias and leakage current with
$250\V$ bias.}\label{tab:5-sensors_dcdc_stave}
\end{center}
\end{table}

The input noise and Double Trigger Noise tests were repeated for comparison,
biasing the sensors with $250\V$. Nitrogen was flushed into the stave 
enclosure and the chiller temperature was set to $13\degC$.

\begin{figure}[!htbp]
 \begin{center}
   \includegraphics[scale=0.6]{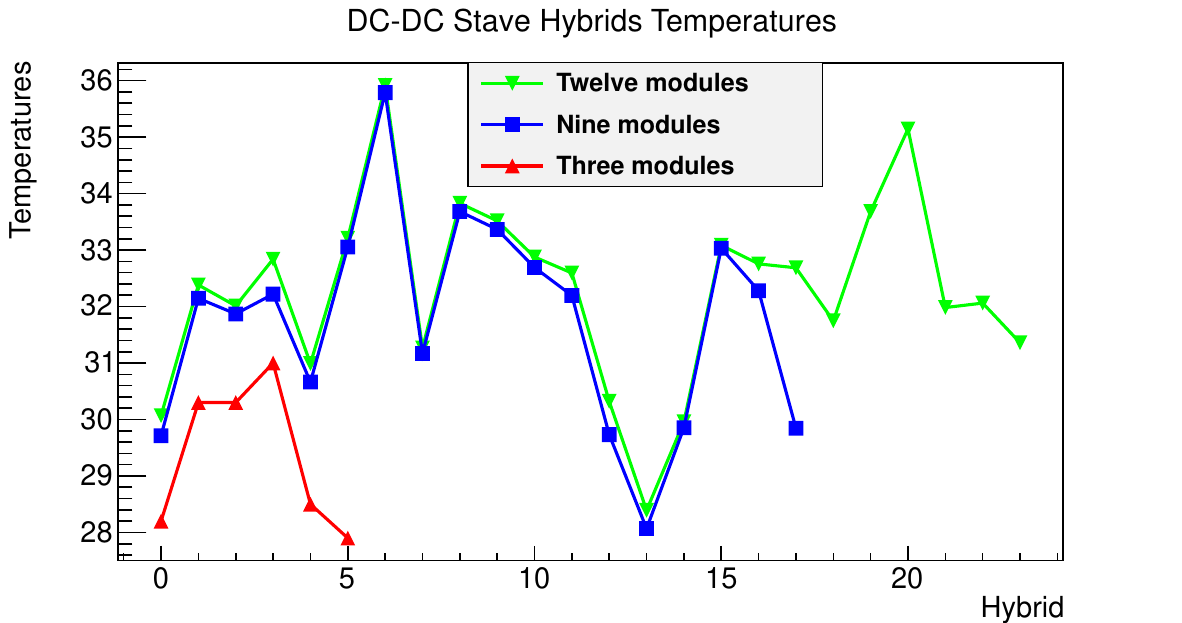}
  \caption{Hybrids temperatures on the DC-DC stave hybrids with 12 
  modules.}\label{fig:5-stave_12modules_temperatures}
 \end{center}
\end{figure}

The hybrids temperatures were also recorded for the 24 hybrids on the stavelet
while in operation. Figure~\ref{fig:5-stave_12modules_temperatures} shows the
temperatures of all the hybrids and the comparison with the first three and
nine modules temperatures.

The tests with 3 and 9 modules were done in October 2013, while the tests with
twelve modules were done in January 2014. The very small difference between 
nine and twelve modules can be explained due to the temperature difference 
during those dates. The nitrogen supply was between $7$ and $15\degC$ colder
in January than in late October. Therefore, despite the increase in heat load
coming from the last three modules, the temperature increase is much lower than
in the comparison between three and nine modules.

\begin{figure}[!htbp]
 \begin{center}
   \includegraphics[scale=0.75]{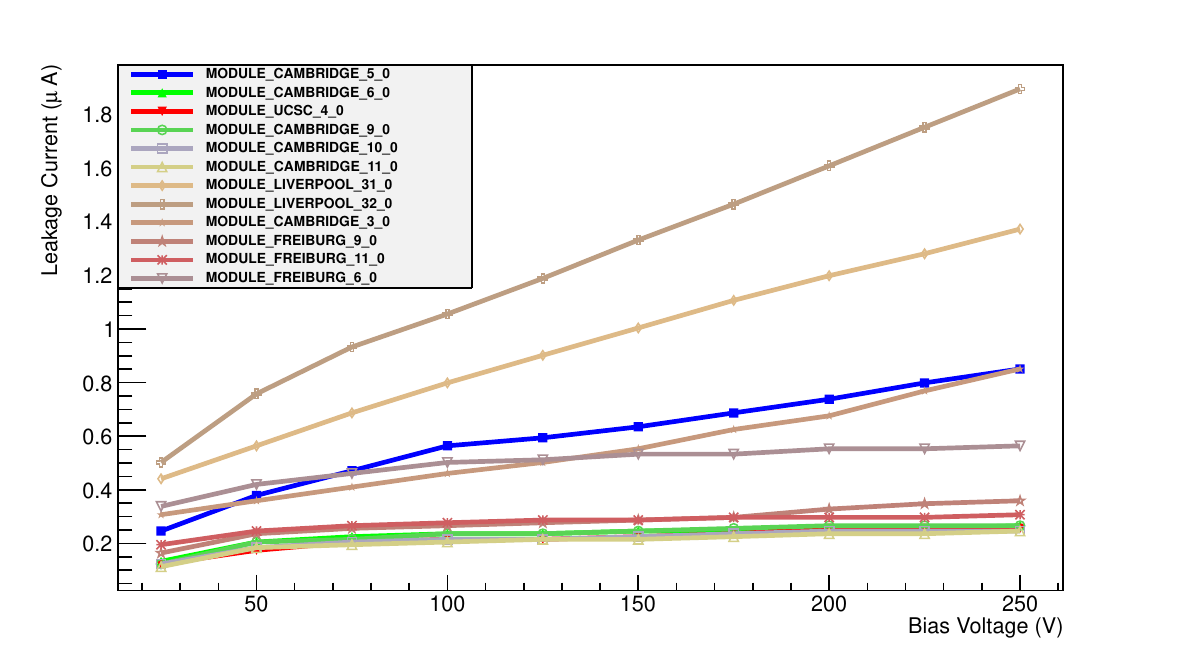}
  \caption{Leakage currents curves for each sensor mounted on the DC-DC stave,
  measured at $22\degC$ in nitrogen.}\label{fig:5-IVcurves_12modules}
 \end{center}
\end{figure}

The leakage currents were measured, performing an I/V scan using SCTDAQ. 
The results are 
shown in Figure~\ref{fig:5-IVcurves_12modules}. All sensors present leakage
current values consistent with the values presented in 
Table~\ref{tab:5-sensors_dcdc_stave}, and they show the stave assembly 
process did not have a negative impact on the sensors. The curves
were measured with the stave inside a light-tight enclosure, with a nitrogen
flow into the enclosure. The temperature of the sensors was $22\degC$ and
the voltage ramp was done in steps of $25\V$, waiting for $10$~seconds between
measuring the current and increasing the voltage.

\begin{figure}[!htbp]
 \begin{center}
   \includegraphics[scale=0.75]{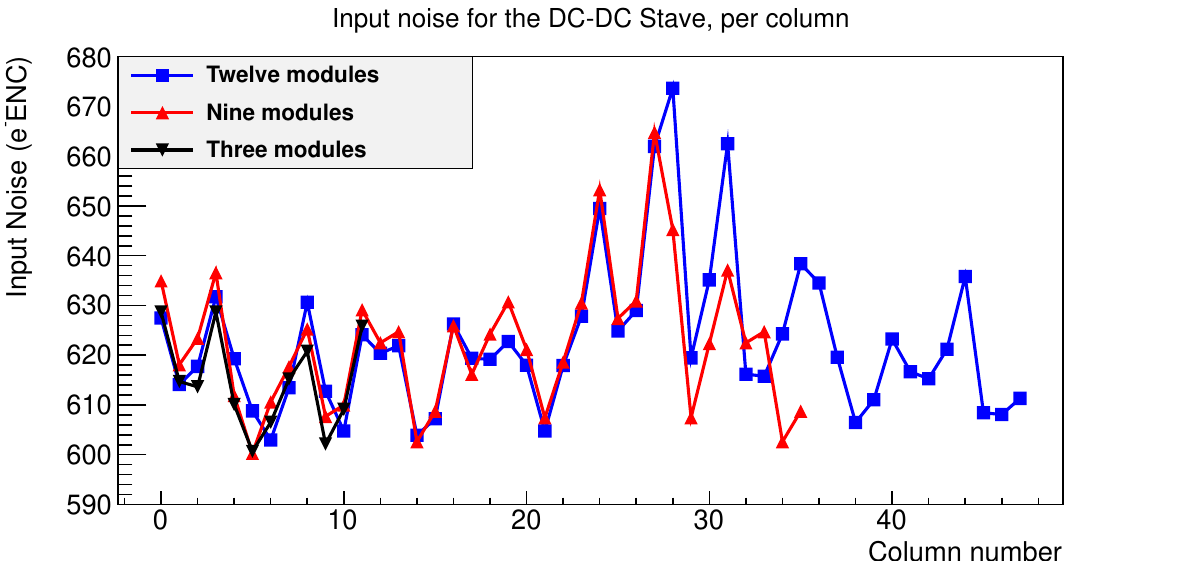}
  \caption{Input noise of the DC-DC stave with twelve modules, compared to
  the noise with three and nine modules.}\label{fig:5-stave_12modules_enc}
 \end{center}
\end{figure}

Figure~\ref{fig:5-stave_12modules_enc} shows the noise results of the stave 
with one complete side. The results with three and nine modules are also
plotted for the corresponding columns. There are no variations on the noise
that differ from the expected increase due to the higher operating temperature
of the hybrids that are further away in the cooling path.

\begin{figure}[!htbp]
 \begin{center}
\subfigure[DTN on the Stave 250 with twelve modules for high thresholds.]{
 \includegraphics[scale=0.65]{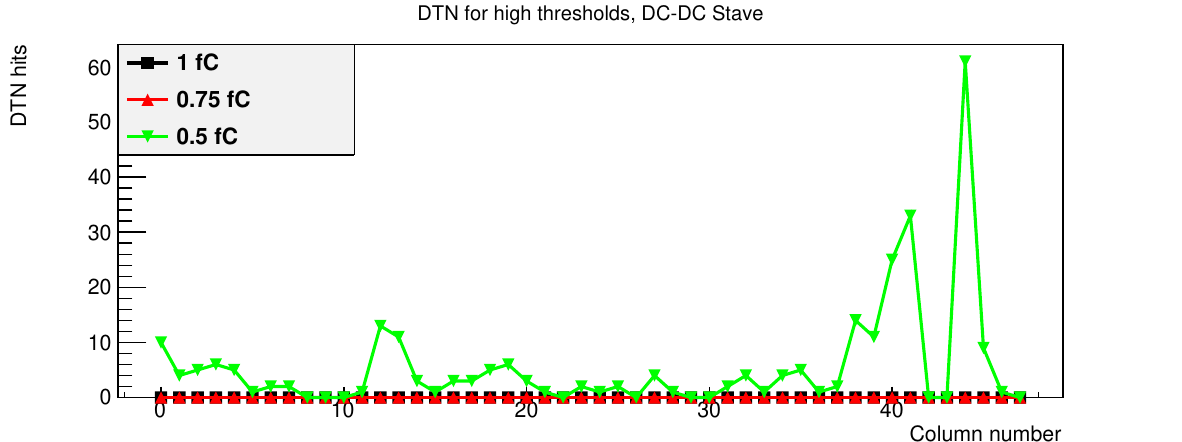}
}

\subfigure[DTN on the Stave 250 with twelve modules for low thresholds.]{
 \includegraphics[scale=0.65]{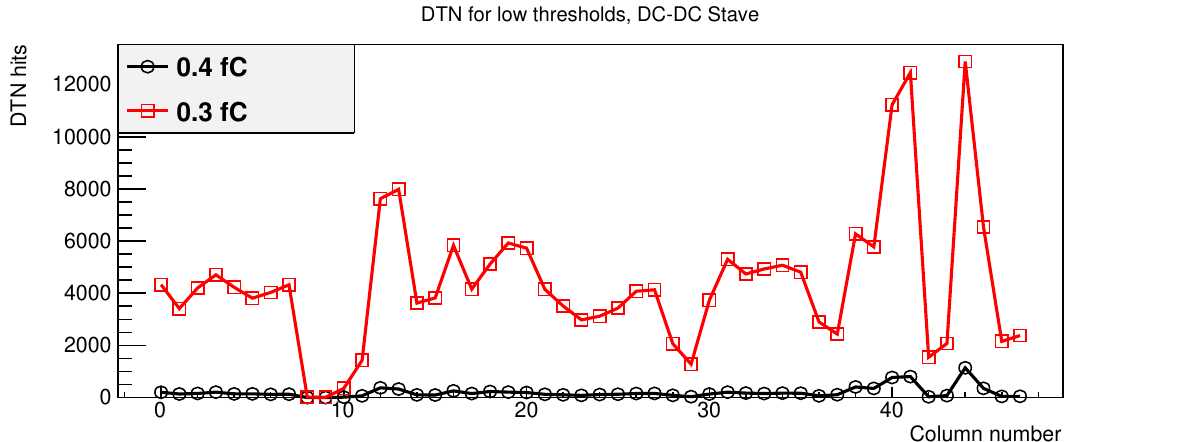}
}
  \caption{Double Trigger Noise of the DC-DC stave with twelve 
  modules.}\label{fig:5-stave_12modules_dtn}
 \end{center}
\end{figure}

Figure~\ref{fig:5-stave_12modules_dtn} shows the Double 
Trigger Noise results on the stave with twelve modules. As before, the X axis 
represents the chip columns (12 modules, 24 hybrids: 48 columns).
The higher threshold Double Trigger Noise results show no hits for the 
$1\fC$ and $0.75\fC$
thresholds, while some hits appear on the $0.5\fC$.

These results show that the single-sided stave with DC-DC conversion has a 
very good performance. The noise of the modules is not influenced by the common
read-out through a long bus. The Double Trigger Noise test also demonstrates 
that the bus does not have a negative effect on the read-out system.

The only drawback of the long bus tape is the signal integrity for distant
hybrids, which can be solved by an adequate choice of termination resistors.

\subsubsection{High Voltage Scan}

After the first side of the stave was finalized, high voltage and low voltage 
scans were performed, in the same way they were done for the stavelets.

Figure~\ref{fig:5-stave_HVscan} shows the input noise evolution on each of
the modules when the sensor bias was scanned in the set 
$V_{\rm bias} = \left\lbrace 10, 25, 50, 75, 100, 125, 150, 200, 250, 
275\right\rbrace$. The scan was performed in both directions, first ramping
up and then ramping down the voltage. The noise shown in the plot is the 
average for each of the modules.

\begin{figure}[!htbp]
 \begin{center}
   \includegraphics[scale=0.7]{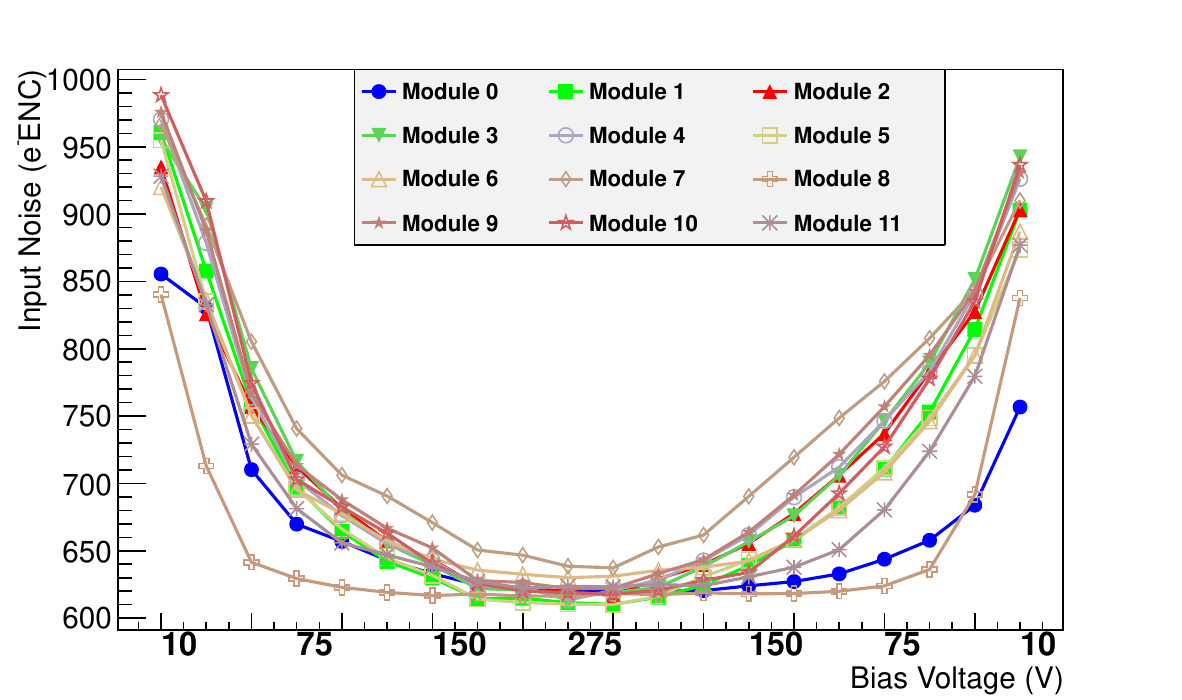}
  \caption{Evolution of the input noise for each module of the DC-DC stave with 
  changing sensor bias.}\label{fig:5-stave_HVscan}
 \end{center}
\end{figure}

The first observation is the different behaviour of the stave modules with 
respect to the stavelets modules. The way down of the bias voltage presents
lower noise than the way up, contrary to what has been observed previously.

Due to the fact that the stave sensors were left unbiased for long periods of 
time, this
may be attributed to the capacitance stabilization with time. The capacitance
of the strips when they are biased decreases with time, at a slow rate with 
time when they are kept in nitrogen.

\subsubsection{Hybrid Powering Variations}

The low voltage power supplies on the stave were programmed to perform noise
measurements when the input voltage to the DC-DC converters was changed.
Due to the long bus tape, the voltage drop along the stave is higher than
on the stavelet. This is shown in Table~\ref{tab:5-stave_voltages} for the
power supplies settings used in the low voltage scans. The three segments are
separated in the table for clarity.

Since the current draw
changes for the different voltage output values on the power supplies, the
voltage drop along the cables and the bus tape is not constant.
For instance, the voltage drop from M0 to M11 at $10.5\V$ setting is $740\mV$
while the drop at $13.5\V$ is $540\mV$.

\begin{table}[!htb]
 \begin{center}
  \begin{tabular}{c|c|c|c|c|c|c|c}
  \toprule
   Voltage on the PSU: & $10.5\V$ & $11.0\V$ & $11.5\V$ & $12.0\V$ & $12.5\V$ & $13.0\V$ & $13.5\V$ \\ \hline
   Module & \multicolumn{7}{c}{Voltage at the DC-DC converter input (${\rm V}$)}\\ \hline
   M0 & $9.04$ & $9.64$ & $10.22$ & $10.78$ & $11.34$ & $11.89$ & $12.44$ \\
   M1 & $9.08$ & $9.67$ & $10.24$ & $10.81$ & $11.36$ & $11.92$ & $12.47$ \\
   M2 & $9.14$ & $9.73$ & $10.30$ & $10.86$ & $11.42$ & $11.97$ & $12.51$ \\
 M3 & $9.25$ & $ 9.83$ & $10.40$ & $10.95$ & $11.50$ & $12.05$ & $12.59$ \\ \hline
M4 & $9.23$ & $ 9.81$ & $10.38$ & $10.93$ & $11.49$ & $12.04$ & $12.58$ \\
M5 & $9.26$ & $ 9.84$ & $10.40$ & $10.96$ & $11.51$ & $12.06$ & $12.60$ \\
M6 & $9.32$ & $ 9.90$ & $10.46$ & $11.01$ & $11.57$ & $12.11$ & $12.65$ \\
M7 & $9.44$ & $10.01$ & $10.57$ & $11.12$ & $11.66$ & $12.20$ & $12.74$ \\ \hline
M8 & $9.58$ & $10.14$ & $10.68$ & $11.22$ & $11.76$ & $12.30$ & $12.83$ \\
M9 & $9.61$ & $10.17$ & $10.71$ & $11.25$ & $11.79$ & $12.31$ & $12.85$ \\
M10 & $9.68$ & $10.23$ & $10.77$ & $11.30$ & $11.84$ & $12.37$ & $12.90$ \\
M11 & $9.78$ & $10.33$ & $10.86$ & $11.39$ & $11.92$ & $12.45$ & $12.98$ \\
\bottomrule
  \end{tabular}
\caption[Voltages at the DC-DC converter inputs with different settings on the
power supplies.]{Voltages at the DC-DC converter inputs with different 
settings on the power supplies. M0 is the module at the far end of the stave, 
whereas M11
is the module closest to the End of Stave card.}\label{tab:5-stave_voltages}
 \end{center}
\end{table}

The voltage on the power supplies was varied in the range $10.5$ to $13.5\V$,
with $0.5\V$ steps. One three point gain scan was run at each setting, both
increasing and decreasing the voltage. Due to the high voltage drop for the 
long bus tape, the results presented here are referenced to the voltage at
the input of the DC-DC converters instead of the power supply setting.

Figure~\ref{fig:5-stave_LVscan_up} shows the results for increasing hybrid
voltage. Each plot contains the four lines of the modules included in each 
power section.

\begin{figure}[!htbp]
 \begin{center}
   \includegraphics[scale=0.77]{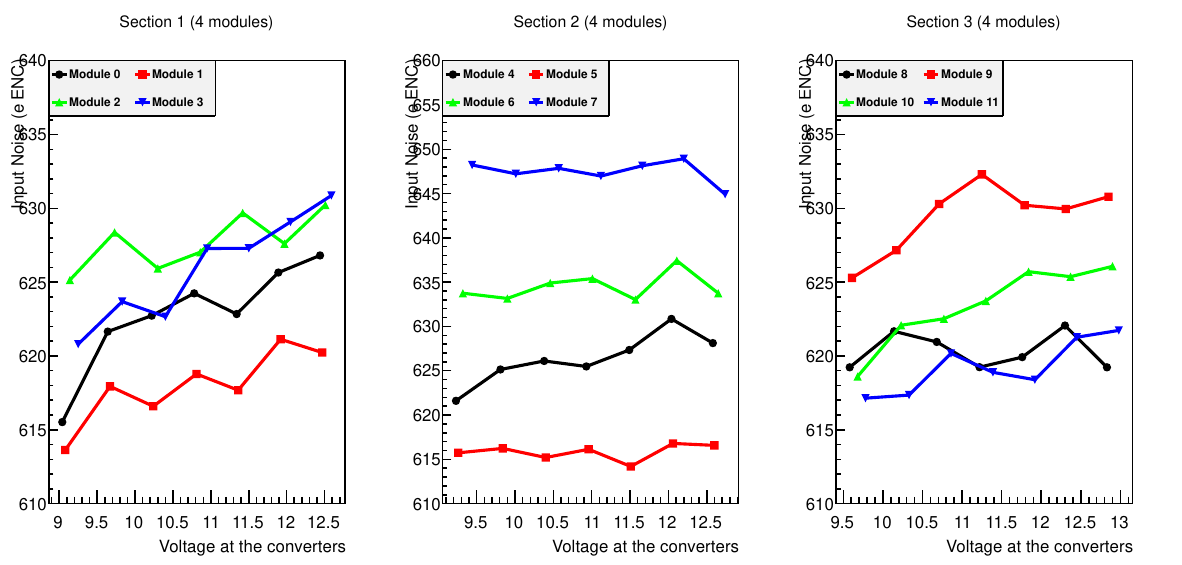}
  \caption{Evolution of the input noise for each module of the DC-DC stave with 
  changing input voltage at the converters, from lower to 
  higher.}\label{fig:5-stave_LVscan_up}
 \end{center}
\end{figure}

\begin{figure}[!htbp]
 \begin{center}
   \includegraphics[scale=0.77]{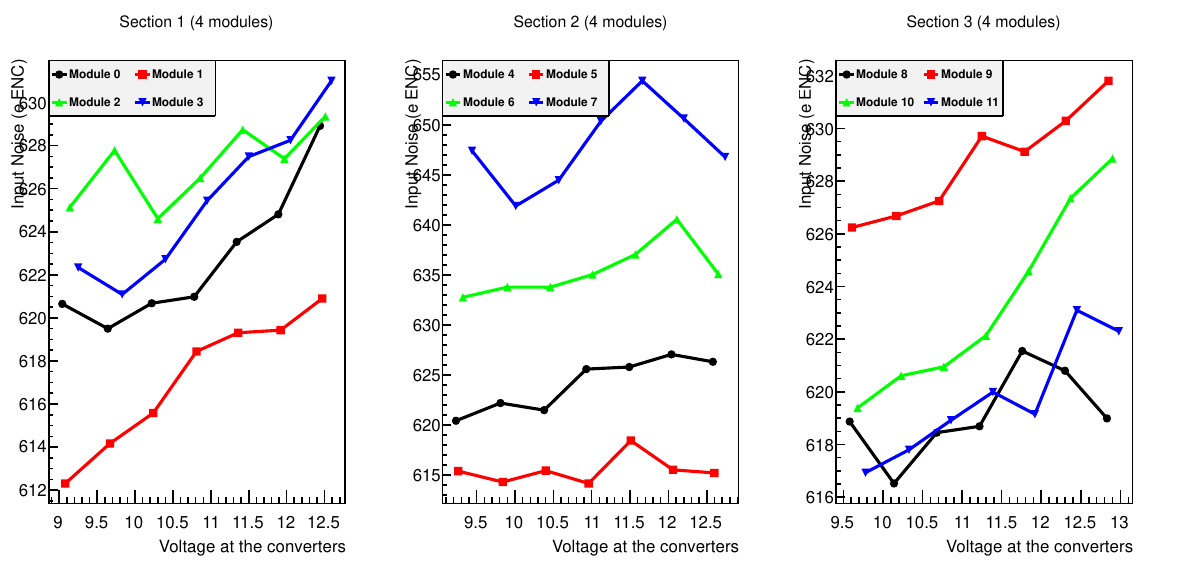}
  \caption{Evolution of the input noise for each module of the DC-DC stave with 
  changing input voltage at the converters, from higher to 
  lower.}\label{fig:5-stave_LVscan_down}
 \end{center}
\end{figure}

Figure~\ref{fig:5-stave_LVscan_down} shows the decreasing voltage results.
The results are compatible with the ones obtained with the DC-DC powered 
stavelet, which showed a noise increase when the voltage at the converters
was set far away from the $10\V$ ``sweet spot'' of the current version of
the converters.

Some of the modules do not show a clear noise increase, while others do. This
does not seem to be related to the sensor strip isolation type, as most of the
modules are of the same type and the results for them are mixed.

\begin{figure}[!htbp]
 \begin{center}
   \includegraphics[scale=0.82,trim=8 0 32 0,clip=true]{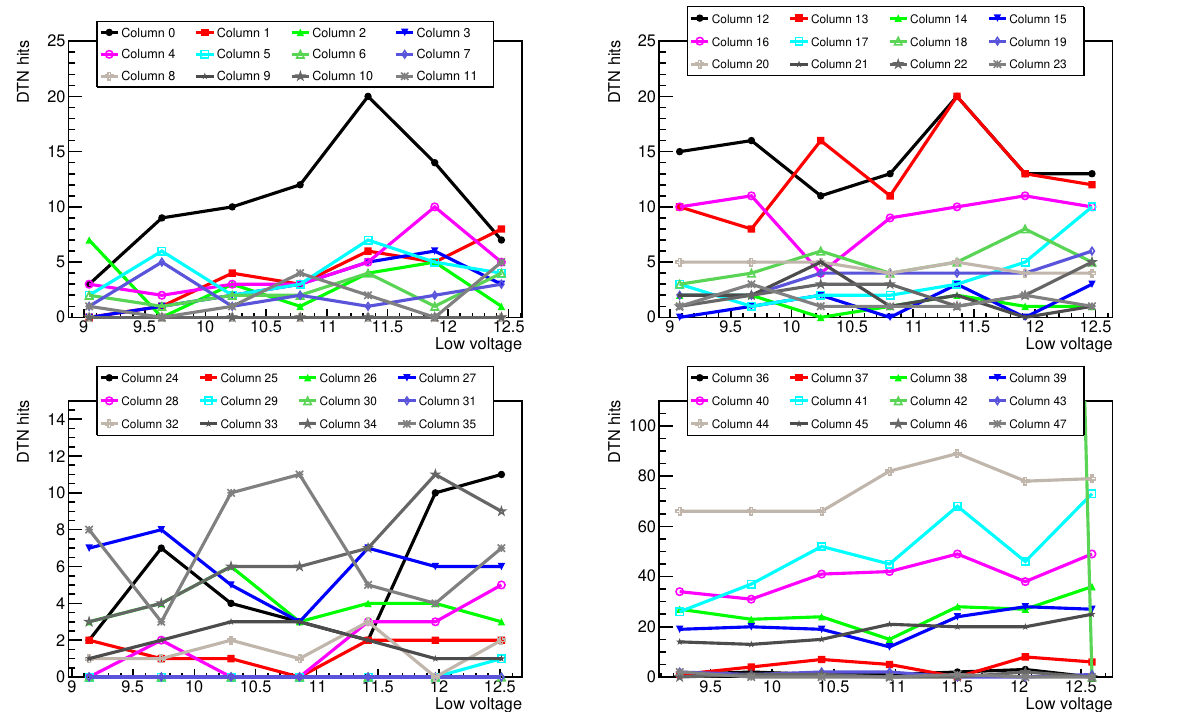}
  \caption{Evolution of the double trigger noise with $0.5\fC$ threshold for each column of the DC-DC stave with 
  changing input voltage at the converters, from lower to 
  higher.}\label{fig:5-staveDTN_LV_up}
 \end{center}
\end{figure}

Double Trigger Noise tests were also performed while changing the low voltage
values. Figure~\ref{fig:5-staveDTN_LV_up} shows the results for every column
with increasing low voltage, with a charge threshold of $0.5\fC$, in order to
have some hits that would allow seeing any trend. All columns show no evidence 
of noise dependence with the low voltage setting of the power supplies.

As was done with the ENC plots, the horizontal axis represents the voltages at
the input of the converters for each module, instead of the setting at the 
power supplies.

The zoom on the bottom 
right plot hides column 42 for the most part. This is due to very high noise
counts at the beginning of the run, caused by some noisy channels that were
eventually masked later in the run.

The Double Trigger Noise results are plotted for each column, in groups of 12
columns (3 modules) per subplot.

Figure~\ref{fig:5-staveDTN_LV_down} shows the plots with decreasing low 
voltage, and we arrive to the same conclusion with respect to the noise 
dependence. In these plots, 
which correspond to the same run as the previous, the noisy channels of column
42 were still masked, so the hits values are more realistic.

\begin{figure}[!htb]
 \begin{center}
   \includegraphics[scale=0.82,trim=8 0 32 0,clip=true]{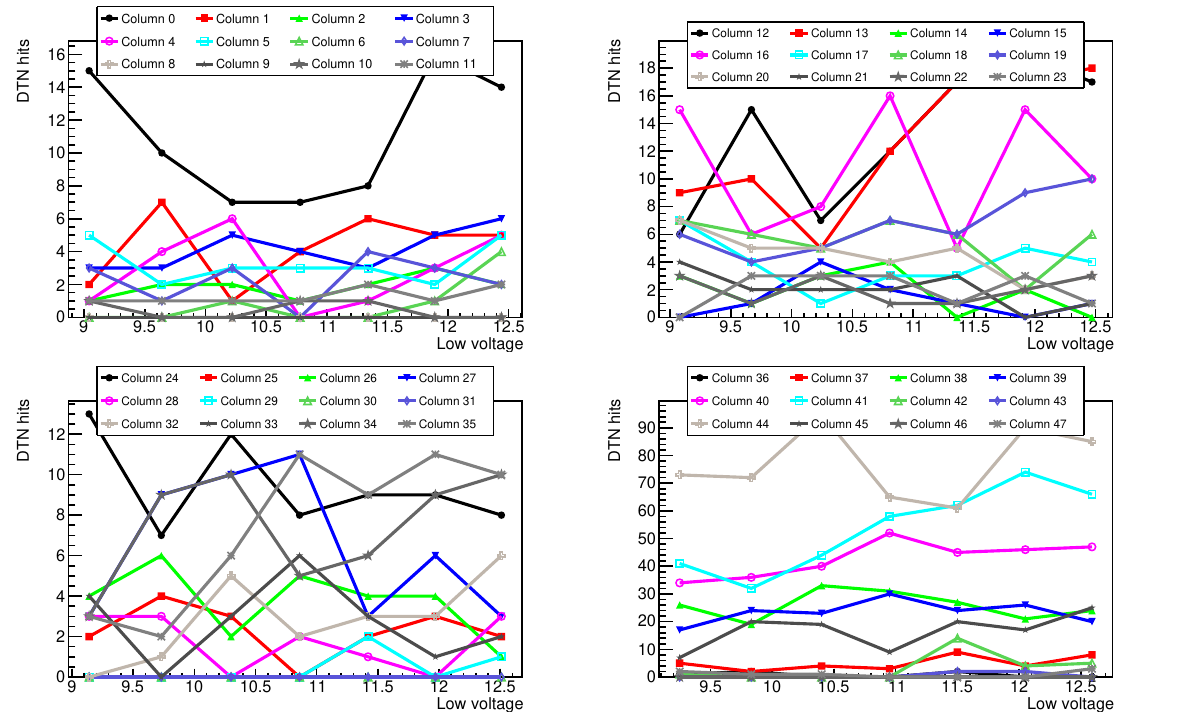}
  \caption{Evolution of the double trigger noise with $0.5\fC$ threshold for each column of the DC-DC stave with 
  changing input voltage at the converters, from higher to 
  lower.}\label{fig:5-staveDTN_LV_down}
 \end{center}
\end{figure}

\section{Discussion}

This chapter has presented the series of tests that have been done with the
upgrade modules and stavelets in the context of this thesis.

The devices produced so far are based on the ABCn25 chip and use a Buffer 
Control Chip (BCC) to multiplex the data coming from the two data links on the
hybrids. Powering of the hybrids is done with DC-DC conversion or Serial 
Powering.

A serially powered module was available for these tests, together with a DC-DC
powered stavelet and another stavelet with serial powering, both holding four
modules each. The tests were done in a testing facility in Building 180 at 
CERN.

Performance of the single module was presented in section~\ref{sec:5-spmodule},
with extensive measurements done varying the ambient and operational 
conditions. Cooling temperature variation was used to extract the noise 
variation with temperature, which was found to be around
$1.5\ENC/{\rm K}$.

Noise evolution with the sensor bias was also analysed in detail, with 
measurements at multiple bias voltages and in both the increasing voltage
direction and the decreasing voltage direction.

Last, low temperature tests were done with a dedicated chiller that is
capable of operating below $-30\degC$. The minimum temperature setting used for
these tests was $-20\degC$ and the electronics worked correctly, with a hybrid
temperature of $2.5\degC$ and ambient temperature around the module of
$8.5\degC$.

In addition, the two stavelets have been tested intensively. Sensor bias scans,
varying the hybrid powering input parameters (voltage or current), using high
voltage multiplexing and trying to interfere using an electromagnetic 
aggressor. The results on the electromagnetic interference tests, presented in 
this chapter were done using a prototype of the optical link that will be used
to read-out the staves.

The high voltage scan tests show the expected decreasing noise results when 
the sensor bias voltage is increased, as the sensors become depleted and 
plateau after over-depletion.

In addition, using a four-way multiplexer designed and built at LBNL to bias 
the stavelets instead of individual high voltage lines resulted in no 
difference on the noise measurements of both stavelets. The motivation for
the use of multiplexing in the high voltage lines is the reduction of cables to
bias the sensors, lowering the material budget by a factor of 13. The results 
shown in this chapter prove that there is no noise increase with any of the two
powering schemes.

The hybrid powering variation shows some performance loss in two of the modules
for the DC-DC conversion case. On the serially powered stavelet, there is a 
noise increase in line with the temperature increase that appears when the input 
current is higher.

An optical link prototype, which has been proposed to be used in the End of 
Stave card to read-out the staves, was used as the electromagnetic aggressor
on the DC-DC stavelet.
There was no evidence of electromagnetic interference that could result in
a performance decrease of the stavelet or the whole DAQ system.

The two stavelets were tested both separately and together on the same DAQ 
setup. The results presented in this chapter show that there is no performance 
loss resulting from the simultaneous operation. The extra noise observed on the
serially powered stavelet is consistent with the temperature increase and the
DC-DC stavelet shows no increase in noise. The Double Trigger Noise test does
not show any additional noise pick-up on any of the two stavelets.

The stavelet concept has been escalated to a full-sized object, the stave. One
such object has been constructed at the time of writing this chapter, a 
one-sided, DC-DC powered stave with twelve modules. The results of the initial
tests of this stave have been presented. The noise performance of the 
single-sided stave is good and a serially powered stave, in chain of modules
configuration, is in construction. The
performance differences between the two will drive the decision of building a
double-sided object.

\chapter{Measurement of the Energy Loss and Particle Identification in the 
ATLAS SCT}\label{chap:de_dx}

\section{Introduction}

The Pixel detector and the Semi-Conductor Tracker (SCT), the innermost parts
of the ATLAS Inner Tracker, were briefly introduced in chapter 1. They are both 
based on silicon detectors and are designed for precise charged particle 
tracking, vertex separation, and precise momentum resolution~\cite{ATLAS}.

\begin{figure}[!htb]
 \begin{center}
  \includegraphics[scale=0.65]{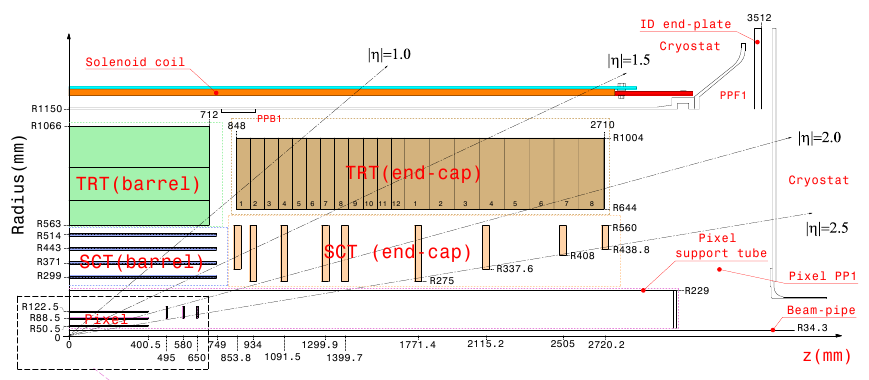}
  \caption{$Rz$ section of ATLAS showing a quarter of the 
  Inner Detector.}\label{fig:7-id_section}
 \end{center}
\end{figure}

However, the SCT and Pixel sub-detectors were not designed to do energy 
measurements. A way to overcome this limitation
is the Time-over-Threshold (ToT) method. It consists in counting the number
of cycles of the $40\MHz$ clock that the charge signal is above the configured 
threshold. These cycles, $25\ns$ long, are named time-bins.

The charge collected in each pixel or strip is approximated by counting these 
time-bins with an 8-bit dynamic range in case of the Pixel detector. In the 
SCT, a 3-bit dynamic range, together with information about the number 
of hit strips is stored.

In the Pixel modules, sixteen front-end (FE) electronics chips are connected 
to each of the sensors through 
bump bonding. Each of the FE-I3 chips contains $2880$ pixel cells of $50\times 
400\mum^2$ size, in an $18\times160$ matrix~\cite{Peric2006178}.

Each cell of the FE-I3 chip has an analogue block, with an amplifier followed 
by a comparator 
that has a programmable threshold. The digital read-out transfers the hit pixel 
address together with  the leading edge and the trailing edge timestamps to the 
buffers of the chip. In these buffers, the Time-over-Threshold is 
calculated, by subtracting the trailing from the leading edge stamps. 

Eight 
bits are used for the ToT calculation, which allows for a charge measurement of 
up to $255$ ToT counts. A typical Minimum Ionizing Particle (MIP) crossing the 
sensor at normal incidence will 
give a ToT count of $30$~\cite{ATLAS-CONF-2011-016}.

The SCT modules have two sides with six ABCD3TA chips each that read-out the 
sensors
strips. There is also a comparator with a programmable threshold which is 
currently set to $1\fC \approx 6242~e^{-}$, for high efficiency and low noise.

For every trigger, the SCT read-out extracts three time intervals from the chip
pipeline. The three intervals sequence is checked against a pattern that is 
determined by the configuration of the SCT read-out. 

The data taken in 2010 used a ``hit'' pattern, which matches 
hits in any of the three time-bins. Data taken in 2011 
and 2012 used a ``level'' pattern, which matches a hit in the central position.
These two patterns are also known as the read-out modes XXX and X1X. In the XXX
mode, only a single or consecutive ones are allowed, that is, patterns ``000'' 
and ``101'' are forbidden.

The specific cluster energy loss $\dedx$ 
(${\rm MeV g^{-1} cm^{-2}}$) is linearly correlated to the cluster charge. 
$\dedx$ can be calculated with the ToT method, counting the number
of time intervals which the signal was above the threshold, for every strip
in a cluster.

The Pixel $\dedx$ calculation exploits the 8 bits resolution of its ToT 
counter and uses a lower threshold ($3.5\ {\rm k}e^{-}$) than the SCT. The 
time-bin information on the SCT uses only 3 bits, what makes the 
Pixel calculation more precise~\cite{ATLAS-CONF-2011-016}. 

The datasets used for this analysis consist of SCT Ntuples, 
where the Pixel track $\dedx$ is provided. With the data contained 
in these Ntuples, it is possible to reconstruct the SCT track $\dedx$ from ToT 
and number of strips, and compare it with the Pixel value. 

The following sections present the study performed on SCT calculations of 
energy loss, $\dedx$. First, the datasets corresponding to the 2010-2012 
running periods are listed. Then, details of how the Time-over-Threshold 
method is used  to calculate the ionization energy $\dedx$ are described, with 
path correction, normalization and comparison with the results from the Pixel.

Next, efficiency and mistag rates of the particle identification process are
presented. An application of the $\dedx$ calculation in tracking radiation 
damage is discussed in section~\ref{sec:7-stability}. 

Part of the work discussed in this chapter has been included as part of an
overview of the SCT performance during the first run of the LHC 
in~\cite{Aad:2014mta}.

\section{Event and Track Selection}

The study used proton-proton collision data taken during 2010 and 2011 with 
$7\TeV$ centre-of-mass energy and during 2012 with $8\TeV$ centre-of-mass 
energy.

The selected SCT tracks were classified according to the location of the hits.
Barrel tracks are those in which all hits associated with the track are located
only in the barrel layers, no hits in endcap modules. Endcap
tracks are those tracks that had some hits in the endcaps.

Apart from the hits locations, tracks can be classified according to the 
origin of the particles:
\begin{itemize}
 \item Primary Vertex (PV): coming from the hard interaction, the tracks have 
 a  small transversal distance to the point of closest approach, $d_0$ (less 
 than $2\mm$). 

\begin{figure}[!htb]
 \begin{center}
  \includegraphics[scale=0.5]{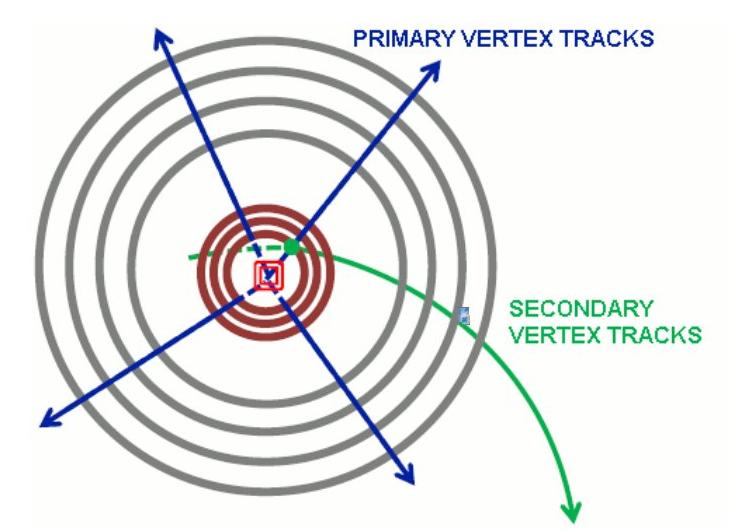}
  \caption{Primary and secondary vertex tracks.}\label{fig:7-pv_sv_tracks}
 \end{center}
\end{figure}

\item Secondary Vertex (SV): apart from the primary vertex, and with the
exception of gas processes (interactions with residual gas molecules), no more 
interactions are expected before the first material layer, which is the beam 
pipe wall. These processes are rich in protons and 
deuterons, because they are taking place in material. They have a larger 
transversal distance, $d_0 > 2\mm$. 

See Figure~\ref{fig:7-pv_sv_tracks} for a 
schematic description of both primary and secondary vertices.
\item Gas interactions: downstream, inside the beam pipe. 
\item Noise.
\end{itemize}

The analysis is focused on tracks originating in primary and secondary 
vertices. Both primary vertex and secondary vertex tracks require a 
longitudinal impact parameter cut of $|z_0| < 100\mm$, to be inside the beam 
spot and to avoid gas interactions. 

The datasets come from SCT Ntuples and for each dataset only events with the 
SCT in ``Ready'' state are analysed.
For every event, a first selection cut requires that they meet the following 
requirements:

\begin{itemize}
\item There has to be at least one primary vertex reconstructed in the event.
\item The primary vertex must have at least $4$ tracks associated to it.
\item At least one of the tracks in the event must have:
\begin{itemize}
 \item Transverse momentum $p_{\rm T} > 500\MeV/c$.
\item At least $6$ SCT hits.
\item At least $1$ Pixel hit.
\item Transverse impact parameter $|d_0| < 1.5\mm$. At least one of the tracks
has to be inside the beam spot.
\item The longitudinal impact parameter and the polar angle have to be such 
that $|z_0\sin\theta| < 1.5\mm$. This requisite complements the previous one, 
in the
$Rz$ plane. 

Tracks with low $d_0$ but large $z_0$ that have a large polar angle
are rejected with this restriction. See 
Figure~\ref{fig:2-track_parameters} for the representation of the parameters.
\end{itemize}
When these conditions are fulfilled, we can say the event originates in a 
proton-proton interaction.
\end{itemize}

Then, for each event, tracks selected for $\dedx$ calculation have to meet the
following requirements:

\begin{itemize}
 \item At least $8$ good SCT hits on track, in order to have a well measured
  track. 
\item Transverse momentum has to be at least $p_{\rm T} > 100\MeV/c$. The 
transverse momentum resolution is better with increasing momentum, but we do
not want to miss low momentum tracks.
\item The momentum of the track has to be lower than $2500\MeV/c$. The energy
loss was calculated up to this momentum, but there is not enough resolution to
perform particle identification after $\approx 800\MeV/c$.

There is no lower momentum cut as the reconstruction already
applies one, $100\MeV/c$ for 2010 and $400\MeV/c$ for 2011 and 2012 data.
\item Longitudinal impact parameter $|z_0| < 100\mm$. These constrains on the
impact parameters select
tracks in the interaction region and also not originating in gas interactions.
\item For Pixel ``true'' particles, at least two good Pixel hits.

The goal of these cuts is to study only tracks that are well measured, with 
enough hits in the SCT and the Pixel detector. In addition, the tracks are
originated in the beam spot, and both primary and secondary vertex tracks are
analysed.
\end{itemize}

\begin{table}[!htb]
	\begin{center}
	\begin{tabular}{c c c c c}
	\toprule
	Run number & Date & Stream type & Read-out mode & Luminosity\\ \hline
   152345 & 3/4/2010 & Minimum bias & XXX & $0.01871\inb$\\
   156682 & 5/6/2010 & Minimum bias & XXX & $1.407\inb$ \\
   159224 & 18/7/2010 & Minimum bias & XXX & $69.16\inb$\\
   165591 & 22/9/2010 & Minimum bias & XXX & $168.9\inb$ \\
    167661 & 25/10/2010 & Minimum bias & XXX & $ 1464\inb$ \\
   179710 & 14/4/2011 & Express  & X1X & $ 6792\inb$ \\
    183407 & 12/6/2011 & Express  & X1X & $ 4.262\times10^4\inb$ \\
    183462 & 14/6/2011 & Express  & X1X & $ 4.693\times10^4\inb$ \\
    186673 & 2/8/2011 & Express  & X1X & $ 3.726\times10^4\inb$ \\
    186729 & 3/8/2011 & Express  & X1X & $ 9.078\times10^4\inb$ \\
    190236 & 1/10/2011 & Express  & X1X & $ 1.037\times10^5\inb$ \\
    191933 & 30/10/2011 & Express  & X1X & $ 197.7\inb$ \\
    201190 & 12/4/2012 & Express &  X1X &  $ 4.726\times10^4\inb$ \\
    204265 & 3/6/2012 & Express & X1X & $ 2.373\times10^5\inb$ \\
   204564 & 7/6/2012 & Express  & X1X & $ 2.014\times10^5\inb$ \\
   204763 & 9/6/2012 & Express & X1X & $ 1.999\times10^5\inb$ \\
   205016 & 15/6/2012 & Express & X1X & $ 1.041\times10^5\inb$ \\
   205017 & 15/6/2012 & Express & X1X & $ 7.403\times10^4\inb$ \\
   205071 & 16/6/2012 & Express & X1X & $ 2.357\times10^5\inb$ \\
   209736 & 3/9/2012 & Express & X1X & $ 1.331\times10^5\inb$ \\
    214086 & 6/11/2012 & Express  & X1X & $ 1.98\times10^5\inb$ \\
    \bottomrule	
	\end{tabular}
	\caption{Datasets used in the $\dedx$ study.}\label{tab:2-datasets}
	\end{center}
\end{table}

The total recorded luminosity of the dataset is $1.76\ifb$. The number of 
recorded events is around $11$ million, summing the minimum bias and the 
express streams that were used in each of the periods. The minimum bias
streams were generated for the SCT Ntuples before April 2011. After that date,
only express streams were available for the SCT. The express stream is normally
used for calibration and data quality, whereas minimum bias streams are used 
for physics. The analysis performed in this thesis does not require information
from the minimum bias streams.

The read-out modes have been explained above: 2010 data were taken in XXX mode,
whereas 2011 and 2012 data were taken in X1X mode. The first records hits in 
any time interval combination of the three selected for a trigger, while the 
second requires that there is one hit in the central time-bin.

The data sets were selected in order to have a good representation of the whole
running period, with some spacing in time. The SCT Ntuples were not available
for all the recorded runs, although there were enough to cover most of the 
first LHC run period.

\section{\texorpdfstring{$\dedx$}{dE/dx} Reconstruction}
A particle passing through matter interacts with electrons and 
nuclei~\cite{Beringer:1481544}. For charged
particles which are not highly relativistic, thus applicable in this study 
case, the main electromagnetic contribution to the 
energy loss is ionization. 

The mean energy loss $\dedx$ due to ionization, as a
function of $\beta$, is given by the Bethe-Bloch formula~\cite{Leo1994}: 

\begin{equation}
 - \dedx = 2\pi N_a r_e^2 m_e c^2 \rho D\frac{Z}{A} \left[\log \left( 
 \frac{2m_e\gamma^2 v^2 W_{\rm max}}{I^2}\right) - 2 \beta^2 - \delta - 
 2\frac{C}{Z} \right]
\end{equation}

The units are ${\rm MeV~cm^{2}~g^{-1}}$ and some of the variables definitions
are:

\begin{itemize}
\item $\beta$ is the projectile velocity in units of $c$.
\item $D$ is a material constant.
\item $Z$ is the atomic number of the medium and $A$ is its atomic mass $[{\rm g~mol^{-1}}]$.
\item $\rho$ is the mass density of the medium $[{\rm g~cm^{-3}}]$.
\item $\delta$, $C$ and $\nu$ are  the density, shell and higher order corrections.
\item $I$ is the mean excitation energy $[{\rm eV}]$.
\item $W_{\rm max}$ is the maximum energy transfer in a single collision.
\end{itemize}

Measurements of energy loss can be used to identify particles if a simultaneous measurement of momentum is 
available. 
In this analysis the goal is to identify protons, kaons and pions.  

Pixel provides their $\dedx$ measurements using the Time-over-Threshold method 
with 8 bits precision. These $\dedx$ values calculated with the Pixel ToT
information, available in the SCT Ntuples, are
used to compare with the ToT approximations for the SCT.

\begin{figure}[!htb]
\begin{center}
 \includegraphics[scale=0.65]{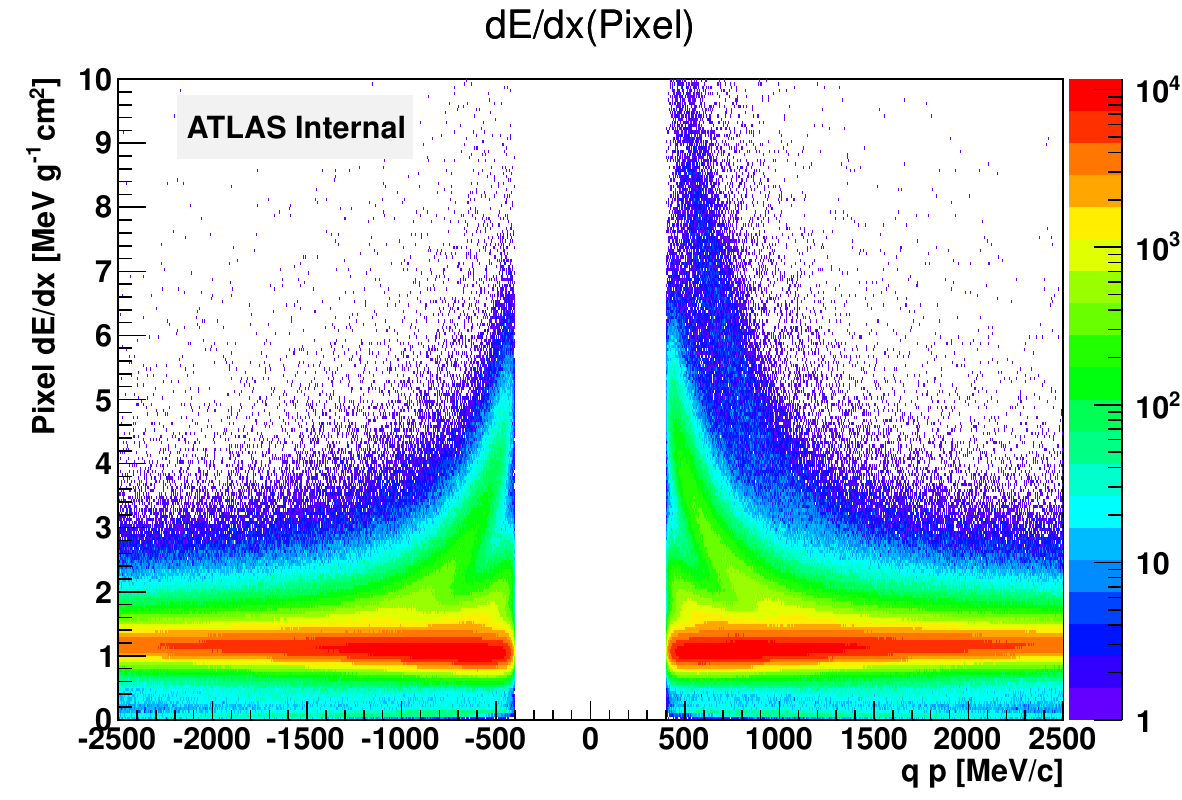}
 \caption{Pixel $\dedx$ extracted for data taken in 2011. At low momenta, the 
 proton  bands are visible for both positive and negative 
 tracks.}\label{fig:7-pixel_dedx}
\end{center}
\end{figure}

Figure~\ref{fig:7-pixel_dedx} shows the $\dedx$ calculated by the Pixel in one 
of the runs in 2011. Particle bands are visible, these are the less resolution 
Bethe-Bloch curves. The representation of $\dedx$ with $qp$ is useful to 
identify the different particles, by looking at the different ionization
energy of the observed bands.

In this figure, the proton, kaon and pion bands are quite visible. 
The lowest $\dedx$ band both for positive and negative tracks corresponds to 
pions, when one looks at the lower momentum region. The next two bands at low
momentum are
kaons and protons, respectively.
There is
a positive deuteron band that
comes from nuclear interactions, while the negative deuteron originates in
fragmentation, therefore the negative one is unstable and less represented.
Protons and deuterons are more visible in secondary vertex tracks, because 
they originate in interactions in the material (beam pipe wall). 

The plot shows that for momentum greater than $800\MeV/c$, all bands converge
to the same $\dedx$ band, making the separation very difficult.

\subsection{SCT Data Acquisition, Hits and Timing}\label{sec:timing}

Each of the 4088 SCT modules are read-out at a rate of $40\MHz$. First, the 
analogue signal is
compared to a $1\fC$ threshold at each time interval and for every channel. 
If the signal is above (below) threshold, a one (zero) is stored in the 
pipeline.

The pipeline is a 132 deep binary FIFO for each of the 128 channels. This is
represented in Figure~\ref{fig:7-abcd_functionality}~\cite{Robinson:1418120}.

\begin{figure}[!htb]
 \begin{center}
  \includegraphics[scale=0.51]{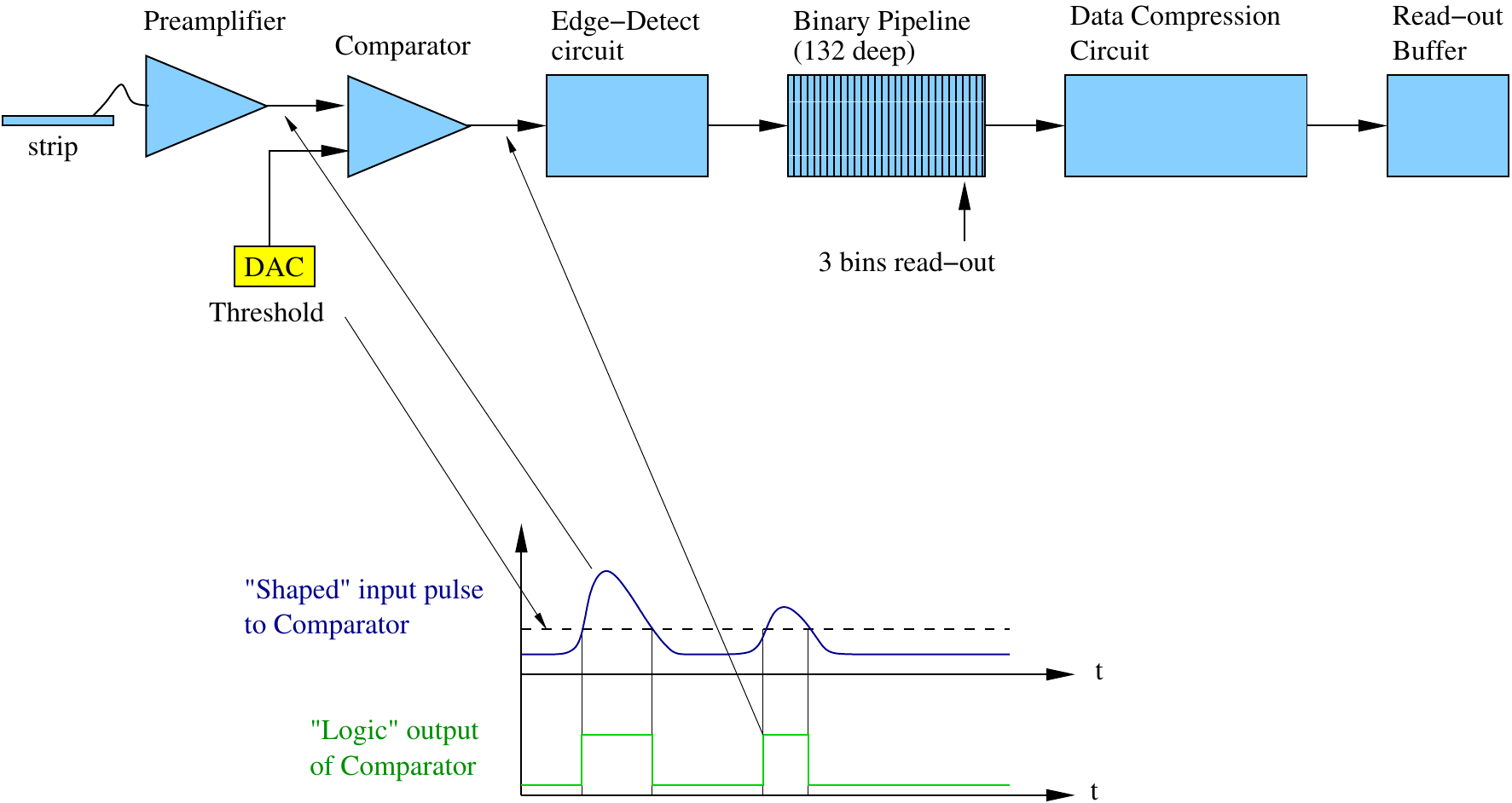}
  \caption{Functionality of the ABCD3TA chip.}\label{fig:7-abcd_functionality}
 \end{center}
\end{figure}

The arrival of the Level 1 trigger from the ATLAS trigger system drives the 
sampling of three pipeline bins. 
The SCT timing is optimized to ensure that the trigger is centred at the 
second bin, resulting in a 01x occupancy
pattern in the three bins. 

The ABCD3TA chip can be configured depending on 
the running conditions, for optimal efficiency:

\begin{itemize}
 \item XXX mode for cosmic rays and $75\ns$ bunch spacing. This was 
 the case for the 2010 runs.
\item X1X mode is used for $50\ns$ bunch spacing, in the 2011 and 2012 runs.
\item 01X mode is designed to be used in runs with $25\ns$ bunch spacing. This 
mode will be the default for operation after 2015.
\end{itemize}

``Timing-in'' the SCT involves setting delays in order to meet these requirements:

\begin{itemize}
 \item The Level 1 trigger arrives at the module when the three bins are at 
 the end of the pipeline.
\item The clock transmitted to the front-end electronics has the correct phase 
relative to the passing of the particles originating in the collisions.
\item The Level 1 trigger is received in the right $25\ns$ time-bin and the 
data from the ATLAS sub-systems are merged into the right events.
\end{itemize}

There are two timing adjustments available, a coarse one with $25\ns$ step 
size and a fine one with $280\ps$ steps~\cite{SCTDAQ}.

\subsection{SCT \texorpdfstring{$\dedx$}{dE/dx} Calculation with the 
Time-over-Threshold Method}

As shown in Figure~\ref{fig:7-tot}, when the pulse signal is larger, it 
results in more time-bins above threshold, that is the definition of 
Time-over-Threshold. As a first approximation, weights $w^i$ are associated to 
the number of time-bins for each channel and summed over all strips 
hit with charge, for all SCT hits associated with a track.

\begin{figure}[!htb]
 \begin{center}
  \includegraphics[scale=0.95]{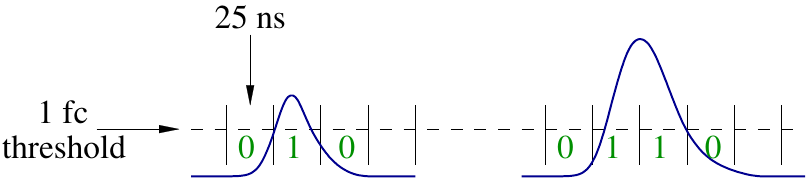}
  \caption{Time-over-Threshold for two different pulse sizes.}\label{fig:7-tot}
 \end{center}
\end{figure}

The obvious choice for assigning weights to the SCT time-bins in XXX read-out 
mode is the following:

\begin{itemize}
 \item Time-bins 101 (illegal time-bin) and 000: $w^i = 0$.
\item Time-bins 001, 010 and 100: $w^i = 1$.
\item Time-bins 011, 110: $w^i = 2$.
\item Time-bin 111: $w^i = 3$.
\end{itemize}

Then, the SCT energy loss, $\dedx_{\rm SCT}$, can be approximated as the sum of
weights of all the strips hits on track:

\begin{equation*}
 \dedx_{\rm SCT} = \sum_{\rm Track\_hits} \sum_{\rm Hit\_strips} w^i_{\rm strip}
\end{equation*}

This is correct for tracks that pass perpendicular to the module, which is
typically not the case. Therefore, a path correction that takes into account 
the incident angle of the particles is needed:

\begin{equation*}
 \dedx_{\rm SCT} = \sum_{\rm Track\_hits} \sum_{\rm Hit\_strips} w^i_{\rm strip} \cos \alpha
\end{equation*}

The angle $\alpha$ is the incident angle on the module. The information 
available in the SCT Ntuple are two local angles, one along the strips 
and another across the strips, $\theta_0^{\rm local}$ and $\phi_0^{\rm local}$.

Redefining the two local angles between $-\pi/2$ and $\pi/2$, leads to 
the local angles $\theta'$ and $\phi'$:

\begin{equation}
 \theta' = \left\{ \begin{array}{c c}
                   \theta_0^{\rm local} - \pi & {\rm ~ if ~} \theta_0^{\rm local} > \pi/2 \\
                    & \\
                   \theta_0^{\rm local} + \pi & {\rm ~ if ~}  \theta_0^{\rm local} < \pi/2
                  \end{array}
\right.
\end{equation}

\begin{equation}
 \phi' = \left\{ \begin{array}{cc}
                   \phi_0^{\rm local} - \pi & {\rm ~ if ~}  \phi_0^{\rm local} > \pi/2 \\
                    & \\
                   \phi_0^{\rm local} + \pi & {\rm ~ if ~}  \phi_0^{\rm local} < \pi/2
                  \end{array}
\right.
\end{equation}

Then, the angle of the particle trajectory with the silicon surface can be defined as

\begin{equation}
  \alpha = \tan^{-1} \left( \sqrt{\tan^2 \theta' + \tan^2 \phi'} \right)
\end{equation}

Finally, the previous $\dedx$ formula has to be normalized with respect to the 
number of hits on the track, leading to the final $\dedx$ formula for the SCT 
(note it is measured in arbitrary units):

\begin{equation}
 \dedx_{\rm SCT} = \frac{\displaystyle \sum_{\rm Track\_hits} \sum_{\rm Hit\_strips} w^i_{\rm strip} \cos \alpha}{\rm Track\_hits} \quad \left[{\rm a.u.}\right]
\end{equation}

With this calculation of the energy loss in the SCT, it is possible to identify 
the bands corresponding to different particle species.

\begin{figure}[!hbt]
 \begin{center}
  \subfigure[Barrel and endcaps.]{
    \includegraphics[scale=0.3]{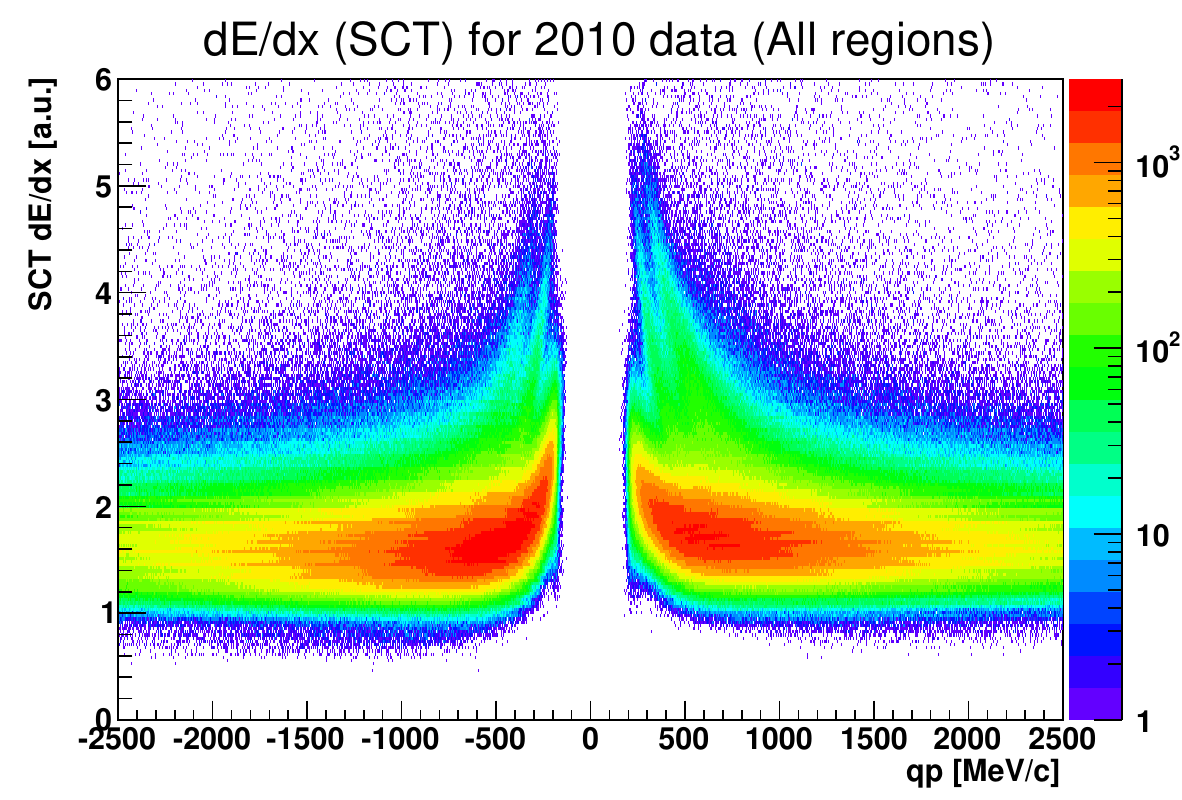}
  }  
   \subfigure[Barrel.]{
     \includegraphics[scale=0.3]{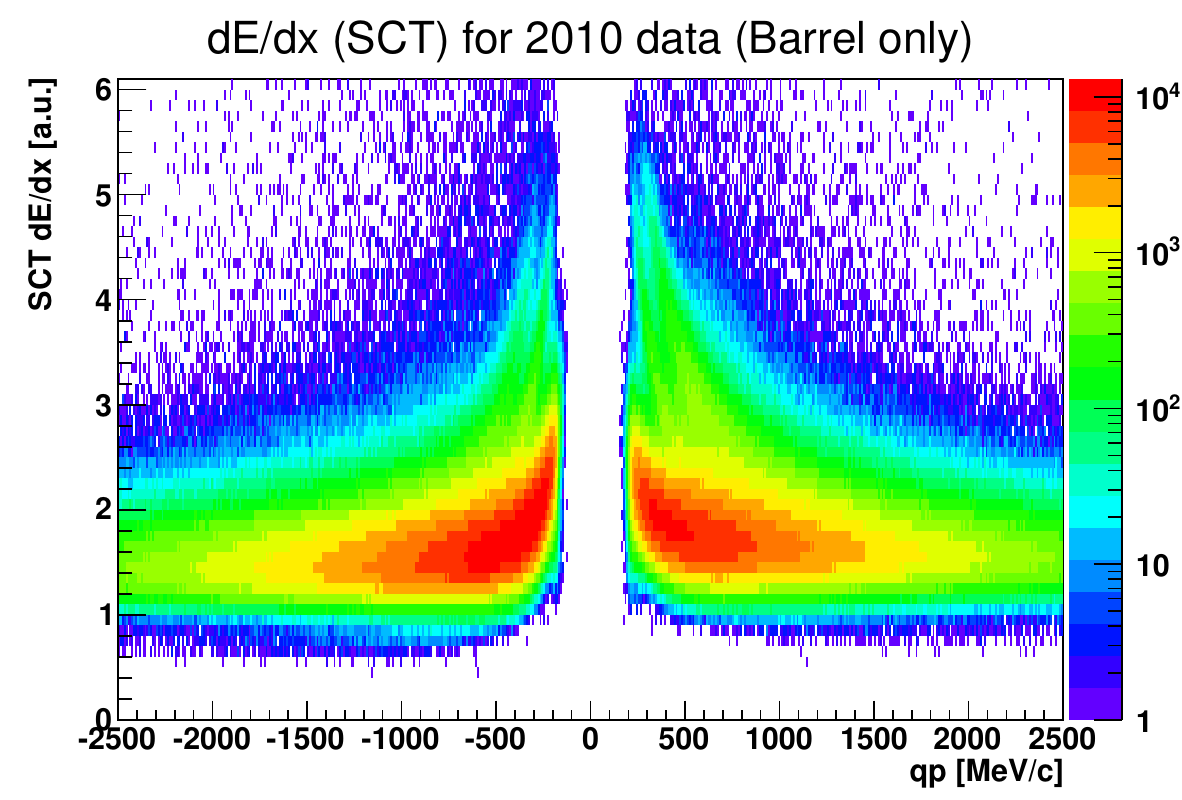}
   }   
   
   \subfigure[Endcap A.]{
     \includegraphics[scale=0.3]{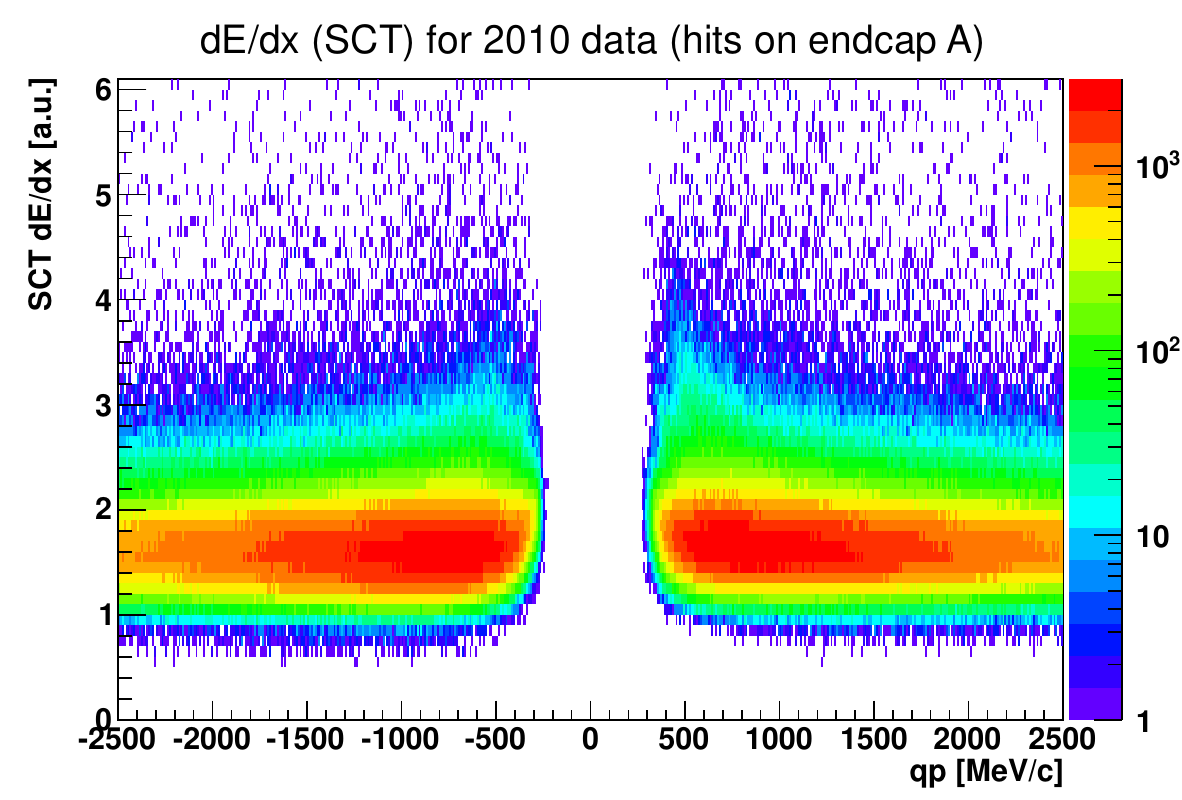}
   }
   \subfigure[Endcap C.]{
     \includegraphics[scale=0.3]{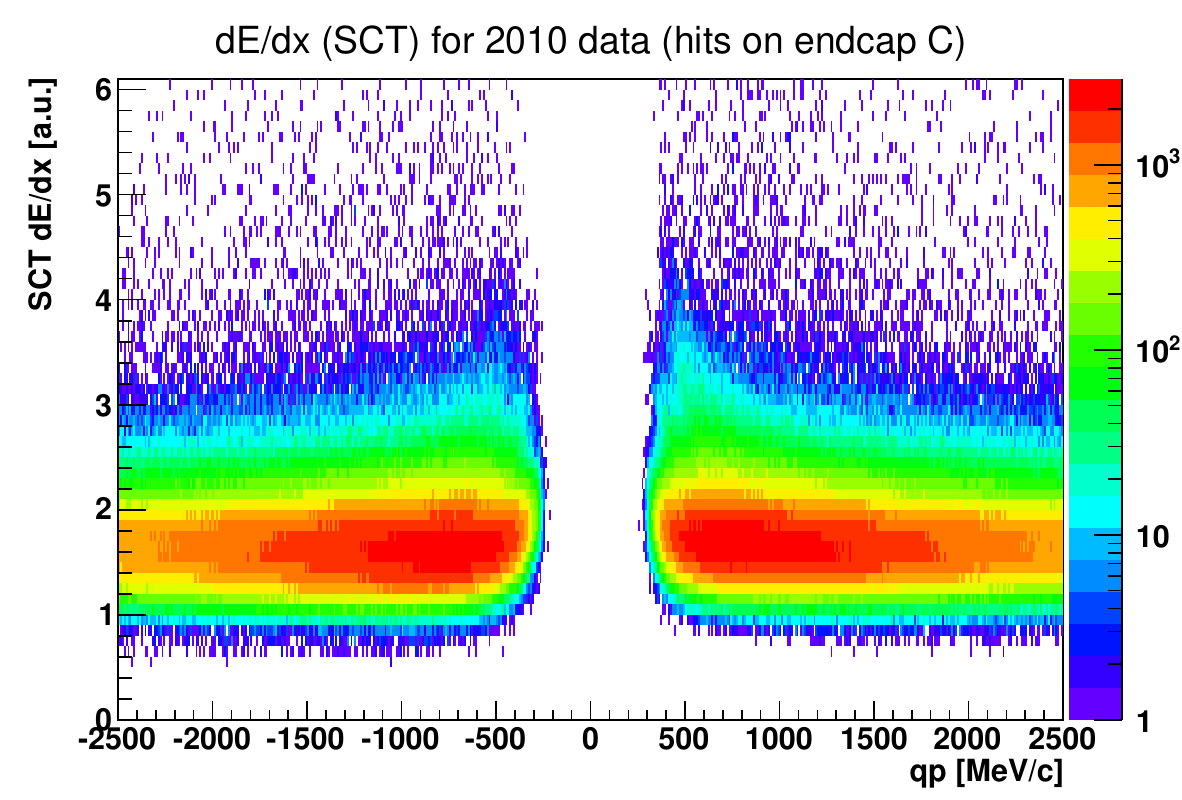}
   }
 \caption{$\dedx$ distributions vs $pq$ in 
 2010.}\label{fig:7-dedx_vs_pq_PV_2010}
 \end{center}
\end{figure}

Figure~\ref{fig:7-dedx_vs_pq_PV_2010} shows the $\dedx$ distribution for the 
SCT in the whole 2010 dataset. The barrel plot shows the calculated $\dedx$ for
tracks that do not leave hits in the endcap, while the endcaps plots correspond
to tracks that have hits in one of the endcaps. 

The band separation for tracks that only hit the barrel is more evident than 
in the tracks that hit the endcaps. Tracks with low total momentum normally do 
not leave hits in the endcap. 

As a result, the endcap $\dedx$ data appears at higher 
momentum than the barrel $\dedx$, around $150\MeV/c$ for the barrel and around 
$250\MeV/c$ for each of the two endcaps. This is primarily caused by the cut on 
the transverse momentum and the minimum number of hits in the SCT, what makes 
the minimum momentum shown in the plots greater than for the barrel tracks.

No primary/secondary vertex separation is done for these plots. Note that the
Y axis is different for the plot showing all regions and the rest.

\section{Particle Identification}\label{sec:7-pid}

The goal of this part of the study is to evaluate if there is any particle 
discriminating power in the 
$\dedx$ calculated with the SCT time-bin information. 

The Monte Carlo 
simulations of SCT Ntuples do not include the realistic time-bin information. 
As a result, it is not possible to use truth information from Monte Carlo 
simulations to 
evaluate the performance of the particle identification using the $\dedx$ 
calculated in the SCT to identify protons, kaons and pions. 

After April 12\textsuperscript{th}, 2011 (run 179579, SCT Ntuple tag f359), the 
SCT Ntuples contain a vector of Pixel likelihoods for pions, kaons and protons. 
Therefore, with 2011 runs after that date it is possible to perform 
efficiency-mistag rate studies, which are shown in the next
section.

The unavailability of the PID information for the Pixel in 2010 data forced
the use of a different approach. A geometrical cut is used on the Pixel 
$\dedx$ to try separate the protons.

This geometrical cut is applied with the following formulas on the
$\dedx$ calculated by the Pixel detector:

\begin{equation}
\left.\begin{array}{r}
  p[{\rm GeV}/c] \geq 0.3 \\
  \\
  \dedx_{\rm Pixel} \geq 1.8 \\
  \\
  1.5\dedx_{\rm Pixel} \leq -6.6\left( p[{\rm GeV}/c] - 1.5\right)\\
  \\
  0.9\dedx_{\rm Pixel} \geq -6.5 \left( p[{\rm GeV}/c] - 0.9\right)
\end{array}
\right\} \Rightarrow \textnormal{Tag as proton}
\end{equation}

or

\begin{equation}
\left.\begin{array}{r}
  0.3 \leq p[{\rm GeV}/c] \leq 0.55 \\
  \\
  4.0\leq \dedx_{\rm Pixel} \leq 8.0
\end{array}
\right\} \Rightarrow \textnormal{Tag as proton}
\end{equation}

Particle identification and resolution of the $\dedx$ reconstruction 
with SCT data taken during 2010 are detailed in~\cite{Garcia-Argos:1557122}. 

The study to find the geometrical cut for the Pixel $\dedx$ is not 
part of this thesis work, but it is presented here because it was used in some
of the studies relevant to this thesis. It is used in the long 
term stability analysis, presented in section~\ref{sec:7-stability}.

\subsection{Particle Identification with 2011-2012 Data}\label{sec:7-pid20112012}

The lack of Monte Carlo simulated Ntuples with full time-bin information and 
truth information limits the scope of the analysis. A way to overcome this 
limitation is the use of the Pixel 
likelihood. A Particle Identification (PID) study with 2011 and 2012 data from the SCT 
was done by extracting pure $i = \pi/K/p$ from the SCT data by using the Pixel 
$\dedx$ likelihood stored in the SCT Ntuple. 
The likelihoods calculated by the Pixel for each particle type, 
$P_{\pi}^{\rm Pixel}$, $P_K^{\rm Pixel}$, 
$P_p^{\rm Pixel}$, are first normalized:

\begin{equation}
 P_{i}^{\rm norm,Pixel} = \frac{P_i^{\rm Pixel}}{P_{\pi}^{\rm Pixel} + P_K^{\rm Pixel} + P_p^{\rm Pixel}}
\end{equation}

Tracks that have normalized Pixel likelihood $P^{\rm norm,Pixel}_{p/K/\pi} > 
0.9$ are considered correctly identified, so we call them ``true particles''. 
The SCT $\dedx$ is calculated for these tracks
to extract the distributions for each particle type in different low momenta 
regions. Unless specified, there are no cuts on secondary or primary vertices, 
and barrel or endcap tracks.

\begin{figure}[!hp]
 \begin{center}
 \subfigure[Pions Pixel $\dedx$.]{
   \includegraphics[scale=0.37]{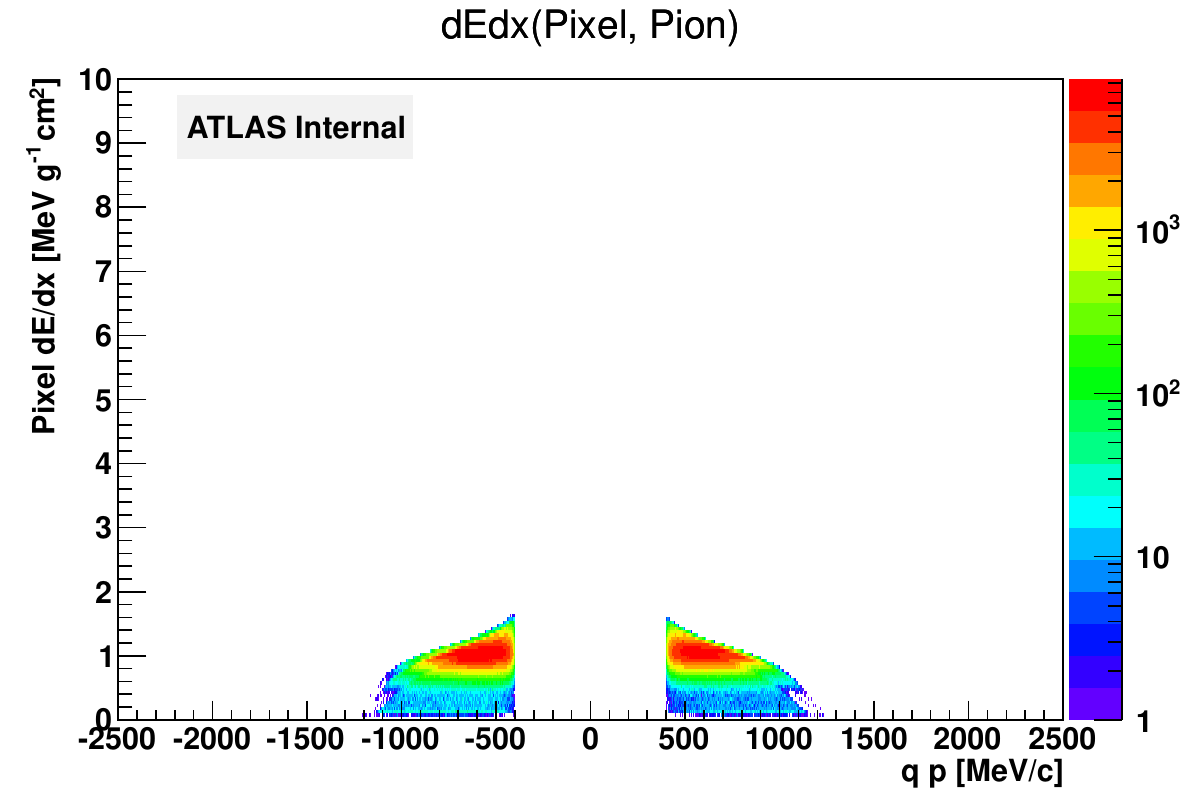}
 }
 \subfigure[Pions SCT $\dedx$.]{
   \includegraphics[scale=0.37]{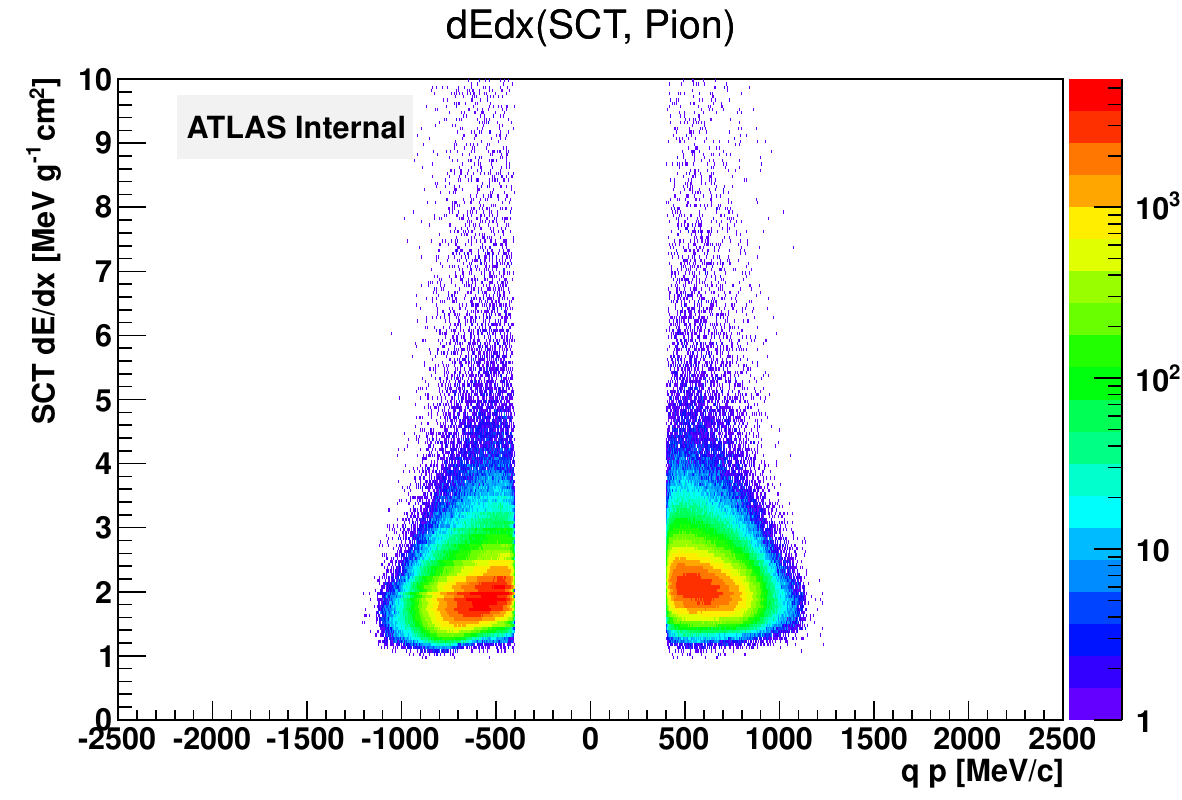}
 }

 \subfigure[Kaons Pixel $\dedx$.]{
   \includegraphics[scale=0.37]{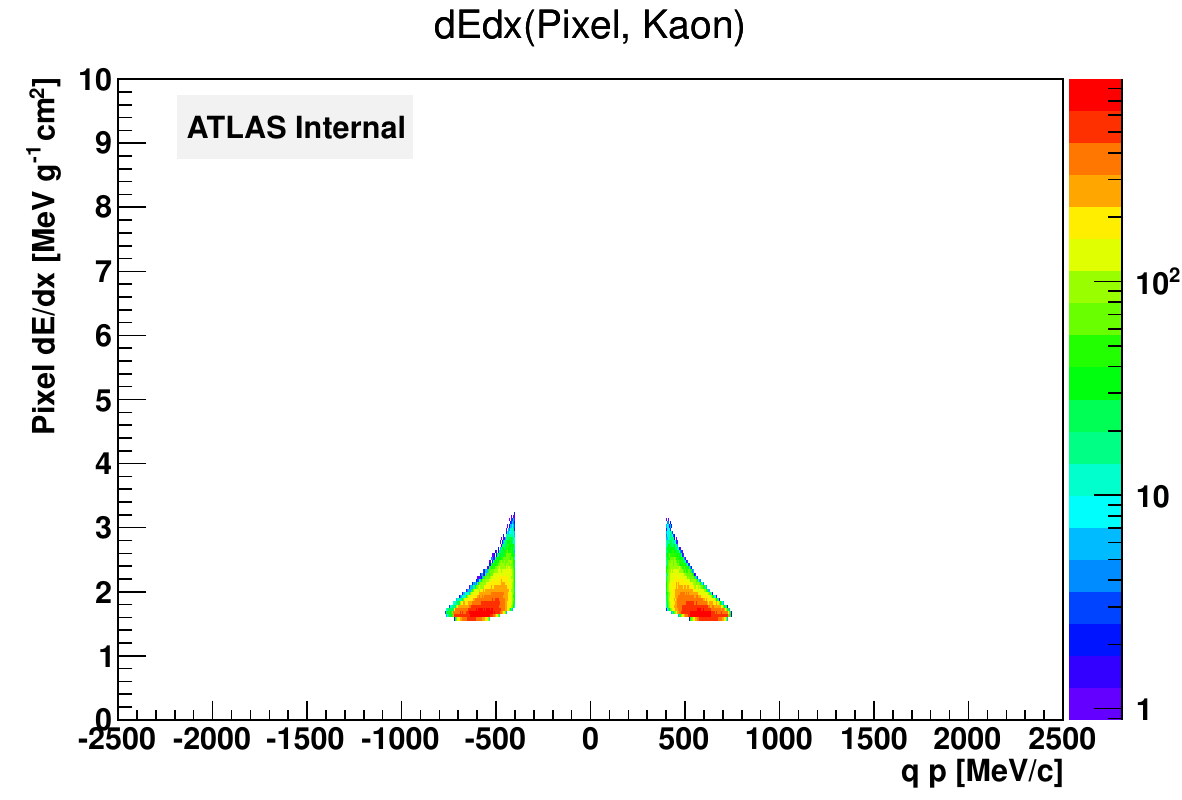}
 }
 \subfigure[Kaons SCT $\dedx$.]{
   \includegraphics[scale=0.37]{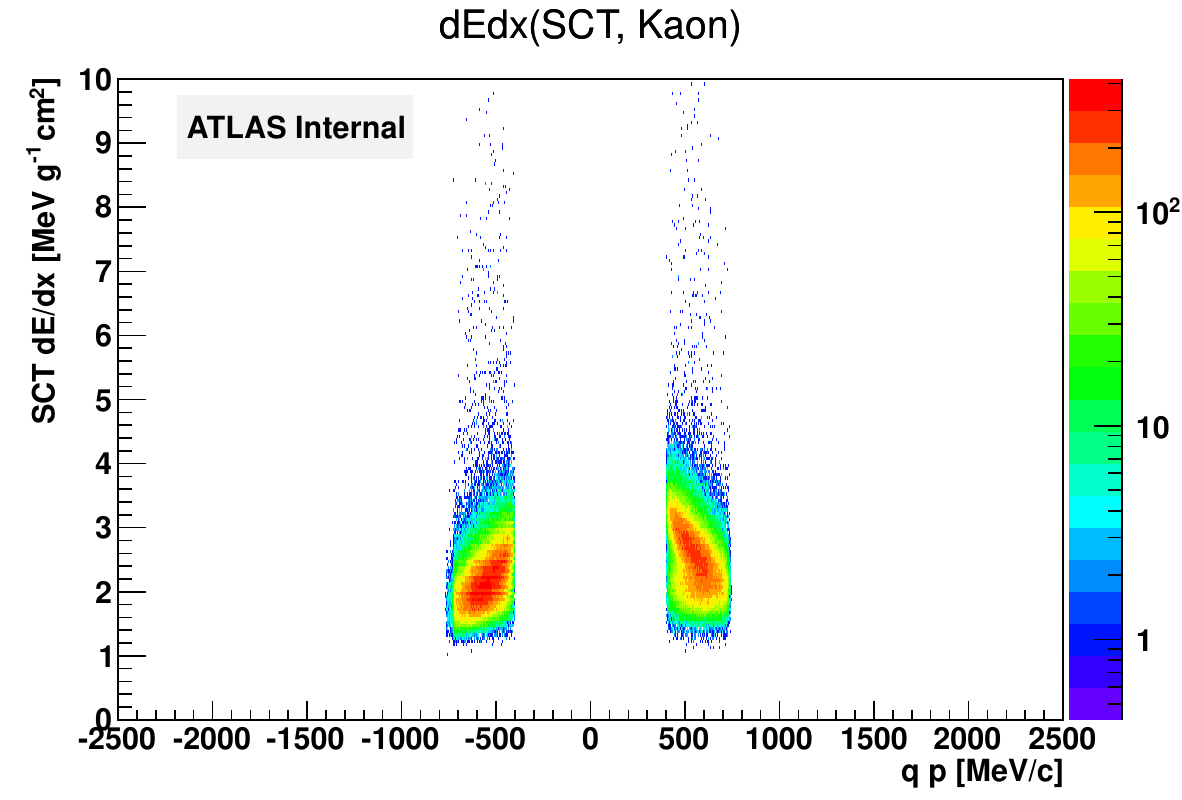}
 }

 \subfigure[Protons Pixel $\dedx$.]{
   \includegraphics[scale=0.37]{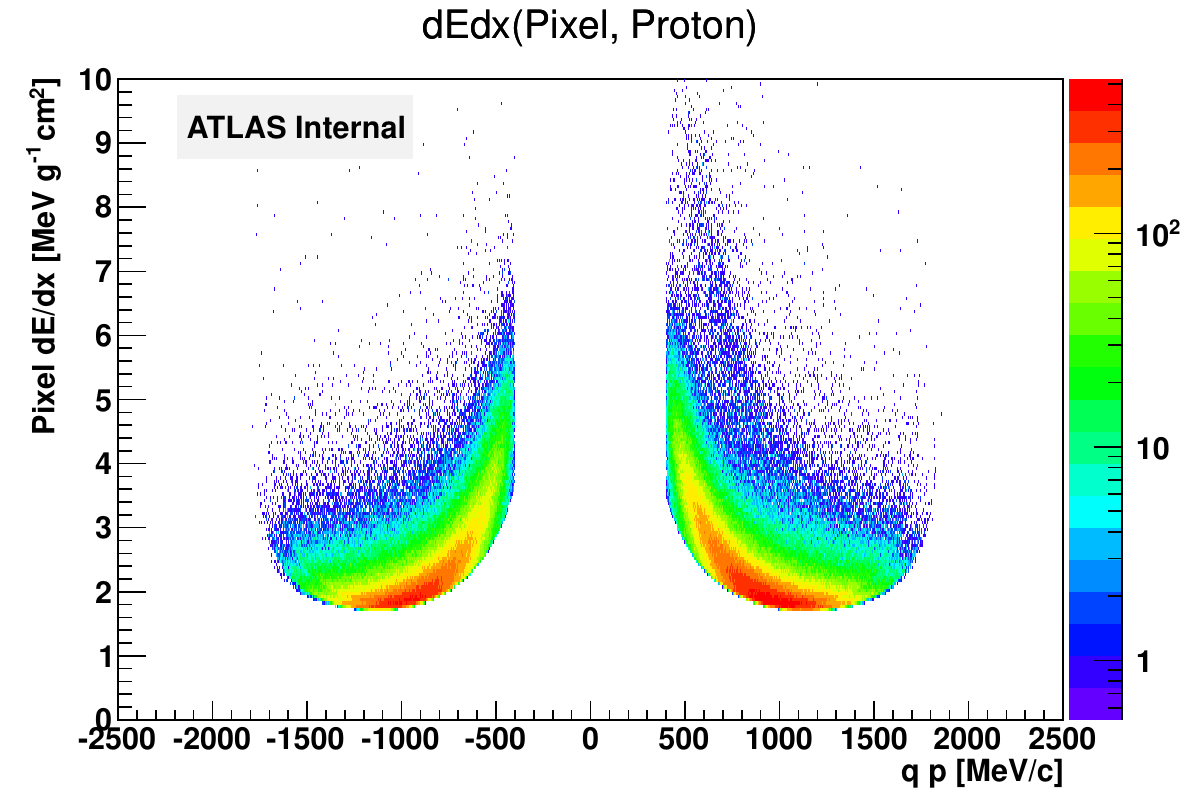}
 }
 \subfigure[Protons SCT $\dedx$.]{
   \includegraphics[scale=0.37]{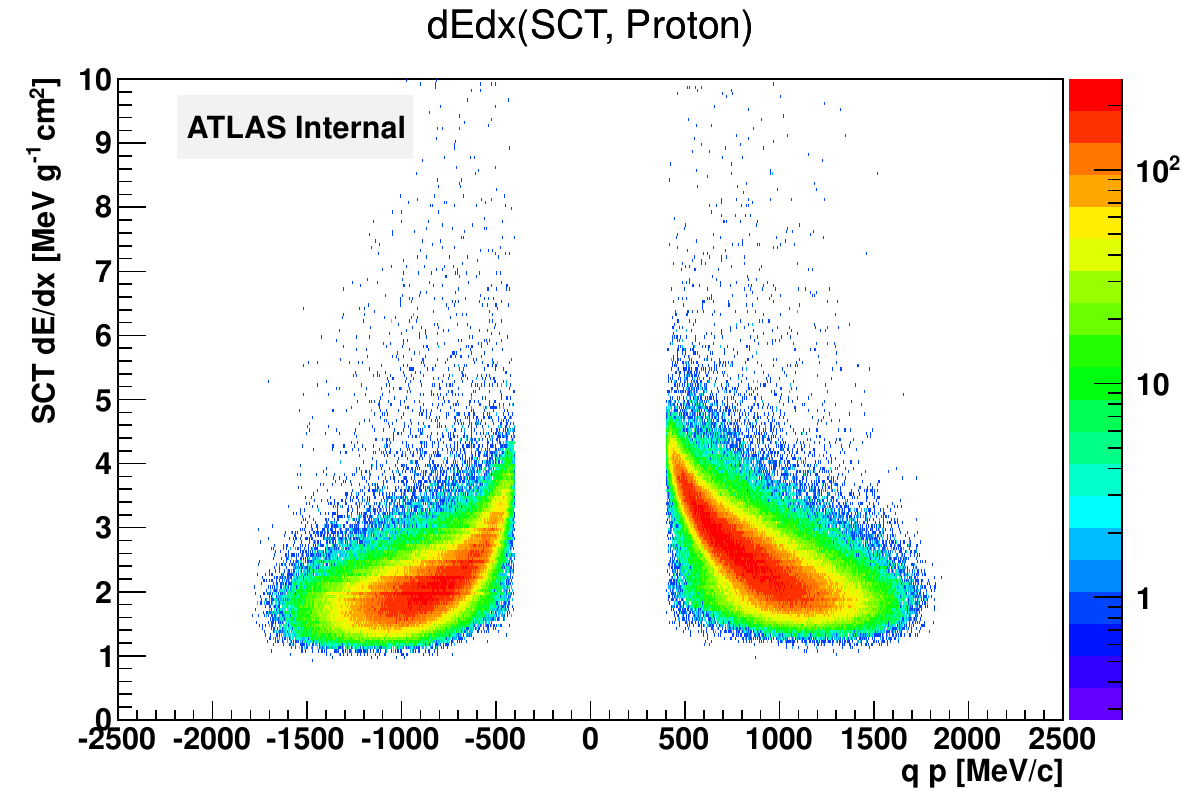}
 }
  \caption{Pixel and SCT $\dedx$ distributions vs $pq$ for the different 
  particle species as tagged by the Pixel  ($P_{i}^{\rm norm,Pixel} > 0.9$). 
  Data taken in 2011.}\label{fig:7-pid_2011_vsqp}
 \end{center}
\end{figure}

Figure~\ref{fig:7-pid_2011_vsqp} shows the various distributions of both the
Pixel and SCT $\dedx$ of each particle species. This plot is generated with 
data taken during 2011.

Tracks in 2010 were reconstructed only for those with momentum greater than
$300\MeV/c$, whereas in 2011 and 2012 the minimum track momentum was 
$400\MeV/c$. This means that for the 2011 and 2012 datasets, there is less
separation between bands. The band separation can be seen when plotting 
the $\dedx$ for each of the particle species in each momentum slice. The 
resolution is better if the distributions are more separated.

Distributions of the individual particle species  for data recorded in 2011 are 
shown in Figures~\ref{fig:7-pid_2011_sct_dedx_distrib} 
and~\ref{fig:7-pid_2011_sct_dedx_distrib_neg}, for positive and negative tracks
respectively. The vertical axis is the fraction of tracks for each momentum
range. Only primary vertex tracks have been selected. The fraction of
secondary vertex tracks is typically $10\%$ of the total tracks for pions and
kaons, and over $20\%$ for protons. 

For low momenta and positive protons, 
nearly $50\%$ of the tracks originate from secondary vertices. 
Figure~\ref{fig:7-SCT_dedx_lowmomentum} shows the different distributions of
positive tracks for
pions, kaons and protons for 2011 data. The plot corresponding to the proton
band shows that the secondary vertex tracks represent almost $50\%$ of all the
proton tracks. The secondary vertex tracks of pions and 
kaons are much less abundant than the primary vertex tracks.

\begin{figure}[!htb]
 \begin{center}
 \subfigure[Primary and secondary vertex pions.]{
  \includegraphics[scale=0.45, trim=70 0 40 0, clip=true]{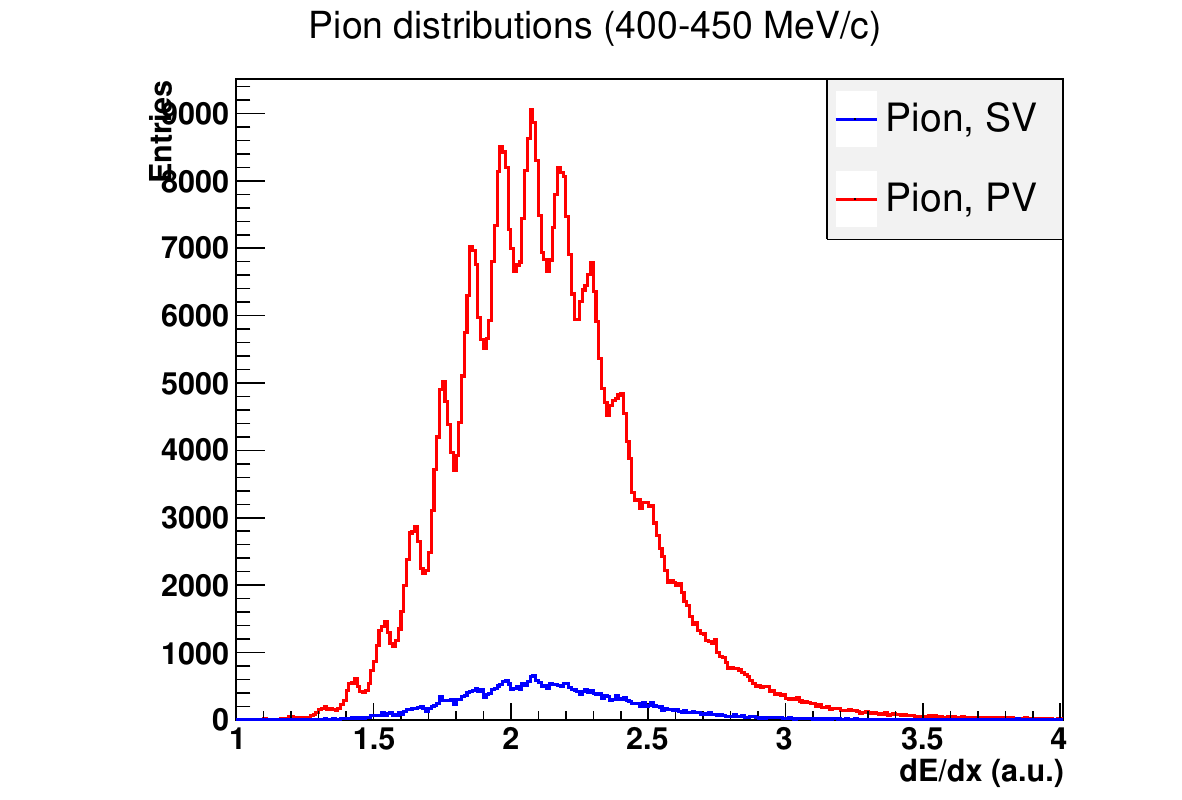}
 }  
 \subfigure[Primary and secondary vertex Kaons.]{
  \includegraphics[scale=0.45, trim=70 0 40 0, clip=true]{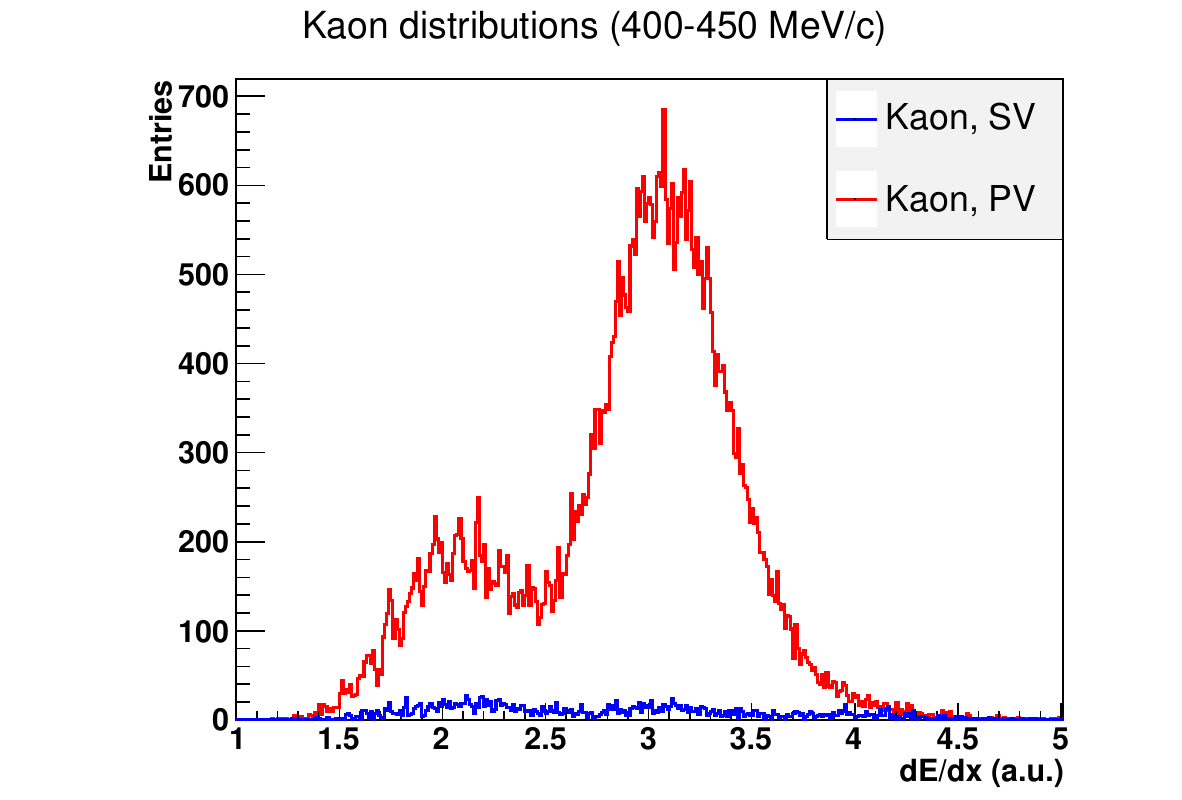}
 }
 
 \subfigure[Primary and secondary vertex protons.]{
  \includegraphics[scale=0.48, trim=70 0 40 0, clip=true]{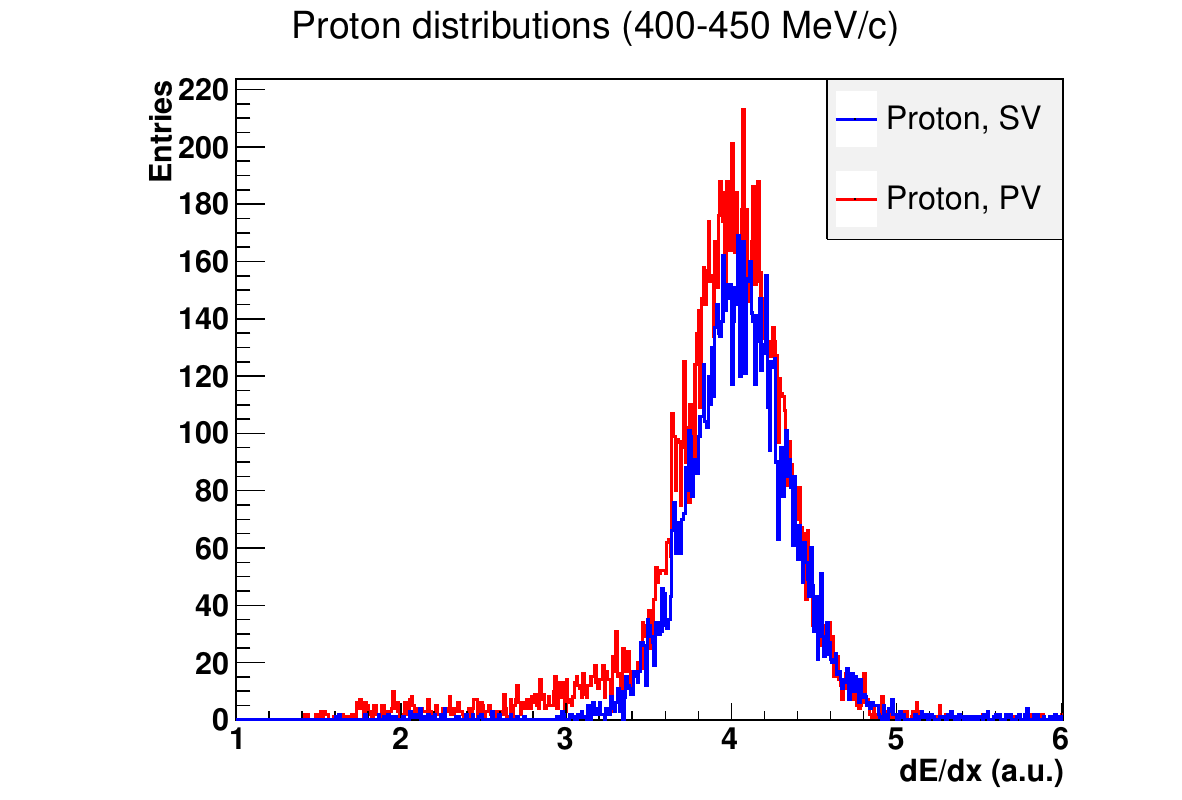}
 }
 \caption[$\dedx$ distributions for positive pions, kaons and protons at low 
 momenta.]{$\dedx$ distributions for positive pions, kaons and protons at low 
 momenta, showing the differences in the rates of appearance for each particle 
 species originating in primary and secondary 
 vertices. The secondary vertex protons represent almost half the total number
 of protons in the sample for momentum in the $(400, 450)\MeV$ 
 range.}\label{fig:7-SCT_dedx_lowmomentum}
 \end{center}
\end{figure}

\begin{figure}[!htbp]
 \begin{center}
 \subfigure[SCT $\dedx$  distributions for each particle for $q>0$ and 2011 data.]{
  \includegraphics[scale=0.75]{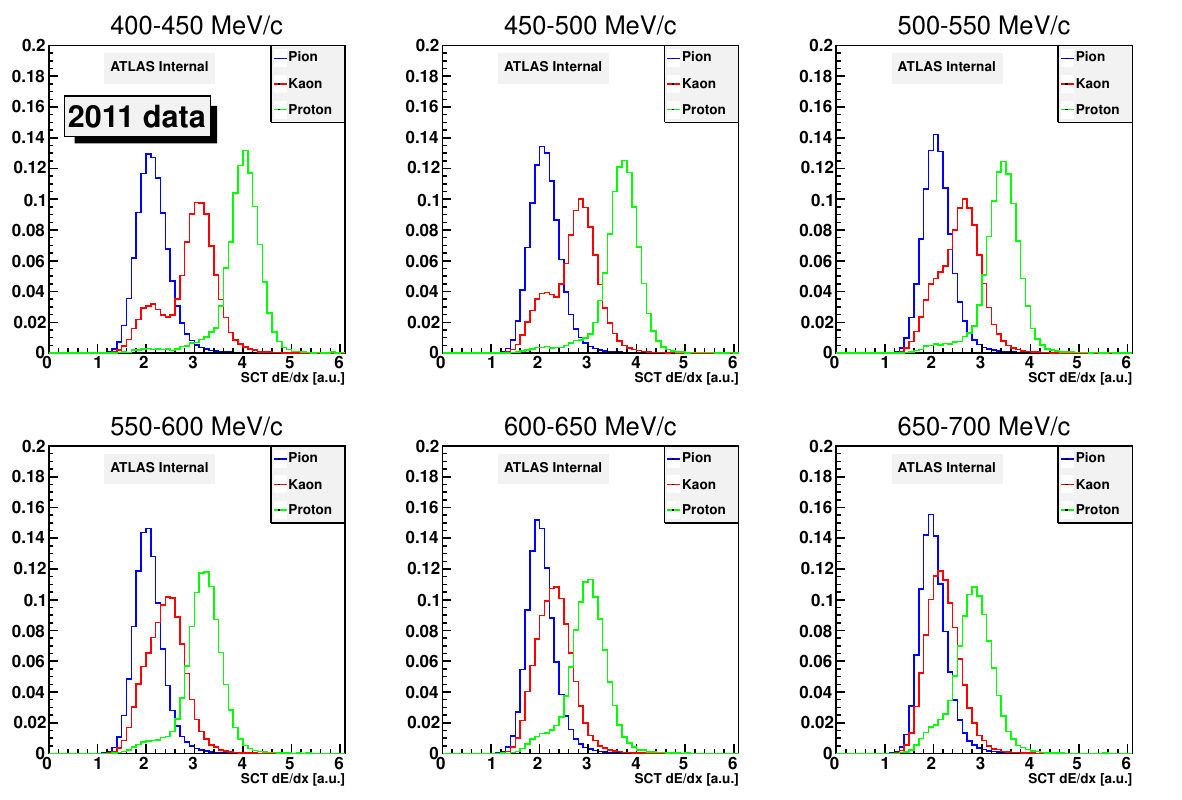}

  \label{fig:7-pid_2011_sct_dedx_distrib}
  }
  
  \subfigure[SCT $\dedx$  distributions for each particle for $q<0$ and 2011 data.]{
  \includegraphics[scale=0.75]{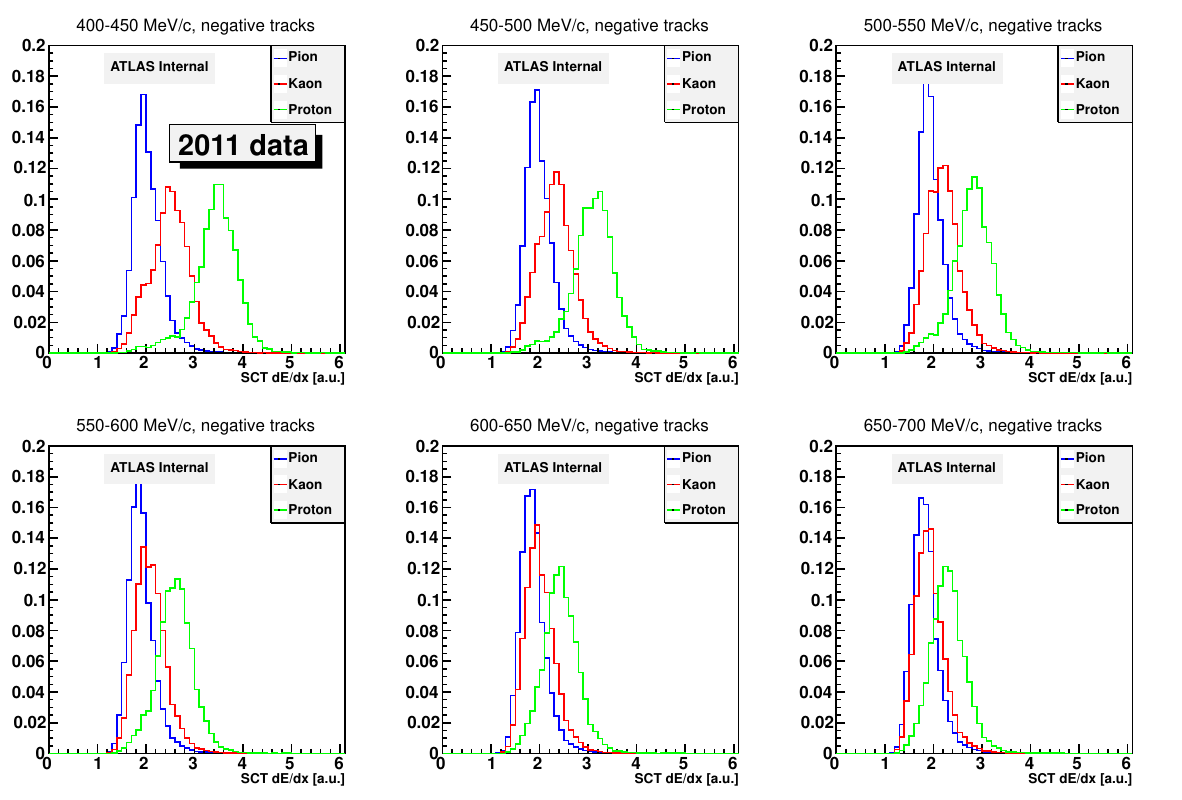}\label{fig:7-pid_2011_sct_dedx_distrib_neg}
  }
  \caption{SCT $\dedx$ distributions for 2011 data, primary vertex tracks
  only.}\label{fig:7-pid_2011_distributions_pv}
 \end{center}
\end{figure}

\begin{figure}[!htbp]
 \begin{center}
 \subfigure[SCT $\dedx$  distributions for each particle for $q>0$ and 2011 data.]{
  \includegraphics[scale=0.75]{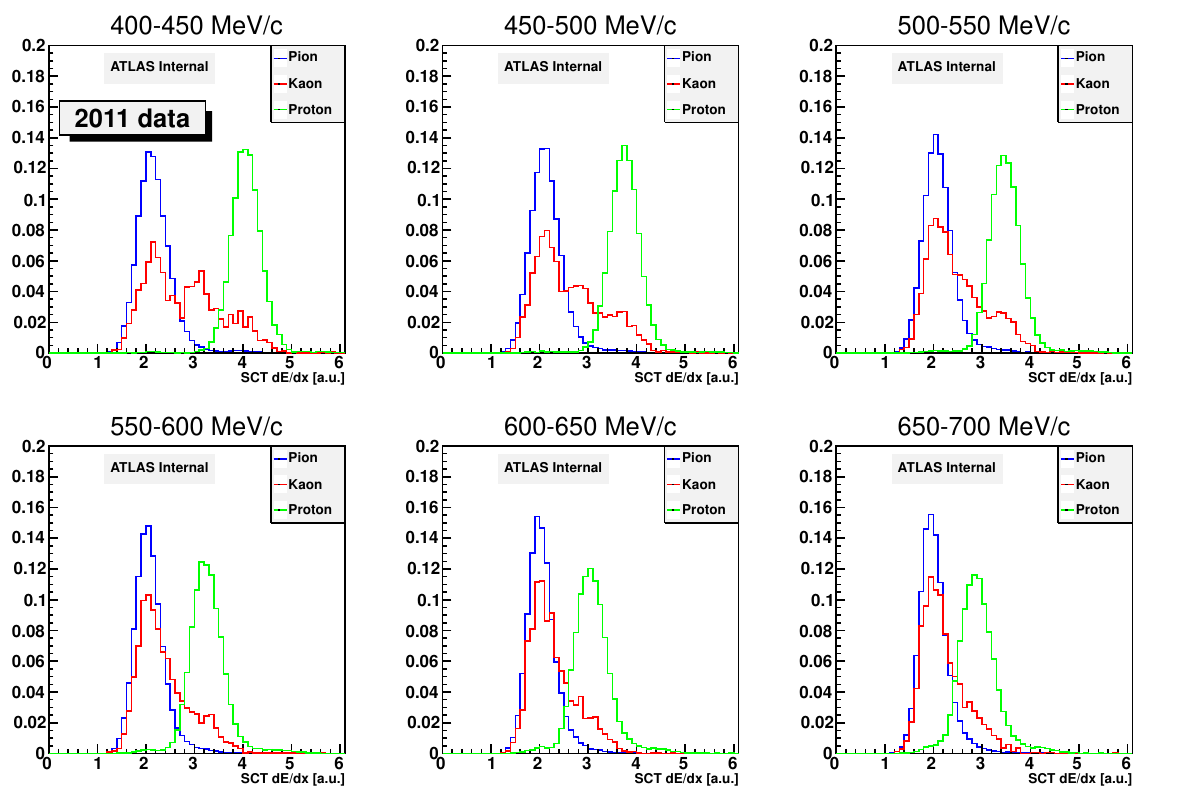}

  \label{fig:7-pid_2011_sct_dedx_distrib_pos_SV}
  }
  
  \subfigure[SCT $\dedx$  distributions for each particle for $q<0$ and 2011 data.]{
  \includegraphics[scale=0.75]{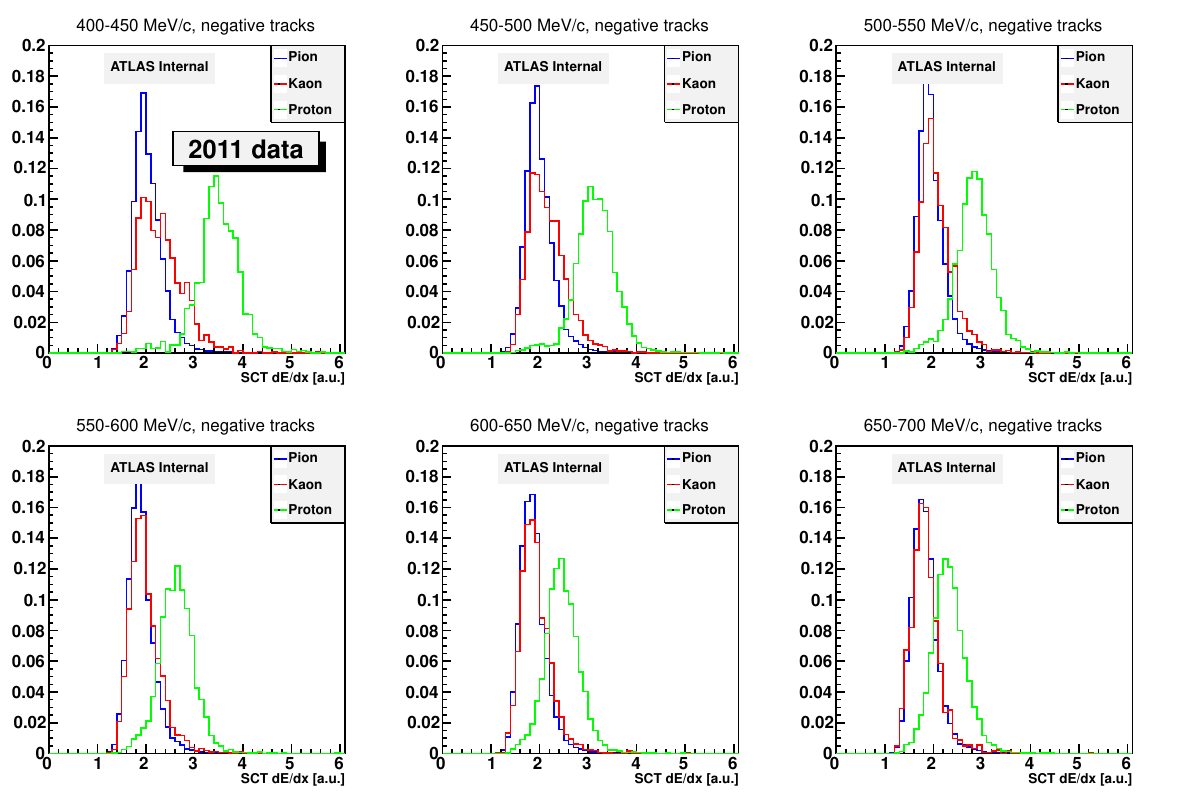}\label{fig:7-pid_2011_sct_dedx_distrib_neg_SV}
  }
  \caption{SCT $\dedx$ distributions for 2011 data, secondary vertex tracks
  only.}\label{fig:7-pid_2011_sct_dedx_distrib_SV}
 \end{center}
\end{figure}

\begin{figure}[!htbp]
 \begin{center}
 \subfigure[SCT $\dedx$  distributions for each particle for $q>0$ and 2012 data.]{
  \includegraphics[scale=0.75]{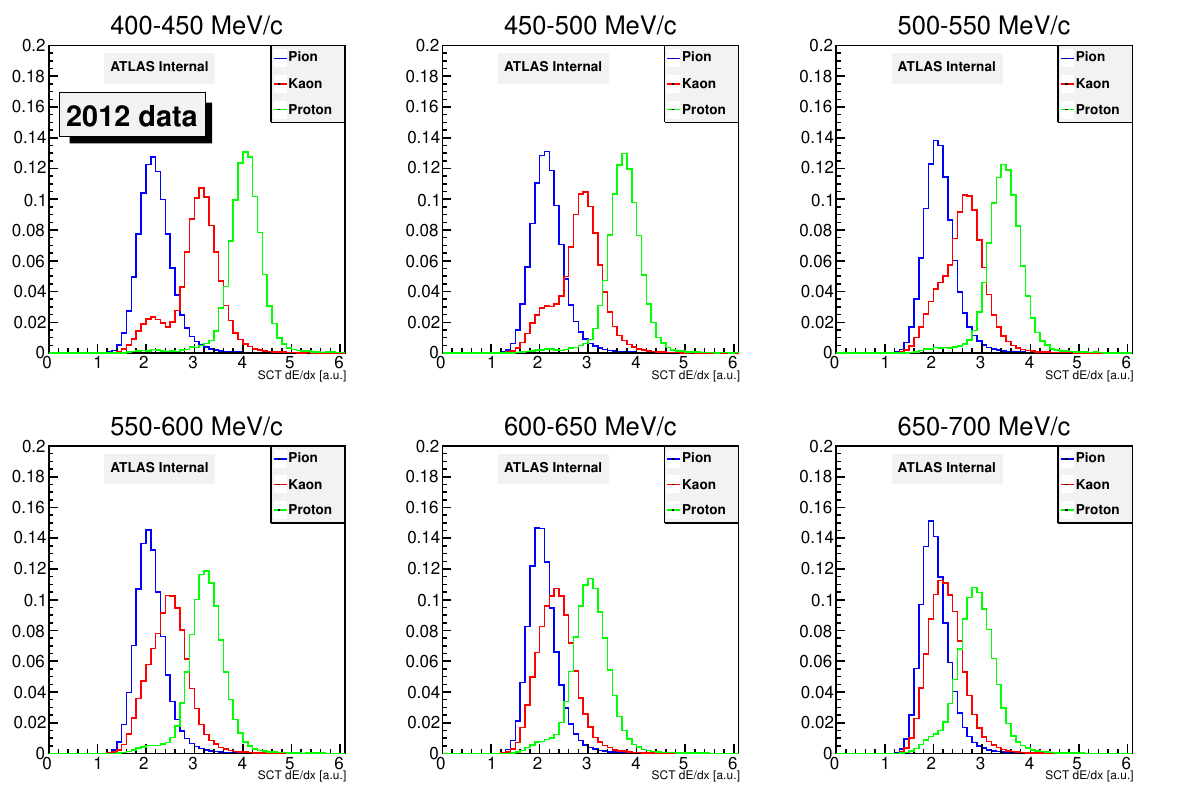}

  \label{fig:7-pid_2012_sct_dedx_distrib}
  }
  
  \subfigure[SCT $\dedx$  distributions for each particle for $q<0$ and 2012 data.]{
  \includegraphics[scale=0.75]{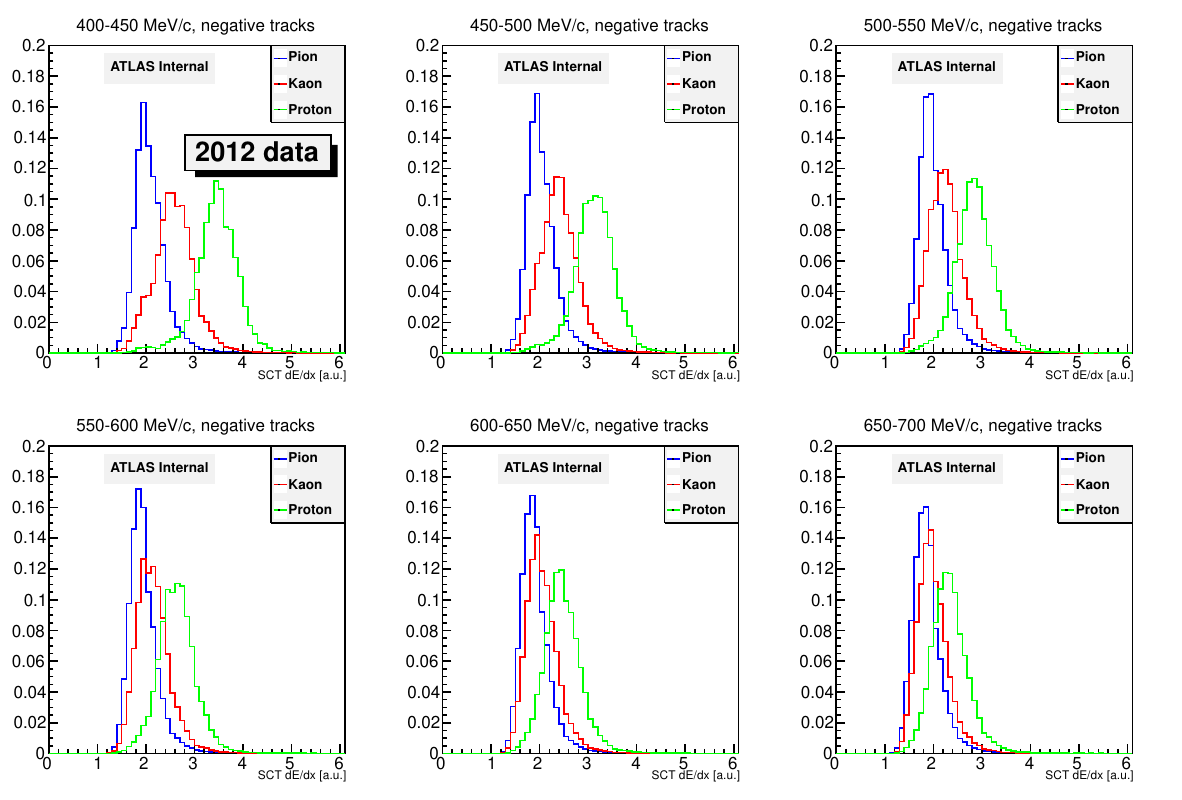}\label{fig:7-pid_2012_sct_dedx_distrib_neg}
  }
  \caption{SCT $\dedx$ distributions for 2012 data, primary vertex tracks
  only.}\label{fig:7-pid_2012_distributions_pv}
 \end{center}
\end{figure}

A sample of the secondary vertex tracks in 2011 is shown in 
Figure~\ref{fig:7-pid_2011_sct_dedx_distrib_pos_SV}, for positive tracks, and
Figure~\ref{fig:7-pid_2011_sct_dedx_distrib_neg_SV}, for negative tracks.
The plots for positive tracks show that the kaon band is mis-identified
and also the proton band is slightly more separated for higher momenta tracks.

The same plots using 2012 data are shown in 
Figures~\ref{fig:7-pid_2012_sct_dedx_distrib} 
and~\ref{fig:7-pid_2012_sct_dedx_distrib_neg} positive and negative tracks 
respectively. Only the primary vertex tracks are shown.

The plots only show a representative sample of the momenta ranges for both
positive and negative tracks, up to 
$700\MeV/c$, in order to illustrate the loss of resolution with increasing
momentum. Although the $\dedx$ has been calculated for tracks up to 
$2.5\GeV/c$, such high momentum does not allow any particle identification 
study due to the lack of band separation at high momentum.
  
The positive tracks have better resolution for low momentum ($< 550\MeV/c$) 
in both periods and worse resolution for higher momentum. Negative tracks, on 
the contrary, show low resolution already at lower momentum, when compared to
the positive tracks.
The reason for this lower resolution of the negative tracks is the tilt angle
of the SCT barrel modules, which is around $11^{\circ}$. 

A negative track hits the modules in the barrel at a greater incidence angle 
than a positive track with the same momentum.
The SCT $\dedx$ calculation relies on the cluster size, and a negative
track will leave smaller clusters for the same momentum as a positive track.
Therefore, the calculation of the charge deposited by
a negative track will result in lower $\dedx$ values than in the case of a
positive track.

Although it is not shown in these plots, there is no kaon band for particle 
momentum above 
$750\MeV/c$ in positive tracks and $800\MeV/c$ in negative tracks. This 
occurs when the Pixel $\dedx$ calculation is such that it is not possible to 
discriminate between pions and kaons. The Pixel assigns a negative likelihood 
to them.

\subsection{Efficiency and Mistag Rate}\label{sec:7-efficiency}

The SCT $\dedx$  distributions of the 3 particle species for positive tracks
are clearly 
separated at low momentum. Protons can be identified in the momentum interval 
$(400,550)\MeV/c$. 
Both data periods have similar distribution of the particles species.

The SCT $\dedx$ likelihood, $P^{\rm SCT}_i$, is defined by Gaussian fitting 
near the first peak each of the SCT $\dedx$  particle distribution from 
Figures~\ref{fig:7-pid_2011_distributions_pv}
and~\ref{fig:7-pid_2012_distributions_pv}, 
in each momentum slice. The fit is done after identifying the particle species
using the Pixel information. The results are the mean $(\dedx)_i$ and 
$\sigma_{\dedx_i}$, with $i=\pi/K/p$: 

\begin{equation}
 P^{\rm SCT}_i = \frac{1}{\sqrt{2\pi} \sigma_{\dedx_i}} \exp \left( - \frac{ \left(\dedx_{\rm SCT} - \dedx_i \right)^2}{2 \sigma_{\dedx_i}^2}\right)
\end{equation}

The same normalization as in the Pixel likelihood is also applied to the SCT
likelihood:

\begin{equation}
 P_i^{\rm norm,SCT} = \frac{P^{\rm SCT}_i}{P^{\rm SCT}_p + P^{\rm SCT}_K + P^{\rm SCT}_{\pi}}
\end{equation}

Since the SCT Particle Identification capability is being compared to that of 
the Pixel, the tagging efficiency is defined as the fraction of tracks that the 
Pixel and the SCT would tag in the same way. For protons, the tagging 
efficiency is

\begin{equation}
 \varepsilon_p = \frac{N({\rm Pixel} = p, {\rm SCT} = p)}{N({\rm Pixel} = p)}
\end{equation}

The mistag rate is calculated under the assumption that the 
Pixel tagging is true for pions and kaons while the SCT would tag as proton:

\begin{equation}
 r_p^{\pi} = \frac{N({\rm Pixel} = \pi, {\rm SCT} = p)}{N({\rm Pixel} = \pi)}\qquad r_p^{K} = \frac{N({\rm Pixel} = K, {\rm SCT} = p)}{N({\rm Pixel} = K)}
\end{equation}

Where $N({\rm Pixel} = i)$ is the number of tracks considered as true $i$ 
particles by the Pixel
($P_i^{\rm norm, PIX} > 0.9$) and $N({\rm Pixel} = i, {\rm SCT} = j)$ is the 
number of tracks 
considered true $i$ particles by the Pixel and tagged as $j$ by the SCT.

With these definitions, the proton tagging efficiency and mistag rate for 
pions and kaons can be checked, using five predefined 
cut values on the SCT normalized likelihood, $P_p^{\rm norm, SCT}$. These cut 
values are $0.1$, $0.3$, $0.5$, $0.7$ and $0.9$.
They are only applied to the SCT likelihood, as the Pixel likelihood is used 
as a means to assign ``truth'' information to the tracks.

\begin{figure}[!htb]
 \begin{center}
 \subfigure[Proton tagging efficiency vs kaon mistag rate.]{
    \includegraphics[scale=0.425, trim=40 0 40 0, clip=true]{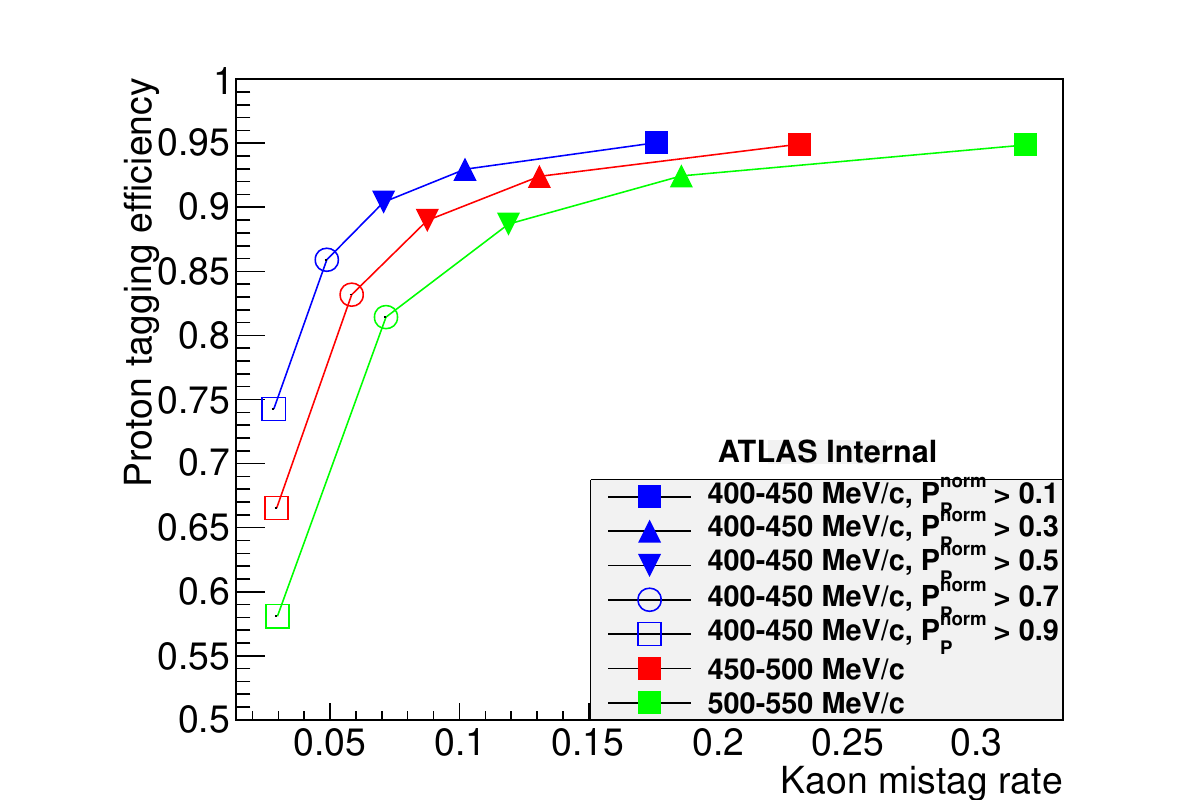}
 } 
 \subfigure[Proton tagging efficiency vs pion mistag rate.]{
    \includegraphics[scale=0.425, trim=40 0 40 0, clip=true]{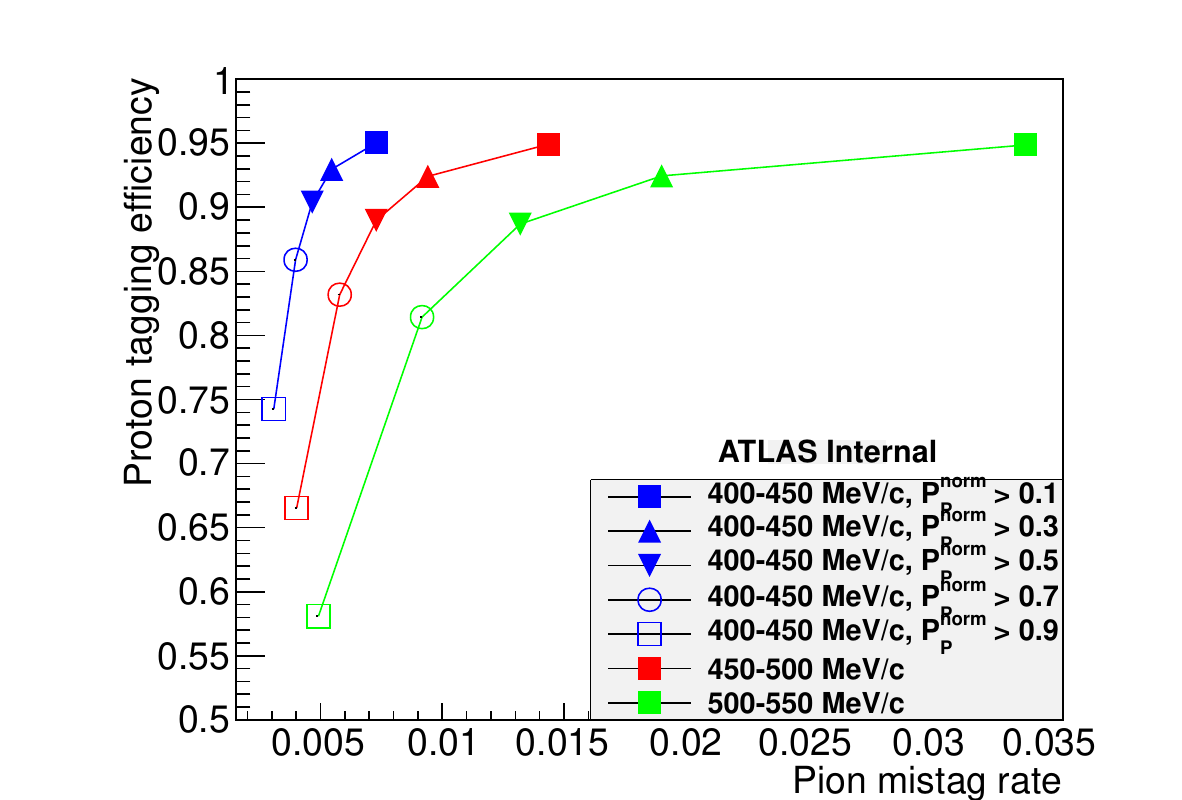}
 }
  \caption[Proton tagging efficiency in the SCT for 2011 data.]{Proton tagging 
  efficiency with respect to the mistag rate  for 
  kaon and pion as a function of 
  the cut on $P_p^{\rm norm, SCT}$, 2011 data. Track momentum in the interval
    $(400, 550)\MeV/c$, with three slices, and for positive particles only.
  }\label{fig:7-tag_eff_2011}
 \end{center}
\end{figure}

\begin{figure}[!hbtp]
 \begin{center}
 \subfigure[Proton tagging efficiency vs kaon mistag rate.]{
    \includegraphics[scale=0.425, trim=40 0 40 0, clip=true]{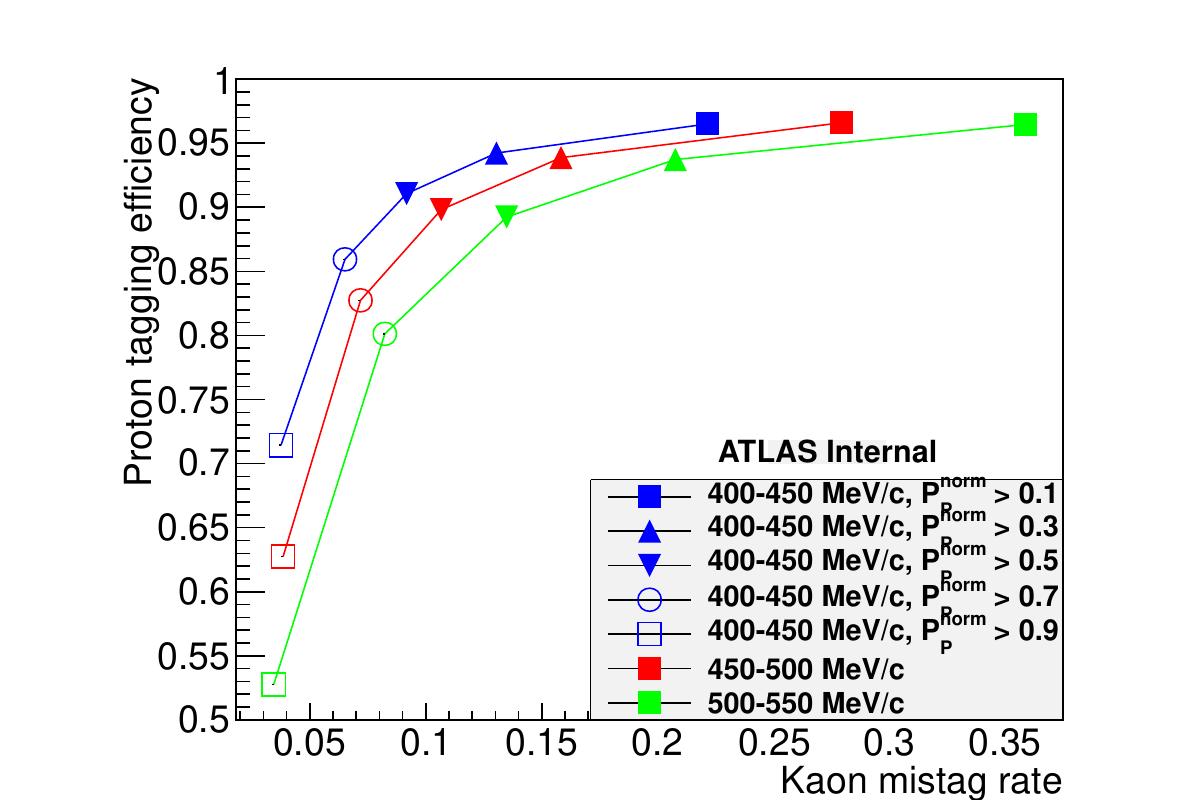}
 }
 \subfigure[Proton tagging efficiency vs pion mistag rate.]{
    \includegraphics[scale=0.425, trim=40 0 40 0, clip=true]{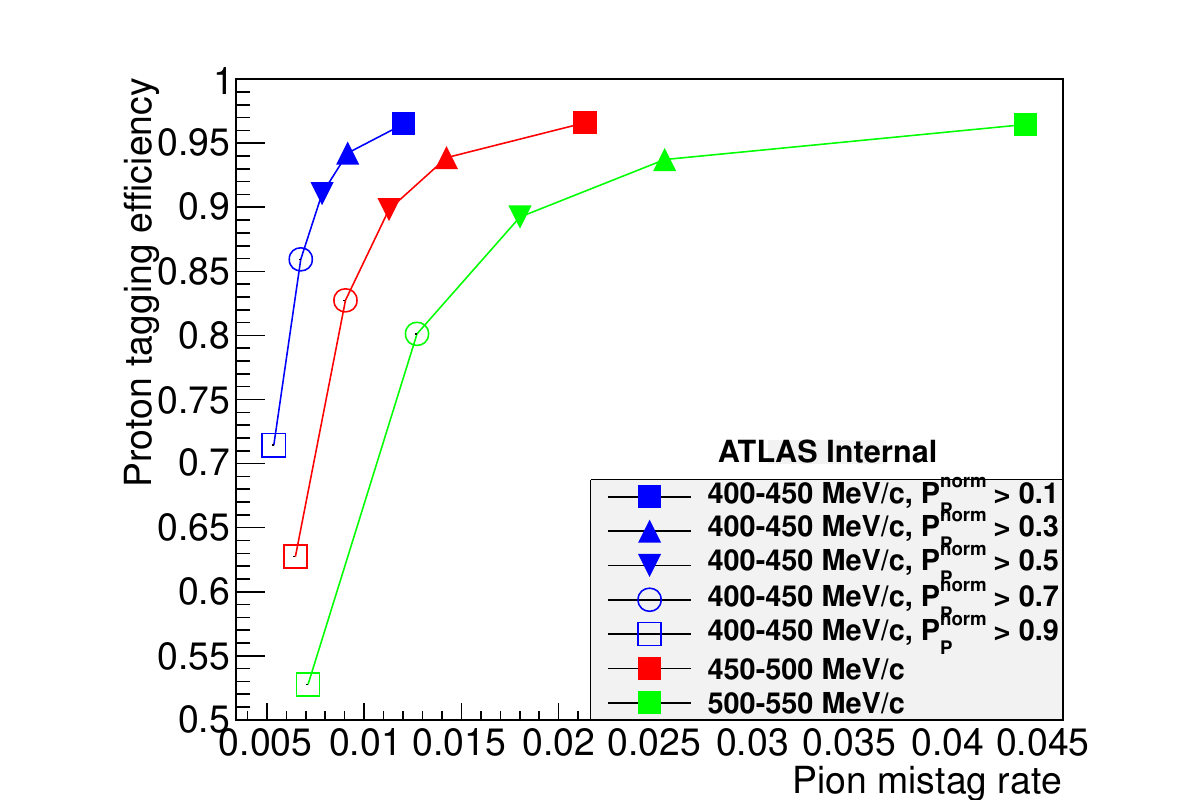}
 }
  \caption[Proton tagging efficiency in the SCT for 2012 data.]{Proton tagging 
  efficiency with respect to the mistag rate  for kaon 
  and pion as a function of 
  the cut on $P_p^{\rm norm, SCT}$, 2012 data. Track momentum in the interval
    $(400, 550)\MeV/c$, with three slices, and for positive particles 
    only.}\label{fig:7-tag_eff_2012}
 \end{center}
\end{figure}

Figures~\ref{fig:7-tag_eff_2011} and \ref{fig:7-tag_eff_2012} show the tagging
efficiency results for low momentum, $400 < p < 550\MeV/c$, both with 2011 and 
2012 data. The results show that the SCT has some independent particle 
discriminating power, with a proton tagging efficiency greater than $95\%$ 
while maintaining a pion mistag rate lower than $5\%$ and a kaon mistag rate 
lower than $40\%$.

The top (blue) line corresponds to $400 < p < 450\MeV/c$, the middle (red) line
to $450 < p < 500\MeV/c$ and the bottom (green) line to $500 < p < 550\MeV/c$.
The filled squares are the cut $P^{\rm norm}_p > 0.1$, upwards triangles are
$P^{\rm norm}_p > 0.3$, downwards triangles are $P^{\rm norm}_p > 0.5$, hollow
circles are $P^{\rm norm}_p > 0.7$ and hollow squares are 
$P^{\rm norm}_p > 0.9$.

For the 2011 data, the kaon mistag rate is lower than $30\%$ and the pion 
mistag rate stays below $4\%$, with a proton tagging efficiency around $95\%$.

During the same period, for a kaon mistag rate less than $20\%$, a tagging 
efficieny greater than $90\%$ is possible, 
whereas for a kaon mistag rate lower than $15\%$, the efficiency is above $95\%$.

In 2012 data, the efficiency is slightly higher, as well as the mistag rates. 
For a proton tagging efficiency of around $96\%$, the kaon mistag rate is lower 
than $40\%$ and the pion mistag rate is below $5\%$.

During this period, the tagging efficiency is close to $94\%$ with kaon mistag 
rates under $22\%$ and pion mistag rates under $3\%$. For a tagging efficiency 
around $90\%$, the mistag rates stay under $15\%$ and $2\%$, for kaons and 
pions respectively.

When comparing the two sets of plots in Figures~\ref{fig:7-tag_eff_2011} and 
\ref{fig:7-tag_eff_2012}, one can observe that the proton tagging efficiency 
does not degrade over time. The difference between the two periods is likely to 
be related to the fit values that are derived from the Pixel ``truth'' particle 
information and are slightly different for each of the two periods.

\section{Long-term Stability}\label{sec:7-stability}

Because the SCT was not designed to do energy measurements, its PID capability 
is limited. However, a very useful study is tracking the radiation damage of 
the silicon tracker over time.

As it was described in section~\ref{sec:2-radiationdamage}, the silicon charge 
collection efficiency drops with accumulated radiation. Due to 
the lower charge collected, the Time-over-Threshold is inevitably reduced. 
Hence, the average time-bin size will drop as well.

One way to track the evolution of the charge collection efficiency is to study 
the change over time of the $\dedx$. For this purpose, the mean $\dedx$ of 
positive protons is tracked in the $(500,550)~{\rm MeV}/c$ range. 
The expected effect is a reduction of the mean value
over time when radiation damage effects are present.

This study involves several components. First, the protons have to be 
identified. The use of Pixel likelihoods is not possible for the 2010 runs 
because they were not calculated. The Pixel $\dedx$, however, is present in 
the data. The geometrical cut described in section~\ref{sec:7-pid} can be 
applied to the Pixel $\dedx$ to separate the protons.

By applying that geometrical cut to the 2010-2012 period it is possible
to get consistent results during the whole period. However, it is necessary to
check for differences between this geometrical cut and the actual Pixel 
likelihood cut that was described in section~\ref{sec:7-pid20112012}.

Calculation of the mean $\dedx$ is done by Gaussian fitting the $\dedx$ 
distribution for the $(500, 550)\MeV/c$ range and extracting the value of the
mean parameter.

\begin{figure}[!htb]
 \begin{center}
  \includegraphics[scale=0.55]{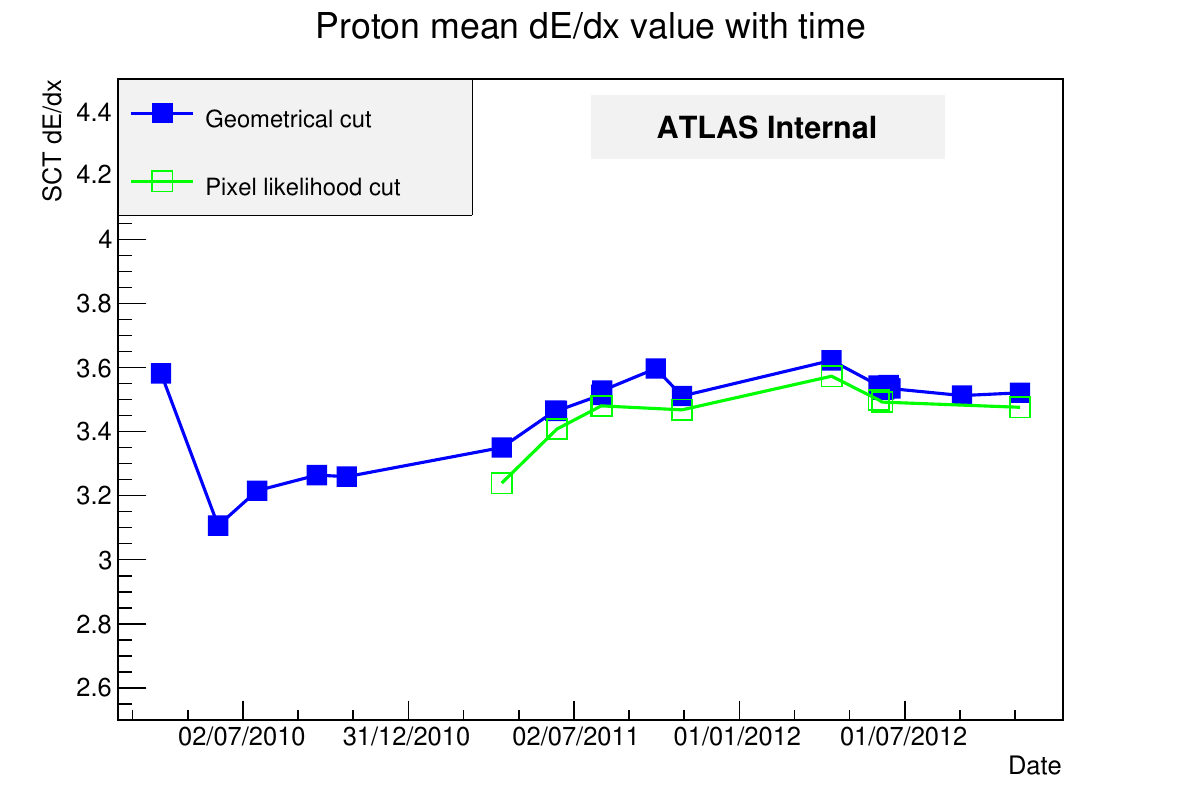}
  \caption{Time evolution of the proton (positive tracks) mean $\dedx$ for track 
  momentum in the 
  $(500,550)\MeV/c$ range, 
  extracting the protons with the geometrical cut and with the Pixel likelihood 
  for 2011-2012.}\label{fig:7-protonmean}
 \end{center}
\end{figure}

Therefore, the first part of the study was to see the actual proton mean
$\dedx$ evolution with time. Figure~\ref{fig:7-protonmean} shows this evolution 
in the whole 2010-2012 period, for the two aforementioned methods of extracting 
the proton band. The geometrical cut is very close to the Pixel likelihood cut 
and the trends are consistent with each other.

It is worth noting that, considering that the goal of this study is to track 
the radiation damage over time, the result shown in 
Figure~\ref{fig:7-protonmean} is initially confusing. The first point is 
much higher than the second one, then the mean $\dedx$ ramps up with time. 
There seems to be no evidence of radiation damage, especially if the first two 
points are ignored. In addition, an increase in the collected charge is 
not something to expect.

From previous studies~\cite{SahalTiming}, it is known that the timing of the 
SCT read-out is not stable with time. The average time-bin size (number of bins 
over threshold) tends to increase with time. This is caused by timing drifts
over time (refer to section~\ref{sec:timing} for a brief review on SCT 
timing). 

A particle traversing one or more strips deposits charge that is collected and 
a signal pulse is generated. When the timing setting is well adjusted with 
respect to the Level 1 trigger, a pulse of width around $20\ns$ is most likely 
to fall within one $25\ns$ time-bin. However, small drifts, of a few 
nanoseconds, on the timing would 
cause this same pulse to fall in two time-bins, hence increasing the average 
measured time-bin size. Even smaller pulses might be above threshold during two 
time-bins if the drift is large.

To prevent this from happening, timing scans are performed in the detector in 
order to calibrate the SCT timing. The goal is to have 01x patterns on the
hits time-bins, as was mentioned in section~\ref{sec:timing}. Next, the effect 
of the timing adjustment in the $\dedx$ calculation on the SCT is studied.

\begin{figure}[htb]
 \begin{center}
  \includegraphics[scale=0.55]{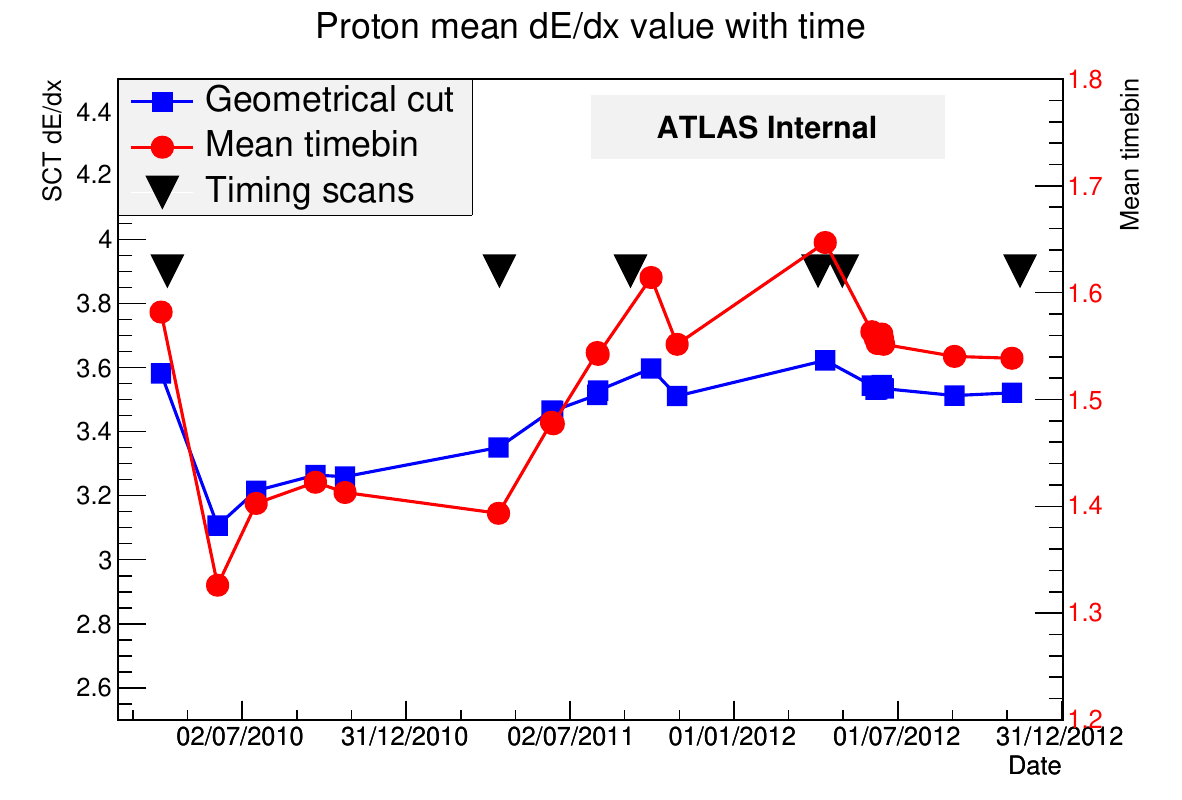}
  \caption{Evolution of the proton mean  $\dedx$ with time, showing the effect
  of the timing misadjustments and the timing scans 
  dates.}\label{fig:7-protonmean_with_timingscans}
 \end{center}
\end{figure}

This part of the study includes a mean time-bin size calculation, for the 
momentum range of interest. The dates when timing scans were performed are also 
known:

\begin{itemize}
 \item 10/4/2010: coarse timing scan. It took place during run 152777, between 
 the first and the second points in the plots.
\item 15/4/2011. It took place during run 179710, one of the runs selected for 
the $\dedx$ study.
\item 8/9/2011. During run 188921.
\item 4/4/2012. Run 200804.
\item 1/5/2012. It took place during run 202660.
\item 15/11/2012. Run 214553.
\end{itemize}

The first scan resulted in a rather large adjustment of the timing. As a 
result, it is not surprising to see the plot in 
Figure~\ref{fig:7-protonmean_with_timingscans}, which shows a large decrease of
the mean time-bin size for the second run of the period. Note that the left Y
axis scale corresponds to the calculated SCT $\dedx$ while the right Y axis 
scale belongs to the mean time-bin size.

This figure shows the mean time-bin size,
together with the mean $\dedx$ of the proton for $(500,550)\MeV/c$ and also 
the dates when timing scans were performed.

In addition, the plot also shows that there is some correlation between the 
mean time-bin size and the calculated $\dedx$. It might be possible to exploit 
this correlation, shown in Figure~\ref{fig:7-timebin_dedx_corr}.

\begin{figure}[!htb]
 \begin{center}
  \includegraphics[scale=0.5]{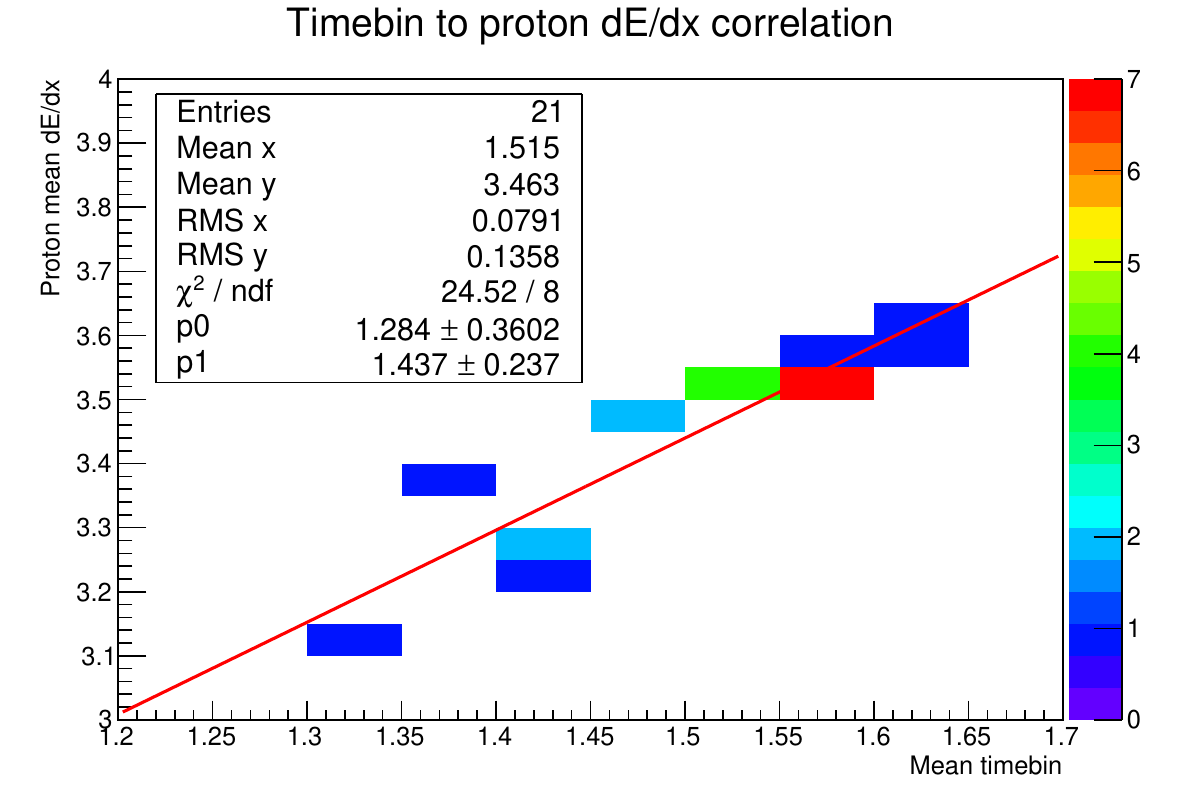}
 \caption{Correlation between the mean time-bin size and the calculated mean 
 proton $\dedx$.}\label{fig:7-timebin_dedx_corr}
 \end{center}
\end{figure}

Although the statistics are very low, the 
correlation is quite clear. A linear fit of the 2010 runs, the first five 
points, gives the following relation:

\begin{equation*}
 p_0 = 1.284 \pm 0.3602 \qquad p_1 = 1.437 \pm 0.237
\end{equation*}

\begin{figure}[!hbt]
 \begin{center}
  \includegraphics[scale=0.55]{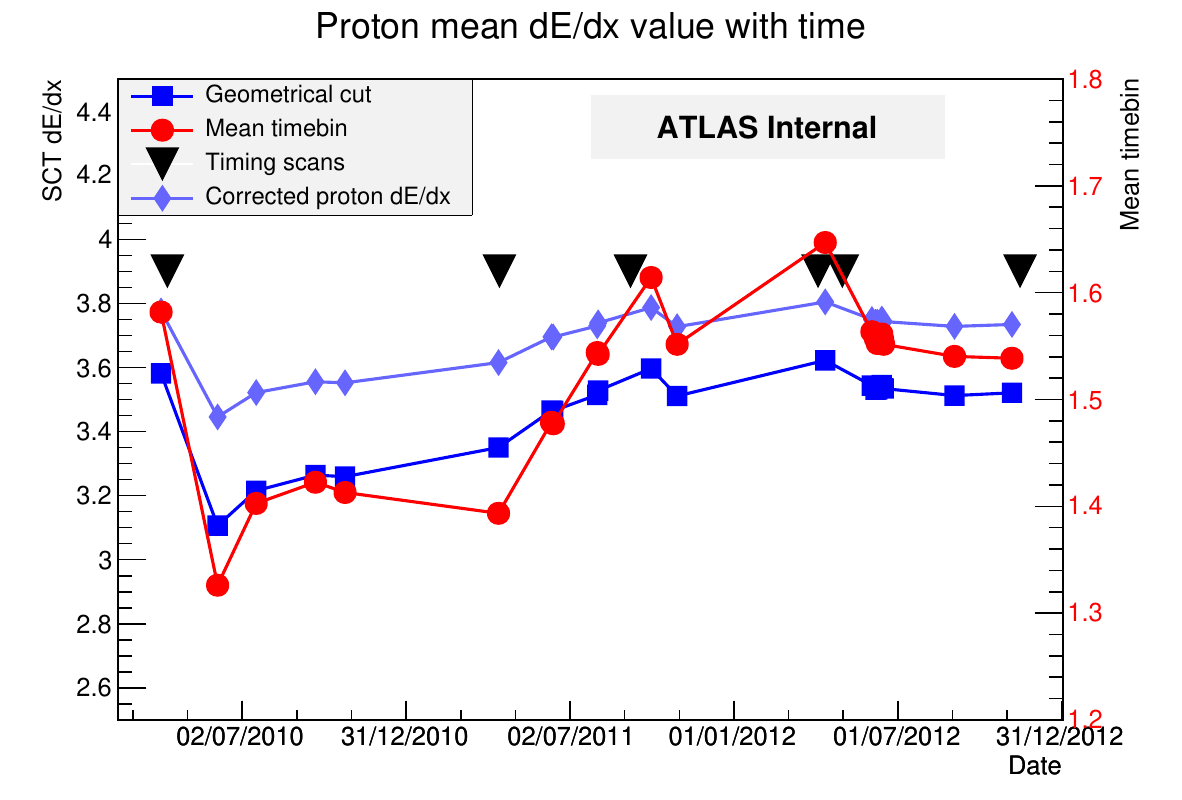} 
  \caption{Evolution of the proton mean  $\dedx$ with time, including a 
  correction for the 2010 runs.}\label{fig:7-protonmean_with_correction}
 \end{center}
\end{figure}

It is possible to use this relationship between the mean time-bin and the 
calculated $\dedx$ to correct for the variations.
Figure~\ref{fig:7-protonmean_with_correction} contains the final correction to 
the $\dedx$ evolution with time.
As before, note that the left Y
axis scale corresponds to the calculated SCT $\dedx$, both corrected and not 
corrected versions, while the right Y axis 
scale belongs to the mean time-bin size.

The corrected proton $\dedx$ in this plot shows a small effect from the 
time-bin size variation over time, although it is not possible to completely 
eliminate it. In
addition, considering the 2011-2012 period is a bit more stable over time, one
can say that there is not just yet any evidence of radiation damage that would 
show up as a reduced $\dedx$ due to a lower charge collection. 

The overall variation of the calculated $\dedx$ is less than $17\%$ in the 
uncorrected numbers, whereas it is less than $10\%$ in the corrected version. 
If the first run is ignored, these variations are even lower. The 
trend since the end of the 2011 run shows a very small variation of around
$2\%$.

\section{Discussion}

This chapter described the results obtained with a Time-over-Threshold method 
for calculating the deposited charge by particles traversing the SCT, thus
reconstructing the energy loss, ${\rm d}E/{\rm d}x$, based on the time-bin
information and the number of affected strips, both available in the SCT 
Ntuples. The results obtained have been compared to the more precise values
provided by the Pixel detector and the performance using efficiency vs mistag 
rate of particle tagging was analysed.

The lack of Monte Carlo simulations for SCT Ntuples that include both truth
information and charge simulation obliged the study to extract ``truth 
information'' from the $\dedx$ calculation performed in the Pixel detector.
A possible follow-up for this study is the generation of Monte Carlo 
simulations with all these characteristics in order to have a proper 
characterization of the SCT $\dedx$ calculation.

The results are encouraging although the SCT read-out has less resolution than 
the Pixel one. It is also important to note that the SCT was never designed for 
this purpose, so this result was not expected from the beginning. The ability
of this method to separate protons from light particles, such as pions, is
shown in the efficiency greater than $90\%$ for low momentum, with a mistag
rate below $10\%$, presented in section~\ref{sec:7-efficiency}.

A possible use of this kind of calculation is to perform long term radiation
monitoring by analysing the ${\rm d}E/{\rm d}x$ variations over time, with
increasing luminosity. 

It was possible to extract the energy loss despite the fact that the SCT was
not designed with that feature in mind. The digitization of the particle hits
causes a great loss of information, what reduces the resolution, but the 
available information allows for some ${\rm d}E/{\rm d}x$ calculation.

It is necessary to point out that, in the future, this calculation will not
be available for day to day data taking. The $13$ and $14~{\rm TeV}$ 
centre-of-mass energy collisions after the first Long Shut-down will 
have a $25~{\rm ns}$ bunch spacing and the SCT will run in 01X mode, with an
additional compression that causes the complete loss of the individual 
time-bins information.

If this analysis is to be used in the future, special runs will need to be 
used, with the appropriate timing and compression settings so the time-bin
information can be recovered from the data.

\chapter{Conclusions and Future Work}

This thesis presents two main topics of research. The first one is the
ATLAS Inner Tracker Phase-II Upgrade, focused on the silicon strips tracker. 
The study shows the results of tests performed on prototypes that have been
built to test the electronics, modules and structures defined for the strips
tracker that will replace the ATLAS Semiconductor Tracker (SCT) and Transition
Radiation Tracker (TRT) sub-systems with an all-silicon tracker when the High 
Luminosity LHC is deployed.

The results shown in chapter 5 prove that the structures to be built for the 
upgrade are feasible. The noise results of a full-size stave with twelve 
modules, and based on the ABCn25 chip, are as good as the noise results for 
the individual modules. The noise pick-up is not affected by the read-out of 
the modules on a $1.2$~metre long bus tape.

In addition, concepts such as high voltage multiplexing and the VersatileLink 
have been tested together with stavelets in order to assess their influence on
the stavelet performance. These devices cause no interference with the modules
read-out and they can be used safely with the strips tracker devices.

Furthermore, a single module has been read-out at the expected working 
temperature setting of the upgraded tracker, $-20\degC$, using a commercial
chiller unit. The tracker will be cooled using a recirculating ${\rm CO_2}$ 
cooling system. Blow-off systems are being tested at CERN and RAL to perform
initial tests with staves.

The future work regarding this topic will consist in testing the next staves
that are being assembled. A serially powered stave with up to 12 modules on one
side will be tested and the results compared to the DC-DC powered stave. The
best performing stave, or both, will be populated on the second side, building
a double sided object for full tests.

Also, the staves will be tested with ${\rm CO_2}$ cooling systems to check 
their behaviour at very low temperatures.

In the near future, the ABC130 chip will be used in the strips upgrade modules,
building staves with up to 13 modules on each side. The tests will be similar 
to those presented here.

The second topic of this thesis is an analysis of the energy loss measurement
using the ATLAS Semiconductor Tracker information. While the SCT was not 
designed to perform this kind of task, chapter 6 shows a method to extract a
way of calculating the $\dedx$, while with a low resolution. It allows some
particle discrimination, compared to the pixel detector, with over $90\%$
proton tagging efficiency. The availability of the time-bin information in
SCT Ntuples enables us to perform this calculation.

This method of calculating $\dedx$ in the SCT has been studied as a means of 
tracking radiation damage in the SCT by measuring the average proton $\dedx$ 
over time, which is expected to decrease as the silicon is damaged by the 
radiation at high luminosities. The timing settings of the SCT have to be
considered in order to correct the deviations that cause inaccurate $\dedx$
calculations.

Further work on this topic is expected as well. The tracking of the proton
mean $\dedx$ can only be performed with the right read-out settings after
the first Long Shut-down of the LHC. After 2015, the SCT will run in a read-out
mode that will make the time-bin information unavailable unless special runs
are used. We expect to be able to do this study in the future to try check 
if this method is appropriate for radiation damage tracking.

\chapter{Resumen}

\section{Introducción}

Esta tesis está centrada en la Fase 2 de la actualización del experimento
ATLAS en el LHC, con énfasis en el futuro detector de trazas de bandas
de silicio. Esta actua-lización supondrá el reemplazo completo del detector
de trazas, para cumplir con los requisitos de granularidad y resistencia
a la radiación necesarios tras la actualización del acelerador. 

Este nuevo acelerador, llamado LHC de alta luminosidad 
(\textit{High Luminosity LHC},
HL-LHC), colisionará protones a una energía en el centro de masas de 
$14\TeV$ y generará alrededor de 140 interacciones por cada cruce de haces
de protones en los detectores.

Además del trabajo relacionado con la actualización de ATLAS, esta tesis 
también presenta un estudio sobre las prestaciones del actual detector de 
bandas de silicio de ATLAS.

Este resumen recoge algunos de los resultados más relevantes,
para una vista más completa puede consultarse la versión en inglés.

\subsection{El experimento ATLAS}

\begin{figure}[!htbp]
\begin{center}
 \includegraphics[scale=0.5]{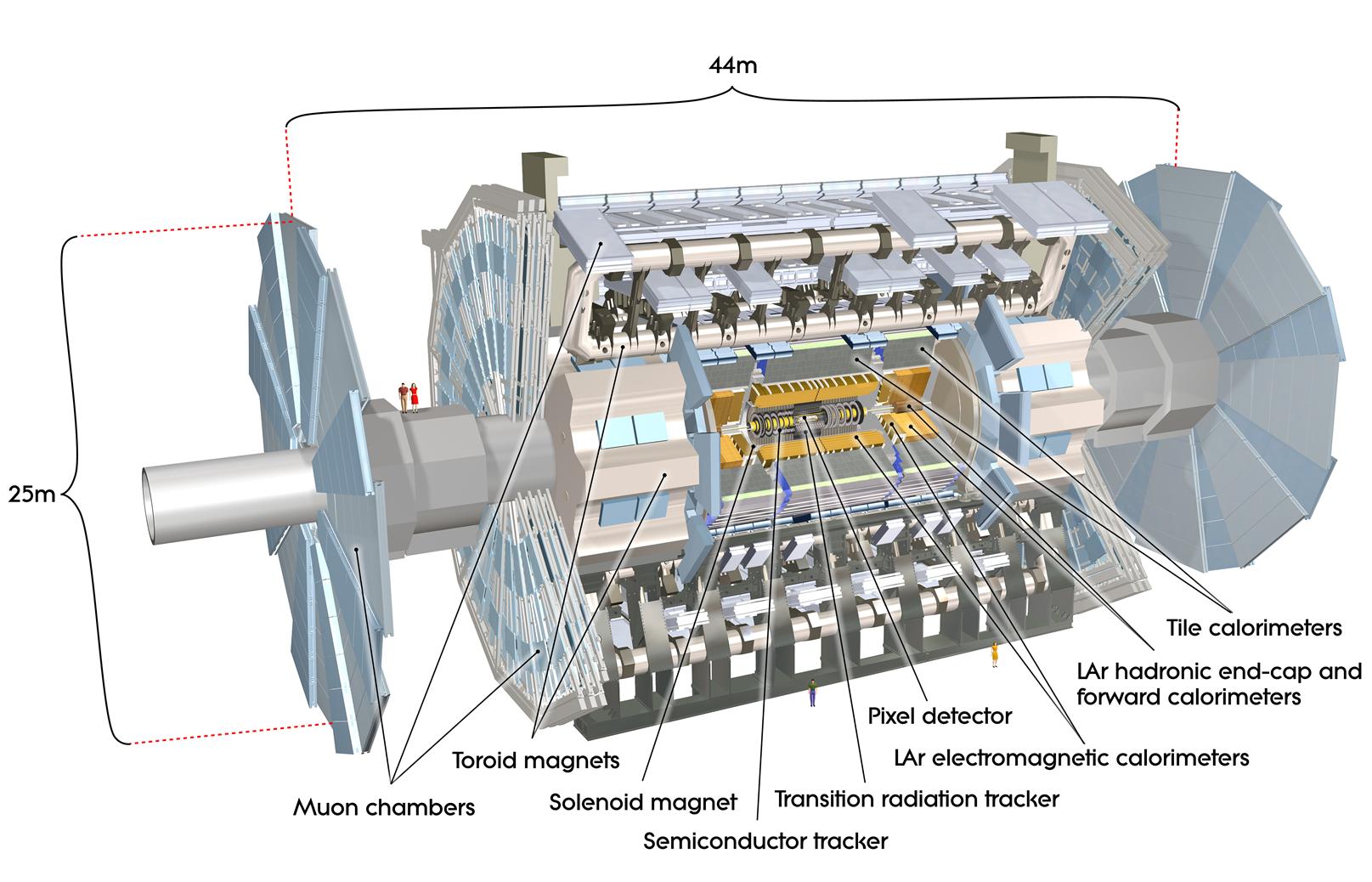}
 \caption{Vista del experimento ATLAS.}\label{fig:8-atlas}
\end{center}
\end{figure}

El experimento ATLAS (\textit{A Toroidal LHC Apparatus}), mostrado en la 
Figura~\ref{fig:8-atlas}, es uno de 
los dos experimentos de propósito general del Gran Colisionador de Hadrones 
(\textit{Large Hadron Collider}, LHC), situado en el ``Punto de 
Interacción 1''. Está formado por múltiples capas, que se 
encargan de detectar diferentes tipos de partículas.

El detector de trazas mide el momento de las partículas cargadas que se 
originan en las colisiones. Se encuentra en la parte más interna del detector, 
rodeando el punto de colisión. 

Está diseñado para medir el momento de las partículas cargadas con gran
resolución y calcular vértices 
primarios y secundarios. Un imán solenoide que genera un campo magnético
de $2$ tesla de intensidad lo rodea para deflectar la trayectoria 
de las partículas.

Tres sub-sistemas componen el detector de trazas: un detector
de píxeles, un detector de bandas de silicio (\textit{Semi-Conductor 
Tracker}, SCT) y un detector de radiación de transición (\textit{Transition 
Radiation Tracker}, TRT). 

Los dos sub-detectores más internos, el de píxeles y el de bandas de 
silicio, están construidos con detectores de silicio. Las partículas 
cargadas que pasan a su través depositan carga eléctrica, la cual puede ser
medida.

\subsection{El detector de trazas de semiconductor en ATLAS}

El detector de trazas de silicio (\textit{Semiconductor Tracker}, SCT)
del experimento ATLAS está formado por 4088 módulos de doble cara, repartidos
en cuatro capas cilíndricas (barril) y dos tapas (endcap) con nueve discos 
cada una.

Los módulos son detectores de bandas silicio de doble cara, con 768 bandas 
activas por cada lado, de $80\mum$ de ancho cada banda. Los sensores son de 
sustrato de silicio tipo n con bandas tipo p. Doce chips leen las 
bandas de cada módulo (seis chips por cada lado), montados sobre un circuito 
híbrido. Los sensores de cada lado están rotados $40\mrad$ para proporcionar
la posición en dos dimensiones del paso de las partículas. 
Hay un total de $6.3$ millones de canales en el detector de bandas de silicio.

Las partículas cargadas que atraviesan el detector depositan carga en el
silicio, la cual es recogida por los chips, amplificada y comparada con un 
umbral. Cuando la carga supera este umbral, se envía a los sistemas que 
procesan los datos fuera del detector. Los chips utilizados en el SCT se 
denominan ABCD3TA~\cite{Campabadal:994402}.

\section{Detector de bandas de silicio en ATLAS Fase 2}

\subsection{Fase 2 del LHC: HL-LHC}

El LHC fue diseñado para ser capaz de entregar una luminosidad integrada de
$400\ifb$, a lo largo de diez años de operación. Ciertos componentes del 
acelerador deberán ser reemplazados tras acumular esa cantidad de 
radiacion~\cite{Rossi:1471000}.

Partes de los experimentos también tendrán que sustituirse por otros debido
a la radiación. Los detectores de trazas de ATLAS y CMS deberán extraerse
para instalar otros nuevos, ya que el silicio de los detectores de píxeles
no podrá soportar más radiación.

El LHC de alta luminosidad (HL-LHC) consiste en un nuevo diseño
del acelerador y de parte de los experimentos. Con este nuevo diseño, se 
pretende alcanzar una luminosidad integrada tras otros diez años de operaciones
entre $2500$ y $3000\ifb$, una cantidad de datos que abre nuevos campos de 
exploración para la física, tanto del Modelo Estándar como más allá del Modelo 
Estándar.

El HL-LHC mantendría la misma energía por haz de protones, pero multiplicaría
por cinco la luminosidad instantánea máxima y casi por seis la luminosidad 
integrada anual. Un mayor número de protones por paquete y un mayor enfoque del
haz serán los mayores cambios que permitan la mayor luminosidad. El incremento 
del enfoque aumentará la probabilidad de colisión por cruce de paquetes, 
pasando de un \textit{pile-up} de $27$ en el diseño del LHC actual a un 
\textit{pile-up} medio de $140$. 

Debido al mayor \textit{pile-up}, los detectores de trazas actuales no podrán
ser utilizados sin una pérdida sustancial de su rendimiento. En ATLAS, el
detector TRT alcanzaría un $100\%$ de ocupación y los detectores de píxeles y el
SCT no podrían resolver trazas cercanas. Se hace necesaria la sustitución del 
detector de trazas por un nuevo diseño con mayor granularidad y se implementará
por completo en silicio.

\subsection{Descripción del detector de bandas}

El requisito para el nuevo detector de bandas de silicio en la Fase 2 de ATLAS
es que soporte una dosis equivalente de neutrones de $1\MeV$ de 
$1.2\times 10^{15}\cm^{-2}$ en las capas más internas, y de 
$5\times 10^{14}\cm^{-2}$ en las más externas. 

Para conseguir sensores que 
puedan soportar la acumulación de esta dosis, se ha cambiado el tipo de sensor
por un sustrato p con implantes tipo n. En este tipo de sustrato, no se produce
inversión de tipo dado que ya es de tipo p.

Las dimensiones de los sensores para los módulos del barril son 
$97.54\times 97.54\mm^2$, con cuatro filas de 1280 bandas cada una en los 
sensores con bandas cortas. En las capas más exteriores, con menor ocupación,
se utilizarán sensores con bandas el doble de largas.

\begin{figure}[!htbp]
\begin{center}
\includegraphics[scale=0.19]{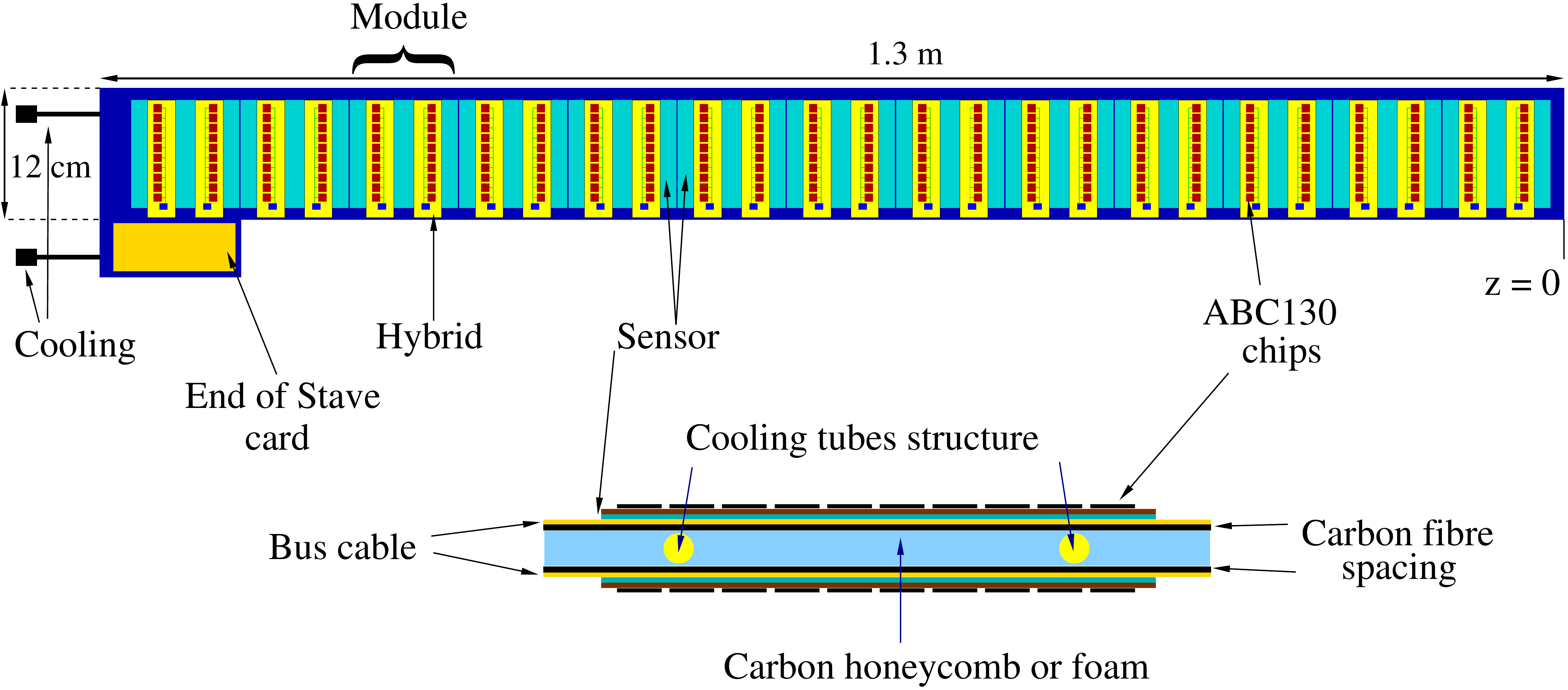}
\caption{Esquema de un stave de barril con 13 
módulos.}\label{fig:8-Stave}
\end{center}
\end{figure}

Los módulos del barril se integrarán en estructuras multi-módulo denominadas
\textit{staves}, conteniendo 26 módulos repartidos en dos lados, para 
proporcionar resolución en dos dimensiones a través de un estéreo-ángulo de
$40\mrad$. Se han producido prototipos de los
staves con un menor número de módulos, llamados \textit{stavelets}, a modo de 
prueba de concepto de los diseños propuestos. Un esquema del stave se muestra
en la Figura~\ref{fig:8-Stave}, con trece módulos por cada lado.

En el caso de los endcaps, los sensores tendrán forma trapezoidal. La 
estructura circular de las tapas requiere que las bandas formen ángulo entre 
ellas, de forma que apunten hacia el centro del punto de interacción.

\subsection{Sistemas electrónicos}

La lectura de los módulos se realiza a través de una tarjeta que en los 
primeros prototipos se implementa con búferes LVDS (\textit{Low Voltage
Differential Signal}) en una placa denominada EoS (\textit{End of Stave}
o \textit{End of Substructure}).

Se han diseñado dos chips de lectura para las bandas, ambos binarios. Un primer
chip con 128 canales, ABCn25, en un proceso de $250\nm$, ha sido utilizado para
construir los módulos que se han probado en el contexto de esta tesis. En el 
futuro, y previsiblemente para la construcción final del detector de trazas
actualizado, se utilizarán chips de 256 canales en un proceso de $130\nm$. Este
chip se denomina ABC130 y su diseño se finalizó durante 
2013~\cite{Affolder:1516555}.

\begin{figure}[!hbt]
 \begin{center}
 \subfigure[Alimentación serie de módulos (``cadena de 
  híbridos'').]{
  \includegraphics[scale=0.6]{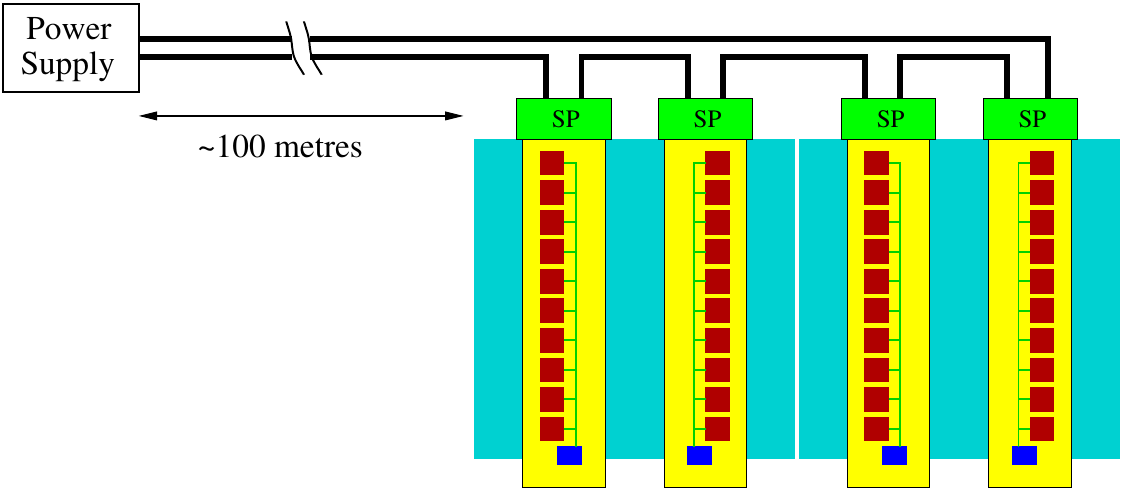}
  \label{fig:8-PoweringModules_SP}
  }
  \subfigure[Alimentación de módulos con conversión DC-DC.]{
  \includegraphics[scale=0.6]{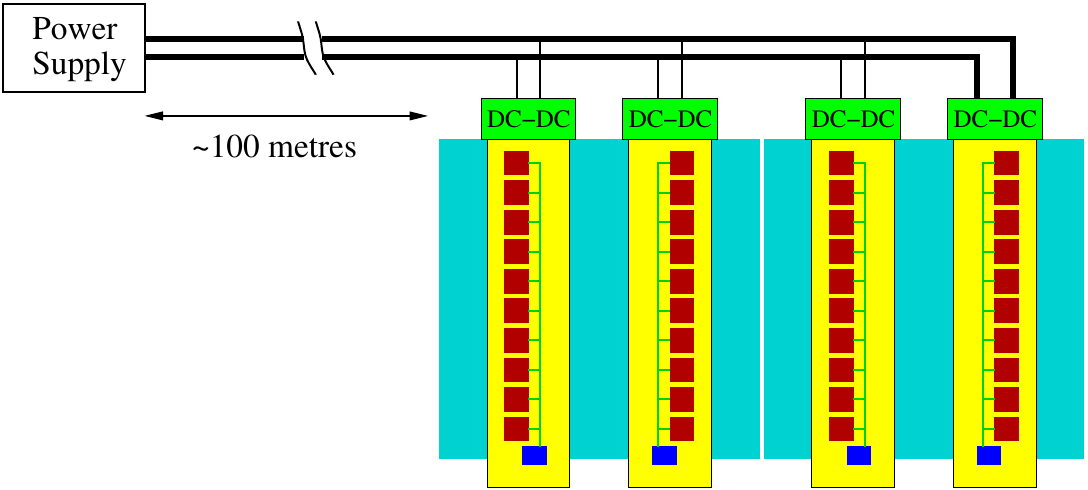}
  \label{fig:8-PoweringModules_DCDC}
  }
  \caption{Esquemas alternativos de alimentación de los 
  módulos.}\label{fig:8-PoweringModules}
 \end{center}
\end{figure}

El consumo de potencia del nuevo detector de trazas será unas 4-5 veces 
superior al actual SCT. Los módulos del SCT se alimentan con cables 
independientes por cada módulo, con fuentes de alimentación que entregan
el voltaje requerido por los chips. Las pérdidas producidas en los cables
debido a la distancia ($100\m$) entre las fuentes de alimentación y el detector
hacen que se pierda una cantidad no despreciable de energía. Esta falta
de eficiencia provoca que se necesite una cantidad de material que afecta al
rendimiento del experimento y eleva los costes.

Para alimentar el detector de trazas actualizado se están investigando sistemas
de alimentación alternativos, como conversión DC-DC (continua-continua) o 
alimentación serie. Parte del trabajo de esta tesis consiste en hacer pruebas
de rendimiento de elementos funcionando con ambos sistemas de alimentación.

Un ejemplo de alimentación serie se muestra en la 
Figura~\ref{fig:8-PoweringModules_SP}, mientras que el equivalente en 
conversión DC-DC se muestra en la Figura~\ref{fig:8-PoweringModules_DCDC}.

Ambas alternativas se basan en reducir la corriente que entregan las fuentes
de alimentación, reduciéndose las pérdidas por resistencia de los cables. En
la conversión DC-DC la fuente entrega una tensión superior a la necesaria para
alimentar los chips, que se reduce a la tensión adecuada en una ubicación 
próxima a los chips.

\subsection{Prestaciones del módulo con alimentación serie}\label{sec:8-modulo}

El estudio de sistemas para la Fase 2 de ATLAS comienza con la construcción
de módulos con un único sensor de bandas de silicio, al cual se pegan dos
híbridos con 20 chips cada uno. 

El módulo medido en el contexto de esta tesis fue fabricado en la Universidad
de Liverpool, con un sensor de grado B (FZ2).
El módulo utiliza alimentación serie, en configuración de ``cadena de 
híbridos'', que implica diferentes niveles de tensión continua para ambos 
híbridos.

En la configuración de ``cadena de módulos'', los dos híbridos de un módulo 
están a la misma tensión de referencia. Se utiliza en estructuras multi-módulo 
como stavelets o staves.

Los tipos de pruebas que se han llevado a cabo con los módulos individuales y
las estructuras multi-módulo son dos.

 El primero, para medir el ruido a la entrada del amplificador del chip, 
 en unidades de electrones (\textit{Equivalent Noise Charge}, ENC).
 Se define como el número de electrones que deben recogerse en el sensor 
 para crear una señal equivalente al ruido del sensor. Se calcula mediante 
 tres barridos del umbral en el comparador del chip inyectando tres cargas
 de magnitud conocida. Normalmente se utiliza un rango de cargas de entrada
 en el cual el amplificador se encuentra en su zona lineal.

El segundo se llama ruido de doble disparo (\textit{Double Trigger Noise}, 
					    DTN). Se utiliza
 para estudiar la susceptibilidad del sistema de adquisición de datos a las
 interferencias electrónicas durante la lectura. Se envían dos disparos 
 espaciados un número de periodos de reloj, variando el espaciado. El segundo
 evento registra la ocupación del módulo al comenzar la lectura del primer
 evento.

Se llevaron a cabo pruebas realizadas sobre el módulo disponible en CERN 
bajo diversas condiciones de funcionamiento, de temperatura y polarizacion
del sensor. Las pruebas se describen en los siguientes
apartados, junto a los resultados obtenidos.

\subsubsection{Variación de temperatura}

El objetivo de esta medida es extraer la dependencia del
ruido con la temperatura. Se tomaron medidas de ruido con diferentes 
temperaturas, realizando 20 repeticiones para cada punto de temperatura.

Las temperaturas de refrigerante con las que se realizó el estudio fueron
$T_{\rm chiller} = \left\lbrace 12, 18, 24\right\rbrace\degC$. Se midió la
temperatura de los híbridos en cada una de las pruebas, como se muestra en
la Tabla~\ref{tab:8-DNDT_temperatures}.

\begin{table}[!htb]
\begin{center}
 \begin{tabular}{c|cccccc}
 \toprule
  $T_{\rm chiller}[\degC]$ &  $12$ &   $18$ &   $24$ \\ \hline
  $T_{\rm hybrid}[\degC]$ & $25.9$ & $30.8$ & $35.7$  \\
  \bottomrule
 \end{tabular}
\caption{Temperaturas del refrigerante y los 
híbridos.}\label{tab:8-DNDT_temperatures}
\end{center}
\end{table}

\begin{figure}[!hb]
\begin{center}
\includegraphics[scale=0.65]{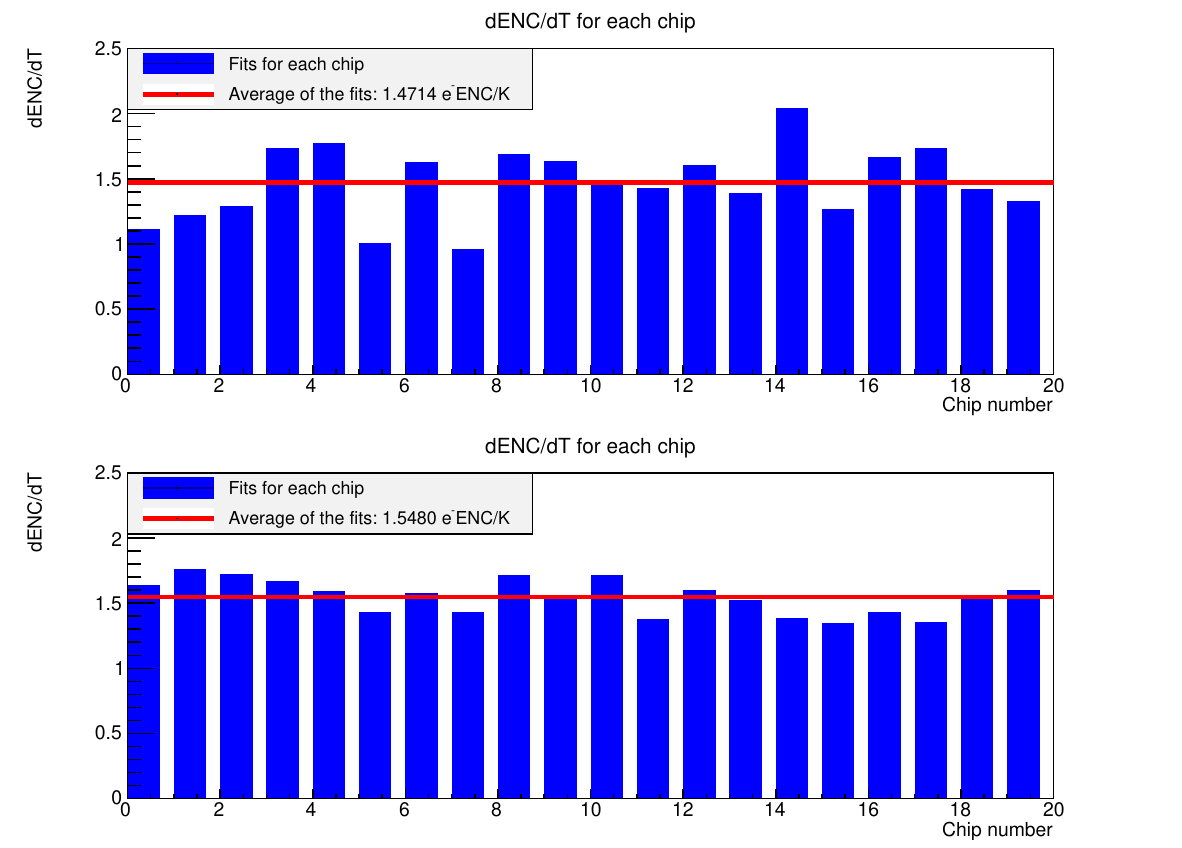}
\caption{Ajuste lineal de la variación del ruido con la temperatura para los
chips de los dos híbridos.}\label{fig:8-spmodule_DNDTfit}
\end{center}
\end{figure}

Para extraer la variación de ruido con la temperatura se tomó la media de las
20 medidas de ruido para cada valor de temperatura de los híbridos y se hizo un
ajuste lineal de los tres puntos. Los valores medios obtenidos para los dos
híbridos fueron los siguientes:

\begin{equation}
\overline{\Delta {\rm ENC}/\Delta T}_{\rm Hybrid~0} = 1.47\pm 0.28~e^{-}{\rm ENC/K} 
\end{equation}

\begin{equation}
\overline{\Delta {\rm ENC}/\Delta T}_{\rm Hybrid~1} = 1.55\pm 0.13~e^{-}{\rm ENC/K} 
\end{equation}

Los resultados de esta prueba se pueden ver en la
Figura~\ref{fig:8-spmodule_DNDTfit}, que muestra el ajuste lineal de cada chip
y la media para los dos híbridos.
La dependencia del ruido se estableció en $1.5\ENC/{\rm K}$.

\subsubsection{Variación de polarización del sensor}

La desertización de la unión p-n varía
en función de esta polarización, y con ella el ruido medido.

\begin{figure}[!htb]
\begin{center}
\includegraphics[scale=0.65]{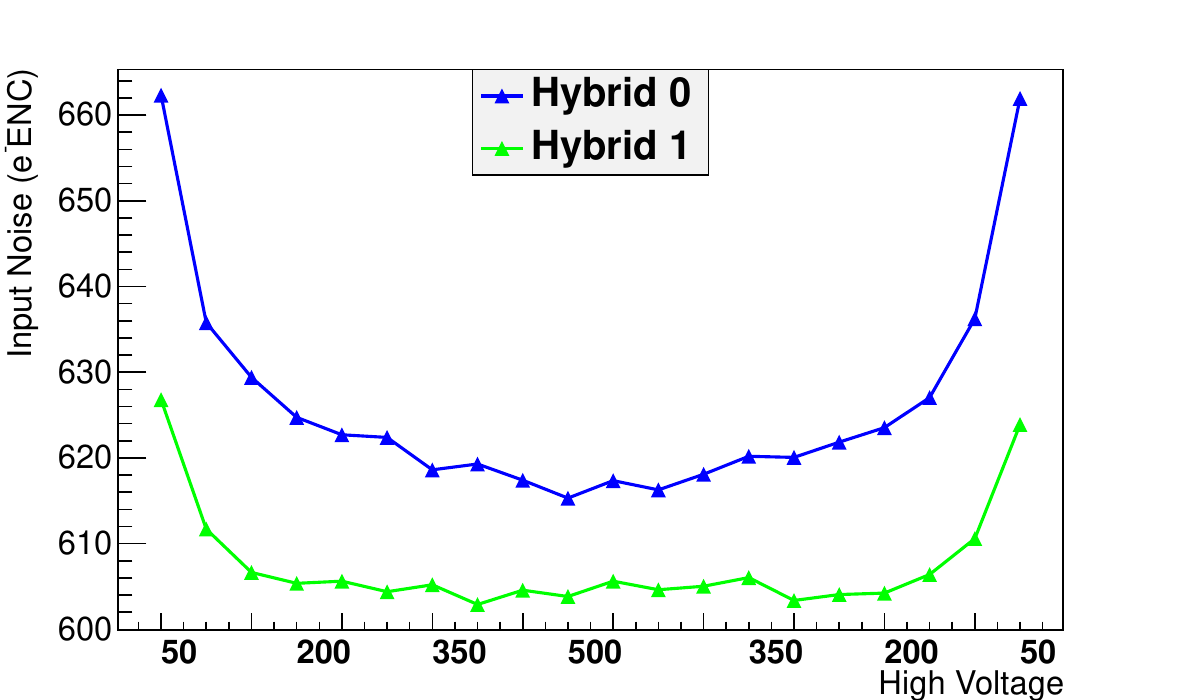}
\caption{Ruido en el módulo serie en función de la polarización 
del sensor.}\label{fig:8-spmodule_hvscan_updown_500V}
\end{center}
\end{figure}

Los resultados se muestran en la 
Figura~\ref{fig:8-spmodule_hvscan_updown_500V}, variación entre $50$ y $500\V$
en sentidos creciente y decreciente.

La razón para realizar el barrido en ambos sentidos es la existencia de
discrepancias en algunos módulos cuando se reduce el voltaje de polarización
del sensor. En ocasiones se observa un aumento del ruido por encima del 
original en los niveles de alto voltaje probados 
anteriormente~\cite{SvenDiplomaThesis}. Este efecto no se observa en este 
módulo. Sin embargo, en algunos módulos de los stavelets sí, como se podrá ver
en la sección siguiente.

\subsubsection{Pruebas a bajas temperaturas}

La electrónica que se incluye en el módulo
y en sus alrededores debe ser capaz de funcionar a bajas temperaturas. El 
detector funcionará con refrigeración por debajo de 
$-20\degC$, utilizando un sistema con ${\rm CO_2}$. 

\begin{figure}[!htb]
\begin{center}
\includegraphics[scale=0.65]{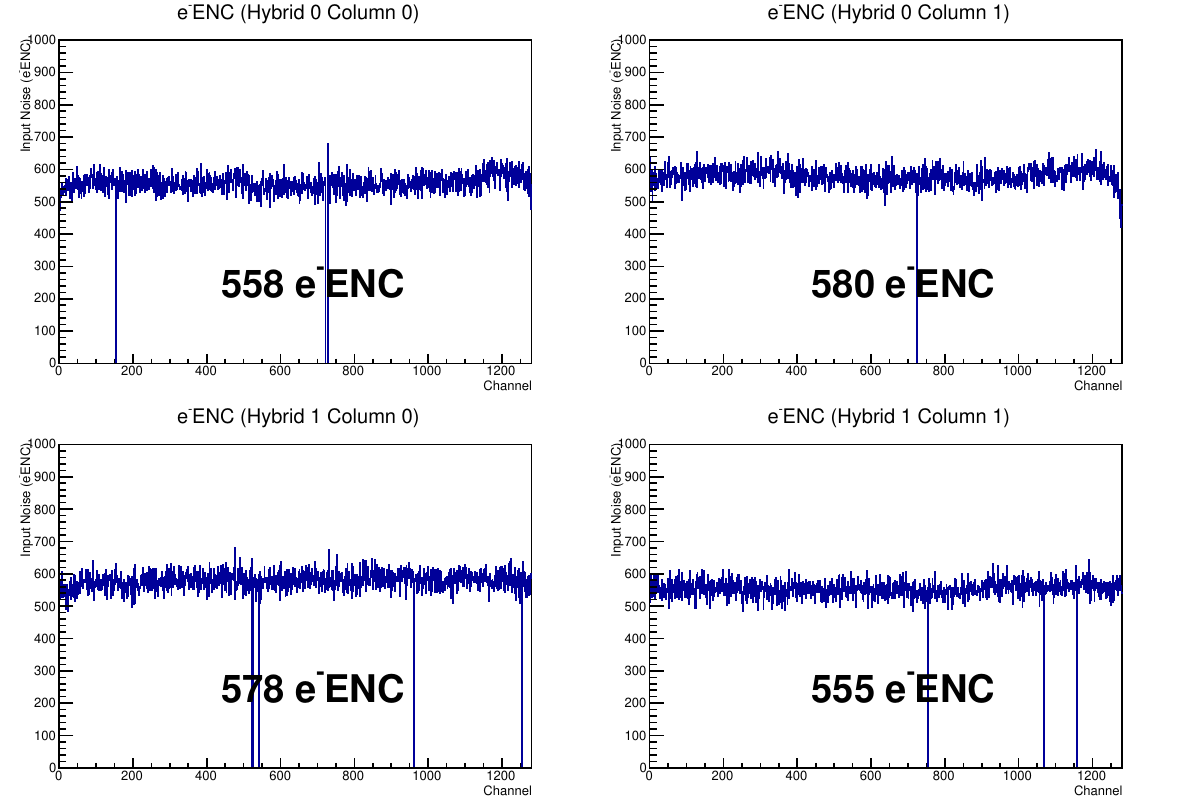}
\caption{Ruido del módulo serie con refrigerante a 
$-20\degC$.}\label{fig:8-SPmodule_m20deg_1fC}
\end{center}
\end{figure}

El módulo fue probado a temperaturas de $-6\degC$, $-12\degC$ y $-20\degC$.
En esta sección se muestra únicamente el resultado para el último valor, por 
tratarse del más relevante. El refrigerante con base de aceite empleado para
estas pruebas permite alcanzar temperaturas alrededor de $-40\degC$. 

Para 
evitar la formación de hielo en las tuberías y en las inmediaciones del módulo
se utilizó aire seco en el contenedor del módulo y revestimientos para las
tuberías.

El resultado de las pruebas a bajas temperaturas se muestra en la
Figura~\ref{fig:8-SPmodule_m20deg_1fC}. Se puede comprobar que el módulo es
capaz de operar a bajas temperaturas. La diferencia de ruido entre la 
temperatura de funcionamiento usual de $12\degC$ y esta prueba a $-20\degC$
es consistente con la variación de ruido con la temperatura presentada
anteriormente.

\subsection{Prestaciones de estructuras con múltiples módulos}

Se han evaluado tres estructuras multi-módulo diferentes en el contexto de 
esta tesis:

\begin{itemize}
 \item Un stavelet con cuatro módulos, utilizando alimentación
 con conversión DC-DC.
 \item Un stavelet con cuatro módulos, utilizando alimentación
 serie.
 \item Un stave, con doce módulos, utilizando alimentación con
 conversión DC-DC.
\end{itemize}

Las tres estructuras fueron ensambladas en el Rutherford Appleton Laboratory
(RAL), y las dos primeras se recibieron en CERN entre 2012 y 2013. La tercera
se probó en RAL durante su construcción. El objetivo de la 
construcción y evaluación de estructuras de cuatro módulos es determinar los
posibles problemas que pueden surgir al alimentar y leer múltiples módulos
dispuestos en un bus común.

\begin{figure}[!htb]
 \begin{center}
 \subfigure[Stavelet con conversión DC-DC.]{
   \includegraphics[scale=0.55]{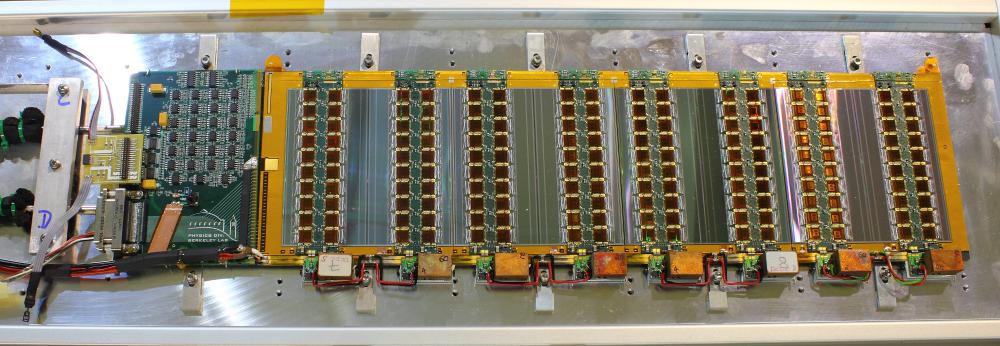}
   \label{fig:8-DCDCstavelet}
 }

 \subfigure[Stavelet con alimentación serie.]{
   \includegraphics[scale=0.55]{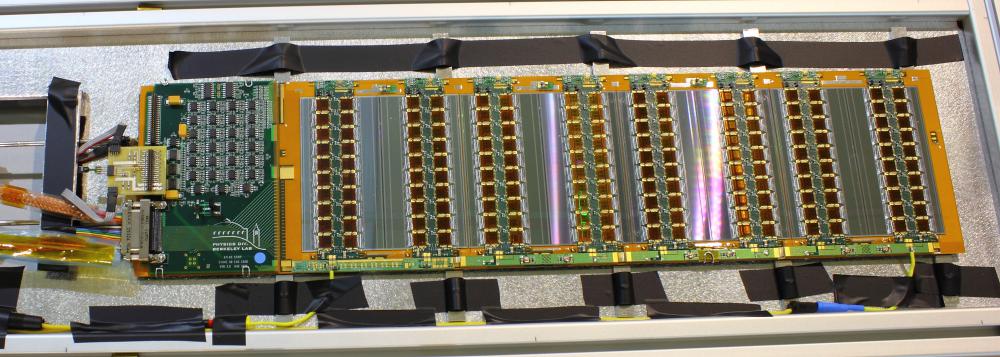}
   \label{fig:8-SPstavelet}
 }
  \caption{Los dos stavelets construidos en RAL y probados en 
  CERN.}\label{fig:8-stavelets}
 \end{center}
\end{figure}

La Figura~\ref{fig:8-DCDCstavelet} muestra una fotografía del stavelet de
cuatro módulos con conversión DC-DC. Cada conversor DC-DC es individual y 
provee la corriente necesaria para alimentar los veinte chips de un híbrido,
proporcionando hasta $5\A$ de corriente a $2.5\V$. La tensión de entrada del
conversor puede ser entre $7\V$ y $15\V$.

Por otro lado, 
el stavelet de cuatro módulos que emplea alimentación serie se muestra en la
Figura~\ref{fig:8-SPstavelet}. Se trata de un stavelet en ``cadena de 
módulos'', donde los dos híbridos de un módulo están referenciados a la misma
tensión. Cada módulo está $2.5\V$ por encima o por debajo de los vecinos. La
corriente de entrada para alimentar el stavelet debe estar entre $9$ y $10\A$
aproximadamente.

Las pruebas de prestaciones de los stavelets de cuatro módulos con conversión 
DC-DC y con alimentación serie fueron las siguientes:

\subsubsection{Barrido de polarización de los sensores}

Se varió el  valor de la tensión de polarización de los sensores,
para comprobar la evolución del ruido en uniones antes
y después de producirse la desertización. Para ambos stavelets se incrementó la
tensión y posteriormente se decrementó.

\begin{figure}[!htb]
\begin{center}
\includegraphics[scale=0.8,trim=7 0 20 0,clip=true]{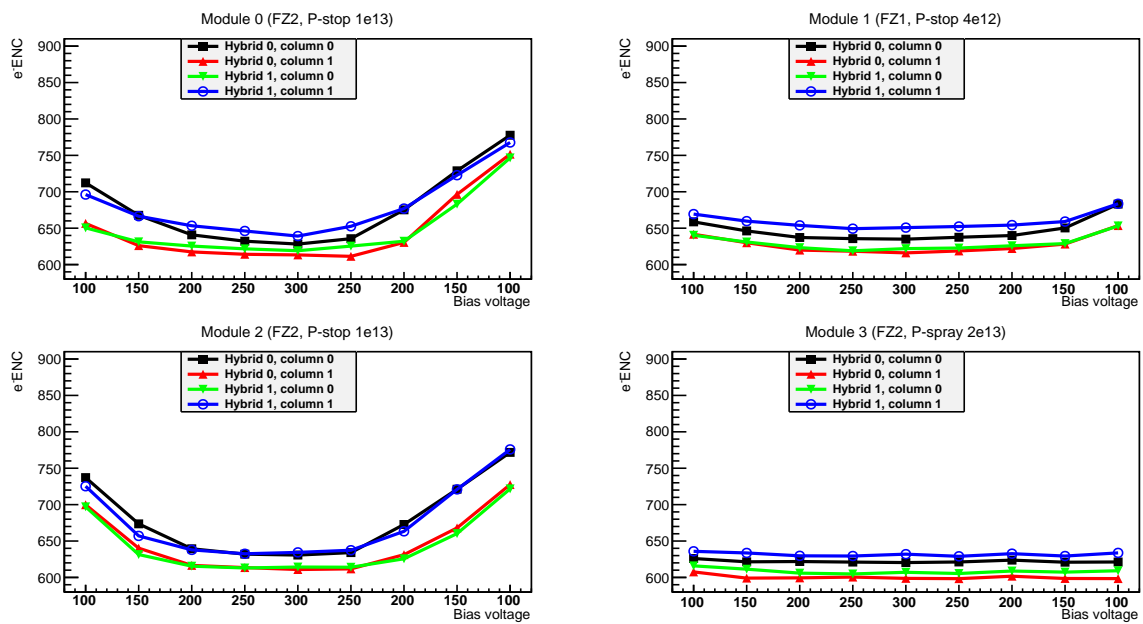}
\caption{Ruido del stavelet DC-DC en función del alto voltaje.}\label{fig:8-DCDC_ENC_vs_HV_UpDown}
\end{center}
\end{figure}

La Figura~\ref{fig:8-DCDC_ENC_vs_HV_UpDown}
muestra la evolución del ruido en el stavelet con conversión DC-DC. Primero
se incrementó de $75\V$ a $300\V$ y posteriormente se disminuyó el voltaje
hasta $75\V$, en pasos de $25\V$.

Algunos sensores presentan un comportamiento 
anómalo, incrementándose el ruido al volver a una polarización inferior desde
una superior. Esto ocurre normalmente con los sensores que utilizan 
aislamiento de tipo P-stop con concentración elevada ($10^{13}~\cm^{-3}$).

\begin{figure}[!hbt]
\begin{center}
\includegraphics[scale=0.8,trim=7 0 20 0,clip=true]{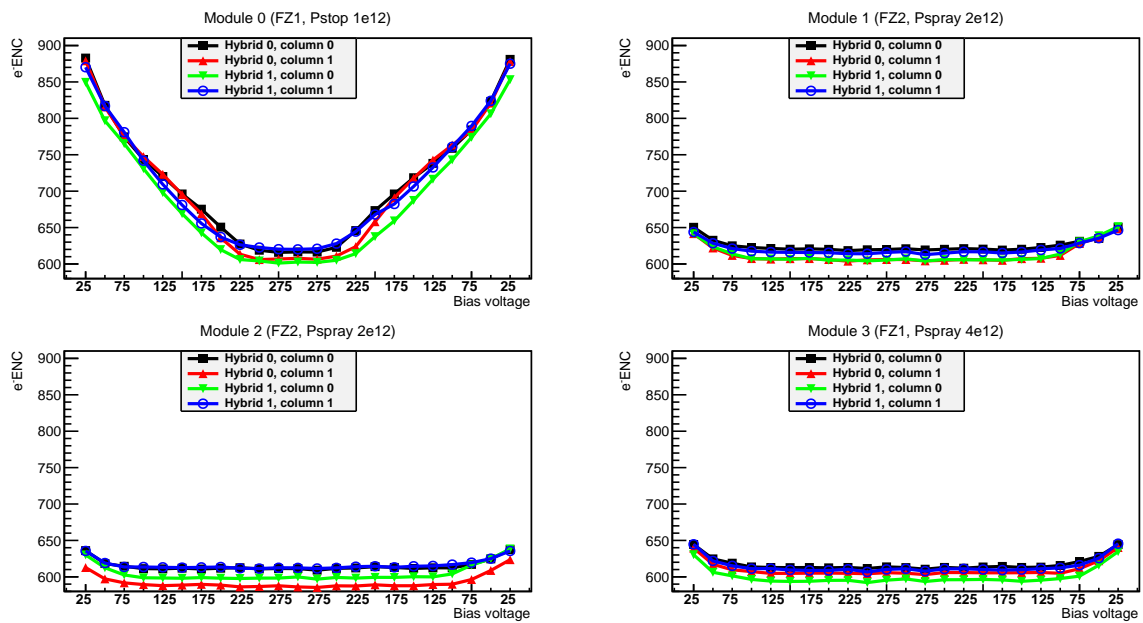}
\caption{Ruido del stavelet serie en función del alto voltaje.}\label{fig:8-SPstavelet_ENC_vs_HV_UpDown}
\end{center}
\end{figure}

 Otro efecto observado en función del tipo de aislamiento es un ruido inicial
 inferior en los que tienen aislamiento P-spray. El ruido permanece bastante 
 por debajo de $700\ENC$ a valores de voltaje muy inferiores al voltaje de
 desertización.

La Figura~\ref{fig:8-SPstavelet_ENC_vs_HV_UpDown}
muestra la misma evolución en el stavelet con alimentación serie. De la
misma manera que en el stavelet con conversión DC-DC, los módulos con sensores
que utilizan aislamiento P-spray parten de un nivel de ruido inferior a baja
polarización. 

Asimismo, el aumento de ruido al volver a un voltaje inferior
es inferior en aislamientos P-stop y P-spray con concentraciones más reducidas 
que en el caso de los sensores del stavelet con conversión DC-DC.

\subsubsection{Variación de la alimentación de los híbridos}

Los conversores DC-DC y los
circuitos de alimentación serie admiten un cierto rango de tensiones y 
corrientes de entrada. Los conversores DC-DC tienen su punto de máxima 
eficiencia en $10\V$, con un rango entre $7$ y $15\V$. Por otro lado, la 
tensión entregada por el circuito de alimentación serie varía en función de la 
corriente. Para el stavelet con ``cadena de módulos'' la corriente típica es
$9.5\A$. 

Se han hecho pruebas variando la tensión de entrada de los conversores DC-DC
entre $10$ y $14\V$. Cuando aumenta la tensión disminuye la corriente, por lo
que se producen menos pérdidas por la resistencia de los cables. Una reducción
de la tensión por debajo de los $9\V$ no es recomendable, debido a la alta
corriente cuando la tensión se reduce.

Sin embargo, es conocido que el punto óptimo de funcionamiento de estos
conversores es con una tensión de entrada de $10\V$, y que tensiones mayores
aumentan el ruido en los módulos.

\begin{figure}[!htb]
\begin{center}
\includegraphics[scale=0.7,trim=7 0 20 0,clip=true]{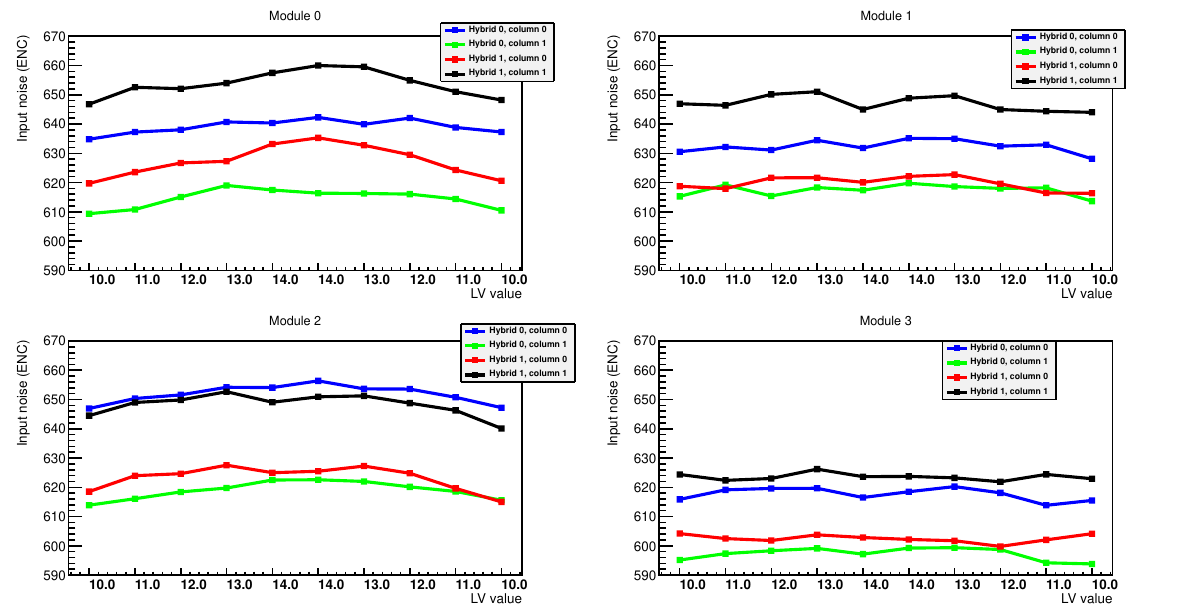}
\caption{Ruido del stavelet DC-DC en función de la tensión de entrada a los
conversores.}\label{fig:8-DCDC_LV_noise_HVfilter}
\end{center}
\end{figure}

\begin{figure}[!htb]
\begin{center}
\includegraphics[scale=0.7,trim=7 0 20 0,clip=true]{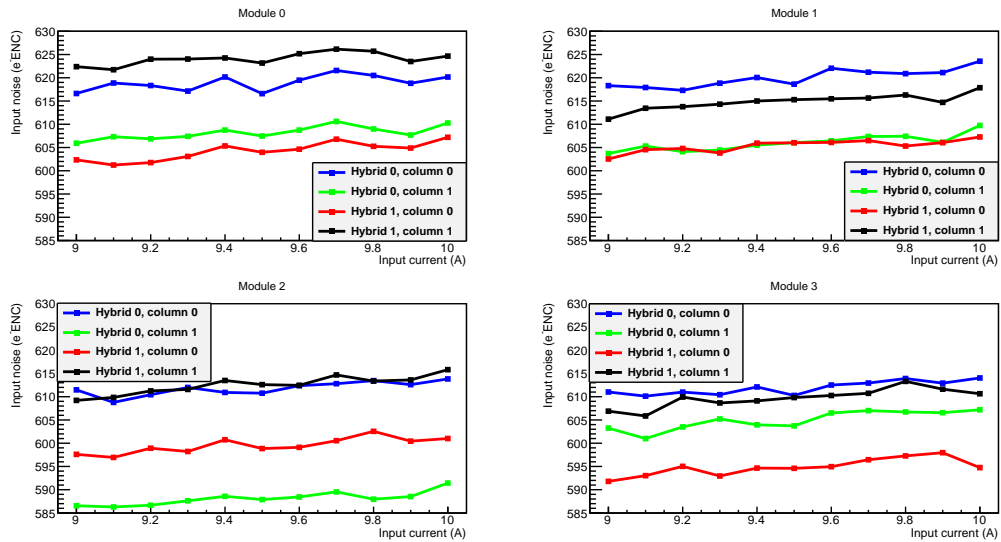}
\caption{Ruido del stavelet con alimentación serie en función de la corriente 
de entrada.}\label{fig:8-SP_LV_noise}
\end{center}
\end{figure}

Los resultados de estas pruebas en el stavelet con conversión DC-DC se muestran 
en la 
Figura~\ref{fig:8-DCDC_LV_noise_HVfilter}. El ruido varía levemente al subir
la tensión hacia $14\V$ en dos de los módulos, mientras que los otros dos 
módulos parecen ser inmunes al aumento de ruido de los conversores al elevar
su tensión de entrada.

En el caso del stavelet con alimentación serie, cuando aumenta la corriente de
entrada también lo hace la tensión. Esto provoca, en última instancia, un 
aumento de la temperatura de los híbridos. Como se ha indicado en la 
sección~\ref{sec:8-modulo}, esto aumenta el ruido de forma proporcional.

La Figura~\ref{fig:8-SP_LV_noise} muestra los resultados de ruido frente a la
corriente de entrada. El aumento de ruido es el esperado en los cuatro módulos
del stavelet. Las variaciones observadas son consistentes con las que se 
producen al repetir una misma medida bajo las mismas condiciones.

\subsubsection{Multiplexión de líneas de alto voltaje}

Con el objetivo de reducir el 
volumen de cables, se ha diseñado un sistema que multiplexa varias líneas de
polarización de los sensores en un único cable, y que se dividen cerca de los
módulos.

\begin{figure}[!htbp]
\begin{center}
\includegraphics[scale=0.8,trim=7 0 20 0,clip=true]{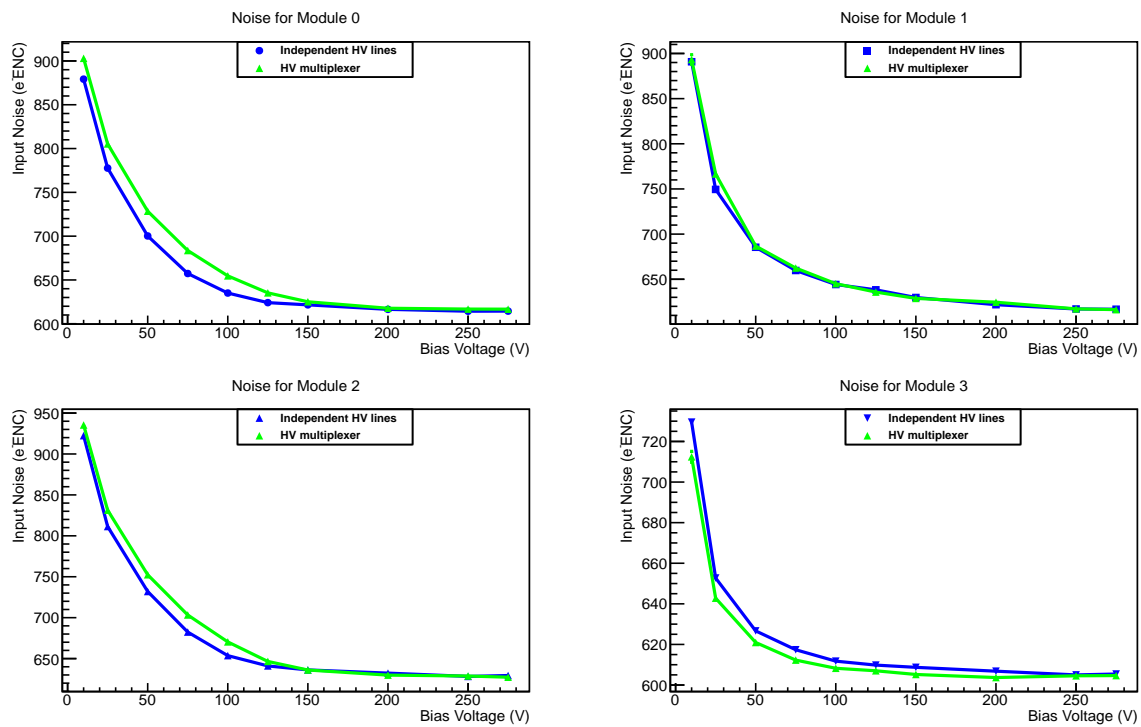}
\caption{Comparación de la dependencia del ruido en el stavelet DC-DC
en función del alto voltaje, 
con multiplexor y con líneas independientes.}\label{fig:8-HVmux_DCDC_12deg}
\end{center}
\end{figure}

\begin{figure}[!htbp]
\begin{center}
\includegraphics[scale=0.8,trim=7 0 20 0,clip=true]{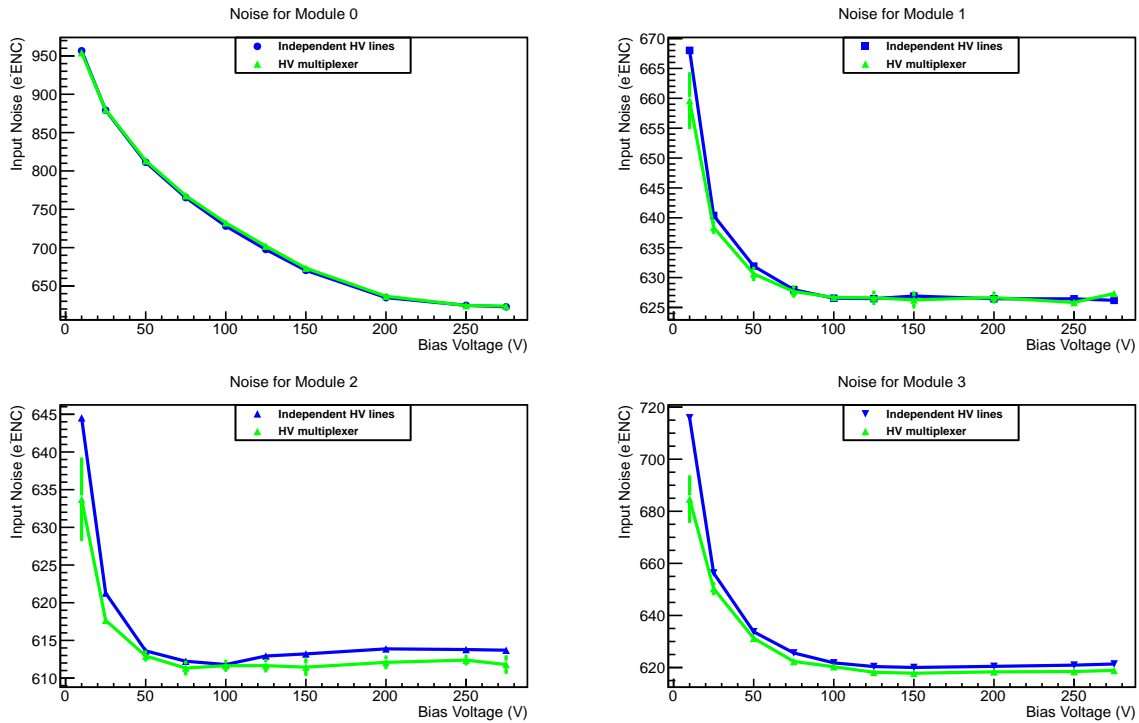}
\caption{Comparación de la dependencia del ruido en el stavelet con 
alimentación serie en función del alto voltaje, 
con multiplexor y con líneas independientes.}\label{fig:8-HVmux_SP}
\end{center}
\end{figure}

En un stave sería posible agregar las 13 líneas de un lado, lo que permitiría
ahorrar
12 pares de cables por cada stave. El objetivo de estas pruebas es verificar 
que 
la multiplexión de líneas de alto voltaje no tiene un efecto negativo en las
prestaciones de los stavelets.

Para las pruebas se ha empleado un multiplexor construido en BNL utilizando
transistores de carburo de silicio (SiC) que pueden conmutar hasta 
$500\V$~\cite{ATLAS:1502664}. Este multiplexor soporta la desconexión 
individual de las líneas de alto voltaje, sin embargo no es posible hacer
medidas de corriente de fugas individuales. 

Para tener una funcionalidad 
completa con multiplexores de alto voltaje, es necesario que los conmutadores
sean suficientemente resistentes a la radiación, capaces de conmutar hasta
$600\V$ y que permitan medir la corriente de fugas de los sensores de forma 
individual.

La Figura~\ref{fig:8-HVmux_DCDC_12deg} muestra los resultados obtenidos con el 
stavelet DC-DC, comparando el ruido con y sin multiplexión. Los resultados del
stavelet con alimentación serie se muestran en la 
Figura~\ref{fig:8-HVmux_SP}. En ambos casos, se observa una variación 
despreciable del ruido en los módulos del stavelet. En el stavelet con 
conversión DC-DC se produce una mezcla entre mayor ruido con el multiplexor
y menor ruido, al igual que en el stavelet con alimentación serie. Estas 
variaciones no representan un aumento sistemático del ruido que pueda 
atribuirse a un efecto causado por el multiplexor.

Estos resultados permiten asegurar que la multiplexión de líneas de alto 
voltaje para polarizar los sensores no produce ningún
efecto negativo sobre el ruido de los stavelets.

\subsubsection{Interferencia electromagnética}

El stavelet con conversión DC-DC se
probó junto a un prototipo de \textit{VersatileLink}, el sistema de 
transmisión óptica que se utilizará en la actualización del detector de 
trazas para comunicarse con los módulos. 

Las pruebas se llevaron a cabo con una temperatura de $12\degC$ en el sistema
de refrigeración y una tensión de polarización de los sensores de $250\V$. En
el momento de realizarse estas pruebas, el sistema de refrigeración no 
funcionaba correctamente y reducía su capacidad de enfriamiento tras un tiempo
de funcionamiento, por lo que la temperatura no se mantuvo estable durante las
pruebas.

La secuencia de pruebas para la medida del ruido ENC fue la siguiente:

\begin{itemize}
 \item Pruebas 1 y 2: prototipo no colocado en las inmediaciones del stavelet. 
 Medida de la referencia de ruido.
 \item Prueba 3: prototipo de VersatileLink situado cerca de la tarjeta EoS, 
 con los cables de alimentación colocados a lo largo del stavelet, encendido.
 \item Prueba 4: igual que el anterior, pero con los cables de alimentación del 
 prototipo pasando desde la tarjeta EoS, como correspondería a la configuración
 realista en el detector. El prototipo está encendido.
 \item Prueba 5: prototipo colocado directamente encima de la tarjeta EoS, 
 encendido.
 \item Prueba 6: VersatileLink colocado sobre un grupo de chips en el módulo
 M3, encendido.
 \item Pruebas 7 y 8: prototipo apagado para volver a ver la referencia de 
 ruido.
 \item Prueba 9: repetición de la prueba 6.
\end{itemize}

\begin{figure}[!hb]
\begin{center}
    \includegraphics[scale=0.5]{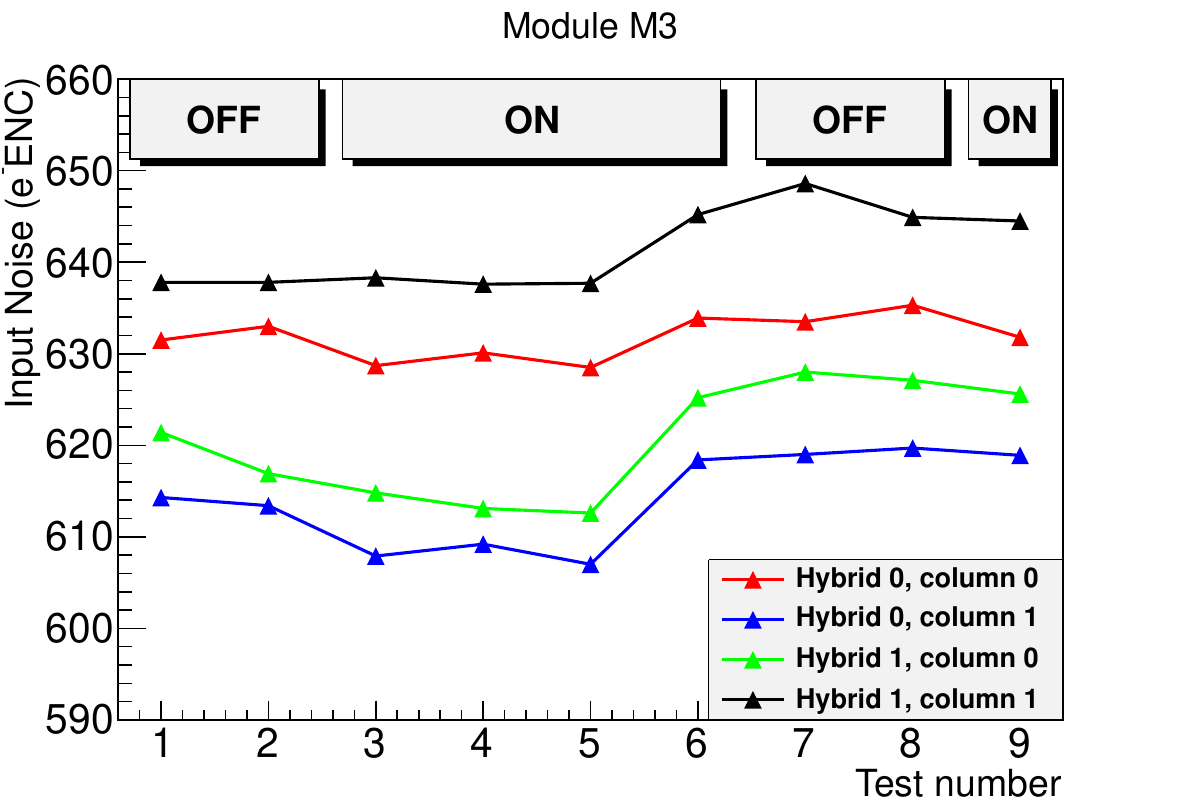}
    \caption{Ruido ENC medido en el módulo M3, con interferencia
 del prototipo de VersatileLink.}\label{fig:8-Versatile_M3}
 \end{center}
\end{figure}

Por otro lado, la medida de ruido de doble disparo (DTN) se hizo en una 
secuencia diferente:

\begin{enumerate}
 \item Sobre el módulo M3, encendido.
 \item Sobre el módulo M3, apagado.
 \item Sobre la tarjeta EoS (orientado longitudinalmente), encendido.
 \item Sobre la tarjeta EoS, apagado.
 \item Sobre la tarjeta EoS, apagado.
 \item Sobre la tarjeta EoS, encendido.
 \item Sobre la tarjeta EoS, girado $90^{\circ}$, encendido.
 \item Sobre la tarjeta EoS, girado $90^{\circ}$, apagado.
\end{enumerate}

\begin{figure}[!htb]
\begin{center}
  \includegraphics[scale=0.775]{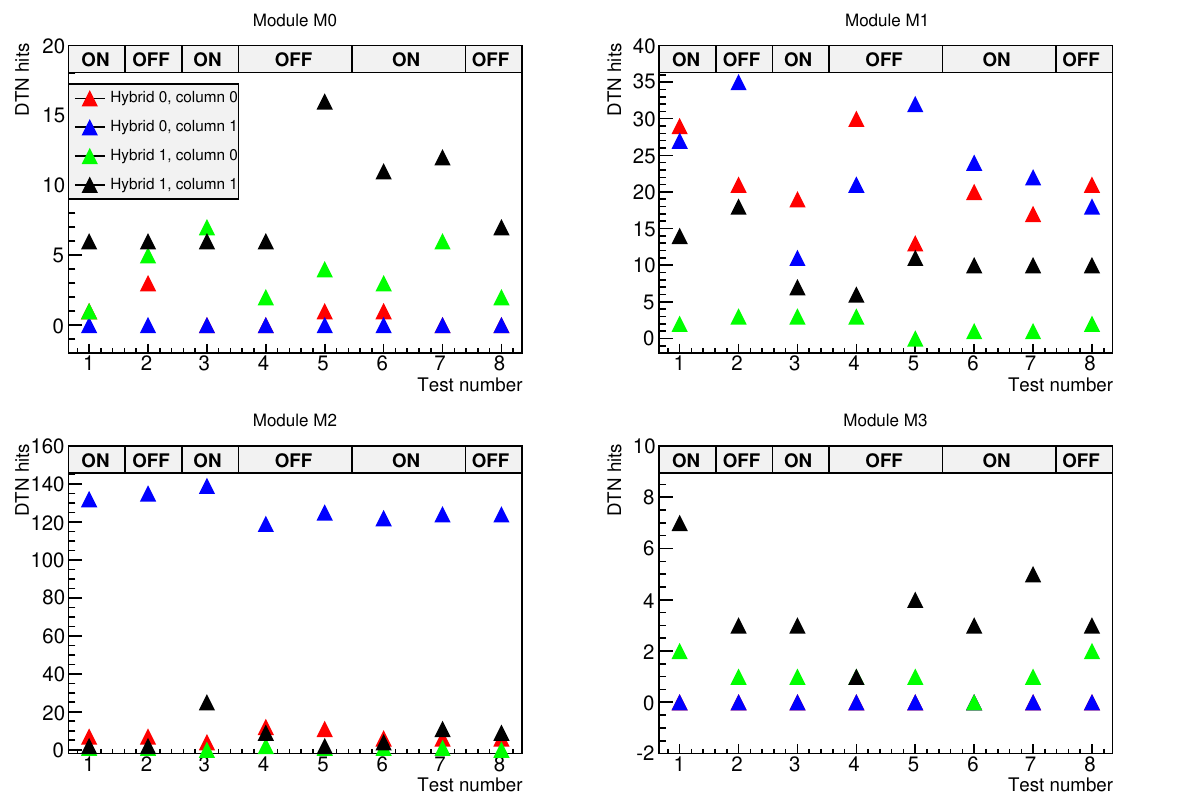}
 \caption{Ruido de doble disparo con umbral a $0.5\fC$, con interferencia
 del prototipo de VersatileLink.}\label{fig:8-versatile_dtnplot}
 \end{center}
\end{figure}

Para la medida de DTN, se activaron todos los módulos del stavelet.

La disponibilidad del prototipo tuvo lugar cuando únicamente el stavelet
con conversión DC-DC se encontraba en el laboratorio, por lo que esta prueba
no se pudo repetir en el stavelet con alimentación serie.

Los resultados mostrados en las
Figuras~\ref{fig:8-Versatile_M3} y~\ref{fig:8-versatile_dtnplot} permiten
asegurar que el prototipo no provoca interferencias en el stavelet con
conversión DC-DC.

Como se mencionó antes, la temperatura del stavelet en el momento de las 
pruebas no se mantuvo estable, debido a un malfuncionamiento del refrigerador. 
El aumento de ruido mostrado en la Figura~\ref{fig:8-Versatile_M3} entre la
prueba 5 y la 6 sigue a un aumento de la temperatura en los módulos del 
stavelet.

La evidencia de que no se trata de un aumento de ruido por 
interferencia del prototipo de VersatileLink se observa en las medidas 7 y 8,
durante los cuales el prototipo permaneció apagado y el ruido se mantuvo 
ligeramente elevado.

En las medidas de ruido ENC se habilitó únicamente el módulo M3 del stavelet,
que es el más próximo a la tarjeta de adquisición y es en sus inmediaciones
donde se situó el prototipo de VersatileLink. 

Los resultados que se pueden ver en la Figura~\ref{fig:8-versatile_dtnplot}
muestran que el ruido de doble disparo se mantiene estable en todo el stavelet
para los periodos en los que el prototipo de VersatileLink está encendido, con
respecto a los periodos en los que está apagado.

El prototipo de VersatileLink probado funciona con en la segunda ventana de 
transmisión óptica, a una longitud de onda
de $1310\nm$, la cual no genera señales en el silicio. Por esta razón, no se 
llevó
a cabo ninguna prueba de fugas de luz, al contrario de lo que se hizo con el
SCT, dado que los láseres funcionan a $850\nm$~\cite{Phillips:2007db}.

\subsubsection{Pruebas individuales y conjuntas de los dos stavelets}

Los dos stavelets
se probaron de forma independiente, y también alimentados y leídos de forma
simultánea y con el mismo sistema de adquisición. 

\begin{figure}[!hbt]
 \begin{center}
  \includegraphics[scale=0.6]{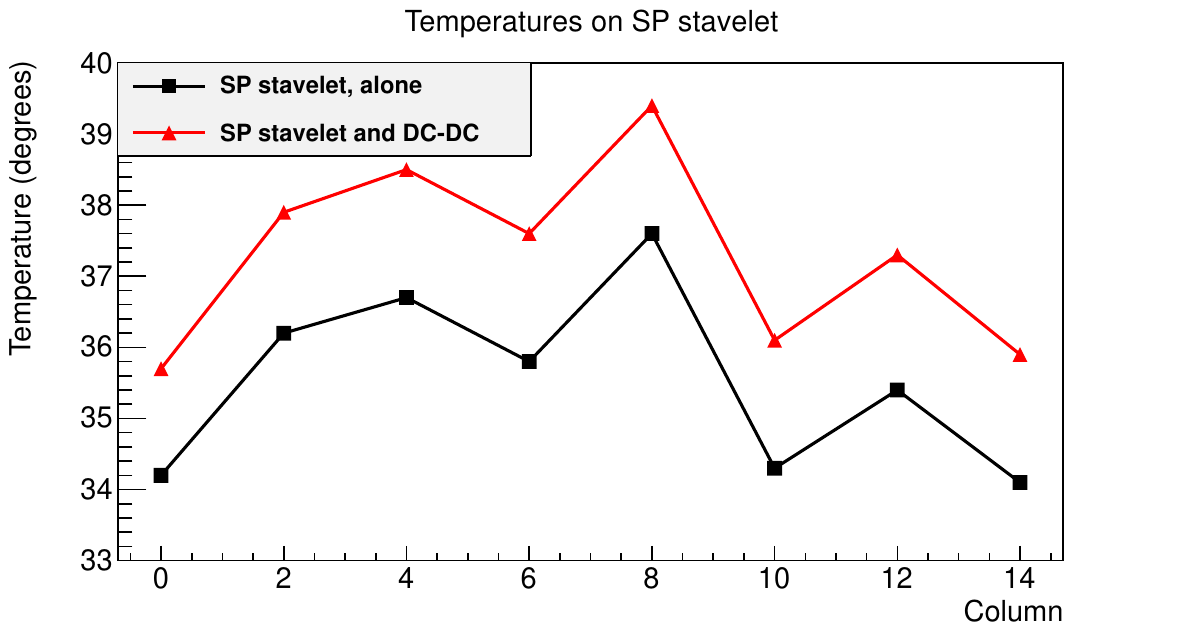}
  \caption{Temperaturas en los híbridos del stavelet serie operado por separado 
  y  junto al stavelet DC-DC.}\label{fig:8-TwoStavelets_Temperatures_SP}
 \end{center}
\end{figure}

La tarjeta de adquisición dispone de dos grupos de conectores que pueden 
emplearse para
medir dos stavelets simultáneamente. La conexión del circuito de 
refrigeración para estas pruebas se hizo en serie, por la falta de 
disponibilidad de conectores para hacerlo en paralelo. El stavelet con 
alimentación serie se situó segundo en la línea de refrigeración, lo que 
provocó que la temperatura de sus módulos fuera superior que en el caso
de operarse por separado. 

La Figura~\ref{fig:8-TwoStavelets_Temperatures_SP}
muestra la comparación entre temperaturas del stavelet con alimentación
serie operado por separado y después de conectarlo después del stavelet con
alimentación DC-DC.

En el stavelet con alimentación DC-DC no se observaron diferencias de 
temperatura entre la operación por separado y la operación junto al stavelet
con alimentación serie.

Los resultados de la 
Figura~\ref{fig:8-TwoStavelets_DCDC} muestran que el stavelet DC-DC no ve
mermadas sus prestaciones al ser medido junto al stavelet con alimentación 
serie.

\begin{figure}[!htbp]
 \begin{center}
  \subfigure[Ruido en el stavelet DC-DC.]{
    \includegraphics[scale=0.65]{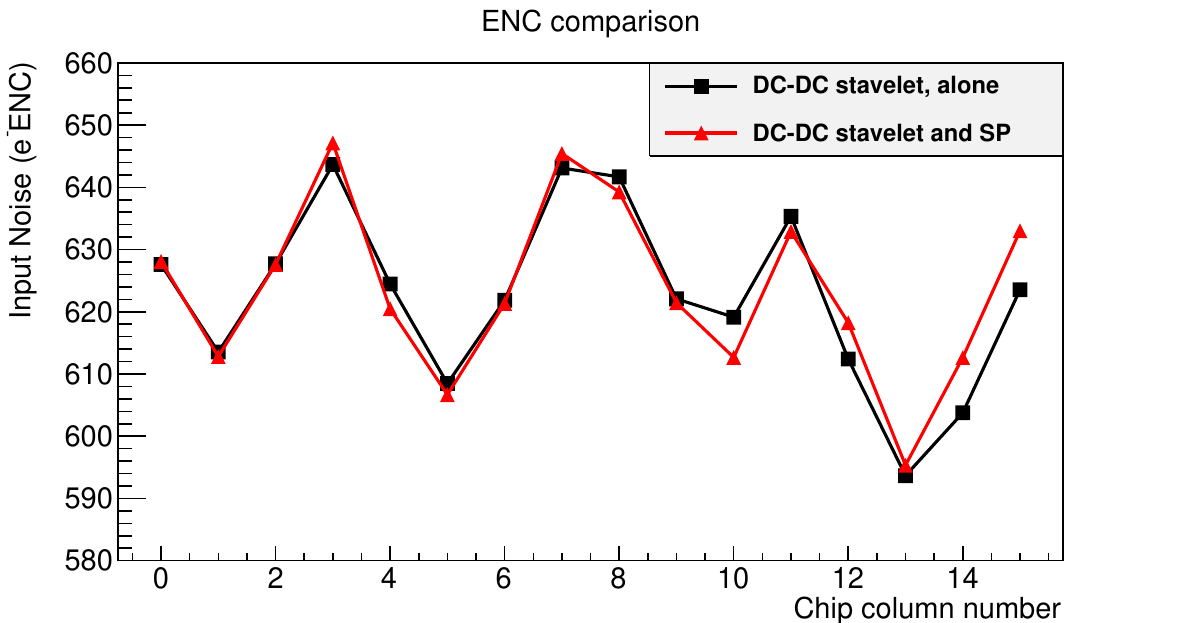}
  }
  \subfigure[Ruido de doble disparo en el stavelet DC-DC.]{
    \includegraphics[scale=0.65]{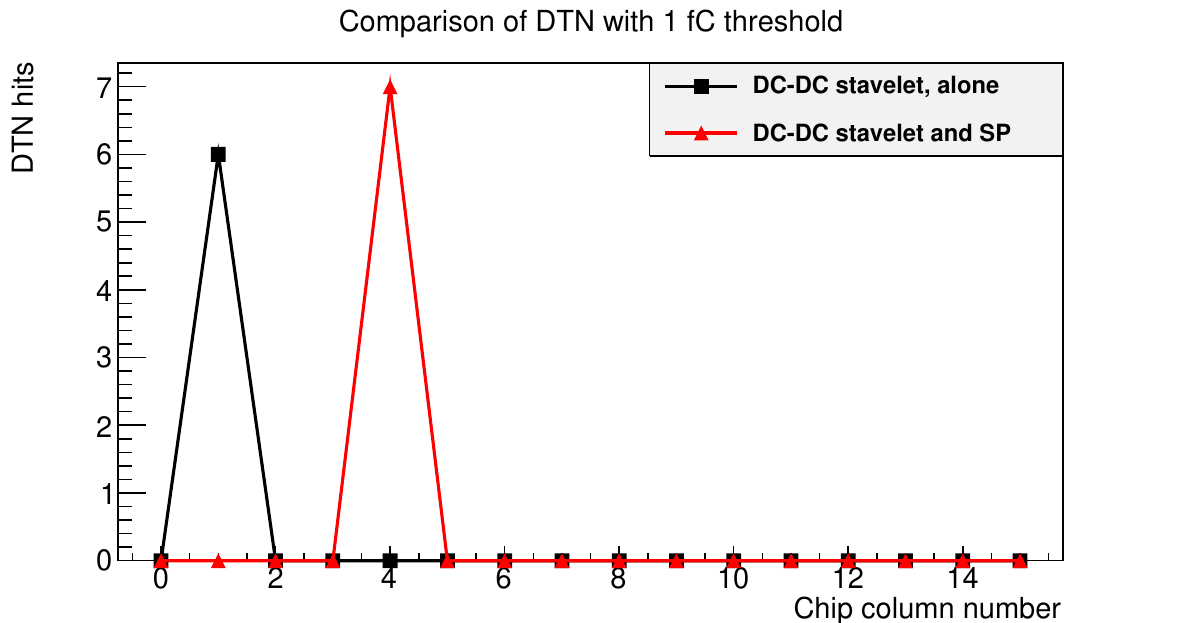}
  }
  \caption{Diferencias de ruido en el stavelet DC-DC cuando funciona 
  independientemente y cuando funciona junto al stavelet 
  serie.}\label{fig:8-TwoStavelets_DCDC}
 \end{center}
\end{figure}

\begin{figure}[!htbp]
 \begin{center}
  \subfigure[Ruido en el stavelet serie.]{
    \includegraphics[scale=0.65]{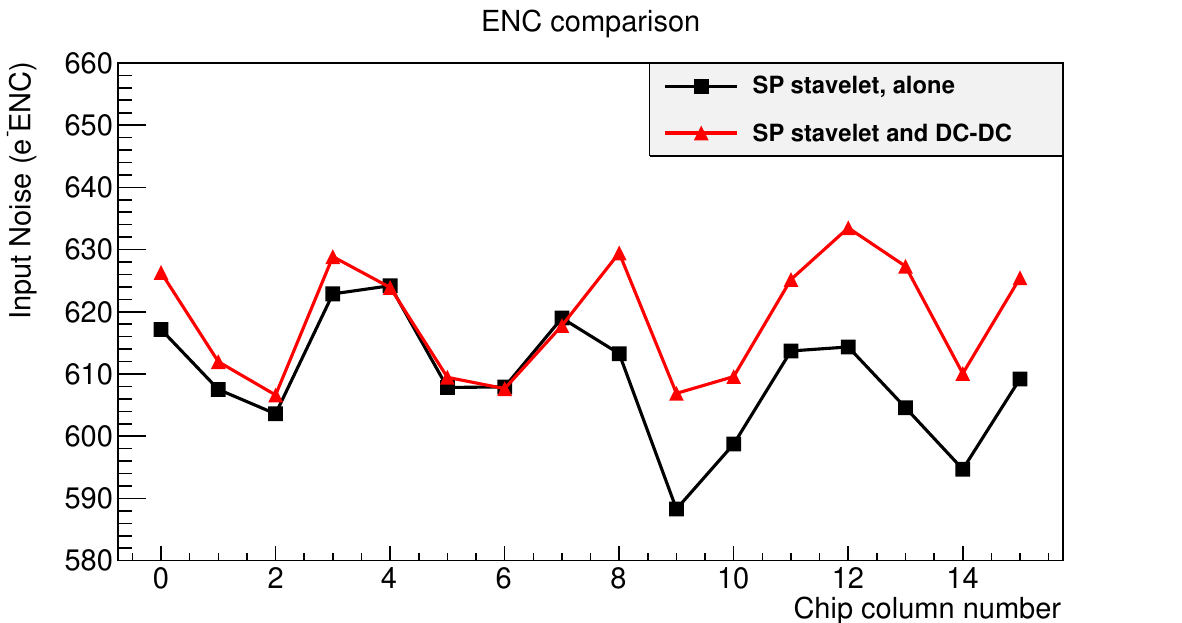}
  }
  \subfigure[Ruido de doble disparo en el stavelet serie.]{
    \includegraphics[scale=0.65]{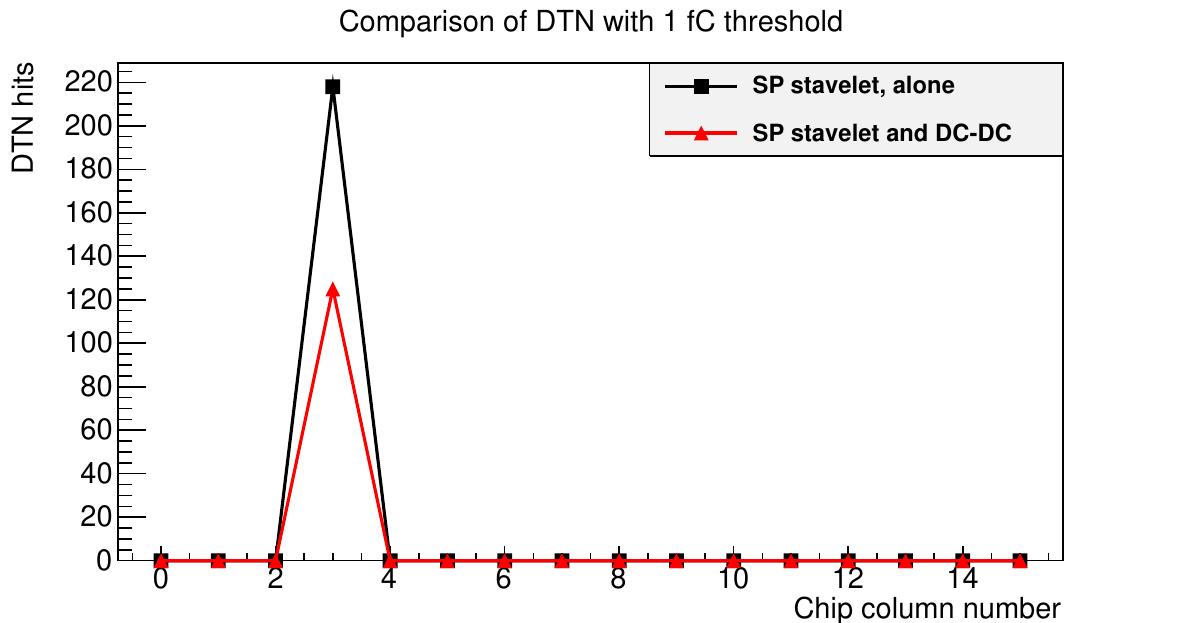}
  }
  \caption{Diferencias de ruido en el stavelet serie cuando funciona 
  independientemente y cuando funciona junto al stavelet 
  DC-DC.}\label{fig:8-TwoStavelets_SP}
 \end{center}
\end{figure}

De la misma manera, la Figura~\ref{fig:8-TwoStavelets_SP} muestra que en el 
stavelet con alimentación serie no se observa aumento del ruido más allá del 
incremento de temperatura de los híbridos, derivado de la configuración del
sistema de refrigeración.

Las únicas precauciones a considerar para probar ambos stavelets de forma
simultánea sin que se produzca aumento del ruido es utilizar un apantallado
adecuado en las líneas de alimentación de los híbridos, reduciendo el ruido
en modo común. En particular, el stavelet con alimentación serie es más 
susceptible al ruido en modo común.

\subsection{Stave de doce módulos con alimentación DC-DC}\label{sec:8-stave}

El primer stave con doce módulos se terminó de construir a finales de 2013 en
RAL. Se montaron módulos procedentes de diversos institutos, todos ellos de
``grado A'', es decir, de la mejor calidad. Los institutos que contribuyeron
con módulos fueron las Universidades de Cambridge (Reino Unido), California 
Santa Cruz (Estados Unidos), Liverpool (Reino Unido), Friburgo (Alemania) y
Glasgow (Reino Unido).

La Figura~\ref{fig:8-stave_12modules_picture} muestra una fotografía del stave
completo con los doce módulos montados sobre la estructura. El diseño con doce
módulos por cada lado es el que se realizó para los módulos con chips de 
$250$ nanómetros, al contrario del diseño con trece módulos para los chips de
$130$ nanómetros~\cite{ATLAS:1502664}.

\begin{figure}[!htb]
 \begin{center}
  \includegraphics[scale=0.58]{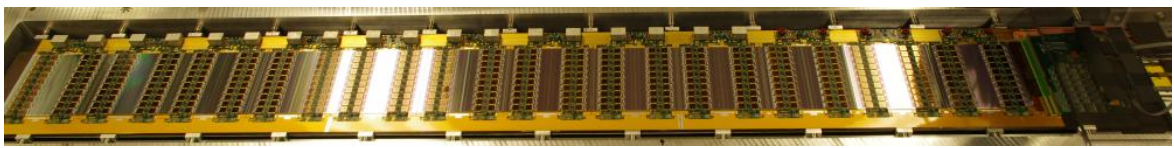}
  \caption{El stave con alimentación DC-DC, con los doce módulos 
  montados.}\label{fig:8-stave_12modules_picture}
 \end{center}
\end{figure}

Los conversores DC-DC utilizados con el stave no son los mismos que en el 
stavelet. Se trata de conversores en tándem, con una entrada de alimentación
y dos salidas, una para cada híbrido de un módulo. 

Las pruebas del stave se hicieron por etapas. En lugar de montar todos los 
módulos de una vez, se montaron en turnos y se probaron en la siguiente 
secuencia: tres, nueve y doce módulos.

\subsubsection{Stave con tres módulos}

Los tres primeros módulos fueron los más alejados de la tarjeta de lectura 
(EoS). Se observó que las temperaturas de los conversores eran demasiado 
elevadas, concretamente en los inductores de cada conversor. 

Para corregirlo, 
se sustituyó el apantallado típico, un soporte de plástico envuelto en pintura 
de cobre, por uno fabricado con cobre estañado. Además, se añadió relleno 
térmico para incrementar la transferencia de calor hacia el exterior.
 
También se  encontraron problemas con la integridad de las señales, por lo que 
se corrigieron las terminaciones de los búferes LVDS en la tarjeta EoS.

\subsubsection{Stave con nueve módulos}

Los seis módulos siguientes se ensamblaron de la misma
manera que los tres primeros, con la salvedad de utilizar pegamento con mayor
conductividad térmica para los conversores DC-DC. Con ello se consiguió 
disminuir la temperatura de los conversores.

Se equiparon los conversores de los nuevos módulos con sensores de temperatura
para comprobar que la refrigeración es correcta. Los sensores se montaron sobre
la placa de circuito impreso y dentro de uno de los apantallamientos, junto
al inductor, de forma alterna entre módulos (en el módulo 3 dentro del 
apantallamiento, el siguiente sobre la placa).

Los problemas de integridad de señal continuaron con estos módulos y hubo que
hacer ajustes en las terminaciones.

\subsubsection{Stave con doce módulos}

Los tres últimos módulos del stave se ensamblaron de la misma forma que los
seis anteriores, incluyendo los conversores DC-DC. 

Las pruebas del stave con tres, nueve y doce módulos se llevaron a cabo
con el refrigerante a $13\degC$, la tensión de polarización de los sensores a
$250\V$ y la humedad se mantuvo por debajo del $10\%$ con nitrógeno.

\begin{figure}[!htbp]
 \begin{center}
   \includegraphics[scale=0.6]{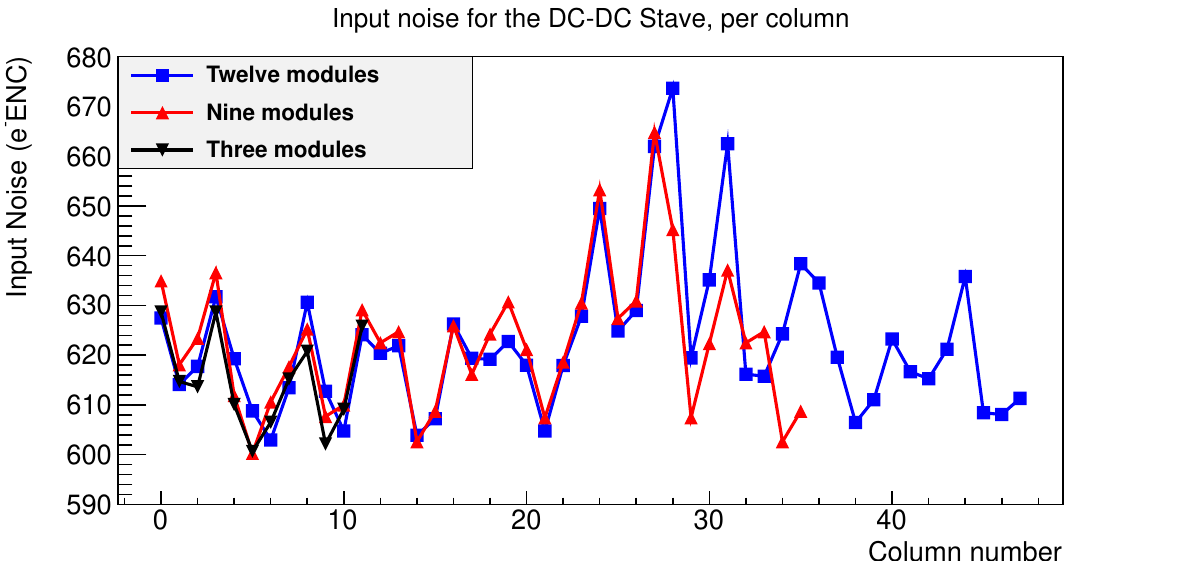}
  \caption{Ruido del stave DC-DC con doce 
  módulos.}\label{fig:8-stave_12modules_enc}
 \end{center}
\end{figure}

Los resultados para todas las pruebas llevadas a cabo en cada etapa de 
ensamblado del stave se presentan en esta sección. El ruido del stave con
doce módulos se muestra en la Figura~\ref{fig:8-stave_12modules_enc},
incluyéndose las medidas realizadas con tres y nueve módulos.

En los resultados se observa que los niveles de ruido ENC se mantienen estables
cuando se incrementa el número de módulos a lo largo del stave. 

Dos módulos
presentan valores de ruido mayores de lo normal, correspondientes a híbridos
que fueron construidos de forma diferente a los demás. Leves defectos en el 
proceso de fabricación provocan que tengan un ruido mayor de lo que les 
corresponde.

\begin{figure}[!htb]
 \begin{center}
   \includegraphics[scale=0.55,trim=3 1 20 20,clip=true]{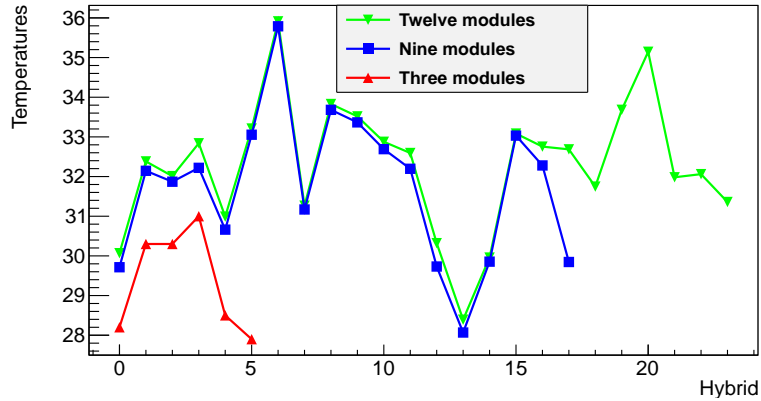}
  \caption{Temperaturas medidas en el stave durante las
  pruebas.}\label{fig:8-stave_12modules_temperatures}
 \end{center}
\end{figure}

La temperatura de los híbridos durante las pruebas con tres, nueve y doce 
módulos se muestra en la Figura~\ref{fig:8-stave_12modules_temperatures}. El
leve aumento entre nueve y doce módulos contrasta con el menor ruido medido en
algunos híbridos con doce módulos. 

El flujo de nitrógeno estaba a una 
temperatura entre $7$ y $15\degC$ inferior en el momento de hacer las pruebas 
con doce módulos respecto de la temperatura cuando se hicieron las pruebas con
nueve módulos, por lo que contribuyó a reducir la temperatura en los chips.

\begin{figure}[!htb]
 \begin{center}
\subfigure[DTN en el stave 250 con doce módulos, umbrales altos.]{
 \includegraphics[scale=0.6]{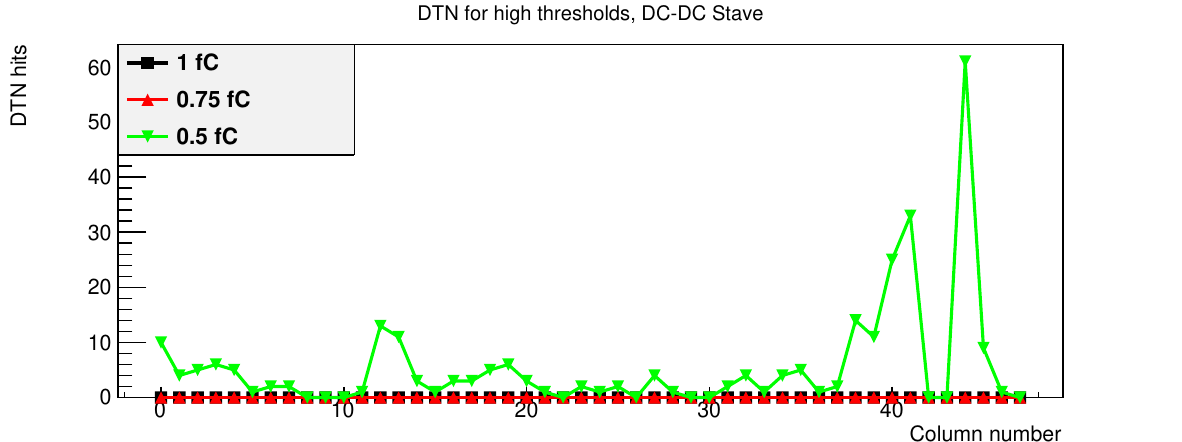}
}

\subfigure[DTN en el stave 250 con doce módulos, umbrales bajos.]{
 \includegraphics[scale=0.6]{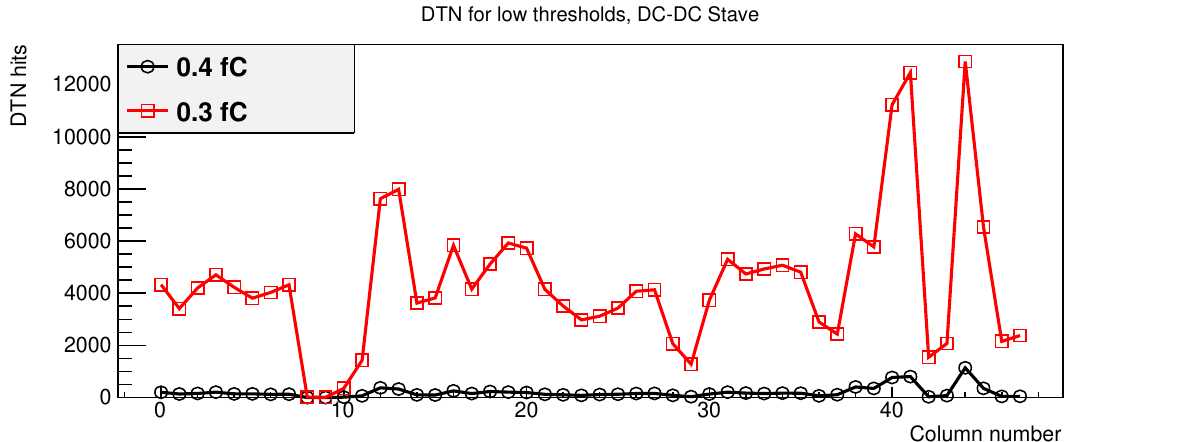}
}
  \caption{Ruido de doble disparo en el stave DC-DC con doce 
  módulos.}\label{fig:8-stave_12modules_dtn}
 \end{center}
\end{figure}

Por otro lado, la medida de ruido de doble disparo se muestra en la 
Figura~\ref{fig:8-stave_12modules_dtn}, para los doce módulos.

Con umbrales elevados, $1\fC$ y $0.75\fC$, no se observa ningún incremento
en la cuenta de ruido. Para el umbral de $0.5\fC$ algunos híbridos presentan
picos, debidos a algunos canales ruidosos que se pueden encontrar en las 
cuentas con umbrales más bajos.

Los resultados muestran que las prestaciones de las estructuras multi-módulo 
con
hasta doce módulos son excelentes, por lo que la viabilidad de los staves como
estructura para implementar los barriles del detector de bandas de semiconductor
está demostrada.

\section{Cálculo del poder de frenada en el SCT}

La pérdida de energía es consecuencia del poder de frenada de un material, o 
$\dedx$, y es un efecto característico del paso de las
partículas cargadas a través de un material. 

En el silicio, la energía perdida por las partículas genera pares 
electrón-hueco, que equivalen a una carga, la cual puede ser medida. Algunos
detectores se construyen con el objetivo de medir esta energía, mientras que
otros se diseñan únicamente para saber si se ha producido el paso de una 
partícula por ellos o no (detectores binarios).

\subsection{Introducción}

Los dos sub-detectores del detector de trazas basados en silicio, 
el detector de píxeles y el de bandas de silicio (SCT) del 
experimento ATLAS no fueron 
diseñados para realizar medidas de energía, sino que son detectores binarios.

Sin embargo, ambos detectores implementan en sus sistemas de adquisición 
temporizadores que miden durante cuánto tiempo ha permanecido la señal generada
por las partículas por encima de un umbral pre-configurado.

El detector de píxeles dispone de un contador de 8 bits, por lo que puede 
contar hasta 256 intervalos de $25\ns$~\cite{ATLAS-CONF-2011-016}. Por otra 
parte, el detector de bandas
de silicio usa un contador de 3 bits, lo que reduce la resolución en partículas
que depositan mayor cantidad de carga.

\subsection{Selección de eventos y de trazas}

Este estudio está basado en los datos recogidos durante las colisiones 
protón-protón entre 2010 y 2012. Las colisiones en 2010 y 2011 se produjeron
a $7\TeV$ de energía del centro de masas, mientras que las colisiones en 2012
fueron a $8\TeV$.

La luminosidad integrada total que se registró en estos conjuntos de datos es
$1.76\ifb$, con un total de $11$ millones de eventos. Los datos están 
almacenados en Ntuplas del SCT (\textit{SCT Ntuples}), que contienen información
de las trazas, tales como el momento, puntos de aproximación al vértice
primario y datos provenientes del detector de píxeles.

Los eventos se seleccionan de entre aquellos que tienen lugar durante el estado
``Ready'' del SCT. Los eventos de interés deben provenir de interacciones 
protón-protón, para lo cual se exige que cumplan los siguientes requisitos:

\begin{itemize}
 \item Al menos un vértice primario reconstruido en el evento. De esta manera
 se pueden descartar eventos que provengan de disparos provocados por rayos 
 cósmicos.
\item El vértice primario debe tener al menos cuatro trazas asociadas.
\item Al menos una de las trazas que formen parte del evento debe cumplir otra 
serie de requisitos:
\begin{itemize}
 \item Momento transversal $p_{\rm T} > 500\MeV/c$.
\item Seis impactos en el SCT.
\item Un impacto en el detector de píxeles.
\item Distancia transversal $|d_0| < 1.5\mm$.
\item Distancia longitudinal tal que $|z_0\sin\theta| < 1.5\mm$.
\end{itemize}
\end{itemize}

De esta forma, se buscan eventos que tengan alguna traza de vértice primario 
con un momento suficientemente elevado.

Para cada evento que pase estos cortes, se seleccionan las trazas para calcular 
su $\dedx$. Las trazas se seleccionan si cumplen además una serie de requisitos 
de calidad, que son los siguientes:

\begin{itemize}
 \item Al menos 8 impactos buenos en la traza. Un impacto que no es ``bueno'' 
 se denomina \textit{outlier}.
 \item Momento transversal $p_{\rm T} > 100\MeV/c$.
 \item Momento total de la traza menor que $2500\MeV/c$.
 \item  $|z_0| < 100\mm$.
\item Dos impactos buenos en el detector de píxeles.
\end{itemize}

De todas las trazas que cumplen con los requisitos se extrae la información
sobre intervalos de tiempo sobre umbral de la Ntupla, para aplicar el método
que se describe a continuación y calcular su $\dedx$. La Ntupla contiene los
datos de intervalos de tiempo sobre umbral de todas las bandas que han generado
señal, para todos los impactos de una traza.

\subsection{Reconstrucción de \texorpdfstring{$\dedx$}{dE/dx}}

En el SCT, el umbral de carga está configurado a $1\fC$ ($6242~e^{-}$) y se
dispone de tres intervalos de $25\ns$ en los cuales se registra la carga
depositada por las partículas que impactan en los sensores. Se calibra la 
temporización del detector tal que se registra un patrón 01x. El modo de
lectura además varía en función del modo de funcionamiento del LHC.

Para la toma de datos en 2010 se usó un patrón de lectura XXX, para un 
espaciado de los paquetes de protones de $75\ns$. Para los datos en 2011 y 2012
se usó un espaciado de $50\ns$ con un patrón de lectura X1X.

\subsubsection{Cálculo de \texorpdfstring{$\dedx$}{dE/dx} en el SCT}

La fórmula para calcular el poder de frenada en el SCT es la siguiente (en
unidades arbitrarias):

\begin{equation}
 \dedx_{\rm SCT} = \frac{\displaystyle \sum_{\rm Track\_hits} 
 \sum_{\rm Hit\_strips} w^i_{\rm strip} \cos \alpha}{\rm Track\_hits} \quad 
 \left[{\rm a.u.}\right]
\end{equation}

Donde $w^i_{\rm strip}$ es el peso asignado según el número de intervalos de
$25\ns$ para cada impacto (por defecto, $1$, $2$ y $3$). El ángulo $\alpha$ es 
la proyección de los ángulos locales $\theta$ y
$\phi$, redefinidos entre $-\pi/2$ y $\pi/2$:

\begin{equation}
  \alpha = \tan^{-1} \left( \sqrt{\tan^2 \theta' + \tan^2 \phi'} \right)
\end{equation}

\begin{equation}
 \theta' = \left\{ \begin{array}{c c}
                   \theta_0^{\rm local} - \pi & {\rm ~ if ~} \theta_0^{\rm local} > \pi/2 \\
                    & \\
                   \theta_0^{\rm local} + \pi & {\rm ~ if ~}  \theta_0^{\rm local} < \pi/2
                  \end{array}
\right.
\end{equation}

\begin{equation}
 \phi' = \left\{ \begin{array}{cc}
                   \phi_0^{\rm local} - \pi & {\rm ~ if ~}  \phi_0^{\rm local} > \pi/2 \\
                    & \\
                   \phi_0^{\rm local} + \pi & {\rm ~ if ~}  \phi_0^{\rm local} < \pi/2
                  \end{array}
\right.
\end{equation}

\begin{figure}[!htbp]
 \begin{center}
    \includegraphics[scale=0.45]{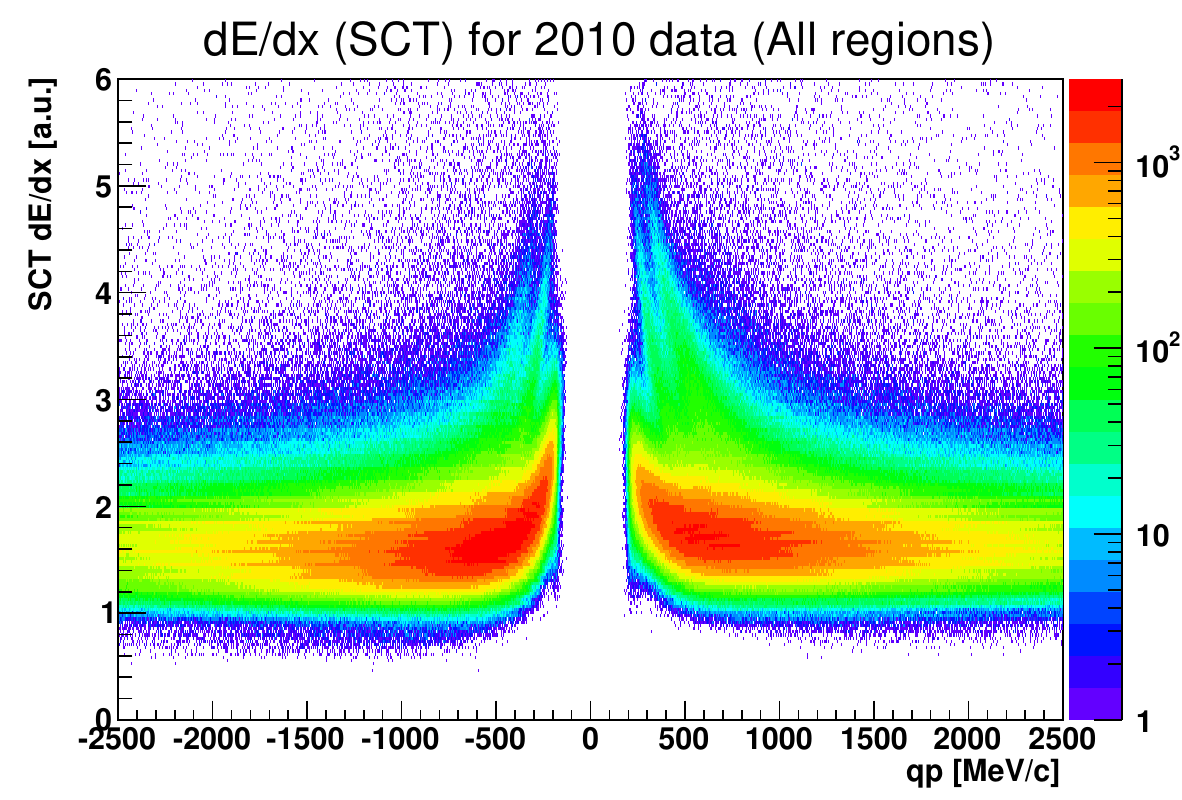}
 \caption{$\dedx$ calculada en el SCT para datos de 
 2010.}\label{fig:8-dedx_vs_pq_PV_2010}
 \end{center}
\end{figure}

La Figura~\ref{fig:8-dedx_vs_pq_PV_2010} muestra el resultado de la 
distribución de la $\dedx$ calculada con los datos del SCT para los datos
recogidos en 2010.

\subsection{Identificación de partículas}

Uno de los objetivos del estudio es evaluar la capacidad de discriminar tipos
de partícu-las mediante el cálculo de $\dedx$ en el SCT. Dado que no existen
simulaciones Monte Carlo para el SCT con informacion realista sobre los 
intervalos de tiempo, no es posible comparar la eficiencia de este cálculo con 
información de verdad.

Por ello, se hace uso del cálculo existente en el detector de píxeles. La 
información de identificación aparece en las Ntuplas del SCT desde 2011. 

En el detector de píxeles se considera que una partícula está bien identificada
(pión, kaón, protón) cuando la verosimilitud normalizada es mayor a $0.9$. Esta
verosimilitud normalizada se define a partir del vector de verosimilitudes para
cada tipo de partícula que está almacenado en la Ntupla:

\begin{equation}
 P_{i}^{\rm norm,Pixel} = \frac{P_i^{\rm Pixel}}{P_{\pi}^{\rm Pixel} + P_K^{\rm Pixel} + P_p^{\rm Pixel}}
\end{equation}

Donde $i = \pi/K/p$.

La Figura~\ref{fig:8-pid_sct_dedx_distrib_ALL} muestra la distribución de $\dedx$
calculada en el SCT para trazas positivas, en los años 2011 y 2012.

\begin{figure}[!htbp]
 \begin{center}
 \subfigure[$\dedx$  del SCT, datos de 2011.]{
  \includegraphics[scale=0.75]{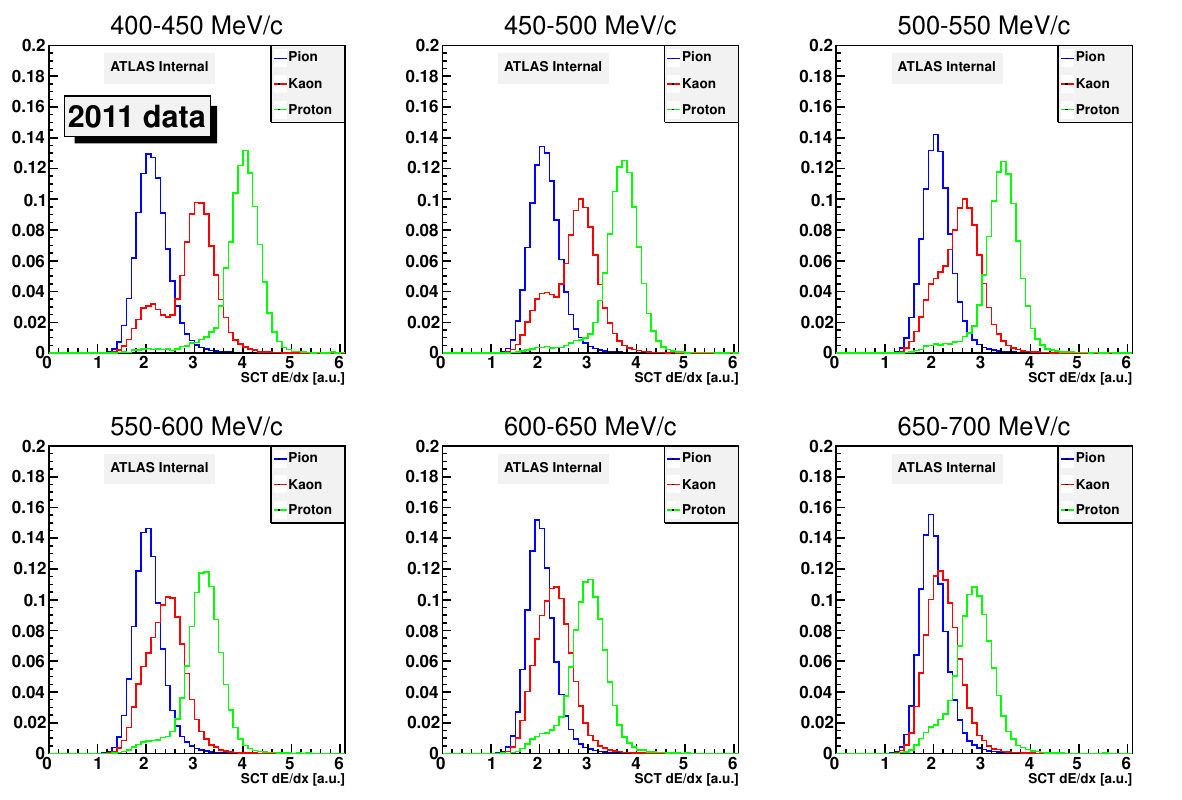}

  \label{fig:8-pid_2011_sct_dedx_distrib}
  }
  
 \subfigure[$\dedx$  del SCT, datos de 2012.]{
  \includegraphics[scale=0.75]{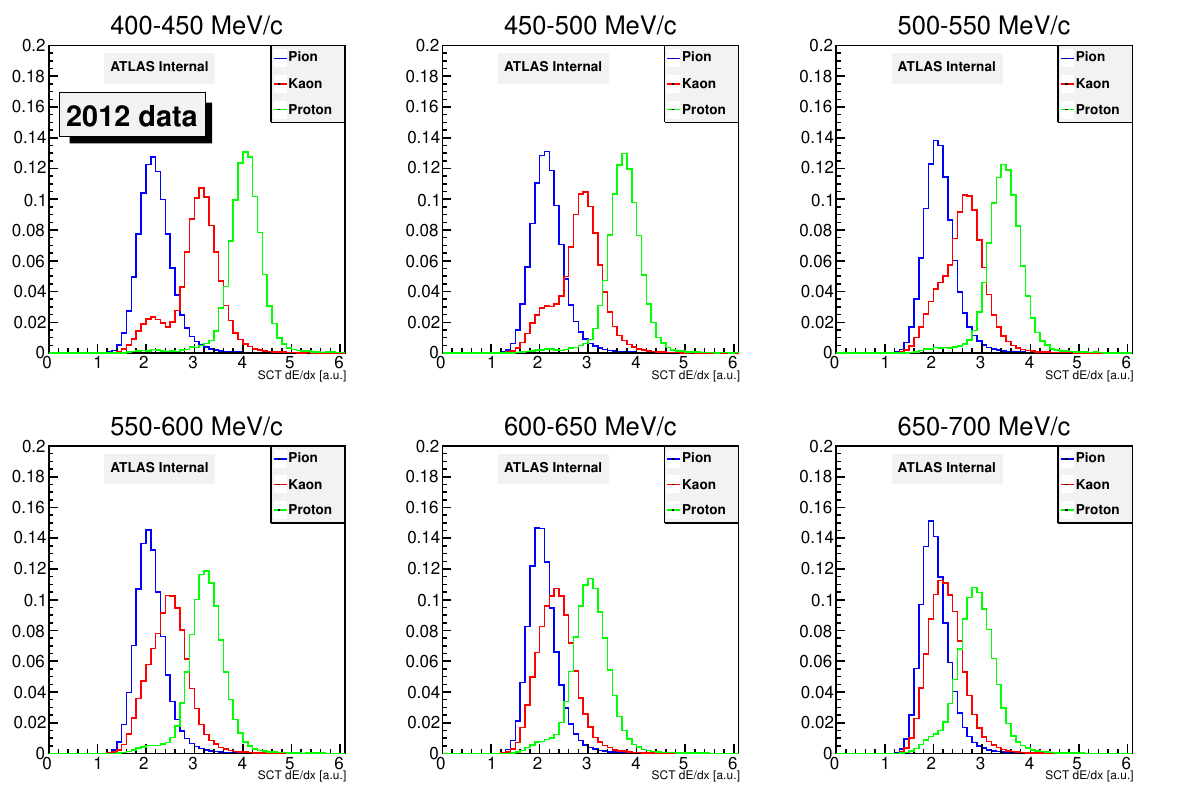}

  \label{fig:8-pid_2012_sct_dedx_distrib}
  }
  
  \caption{Distribuciones de $\dedx$ en el SCT para datos de 2011 y 2012, 
  trazas positivas y vértices primarios.}\label{fig:8-pid_sct_dedx_distrib_ALL}
 \end{center}
\end{figure}

Se observa en las gráficas que existe cierta capacidad para separar las bandas,
por lo que es posible en principio explotar esta separación. Haciendo uso de 
esta información, se hace un ajuste a una gaussiana de las curvas, de manera
que se puede calcular la verosimilitud de una $\dedx$ dada para los tres tipos
de partículas:

\begin{equation}
 P^{\rm SCT}_i = \frac{1}{\sqrt{2\pi} \sigma_{\dedx_i}} \exp \left( - \frac{ \left(\dedx_{\rm SCT} - \dedx_i \right)^2}{2 \sigma_{\dedx_i}^2}\right)
\end{equation}

Siendo $(\dedx)_i$ y $\sigma_{\dedx_i}$ la media y varianza del ajuste para 
cada tipo de partícula en cada rango de momento, y $i=\pi/K/p$.

La verosimilitud de cada tipo de partícula se normaliza de la misma forma en
la que se normaliza la calculada por el detector de píxeles y se calculan las
tasas de eficiencia y de fallo de etiquetado para protones:

\begin{equation}
 r_p^{\pi} = \frac{N({\rm Pixel} = \pi, {\rm SCT} = p)}{N({\rm Pixel} = 
 \pi)}\qquad r_p^{K} = \frac{N({\rm Pixel} = K, {\rm SCT} = p)}{N({\rm Pixel} = K)}
\end{equation}

Siendo $N({\rm Pixel} = i)$ el número de trazas que el detector de píxeles 
etiquetaría como verdaderos $i$ (pión, kaón, protón) y 
$N({\rm Pixel} = i, {\rm SCT} = j)$ es el número de trazas que el detector de
píxeles considera verdaderas $i$ y que el SCT etiquetaría como $j$.

\begin{figure}[!htb]
 \begin{center}
 \subfigure[Eficiencia de etiquetado de protones frente a tasa de
 fallos de etiquetado de kaones.]{
    \includegraphics[scale=0.37]{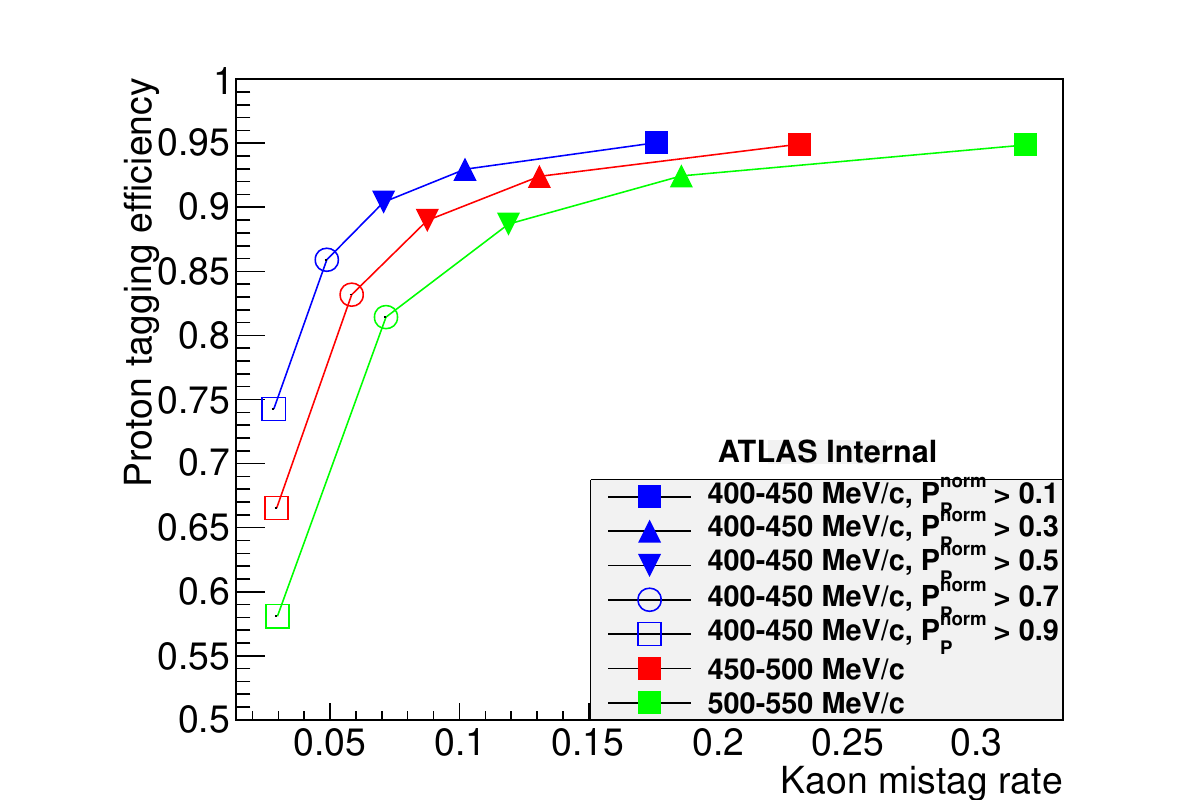}
 }
 \subfigure[Eficiencia de etiquetado de protones frente a tasa de
 fallos de etiquetado de piones.]{
    \includegraphics[scale=0.37]{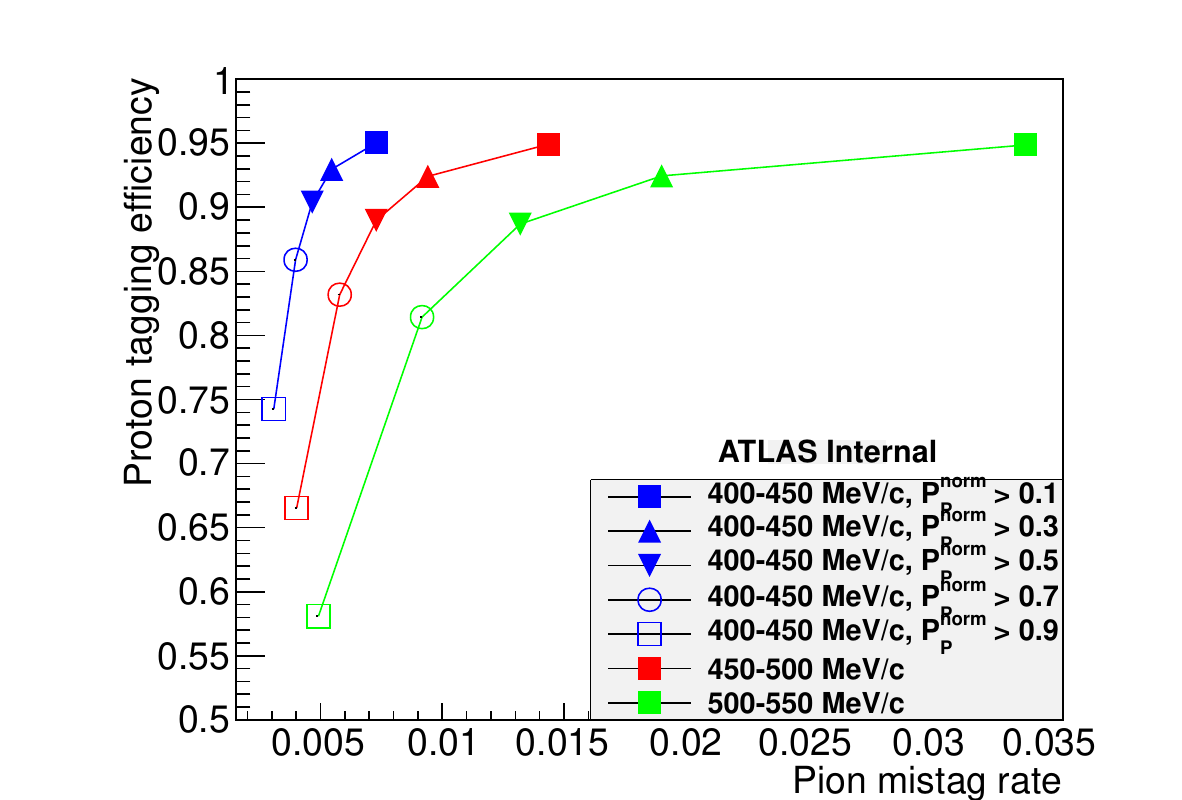}
 }
  \caption{Eficiencia de etiquetado de protones en función del corte de
  verosimilitud $P_p^{\rm norm, SCT}$, datos de 
 2011.}\label{fig:8-tag_eff_2011}
 \end{center}
\end{figure}

\begin{figure}[!htb]
 \begin{center}
 \subfigure[Eficiencia de etiquetado de protones frente a tasa de
 fallos de etiquetado de kaones.]{
    \includegraphics[scale=0.37]{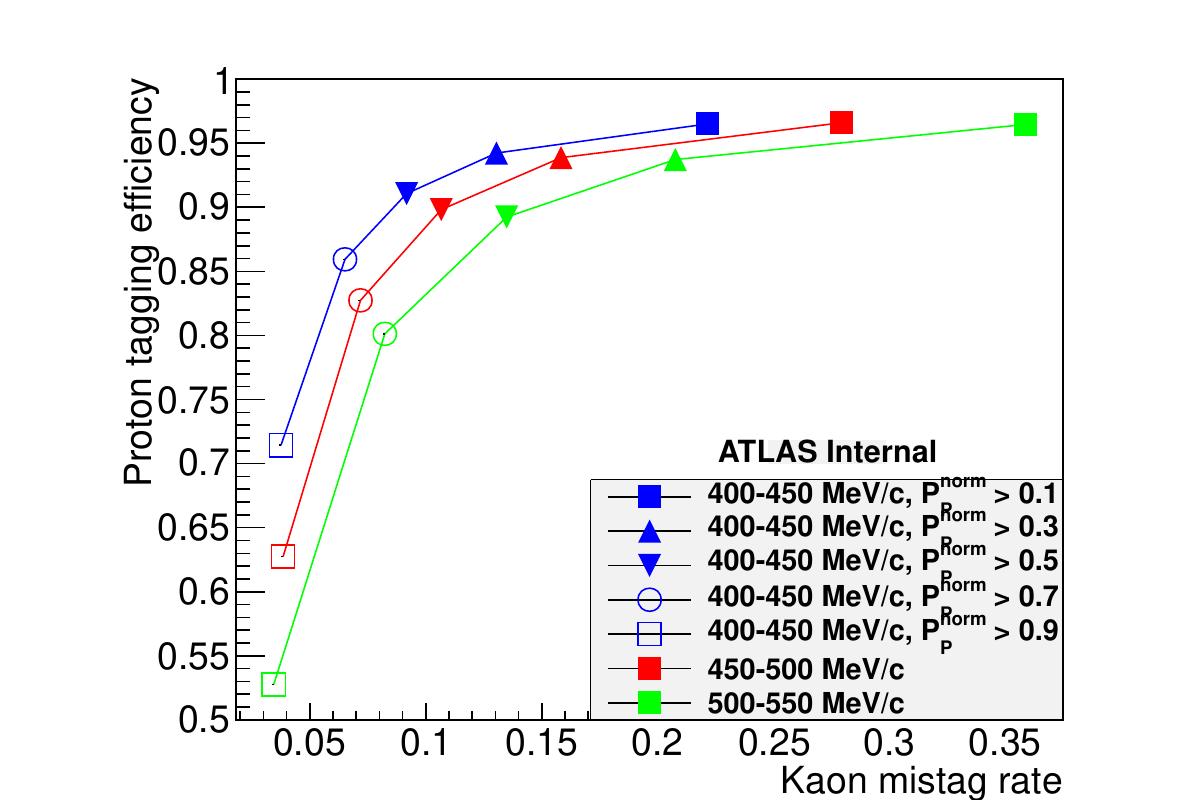}
 } 
 \subfigure[Eficiencia de etiquetado de protones frente a tasa de
 fallos de etiquetado de piones.]{
    \includegraphics[scale=0.37]{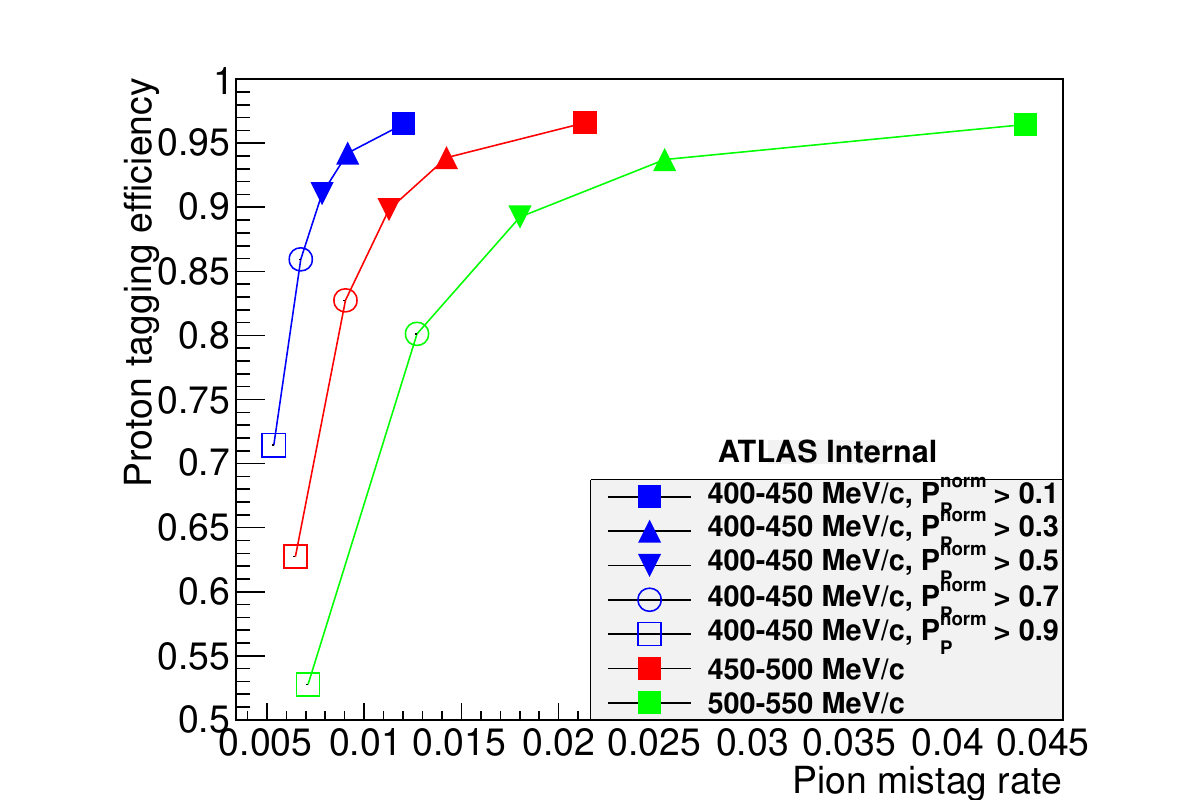}
 }
  \caption{Eficiencia de etiquetado de protones en función del corte de
  verosimilitud $P_p^{\rm norm, SCT}$, datos de 
 2012.}\label{fig:8-tag_eff_2012}
 \end{center}
\end{figure}

Con estas definiciones, se hicieron varios cortes de verosimilitud normalizada
en el SCT, $0.1$, $0.3$, $0.5$, $0.7$ y $0.9$. Para estos cortes, se observó la 
tasa de eficiencia frente al fallo de etiquetado de los protones. Los 
resultados se
muestran en las Figuras~\ref{fig:8-tag_eff_2011} y~\ref{fig:8-tag_eff_2012}, 
para los datos de 2011 y 2012, respectivamente. 

Se obtiene una eficiencia máxima del $95\%$ con un fallo de etiquetado de 
piones por protones inferior al $5\%$ y de kaones por protones inferior al
$40\%$, en 2012.

\subsection{Estabilidad a largo plazo}

Dado que el SCT no se diseñó para efectuar medidas de energía, la capacidad
de hacer identificación de partículas es limitada. Sin embargo, existe una 
aplicación del cálculo de $\dedx$ que resulta interesante desde el punto de 
vista del rendimiento del detector. Dado que la $\dedx$ está directamente
relacionada con la carga depositada por las partículas y la carga recogida por
los detectores se reduce con el daño por radiación, se puede analizar el
valor de $\dedx$ a lo largo del tiempo, como forma de monitorizar el daño
por radiación en el detector.

Se ha optado por monitorizar la media de $\dedx$ calculada en protones 
positivos con momento entre $500$ y $550\MeV/c$. La identificación de protones
se ha realizado usando un método de corte geométrico de la $\dedx$ calculada
por el detector de píxeles, para poder extraer los protones desde el comienzo
de las operaciones del detector. 

Asimismo, es necesario tener una buena temporización del detector, por lo que
se efectuaron correcciones en ciertas fechas señaladas. El intervalo de tiempo
medio en el SCT para los protones identificados por el detector de píxeles 
varía con el tiempo, y afecta al cálculo de $\dedx$ efectuado en el SCT.  Es
posible corregir levemente estas variaciones, pero lo más importante es conocer
que estas variaciones existen para poder interpretar correctamente los datos.

\begin{figure}[!hbt]
 \begin{center}
  \includegraphics[scale=0.5]{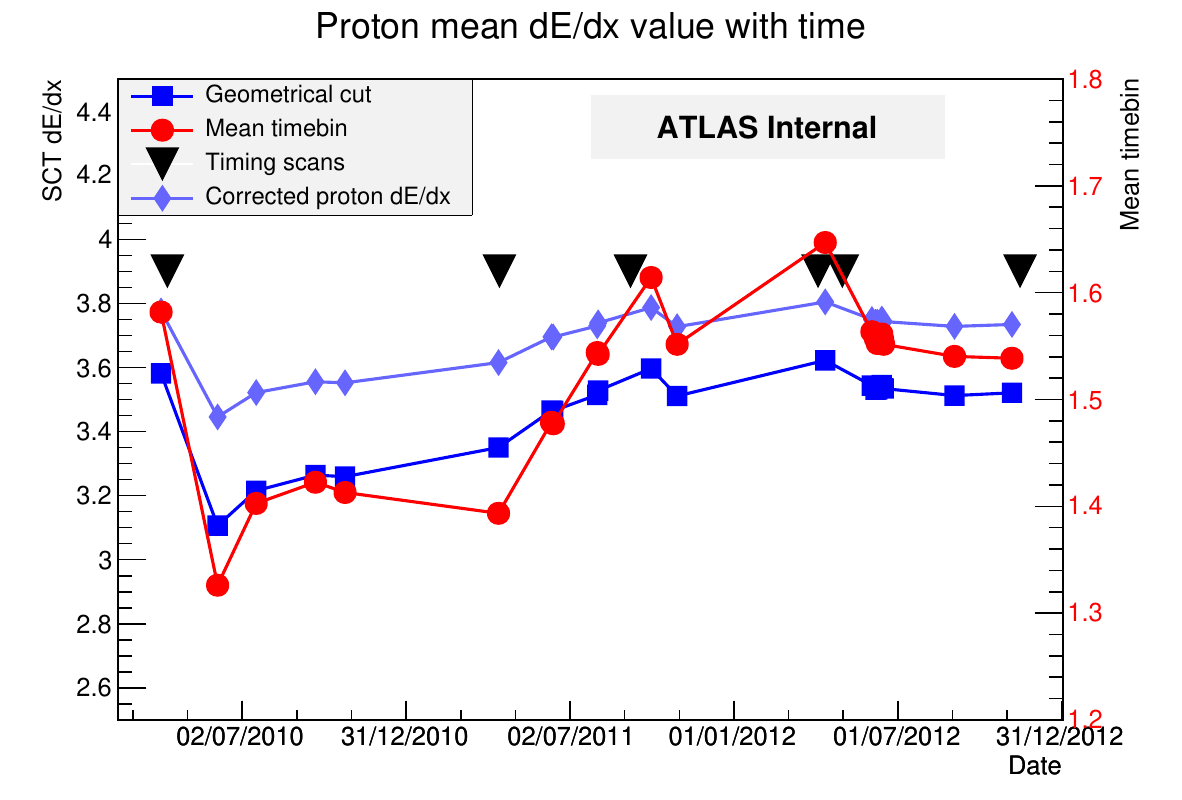} 
  \caption{Evolución de la $\dedx$ media de protones con el 
  tiempo.}\label{fig:8-protonmean_with_correction}
 \end{center}
\end{figure}

La Figura~\ref{fig:8-protonmean_with_correction} muestra la evolución de la 
$\dedx$ media de los protones con el tiempo, incluyendo el ajuste del tamaño
medio de los intervalos de tiempo. Los incrementos de $\dedx$ se deben a 
leves desajustes de la temporización, que provocan que un pulso pueda ser
detectado por encima del umbral durante un intervalo de tiempo más del que le
correspondería.

Las variaciones observadas son inferiores al $17\%$ en la gráfica no corregida,
mientras que son inferiores al $10\%$ en la versión corregida. La tendencia
desde el final de 2011 es de una variación alrededor del $2\%$.

Dado que el SCT aún no padece los problemas derivados del daño por radiación
que reducen la carga recogida por los detectores, no es posible observar 
variación en ese sentido. Sin embargo, esperamos poder aplicar este método en
el futuro para monitorizar el daño por radiación, cuando la luminosidad 
acumulada sea mucho mayor.

\section{Conclusiones}

En esta tesis se ha presentado el trabajo realizado con detectores de bandas de
silicio que se han prototipado para la Fase 2 de la actualización del 
experimento ATLAS. Esta actualización tendrá lugar para que el detector de
trazas del experimento sea capaz de funcionar con los requisitos de 
luminosidad, granularidad y resistencia a la radiación del LHC de alta
luminosidad.

Los resultados sobre estructuras multi-módulo son prometedores y permiten 
verificar que los diseños propuestos para la lectura agregada de módulos y su
alimentación son factibles. El aumento en el número de canales de lectura, con
el consiguiente incremento en la potencia eléctrica consumida, hace necesario
un cambio importante en la forma de alimentar la electrónica en los detectores.

La forma de adquisición de datos también cambiará para reducir el número de 
líneas de datos que van desde los módulos a la sala en la que se procesa la
información. Para ello se agregarán trece módulos en una línea de datos 
compartida. 

En el contexto de esta tesis se han prototipado y evaluado varios componentes,
que implementan estas reducciones de líneas. Los resultados muestran que el
rendimiento de los módulos no se ve perjudicado por la multiplexación de líneas
de alimentación, lectura de los módulos o líneas de alto voltaje.

El primer prototipo de estructura del barril con doce módulos también ha sido 
medido con éxito.

El trabajo presentado en esta tesis continuará con la evolución de los chips
en proceso de $130\nm$, y se deberán probar las mismas estructuras con módulos
que incluyan este chip.

Asimismo, se ha presentado un segundo tema de investigación, consistente en 
estudiar la posibilidad de calcular el poder de frenada, $\dedx$, en el 
detector
de bandas de silicio (SCT) del experimento ATLAS. A pesar de que no fue diseñado
para realizar esta tarea, es posible realizar un cálculo aproximado de este
parámetro. Se ha estudiado la aplicación de este cálculo como forma de hacer
identificación de partículas en el SCT y para hacer una monitorización del
daño debido a la radiación en los sensores de silicio.

\backmatter

\renewcommand{\chaptermark}[1]{\markboth{#1}{}}
\renewcommand{\sectionmark}[1]{\markright{\thesection.\ #1}}

\chapter{Acronyms}

\begin{acronym}
\acro{ABC}{ATLAS Binary Chip}
\acro{ALICE}{A Large Ion Collider Experiment}
\acro{ASIC}{Application-Specific Integrated Circuit}
\acro{ATLAS}{A Toroidal LHC Apparatus} 

\acro{BCC}{Buffer Control Chip}
\acro{BSM}{Beyond Standard Model}

\acro{CERN}{Organisation Européenne pour la Recherche Nucléaire (European Organization for Nuclear Research)}
\acro{CMS}{Compact Muon Solenoid}

\acro{DAQ}{Data Acquisition}
\acro{DTN}{Double Trigger Noise}

\acro{ENC}{Equivalent Noise Charge}
\acro{EoS}{End of Stave or End of Substructure}

\acro{FCal}{Forward Calorimeter}
\acro{FE}{Front-End}
\acro{FSI}{Frequency Scanning Interferometry}
\acro{FZ}{Float-Zone}

\acro{GaAs}{Gallium Arsenide}
\acro{GBT}{GigaBit Transceiver}
\acro{Ge}{Germanium}

\acro{HCC}{Hybrid Control Chip}
\acro{HEC}{Hadronic Endcap Calorimeter}
\acro{HEP}{High Energy Physics}
\acro{HL-LHC}{High Luminosity-Large Hadron Collider}
\acro{HSIO}{High Speed Input/Output}
\acro{HV}{High Voltage}

\acro{IP}{Interaction Point}

\acro{LAr}{Liquid Argon}
\acro{LEP}{Large Electron-Positron collider}
\acro{LHC}{Large Hadron Collider}
\acro{LHCb}{Large Hadron Collider beauty}
\acro{LHCf}{Large Hadron Collider forward}
\acro{LS1}{Long Shut-down 1}
\acro{LS2}{Long Shut-down 2}
\acro{LS3}{Long Shut-down 3}
\acro{LV}{Low Voltage}
\acro{LVDS}{Low Voltage Differential Signal}

\acro{MIP}{Minimum Ionizing Particle}

\acro{NIM}{Nuclear Instrumentation Module}
\acro{NO}{Noise Occupancy}

\acro{PCA}{Point of Closest Approach}
\acro{PID}{Particle IDentification}

\acro{SCT}{Semi-Conductor Tracker}
\acro{SD}{Strobe Delay}
\acro{Si}{Silicon}
\acro{SiC}{Silicon Carbide}
\acro{SM}{Standard Model}
\acro{SP}{Serial Powering}

\acro{ToT}{Time over Threshold}
\acro{TOTEM}{TOTal Elastic and diffractive cross section Measurement}
\acro{TRT}{Transition Radiation Tracker}

\acro{VME}{VERSAmodule Eurocard}

\end{acronym}

\bibliography{../bibliography/refs_cosmic,../bibliography/refs_silicon,../bibliography/refs_powering,../bibliography/refs_tracking,../bibliography/refs_misc,../bibliography/refs_physics,../bibliography/refs_upgrade,../bibliography/refs_vcsels,../bibliography/refs_dedx}


\end{document}